\documentclass{aa}  

\usepackage{graphicx}
\usepackage[varg]{txfonts}
\usepackage[colorlinks,citecolor=blue]{hyperref}
\usepackage{float}
\usepackage{placeins,afterpage}
\usepackage[caption = false]{subfig}

\usepackage{booktabs}
\usepackage{pdflscape}
\usepackage{float}
\setcitestyle{notesep={ }}

\begin{document} 

   \title{CRIRES high-resolution near-infrared spectroscopy of\\ diffuse interstellar band profiles}
   \subtitle{Detection of 12 new DIBs in the YJ band and\\ the introduction of a combined ISM sight line and stellar analysis approach}

   \author{A.~Ebenbichler\inst{1}
          \and
          A.~Postel\inst{1}
          \and
          N.~Przybilla\inst{1}
          \and
          A.~Seifahrt\inst{2}
          \and
          D.~We{\ss}mayer\inst{1}
          \and 
          W.~Kausch\inst{1}
          \and
          M.~Firnstein\inst{3}
          \and 
          K.~Butler\inst{4}
          \and
          A.~Kaufer\inst{5}
          \and 
          H.~Linnartz\inst{6}
          }

   \institute{Institut f\"ur Astro- und Teilchenphysik, Universit\"at Innsbruck,
              Technikerstr. 25/8, 6020 Innsbruck, Austria\\
              \email{Alexander.Ebenbichler@uibk.ac.at}
             \and
                 Department of Astronomy and Astrophysics, University of Chicago, 5640 S Ellis
                 Avenue, Chicago, IL 60637, USA
             \and
             Dr. Karl Remeis-Sternwarte \& Erlangen Centre for Astroparticle Physics, Friedrich-Alexander-Universit\"at Erlangen-N\"urnberg, Sternwartstr. 7, 96049 Bamberg, Germany
             \and
                 LMU M\"unchen, Universit\"atssternwarte, Scheinerstr. 1, 81679 M\"unchen, Germany
             \and
                 European Southern Observatory, Alonso de Cordova 3107, Vitacura, Casilla 19001,
                 Santiago, Chile
             \and
                 Laboratory for Astrophysics, Leiden Observatory, Leiden University, 
                 P.O.~Box 9513, 2300 RA Leiden, The Netherlands
             }

   \date{Received ....; accepted ....}

 
  \abstract
   {}
   {A high spectral resolution investigation of diffuse interstellar bands (DIBs) in the
   near-infrared ($YJ$ band) is conducted to test new methods, to confirm and improve existing parameters, and to search for new DIBs.}
   {The CRyogenic high-resolution InfraRed Echelle Spectrograph (CRIRES) on the European Southern Observatory's Very 
   Large Telescope was
   employed to obtain spectra of four reddened background supergiant stars (HD~183143, HD~165784, HD~92207, 
   HD~111613) and an unreddened
   comparison star (HD~87737) at the highest resolution of $R$\,$\approx$\,100\,000 currently achievable
   at near-infrared wavelengths, more than twice as high as accomplished in previous near-infrared DIB studies. The correction for telluric absorption was performed by a 
   modelling approach. Non-local thermodynamic equilibrium spectral modelling of available optical and the new
   near-infrared stellar spectra facilitated a comprehensive characterisation of the
   atmospheric properties of the background stars. As a consequence, a more precise and accurate
   determination of the reddening and the reddening law along the respective sight lines could be achieved than feasible before by 
   comparison of the observed and  model spectral energy distributions. 
   For DIBs that overlap with stellar lines the DIB profile shapes could be
   recovered.}
   {Seventeen known near-infrared DIBs were confirmed, and 12 previously unknown and generally weaker DIBs were
   identified in the $YJ$ band. Three DIBs that show uniform profiles along all 
   sight
   lines were identified, possibly connected to transitions from a common lower state of the
   same  carrier. The divergent extinction curve towards the  frequently discussed DIB standard 
   star HD~183143 could
   be reproduced for the first time, requiring extra absorption by $\sim$3.5\,mag
   due to polycyclic aromatic hydrocarbons (PAHs) to match the ultraviolet extinction bump. 
   This extra absorption probably stems from a circumstellar bubble lying in front of the
   star which is intersected tangentially by the  line of sight, making this
   particular sight line  more peculiar than standard.}
   {}

   
   \titlerunning{CRIRES high-resolution near-IR spectroscopy of DIB profiles}
   \authorrunning{Ebenbichler et al.}

   \maketitle
%

\section{Introduction}

   Diffuse interstellar bands (DIBs) are absorption features that have been reported in the 
   spectra of reddened stars in the Milky Way since their first detection by \citet{Heger22}. They are caused by 
   material populating
   the diffuse interstellar medium (ISM) and grow in strength with increasing extinction.
   The DIBs are ubiquitous in sight lines towards background sources within the Galactic
   disc,  except for nearby stars \citep[for a review of the development of the field see][]{Snow14}. They have also been found in nearby dust-harbouring 
   irregular and
   spiral galaxies \citep[e.g.][]{Ehrenfreundetal02,Cordineretal08a,Cordineretal08b}, more
   distant dusty starburst galaxies \citep{HeLe00}, and a damped Ly$\alpha$ system at
   redshift 0.524 \citep{Junkkarinenetal04}.
   
   The carriers of DIBs are thought to be large carbonaceous molecules (from several tens   up to 
   $\sim$100 atoms) in the gas phase \citep{1995ARA&A..33...19H,Tielens14}.
   It is challenging to identify the carriers of DIBs given the
   overwhelming number of potential candidates \citep[see e.g.][for a discussion]{Omont16,Omontetal19}.
   Over 550 DIBs have been found so far \citep{2019ApJ...878..151F}, and only 4 DIBs ($\lambda\lambda$  
   9365, 9428, 9577, 9632\,{\AA}) 
   have an unambiguously identified carrier, which is the buckminsterfullerene cation C$_{60}^+$ 
   \citep[e.g.][]{2015Natur.523..322C,2017ApJ...846..168S,2019ApJ...875L..28C}. 
   
   In the absence of direct identifications of DIB carriers, many indirect clues have been
   collected, providing numerous constraints on their nature.
   Many studies have established DIB families,  groups of DIBs that behave in a similar
   manner,  based on correlations of their respective equivalent widths \citep[$EW$s; 
   e.g.][]{1987Krelowski,1997Cami}. The $\lambda\lambda$6614 and 6196\,{\AA} DIBs show the 
   highest $EW$ correlation of any optical DIB pair known to date \citep{McCall_2009}. There have been
   several studies on both members of this pair including profile studies at very high 
   spectral resolving power
   \cite[$R$\,=\,$\lambda/\Delta\lambda$\,=\,600\,000,][]{1995MNRAS.277L..41S}, finding 
    substructures. 
   \cite{McCall_2009} conclude that the DIB pair $\lambda\lambda$6614, 6196\,{\AA} is not
   certain to have the same carrier because they have different band profiles, which should be 
   very similar if the DIBs arise from transitions from the same molecular ground state.
   However, the carriers should share a common chemical history.
   Substructures in DIB profiles are believed to arise from unresolved P-, R-, and Q-branches of
   molecular transitions \citep{McCall_2009}.
   
   A study of the special subclass of C$_2$-DIB profiles in the optical 
   was conducted by the ESO Diffuse Interstellar Bands Large Exploration Survey
   \citep[EDIBLES,][]{Coxetal17,2018A&A...616A.143E} at 
   $R$\,$\approx$\,70\,000-100\,000. 
   They resolved DIBs with substructure, interpreted as
   rotational P(Q)R-branches of molecular carriers. Differences in the relative amplitudes of the 
   peaks and peak substructure separations for the same sight lines imply that the C$_2$-DIBs
   arise from different molecular carriers of different sizes.
   Some of the DIBs showed constant separation of the absorption subpeaks, while others
   potentially showed variations or global broadening in response to varying C$_2$ 
   temperatures. 

   Most of the DIB analyses to date have concentrated on the optical wavelength range.
   The discoveries of DIBs in the near-infrared
   \citep[NIR,][]{1990Natur.346..729J,1994Natur.369..296F, 2007A&A...465..993G,geballe2011, 
   2014A&A...569A.117C, 2015ApJ...800..137H, 2017MNRAS.467.3099G} opened up new possibilities 
   for DIB studies because of the significantly lower transition energies than in the optical,
   shifting the focus from electronic to ro-vibrational transitions.
   This wavelength region is also less dense in DIBs which decreases the chance of coincidental overlaps.
   At the same time, the survey of the 15273\,\AA\ DIB within the Apache Point Observatory 
   Galactic Evolution Experiment (APOGEE) by \citet{2019apogee} produced a 
   Galactic overview of DIB absorption at high spectral resolution ($R$\,=\,22\,500) for the first time,
   covering over 124\,000 sight lines.
   Studies of NIR DIBs in the past decade profited from the rapidly increasing spectral
   resolution of the available instrumentation, from $R$\,$\approx$\,10\,000 with 
   X-shooter in the mini-survey of \citet{2014A&A...569A.117C}, over $R$\,$\approx$\,28\,000
   with the Warm INfrared Echelle spectrograph to Realize Extreme Dispersion and sensitivity 
   (WINERED) in the study of \citet{2015ApJ...800..137H} to $R$\,$\approx$\,45\,000 with the
   Immersion Grating INfrared Spectrograph (INGRIS) achieved in the work of
   \citet{2017MNRAS.467.3099G}.

   Besides correlations of DIB $EW$s with each other, relations of DIB $EW$s to other
   ISM tracers have been investigated, including the \ion{H}{i} or H$_2$ column density, the presence 
   of simple diatomic molecules (like C$_2$, see above), and the dependency on the ultraviolet (UV) radiation 
   field along the sight 
   lines. Most commonly discussed is the general correlation with dust column density, 
   typically reduced to extinction $A_V$ or colour excess $E(B-V)$ for practical purposes, 
   with the total-to-selective extinction ratio $R_V$\,=\,$A_V/E(B-V)$ parametrising an 
   average extinction law \citep{cardelli1989}. The accurate and precise determination of the 
   wavelength-dependent extinction is consequently of great importance for DIB studies. 
   Traditionally, the colour excess and extinction law are derived by comparing colours and the 
   spectral energy distribution (SED) of reddened objects with those of unreddened identical
   standard stars,  known as  the pair method \citep[e.g.][]{Massaetal83}.
   This possibly introduces systematic uncertainties in the analyses, for example because of  spectral-type 
   mismatch or unrecognised binarity of objects, which is particularly high among early-type 
   stars (with a high 
   fraction of similar-mass binary components). Thus, the use of 
   stellar atmosphere models and synthetic SEDs to derive extinction curves
   \citep[e.g.][]{FeMa05} is to be preferred.
   
   Typically, OB-type stars are employed as background stars for studying DIBs. The reasons are their high luminosities, near-continuous spectra with only few stellar features, and fast rotation, which allow DIBs to be easily identified and measured. In contrast, the focus here is on supergiants of spectral  types late B to early A. Compared to OB-type stars they have low bolometric corrections and rather flat SEDs throughout the optical and near-IR (i.e. their very high luminosity is concentrated at wavelengths with DIBs). BA-type supergiants therefore facilitate longer sight lines to be covered at the same background star magnitudes. This allows higher reddening values to be reached, which is an advantage for the search for new (intrinsically weaker) DIBs. On the other hand, they are slow rotators, and show spectra richer in stellar lines than OB-stars;  however, this is not problematic for DIB studies if the stellar spectra are understood well. As BA-type supergiants are among the visually brightest stars in actively star-forming galaxies \citep[e.g.][]{Kudritzkietal08} they will become primary targets for future DIB studies in nearby galaxies, already accessible today  to intermediate-resolution spectroscopy with 8--10m class telescopes  out to distances of 7 to 8\,Mpc \citep{Bresolinetal01,Kudritzkietal13,Kudritzkietal14}.
   
       \begin{table*}
        \caption{The observational star sample.}\vspace{-1mm}
        \setlength{\tabcolsep}{1.1mm}
        \centering
        {\small
\begin{tabular}{llcrccrccccc}
\hline\hline
\multicolumn{5}{c}{} & \multicolumn{3}{c}{CRIRES Observations} & & \multicolumn{3}{c}{Optical Observations} \\
\cline{6-8} \cline{10-12}
Object & Sp.Type & $V$ \tablefootmark{a} & $B-V$ \tablefootmark{a} & $J$ \tablefootmark{b} & Date & Exp.time & S/N & & Date & Exp.time & S/N \\
\multicolumn{2}{c}{} & mag & mag & mag & YYYY-MM-DD & s & & & YYYY-MM-DD & s & \\
\hline
HD 183143 & B7\,Iae & 6.839 & 1.185    & 4.179 & 2014-05-27, 2014-06-13, 2014-06-23 & 60$-$90 & 200$-$500 & & 2013-08-18 & 720 & 500 \\
          &         &       &          &       & 2014-06-24, 2014-06-28\\
HD 165784 & A2\,Iab & 6.538 & 0.865    & 4.643 & 2014-05-26, 2014-06-17, 2014-06-23 & 60$-$90 & 200$-$500 & & 2007-07-09 & 140 & 180 \\
          &         &       &          &       & 2014-06-24, 2014-07-04, 2014-07-06\\
          &         &       &          &       & 2014-07-07, 2014-07-08, 2014-07-14\\
HD 92207  & A0\,Iae & 5.476 & 0.500    & 4.129 & 2007-05-09, 2007-05-10 & 8$-$12  & 180$-$280 & & 2013-05-10 & 500 & 400 \\
HD 111613 & A1\,Ia  & 5.741 & 0.318    & 4.737 & 2007-05-10 & 15$-$25 & 200$-$300 & & 1999-01-23 & 600 & 400 \\
HD 87737  & A0\,Ib  & 3.486 & $-$0.026 & 3.499 & 2007-05-10 & 10      & 200$-$300 & & 1999-01-21 & 120 & 440 \\
\hline
\end{tabular}        \label{tab:targets}
        }\vspace{-2mm}
        \tablefoot{
         The first four stars are employed for DIB measurements, the fifth is an unreddened standard star.
        \tablefoottext{a}{\cite{Mermilliod97}}
        \tablefoottext{b}{\cite{2003cutri}.}
        }
    \end{table*}

   \begin{figure*}
   \centering
      \subfloat{
      \includegraphics[width = .245\linewidth]{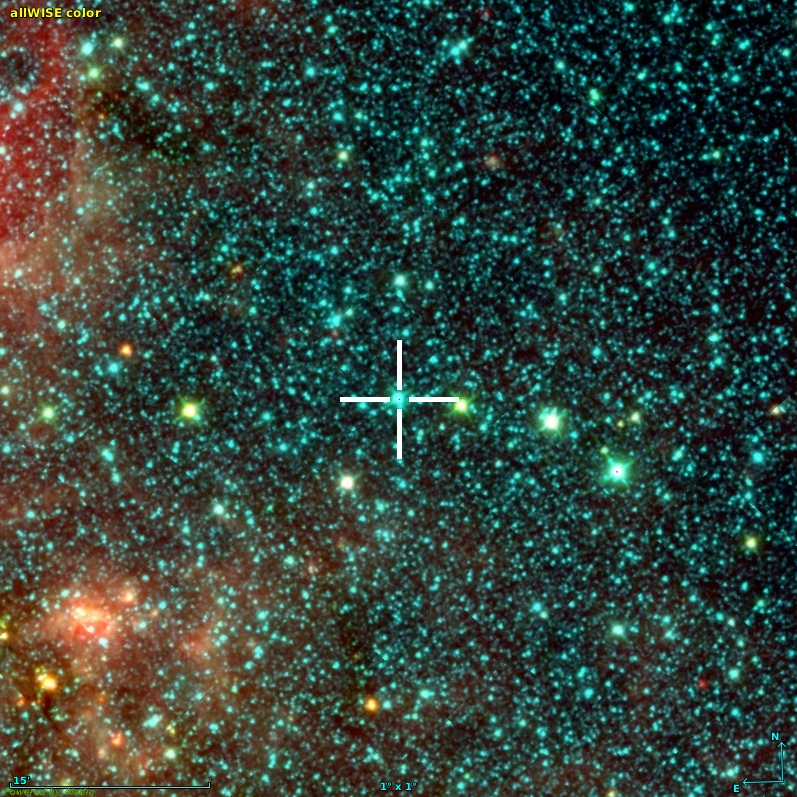}}
   \subfloat{
      \includegraphics[width = .245\linewidth]{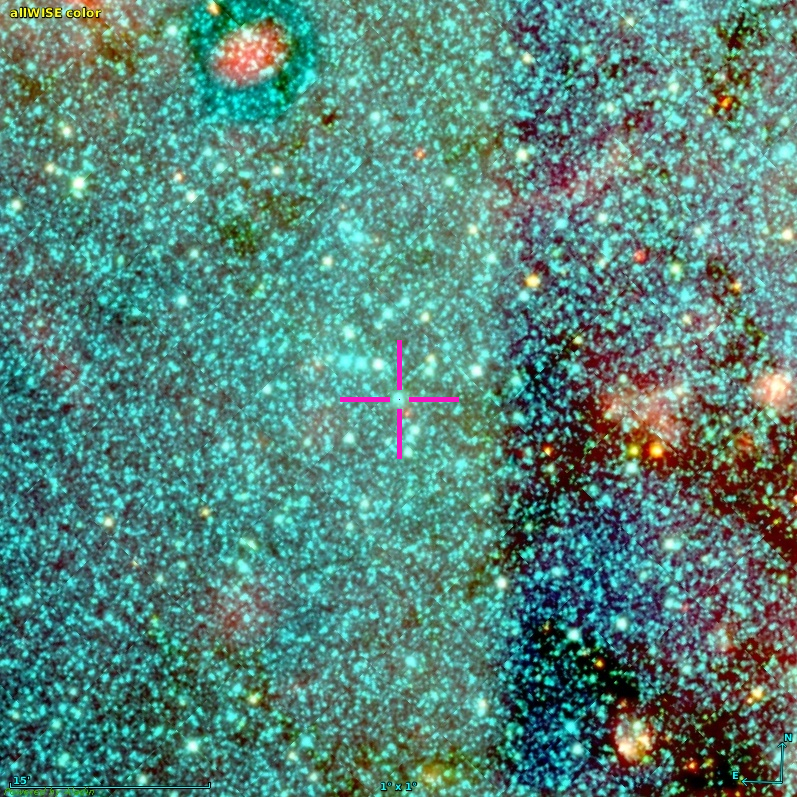}}
   \subfloat{
      \includegraphics[width = .245\linewidth]{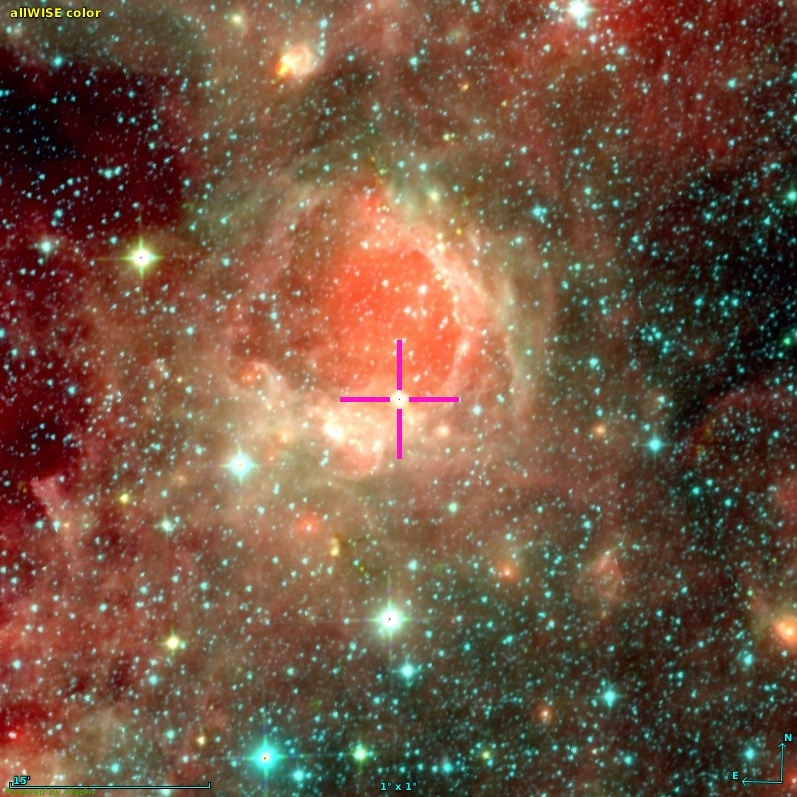}}
   \subfloat{
      \includegraphics[width = .245\linewidth]{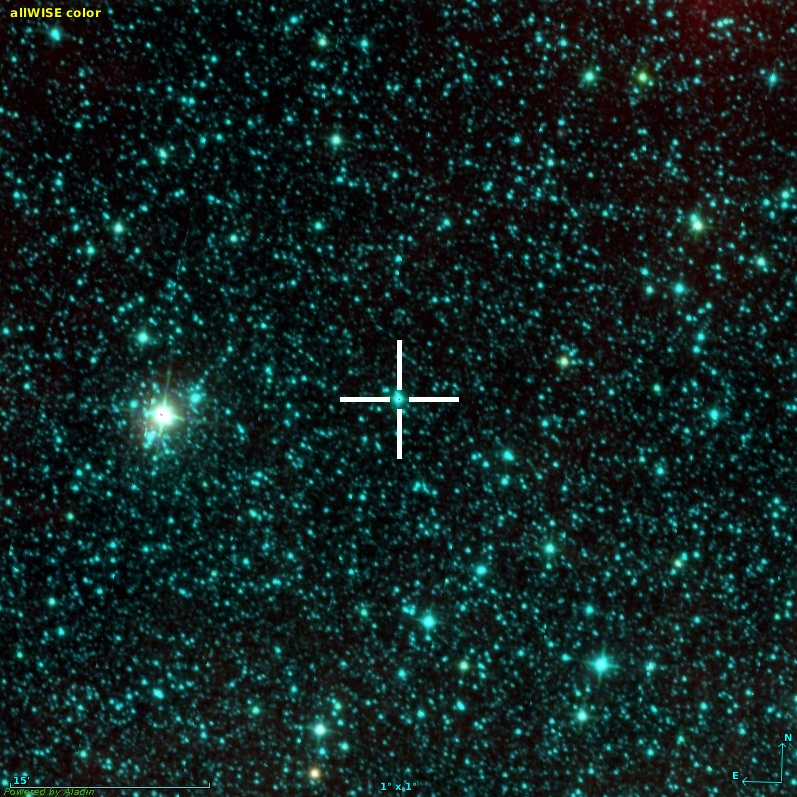}}
   \caption{WISE imaging ($1^\circ \times 1^\circ) $ of the ISM in the sight lines towards the
   target stars, from left to right: HD~183143, HD~165784, HD~92207 and HD~111613.
   Colour code: Red (W4, 22\,$\mu$m), green (W2, 4.6\,$\mu$m), blue (W1, 3.4\,$\mu$m). The
   target stars are centred and indicated by a cross.}
   \label{fig:thumbnails}
   \end{figure*}
   
   In the present work a study of NIR DIBs at more than double the spectral resolution 
   reported so far is presented, finally reaching the contemporary standard set at optical 
   wavelengths. Likewise, an improved characterisation is presented of the background stars' properties 
   based on models that account for deviations from the standard assumption of local 
   thermodynamic equilibrium (LTE), known as  non-LTE models,  
   which in turn allows properties of 
   the ISM sight lines like reddening and reddening law to be constrained to unprecedented accuracy and precision.  
   We describe the observation and reduction of our NIR and optical spectra in
   Sect.~\ref{sec:data_reduction} and the quantitative analysis of the background stars in
   Sect.~\ref{sec:stellar_analysis}, including the comparison of the model with the observed SEDs. 
   The methods for the analysis of
   the velocity components of the ISM along the sight lines and for DIB measurements (plus a brief comparison with previous measurements) are discussed in 
   Sects.~\ref{sec:atomic_lines} and \ref{sec:dib_measurements}. New NIR DIBs are introduced in   
   Sect.~\ref{sec:dib_candidates} and we present a set of DIBs with common absorption profiles
   in Sect.~\ref{sec:profile_sets}. 
   Finally, a brief summary is given and conclusions are drawn.

\section{Observations and data reduction \label{sec:data_reduction}}

Basic information on the observational sample is summarised 
in Table~\ref{tab:targets}: the spectral type, Johnson $V$ and $J$ magnitudes from the Two 
Micron All Sky Survey (2MASS), 
the Johnson $B-V$ colour,  the observing logs, observation dates, exposure times, and 
signal-to-noise ratio per pixel $S/N$ achieved for the spectra, both for the
NIR  and our complementary optical observations. 
The table is ordered by colour excess of the target stars (see below), from high to low.
The first four (reddened) stars were
employed for DIB measurements in the present work. The nearby supergiant HD~87737 is located at a high Galactic latitude and
is thus practically unreddened with a colour excess of $E(B-V)$\,=\,0.02. It therefore
served as an unreddened reference star in the present work for the confirmation of DIBs 
and the identification of stellar absorption lines. This was possible because of the
relatively narrow spectral range of B7 to A2 covered here, with HD~87737 an A0-type star 
conveniently situated in the middle.

Colour composites of images obtained by the Wide-field 
Infrared Survey Explorer (WISE) are shown in Fig.~\ref{fig:thumbnails} that allow 
the ISM environment of the DIB target stars to be visualised. The star 
HD~92207 lies on a sight line towards the \ion{H}{ii} region IC~2599, 
which is 
powered by OB stars in the young open cluster NGC~3324, whereas the other three stars appear 
to be located on diffuse ISM sight lines (but see below for further discussion).

The NIR spectra analysed in this work were obtained with the CRyogenic high-resolution 
InfraRed Echelle Spectrograph \citep[CRIRES,][]{Kaeufletal04} 
on the European Southern Observatory (ESO) Very Large Telescope (VLT) Unit Telescope~1 (UT1) 
at Cerro Paranal, Chile.
The spectrograph provided a resolving power of up to $R$\,$\approx$\,100\,000, depending on 
slit width, and made observations within a wavelength range $\sim$0.95--5.2\,$\mu$m possible. 
A single instrumental setting covered a narrow wavelength range ($\sim\lambda$/70),
and was recorded on four detector chips, yielding four disconnected spectral pieces per 
exposure. However, because of order overlap on chips 1 and 4 in the $YJ$ bands, only spectra recorded on chips 2 and 3 were employed for the further analysis. A near-continuous wavelength coverage was achieved by choosing adjacent instrumental settings. 
Coverage of an entire spectroscopic NIR band therefore required approximately 70
wavelength settings to be taken.

The data were obtained during two observing runs, in one of the first visitor-mode runs in 
2007 when CRIRES was offered to the community by ESO for the first time and in one of the 
last service-mode observing runs in 2014 before CRIRES was removed from the VLT-UT1 for 
upgrade. The data reduction was performed close to the time of the observations for both datasets, 
thus it is optimised to the status of the instrument at the time of observation; the 
instrument was subject to many smaller interventions in between.
    
Data for \object{HD 87737}, \object{HD 92207,}
and \object{HD 111613} were obtained without adaptive optics (AO) support within a programme on benchmark 
spectroscopy of NIR stellar features of bright A-type supergiants with a 0.2{\arcsec} slit (i.e. at a nominal $R$\,$\approx$\,100\,000). 
The wavelength coverage of the $YJ$ band was incomplete.
As a consequence, several previously reported NIR DIBs are not covered by these data.
The signal-to-noise ratio per pixel was $S/N$\,$\approx$\,200--300.

Basic  data reduction followed the standard recipes for long-slit IR
spectrographs. Observations were always performed in a single AB nodding pattern for all 
wavelength settings. All raw frames were treated with a non-linearity correction
before pairwise subtraction removed the atmospheric emission
features. The 
individual A$-$B and B$-$A frames were then divided by a normalised flat field. The flat-fielding 
step corrected the pixel-to-pixel gain variations of the chips, but due to repeatability 
problems of the intermediate slit of CRIRES this step influenced the slope of the continuum in
most spectra, which had to be corrected in a final step of the data reduction. An ESO Recipe 
Execution Tool (EsoRex)  of the CRIRES data reduction pipeline was used for optimal
extraction of the 1D spectrum in each nodding position. The two 1D spectra of each set-up and 
target combination were then cross-correlated to identify offsets in wavelength due to the 
slit curvature, shifted accordingly and co-added. In this step the imprints of optical ghosts 
that were present in one of the two nodding position in most $YJ$ band settings were 
removed, and cosmetic artefacts and outliers were filtered by an iterative procedure.

The telluric absorption lines still contained in the processed 1D spectra could not
be removed by dividing by the spectrum of an early-type standard star since this
would have compromised the intrinsic features of the targets. Instead, the
telluric features were modelled using the Fast Atmospheric Signature CODE 
\citep[FASCODE,][]{1981SPIE..277..152C,1992JGR....9715761C}, a line-by-line
radiative transfer code for the Earth's atmosphere, and the HIgh-resolution TRANsmission 
molecular absorption database \citep[HITRAN,][]{2005JQSRT..96..139R}. Global Data Assimilation
System (GDAS) atmospheric profiles were used as input for FASCODE and delivered the 
necessary information for temperature, pressure, and humidity as a function of height in the
atmosphere. The profiles were retrieved from the National Oceanic and Atmospheric
Administration's (NOAA) website\footnote{\url{http://www.ready.noaa.gov/ready/amet.html}}
for the two nights of observations. The amount of precipitable water vapour predicted
by the models was adapted to achieve an optimal fitting in comparison to
the measurements. The resulting spectra were smoothed with a Gaussian kernel to
match the resolution of the science spectra. In a final step the telluric absorption features
were then removed by dividing the science spectra by the model spectra. A detailed discussion
of the performance and limitations of this particular approach for telluric correction is given
by \citet{Seifahrtetal10}.
    
The significantly reddened supergiant targets \object{HD 165784} and \object{HD 183143} were
observed at a nominal $R$\,$\approx$\,100\,000
covering a spectral range between $9550$ and $13450\,${\AA}. A 
$S/N$\,$\approx$\,200--500 per pixel  was achieved in no-AO mode. The raw data were reduced using the CRIRES data
reduction pipeline \citep{crires_pipeline_man} in a way similar to that described above, but using the version that was adapted to the
instrument at the time of observation. The spectra were corrected for telluric absorption 
using Molecfit \citep{molecfit,Kauschetal15}, which follows a  modelling approach similar to
FASCODE.
The precipitable water vapour was determined between 1 and 2\,mm by Molecfit for most settings.

    \begin{table*}
        \setlength{\tabcolsep}{.01mm}
        \centering
        \caption{Parameters of the sample stars and ISM sight line
        characterisation.}
        {\small
        \setlength{\tabcolsep}{1.mm}
\begin{tabular}{llccccccccccl}
\hline\hline
Object & Sp.Type  & $T_\mathrm{eff}$  & $\log g$  & $y$  & $\xi$  & $v \sin i$  & $\zeta$  & $d_\mathrm{spec}$ & $d_\mathrm{EDR3}$ & $E(B-V)$  & $R_V$  & ISM Type  \\\cline{6-8}
\multicolumn{2}{c}{} & K & cgs &  & \multicolumn{3}{c}{km\,s$^{-1}$} & kpc & kpc & \\
\hline
HD 183143 & B7\,Iae  & 12800$\pm$200 & 1.76$\pm$0.05 & 0.099$\pm$0.005 & 7$\pm$2 & 37$\pm$4 & 27$\pm$5 & 2.43$\pm$0.24 & 2.17$\pm$0.12 & 1.22$\pm$0.02 & 3.3$\pm$0.05  & diffuse+shell\\
HD 165784 & A2\,Iab & 9000$\pm$200  & 1.50$\pm$0.10 & 0.119$\pm$0.007 & 7$\pm$1 & 18$\pm$3 & 35$\pm$5  & 1.77$\pm$0.33 & 1.59$\pm$0.09 & 0.86$\pm$0.02 & 3.1$\pm$0.2 & diffuse \\
HD 92207  & A0\,Iae & 9500$\pm$200  & 1.20$\pm$0.10 & 0.120$\pm$0.020 & 8$\pm$1 & 30$\pm$5 & 20$\pm$5 & 3.06$\pm$0.47 & 2.14$\pm$0.28 & 0.48$\pm$0.03 & 3.9$\pm$0.2 & diffuse+\ion{H}{ii}\\
HD 111613 & A1\,Ia  & 9150$\pm$150  & 1.45$\pm$0.10 & 0.119$\pm$0.014 & 7$\pm$1 & 19$\pm$3 & 21$\pm$3 & 2.47$\pm$0.44 & 2.21$\pm$0.14 & 0.39$\pm$0.03 & 3.5$\pm$0.2 & diffuse \\
HD 87737  & A0\,Ib  & 9600$\pm$200  & 2.05$\pm$0.10 & 0.129$\pm$0.013 & 4$\pm$1 & 0$\pm$3 & 16$\pm$2 & 0.68$\pm$0.13 & 0.56$\pm$0.09 & 0.02$\pm$0.02 & 3.1$\pm$0.2 & diffuse \\
\hline
\end{tabular}        \label{tab:stellar_parameters}
        }
    \end{table*}

\begin{table*}[t!]
\centering
\footnotesize
\caption[]{Non-LTE abundances of the sample stars. \label{tab:abundances}}
\begin{tabular}{lcccccl}
\hline
\hline
\footnotesize
Ion            & \multicolumn{5}{c}{$\log \varepsilon_X$} & model atom\\
\cline{2-6}\\[-3mm]
               & HD 183143 & HD 165784 & HD 92207 & HD 111613 & HD 87737\\
\hline\\[-3mm]
\ion{C}{i}     & 8.31$\pm$0.07 & 8.38$\pm$0.04 & 8.33          & 8.29$\pm$0.10 & 8.25$\pm$0.06 & \citet{Przybillaetal01b}\\
\ion{N}{i}     & 8.69$\pm$0.06 & 8.62$\pm$0.05 & 8.25$\pm$0.04 & 8.45$\pm$0.04 & 8.53$\pm$0.07 & \citet{PrBu01}\\
\ion{O}{i}     & 8.91$\pm$0.08 & 8.80$\pm$0.04 & 8.79$\pm$0.07 & 8.72$\pm$0.04 & 8.73$\pm$0.06 & \citet{Przybillaetal00}\\
\ion{Mg}{i/ii} & 7.69$\pm$0.09 & 7.56$\pm$0.08 & 7.49$\pm$0.11 & 7.48$\pm$0.05 & 7.53$\pm$0.04 & \citet{Przybillaetal01a}\\
\ion{Fe}{ii}   & 7.66$\pm$0.10 & 7.63$\pm$0.08 & 7.34$\pm$0.07 & 7.53$\pm$0.06 & 7.55$\pm$0.07 & \citet{Becker98}\\
\hline\\
\end{tabular}
\end{table*}

In addition, complementary optical spectra observed with the 
Fibre Optics Cassegrain Echelle Spectrograph \citep[FOCES,][]{Pfeifferetal98} on the Calar 
Alto 2.2\,m telescope in Spain ($R$\,=\,40\,000) for HD~183143 and with the Fibre-fed Optical
Echelle Spectrograph \citep[FEROS,][]{Kauferetal99} on the ESO/MPG 2.2\,m telescope at 
La Silla/Chile ($R$\,=\,48\,000) for the other targets were used (see Table~\ref{tab:targets}). Data reduction for HD~183143 was
performed in the same way as for the FOCES observations discussed by \citet{2012A&A...543A..80F},
and the reduction of the raw data for the other stars was performed as described by
\citet{2006A&A...445.1099P}. These spectra were used for optical DIB measurements and for
measurement of the velocity dispersion of interstellar absorbers based on \ion{K}{i} lines.

To construct the stellar SEDs, flux-calibrated, low-dispersion, and large aperture (10{\arcsec}$\times$20{\arcsec})  
spectrophotometry by the International Ultraviolet Explorer (IUE) were extracted from the 
Mikulski Archive for Space Telescopes (MAST\footnote{\url{http://archive.stsci.edu/}}):
data IDs SWP06550 and LWR05637 for HD~183143, SWP43216 and LWP21843 for HD~92207, and
SWP08566 and LWR07305 for HD~87737. 
If IUE data were not available (as was the case for HD~165784 and HD~111613) UV magnitudes 
from the Astronomical Netherlands Satellite (ANS) UV Catalogue of Point Sources 
\citep{1982A&AS...49..427W} were adopted. These data were supplemented by
Johnson $UBV$ magnitudes from \cite{Mermilliod97}, $JHK$ magnitudes from 2MASS 
\citep{2003cutri,2006Skrutskie}, and infrared (IR) magnitudes from the WISE All-Sky Data Release 
\citep{2012cutri}. 

Contamination of the SEDs by other nearby objects can be excluded. An OB-type main-sequence star within the IUE aperture would reveal its presence by an unaccounted UV rise, anything else is outshone by the BA-type supergiant that concentrates the integrated luminosity of a globular cluster or even of a dwarf galaxy into a point source.

\section{Quantitative analysis of background stars \label{sec:stellar_analysis}}
    A thorough physical characterisation of the background stars of interstellar sight lines 
    including the calculation of complete synthetic spectra is highly beneficial, not to say a prerequisite, for DIB 
    studies. First, the synthetic SEDs allow extinction curves to be determined in the 
    best possible way, thus minimising systematic errors. Second, this allows the pure DIB 
    profiles of features that are (fully) blended 
    with stellar lines to be recovered. Background sources are typically selected among 
    early-type stars for various reasons:  relatively line-free spectra; continuum fluxes
    at optical wavelengths that reflect the Rayleigh-Jeans tail of the SEDs; high luminosities
    that render the stars bright even over kiloparsec distances, even more so when   considering giants and supergiants in addition to  the more frequent main-sequence stars. However, 
    quantitative analyses of early-type stars require non-LTE techniques to be applied because
    of the intense radiation fields, and hydrodynamical
model 
    atmospheres for the most luminous objects  to account for strong mass outflow.

    \subsection{Model atmospheres and spectral analysis}
    A determination of a number of atmospheric parameters is required for a thorough description of the background
    star: the effective temperature $T_\mathrm{eff}$, surface gravity $g$, microturbulence $\xi$, macroturbulence
    $\zeta$, projected rotational velocity $v \sin i$, helium abundance (by number) $y$, and elemental abundances 
    $\log \varepsilon_\mathrm{X}$\,=\,$\log N(X)/N(\mathrm{H})$\,+\,12, where $N(X)$ is the number density of 
    element $X$ and $N$(H) the number density of hydrogen. To do this, an inverse problem needs to be solved: 
    finding the correct parameters that allow the observed spectrum to be reproduced. 
 
    A hybrid non-LTE approach was employed. Line-blanketed hydrostatic model atmospheres were 
    calculated with the {\sc Atlas9} code \citep{kur1993b} under the assumption of LTE.
    These were complemented by non-LTE line-formation calculations performed with updated and
    extended versions of {\sc Detail} and {\sc Surface} \citep{giddings1981, daresbury1985}.
    The coupled radiative transfer and statistical equilibrium equations were solved iteratively with {\sc Detail}, which
    provided non-LTE level populations, and the formal solution was computed based on these with {\sc Surface}, 
    which furthermore employed refined line-broadening theories.
    Synthetic spectra were compared to observations aiming at $\chi^2$ minimisation,  and the physical stellar 
    parameters were adjusted in an iterative way to improve the fit. The main spectral diagnostics were
    temperature-sensitive metal ionization equilibria (requiring that abundances derived from different 
    ionization stages of an element agree) and gravity-sensitive Stark-broadened hydrogen lines. Further constraints 
    came from the match between the observed and model SEDs, and from luminosity constraints based on distances.
    Further details of the analysis methodology for applications to BA-type supergiants are discussed by 
    \citet{2006A&A...445.1099P}, with some refinements introduced later by \citet{2012A&A...543A..80F}. 

    Results for the basic atmospheric parameters of the sample stars are summarised in Table~\ref{tab:stellar_parameters}.
    For each object the HD number is listed, along with its spectral type; the effective temperature; the logarithm of the surface 
    gravity; the surface helium abundance; and the microturbulent, macroturbulent,
    and projected rotational velocities, all with their uncertainties. 
    Furthermore, spectroscopic distances $d_\mathrm{spec}$ \citep[calculated according to Eq.~1 
    of][]{NiPr12}, distances from the inversion of the Gaia Early Data Release 3 
    \citep[EDR3,][]{Gaia16,Gaia21} parallax, and the derived $E(B-V)$ and $R_V$ values from a 
    comparison of observed and model SEDs (see below) are also summarised in 
    Table~\ref{tab:stellar_parameters}.  In addition,  the ISM sight line type is stated.
    Table~\ref{tab:abundances} lists non-LTE elemental abundances for
    those chemical species that contribute almost all observed stellar lines in the analysed NIR bands. 
    References to the model atoms used for NIR line-formation calculations are also tabulated. 
    The model atoms of \citet{PrBu04} and \citet{Przybilla05} were employed for hydrogen and helium, respectively.  

\begin{figure}
    \centering
    \includegraphics[width=.95\linewidth]{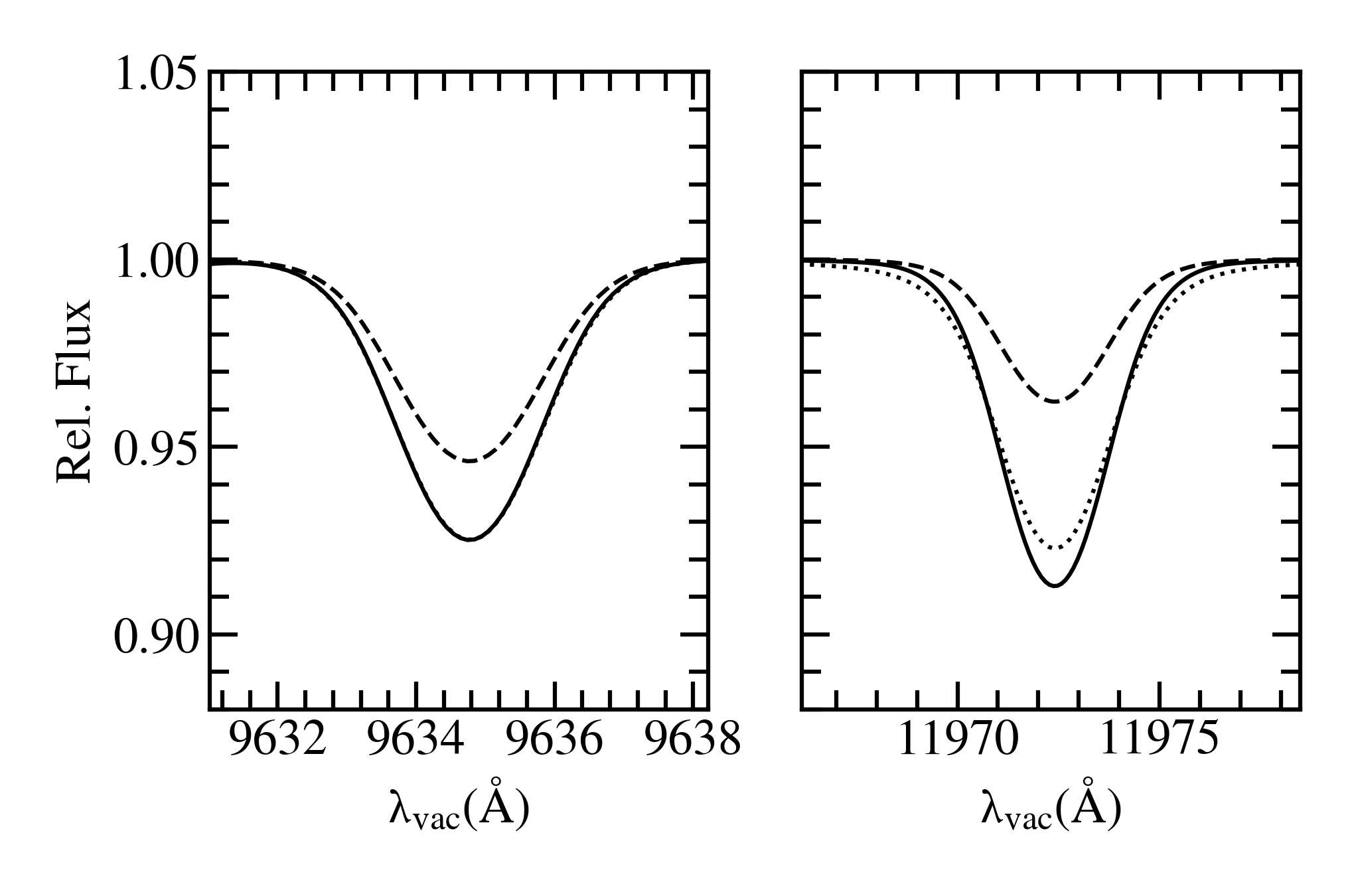}
    \caption{Importance of non-LTE effects. Left panel: Comparison of synthetic line 
    profiles for the \ion{Mg}{ii} $\lambda$9632\,{\AA} blend to DIB $\lambda$9632{\AA} in 
    HD~165784, for 
    the average magnesium abundance derived for the star: non-LTE (solid line), LTE (dashed line), LTE profile for an 
    adjusted abundance that allows the non-LTE $EW$ to be reproduced (dotted line: identical to the solid line in this case). 
    The profiles were convolved with the appropriate broadening functions for the star.
    Right panel: Same as in the left panel, but for \ion{He}{i} $\lambda$11970\,{\AA} blend to DIB $\lambda$11970\,{\AA}, for HD~183143 (see text for  discussion).}
    \label{fig:nlte_lte}
\end{figure}

    It should be noted that the modelling approach presented here is applicable to a wide variety of early-type 
    stars that are used as background objects to investigate interstellar lines of sight. It was successfully applied
    for the analysis of normal early B-type stars \citep{NiPr07,NiPr12,NiPr14} as well as chemically peculiar B-type
    stars \citep{Przybillaetal08,Przybillaetal16,Przybillaetal21,Mazaetal14}, and for spectroscopic binary or multiple
    early-type systems
    \citep{Gonzalezetal17,Gonzalezetal19}. Limitations are set at high luminosities, for O-type stars and early B-type 
    supergiants above $\sim$30\,$M_\odot$, where pronounced mass outflow occurs and the intense radiation field gives 
    rise to significant non-LTE effects on the atmospheric structure, both of which cannot be realistically handled by a hydrostatic hybrid non-LTE approach. 
 
 \begin{figure}
        \centering
           \subfloat[DIB 9632 -- HD~165784]{
        \includegraphics[width=.95\linewidth, trim=0 15 0 0, clip]{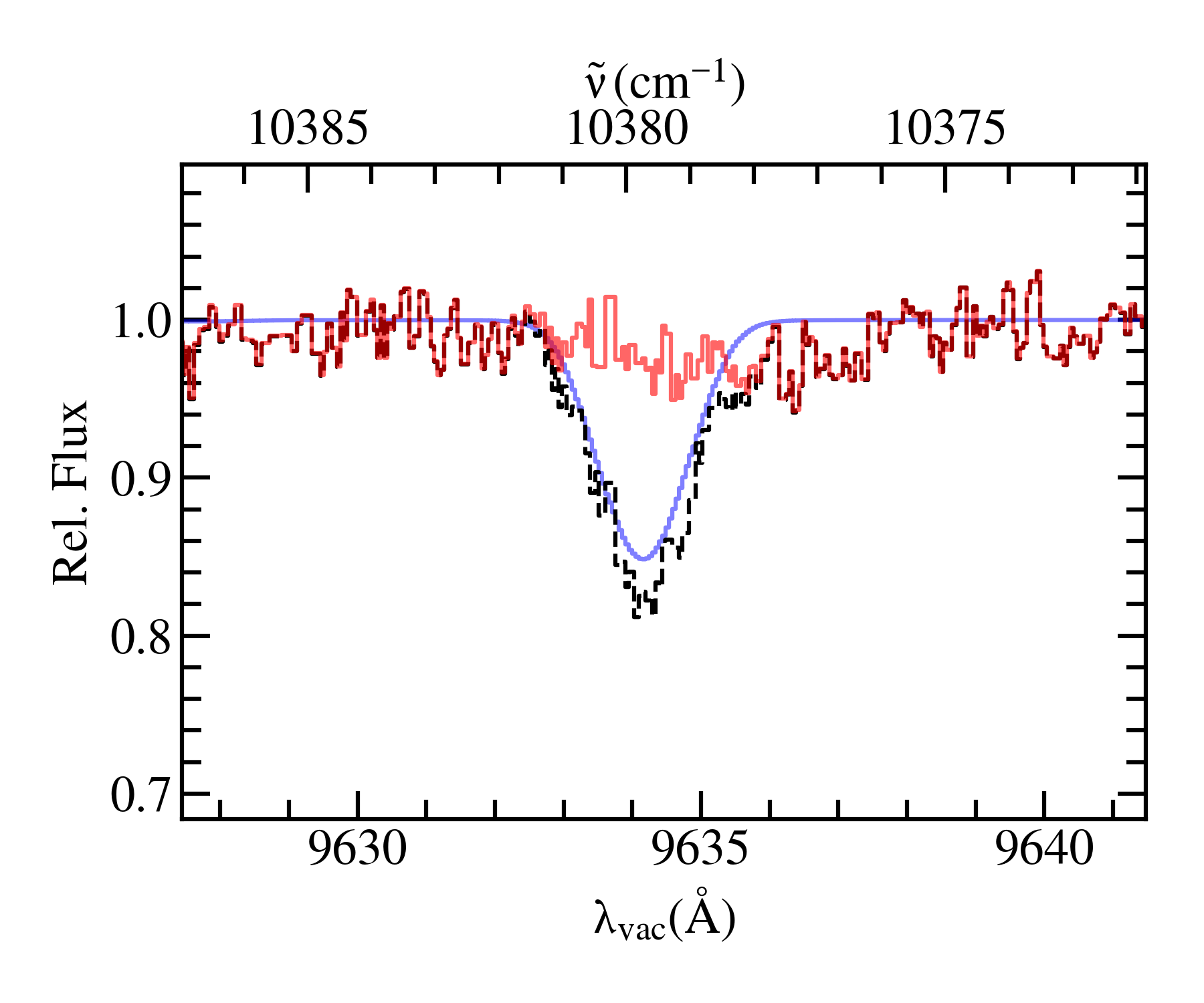}}\\[-1 mm]
           \subfloat[DIB 11970 -- HD~183143]{
        \includegraphics[width=.95\linewidth, trim=0 15 5 0, clip]{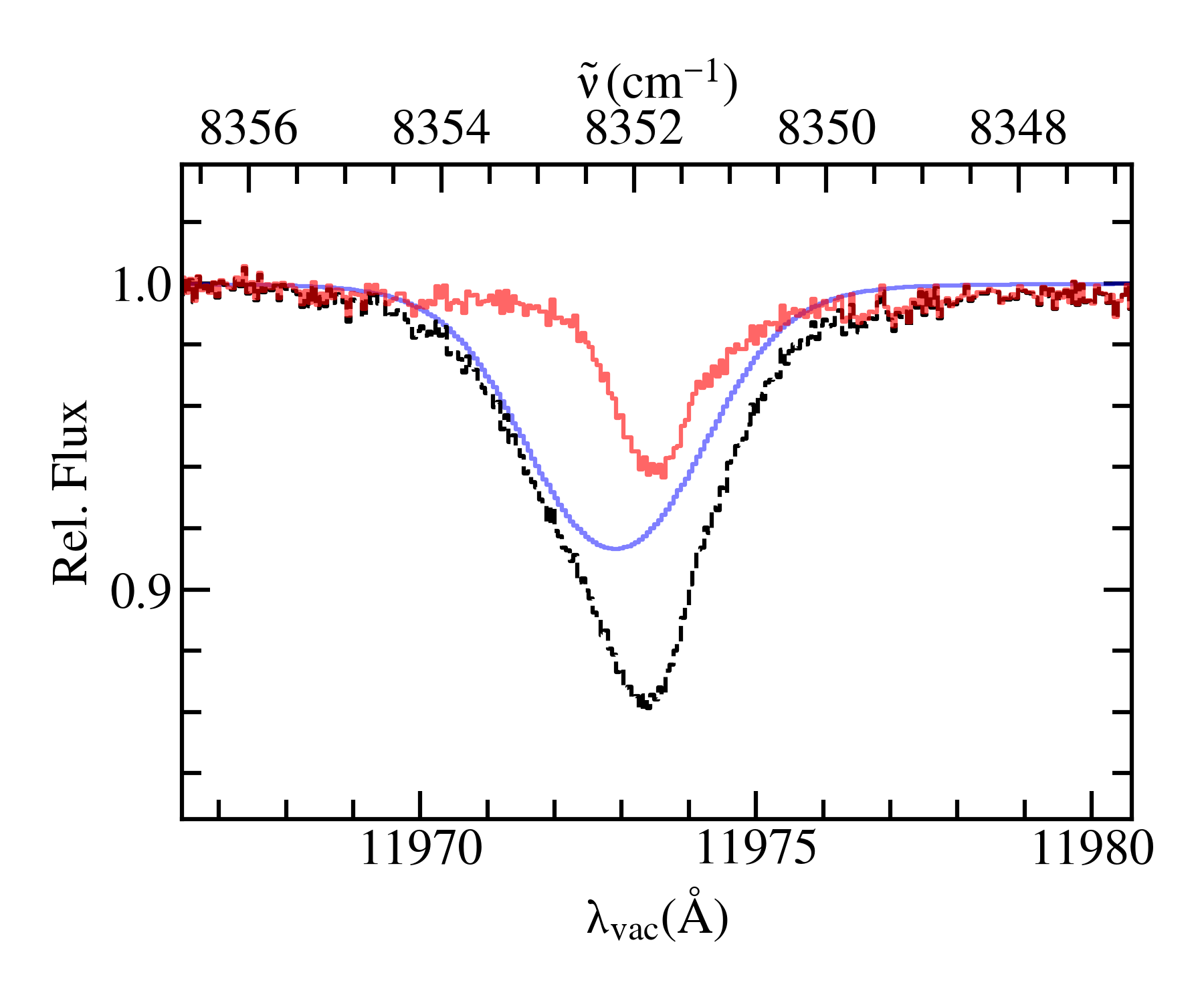}}
        \caption{Correction of the band profile of DIB\,$\lambda$9632 in HD~165784 and DIB\,$\lambda$11970 in HD~183143. Displayed are the 
        observed spectrum (dashed black line); the synthetic spectrum (blue line) of the \ion{Mg}{ii} line of the background star 
        HD 165784; the \ion{He}{i} line of the background star 
        HD 183143 and the corrected spectrum, i.e. the observed spectrum divided by the
        stellar spectrum, showing the DIB (red line). Wavelengths and wavenumbers are given in the barycentric rest frame.}
        \label{fig:divide_stellar}
\end{figure}
 
    \subsection{Stellar blends}\label{sec:stellar_blends}
    Diffuse interstellar bands may be blended  with telluric lines and with stellar absorption lines. A correction can be 
    performed in analogy to the telluric case, by dividing the observed spectra with appropriate synthetic spectra.
    The importance of accounting for non-LTE effects in the line formation calculations is shown in 
    Fig.~\ref{fig:nlte_lte} for two exemplary cases of stellar lines that typically blend with DIBs: a \ion{Mg}{ii} line
    that overlaps with one of the stronger C$_{60}^+$ DIBs \citep{2018A&A...614A..28L} and a \ion{He}{i} line that overlaps with one of the DIBs reported here for 
    the first time. Non-LTE effects are very strong in both cases, increasing the $EW$s of the lines by factors of 2--3, with our non-LTE models being able to reproduce observations.
    While the profile of the \ion{Mg}{ii} line could in principle be closely reproduced in LTE by increasing the magnesium 
    abundance until the $EW$s match, this is not possible in the case of the \ion{He}{i} line. Reliable and well-tested 
    non-LTE models are therefore mandatory to recover the intrinsic DIB profiles. 
    Two examples of the procedure on the observed 
    spectrum of HD~183143 and HD~165784 are shown in Fig.~\ref{fig:divide_stellar}. 
    The broad C$_{60}^+$ DIB $\lambda9632$ has a shallow but symmetrical profile after correction for the stellar \ion{Mg}{ii} line.
    The \ion{He}{i} line is much stronger than the DIB, and 
    it will dominate even more for earlier spectral types, which explains why the DIB has not been reported previously.
    This \ion{He}{i} blend is important only for HD~183143 and HD~92207 among our reddened sample stars; the two other 
    supergiants (and HD~87737) are too cool to excite this line, thus unambiguously confirming the new DIB at this wavelength.   
    
    The correction for stellar blends with hydrostatic non-LTE models is also limited for objects like the sample
    supergiants. Some NIR hydrogen lines are affected by the stellar winds, similarly to some of the Balmer lines,
    such that the feasibility of corrections of DIB profiles for hydrogen line blends needs to be decided on a case-by-case 
    basis.

\subsection{Extinction curves\label{sec:extinction_curve}}

    The ISM is not homogeneous in the Milky Way and shows variations in gas and grain 
    composition, the local radiation field and turbulence.
    These variations can manifest themselves in modifying DIB characteristics and also in the shape of
    the extinction law, which depends on the interstellar dust composition.

\begin{figure}
      \centering
      \includegraphics[width=\linewidth]{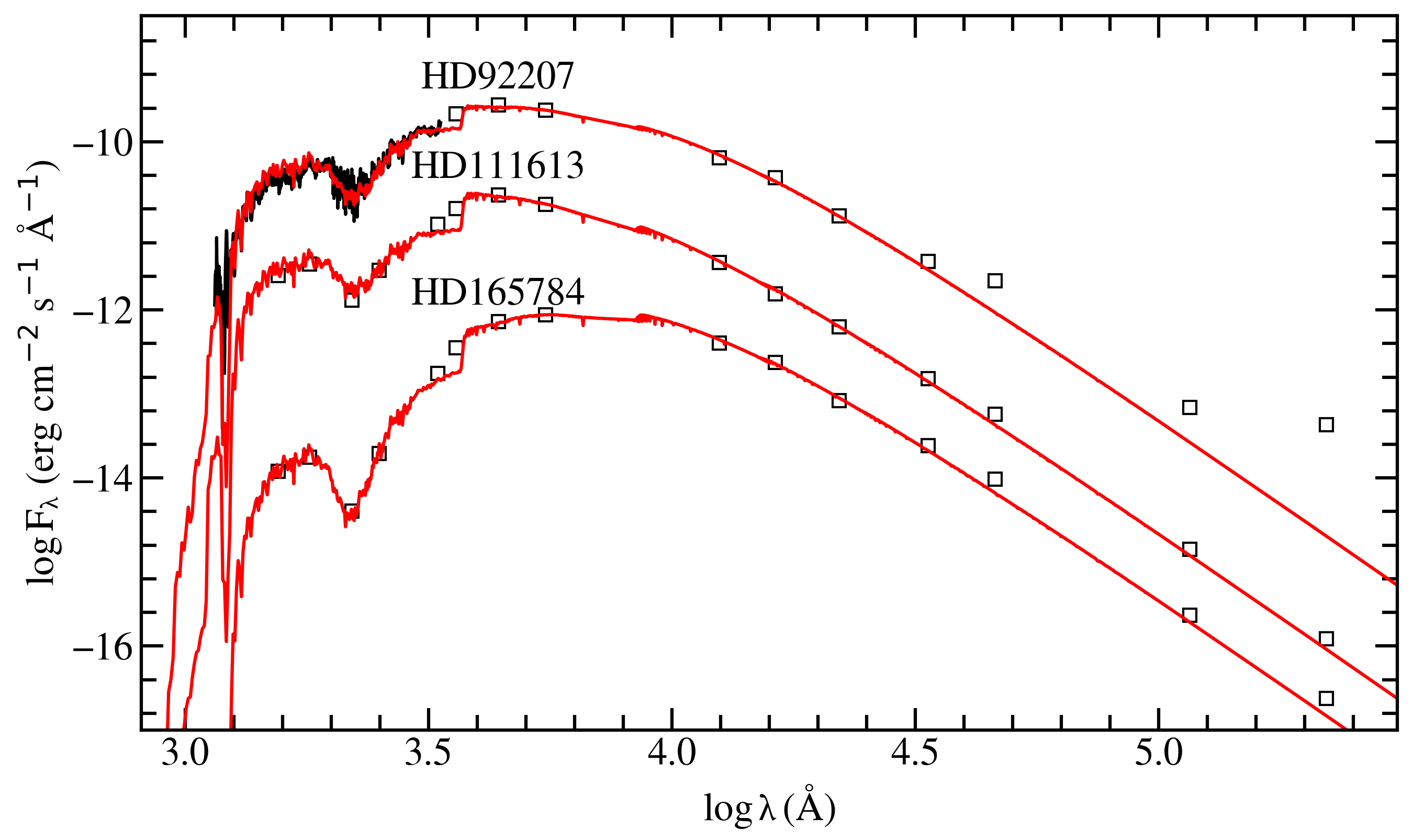}
      \caption{Measurements of three sample star SEDs, compiled from IUE data (black curve) and photometric data (boxes). The
      CCM extinction law was employed to redden the synthetic {\sc Atlas9} SEDs of the background stars (red); see  text for  
      discussion.}
      \label{fig:multi_sed}
  \end{figure}

    The total-to-selective extinction ratio $R_V$ was shown to correlate with some normalized DIB $EW$s
    \citep[$EW$ per unit extinction $A_V$;][]{2018A&A...620A..52R} because it depends on the composition of the 
    diffuse ISM. Therefore, a meaningful characterisation of an interstellar sight line requires  the individual determination 
    of  $E(B-V)$ and of $R_V$, instead of adopting a standard value $R_V$\,=\,3.1. 
    For the present work, an approach 
    was selected in which the model SED for the parameters determined from the quantitative spectroscopic analysis was
    reddened using the Cardelli-Clayton-Mathis (CCM) extinction law \citep{cardelli1989} for a comparison with the observed 
    SED. Best fits for the targets HD~165784, HD~92207, HD~111613, and HD~87737 were obtained for the $E(B-V)$ and $R_V$ 
    values  summarised in Table~\ref{tab:stellar_parameters}.
    The fits are displayed for the three reddened stars in Fig.~\ref{fig:multi_sed}. 
    Reddened SEDs are compared to photometric magnitudes, and if available IUE data are
    used for the UV-bump.
    Only the cases of HD~92207 and HD~183143 require some further discussion.
    
    \begin{figure}
       \centering
       \includegraphics[width=\linewidth]{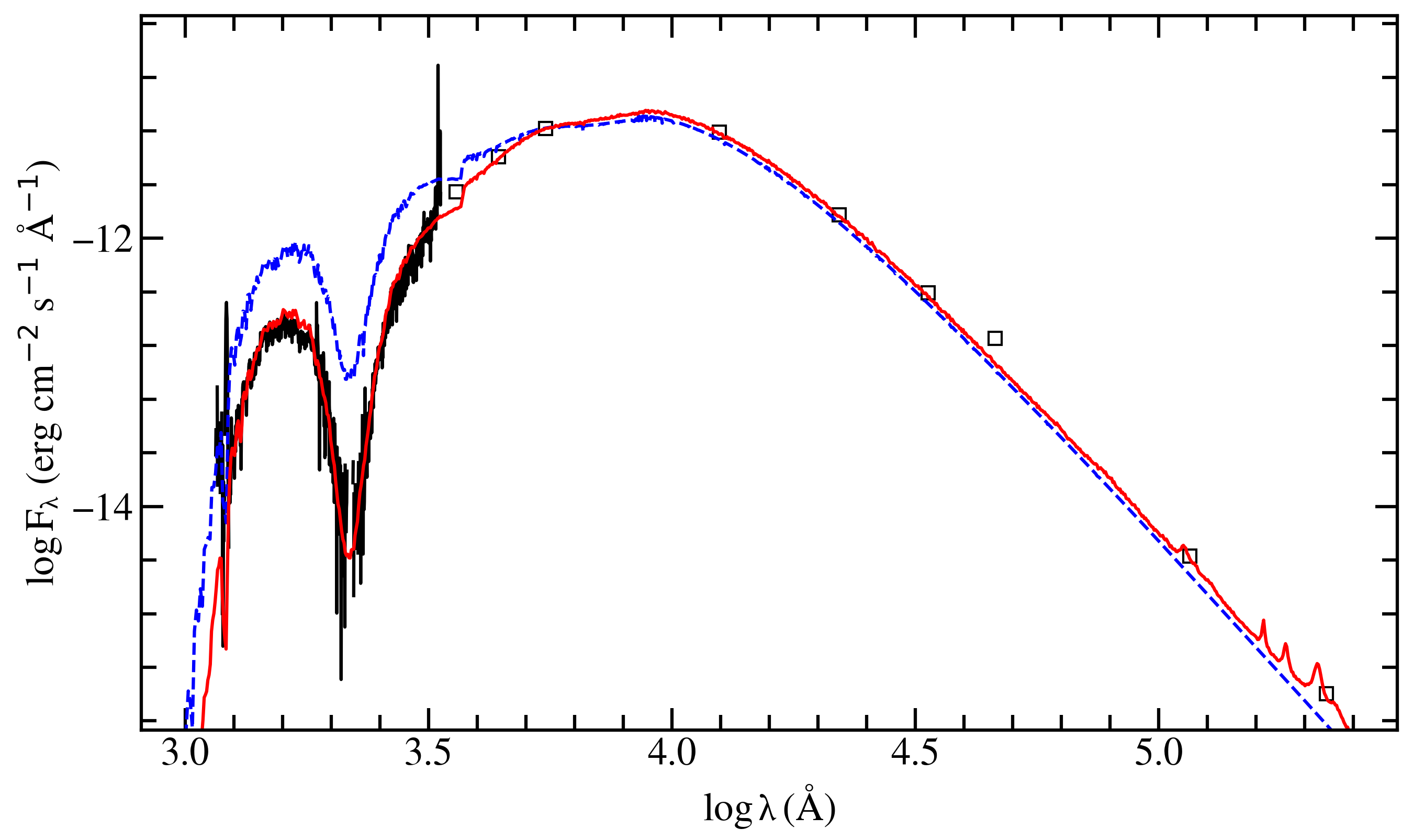}
       \caption{Comparison of the measured (black) and model SEDs for HD~183143. Two model solutions are shown: Application 
       of the CCM extinction law to the {\sc Atlas9} SED with $E(B-V)$\,=\,1.02 and $R_V$\,=\,3.8 
      (dashed blue line) and a modification by employing {\sc Skirt} to introduce additional extinction by a PAH component, 
      including secondary emission (red line). With the additional PAH component extinction, values of $E(B-V)$\,=\,1.22 and 
      $R_V$\,=\,3.3 are required to obtain a near-perfect fit to the observed SED.}
       \label{fig:sed_pah}
   \end{figure}    
 
    \begin{figure*}
   \centering
      \includegraphics[width = 0.41\linewidth]
      {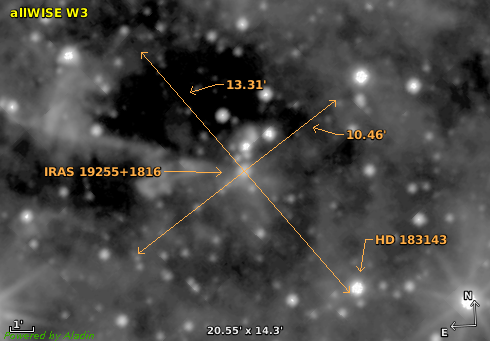}\hspace{3mm}
      \includegraphics[width = 0.41\linewidth]
      {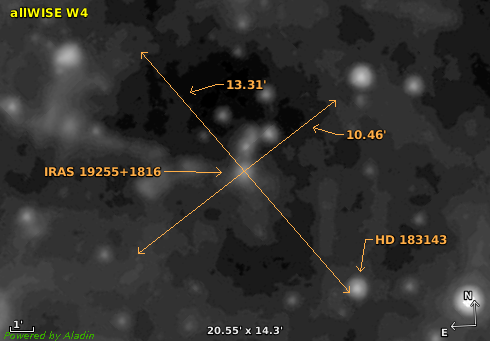}
   \caption{Images of the ISM bubble on the sight line towards HD~183143 in the WISE W3
   ($12\,\mu$m, left) and W4 ($22\,\mu$m, right) bands.
   The orange lines give the extent of the major and minor axes of the bubble.}
   \label{fig:ir_bubble}
   \end{figure*}
   
    \paragraph{HD 92207.}
    The sight line towards HD~92207 appears complicated even at a first glance (see
    Fig.~\ref{fig:thumbnails}). The crucial question is whether the star is a foreground
    object to the open cluster NGC~3324 and its surrounding \ion{H}{ii} region IC~2599, a 
    member, or even a background object. A recent analysis of cluster members based on Gaia 
    EDR3 
    data finds a distance of 2.8$\pm$0.2\,kpc to NGC~3324 \citep{Bishtetal21}. The distance to
    HD~92207 is more difficult to constrain; the Gaia EDR3 parallax implies a short 
    distance of 2.1$\pm$0.3\,kpc, while the spectroscopic distance of 3.1$\pm$0.5\,kpc (see 
    Table~\ref{tab:stellar_parameters}) is compatible with the open cluster distance. This
    would correspond to bolometric magnitudes $M_\mathrm{bol}$ of about $-$8.3\,mag for the 
    short distance and about $-$9.1\,mag for the long distance. However,
    HD~92207 is one of the most luminous BA-type supergiants known in the Milky Way, more 
    luminous 
    than Deneb \citep[$M_\mathrm{bol}$\,$\approx$\,$-$8.5\,mag,][]{SchPr08} because of the 
    much higher mass-loss rate, as evidenced by the highly pronounced H$\alpha$ P Cygni 
    profile \citep{Kauferetal96}. It therefore appears that HD~92207 may indeed be spatially 
    close to the cluster NGC~3324
    (implying that an unaccounted systematic error may trouble the Gaia parallax), 
    which is further corroborated by a consistent colour excess, 0.48$\pm$0.03\,mag versus 
    0.45$\pm$0.05\,mag for the cluster \citep{Bishtetal21}. Our $R_V$ value of 3.9 is also 
    consistent with a sight line traversing an \ion{H}{ii} region, where values larger than 
    the canonical value of 3.1 are typically realised. The observed flux excess for this star in the WISE bands
    can be attributed to the high infrared luminosity of the \ion{H}{ii} region (see Fig.~\ref{fig:thumbnails}).

    \paragraph{HD 183143.\label{sec:shell_section}}
    Unlike the other four stars, no reasonable fit of the observed SED could be achieved using the CCM extinction law 
    in the case of the DIB standard star HD~183143. As shown in Fig.~\ref{fig:sed_pah}, a fit with $E(B-V)$\,=\,1.02 and 
    $R_V$\,=\,3.8, which matches the SED for $\gtrsim$5000\,{\AA}, fails to reproduce the 
    strong extinction shortwards, overpredicting the flux by about two orders of magnitude in the 2175\,{\AA} 
    extinction bump. The colour excess is also incompatible with a value derived from considering an intrinsic $(B-V)$ colour 
    from the spectral-type, which yields $E(B-V)\,=\,1.27$ for HD~183143 \citep{Hobbsetal09}.
    It is well documented that the sight line towards HD~183143 shows two roughly equally strong
    interstellar absorption components \citep[see also][]{Hobbsetal09}. A tentative approach to modelling the extinction curve by overlapping two distinct 
    CCM-type extinction laws also yielded no success. In order to solve this discrepancy for this highly important object
    in the context of DIB investigations, a more refined approach is required.
    
   \begin{figure*}
  \centering
  \subfloat[]{
      \includegraphics[trim=25 0 0 0, width = .24\textwidth]
      {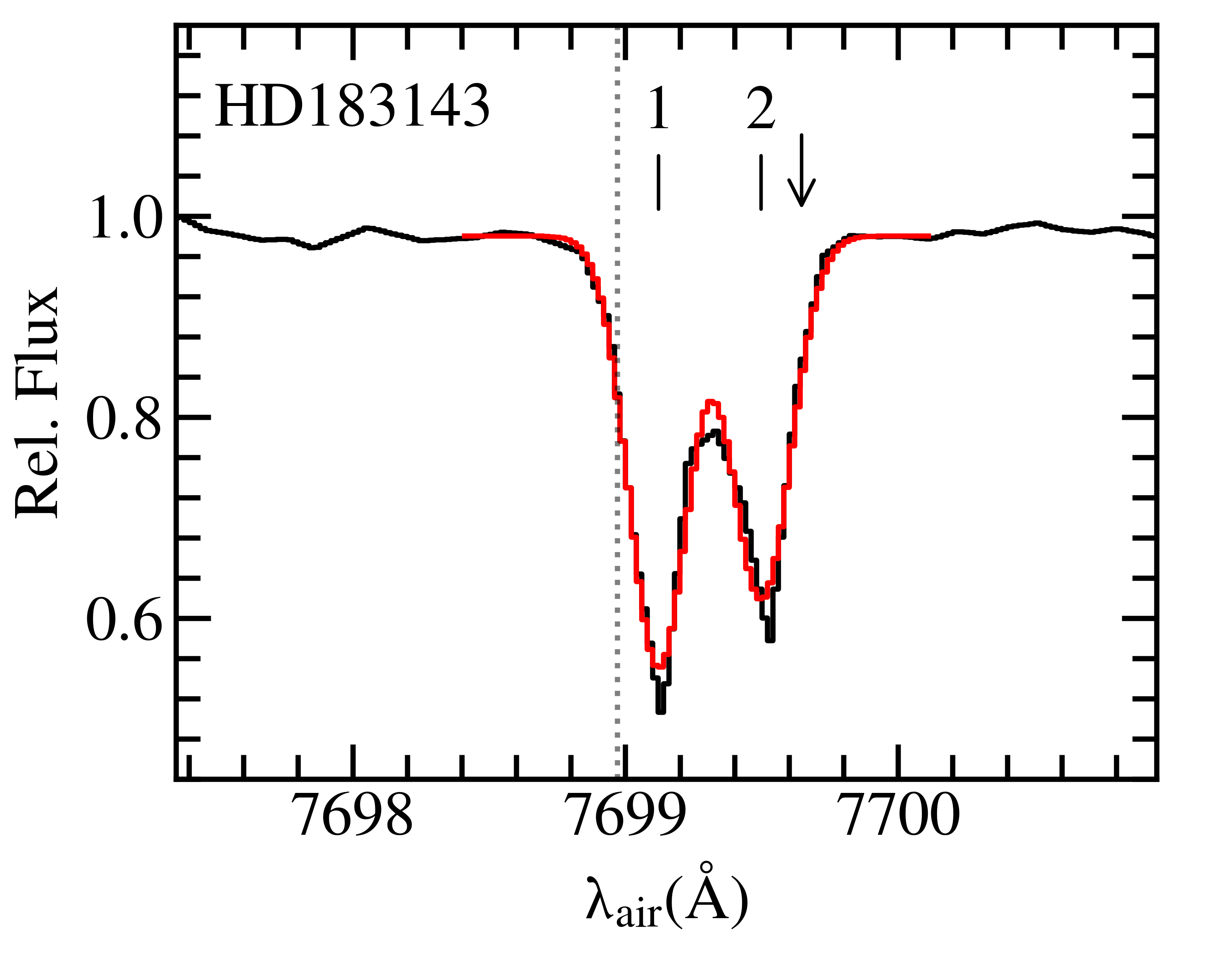}\label{fig:KI_HD183143}}
  \subfloat[]{
      \includegraphics[trim=25 0 0 0,clip, width = .24\textwidth]
      {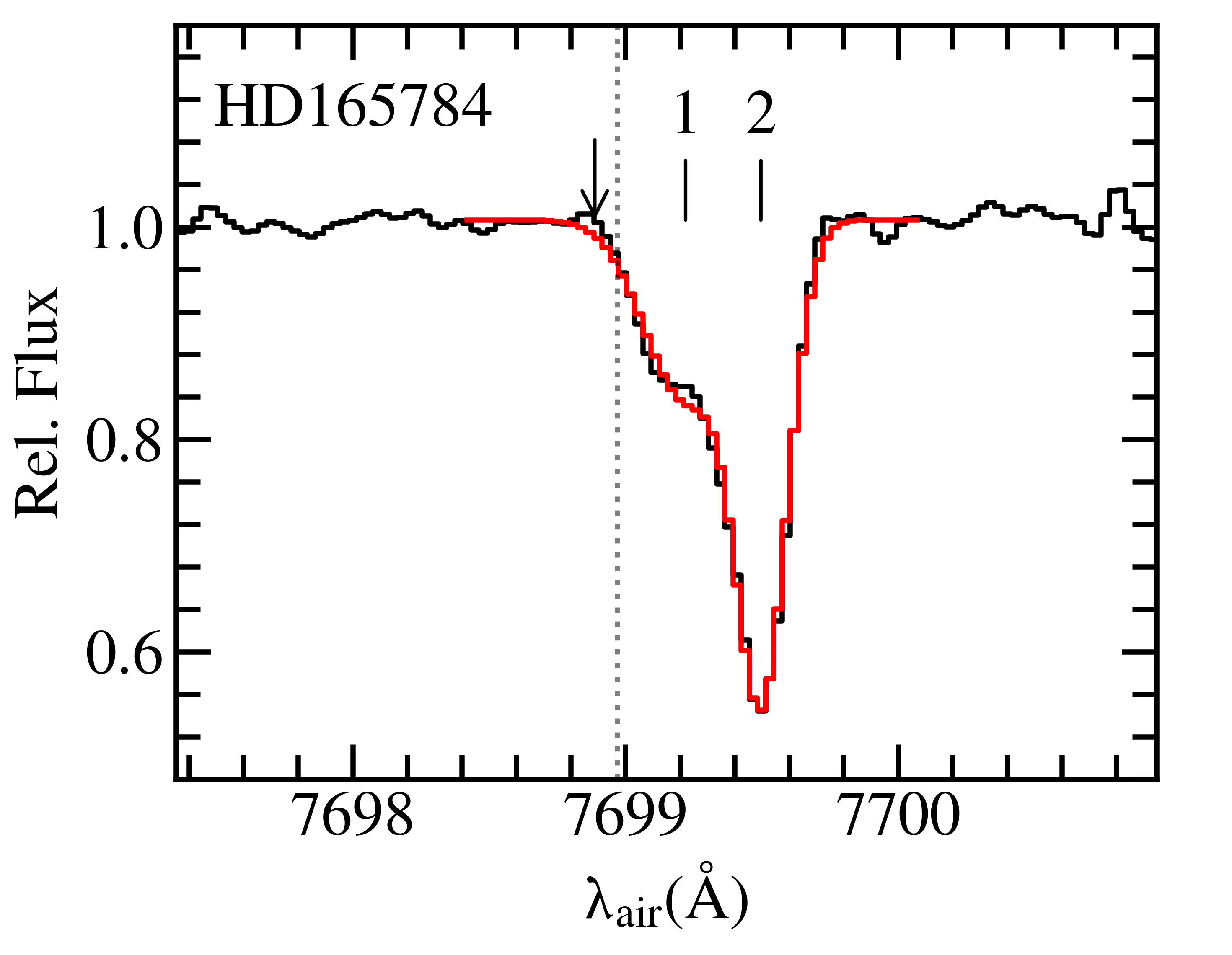}\label{fig:KI_HD165784}}
  \subfloat[]{
      \includegraphics[trim=25 0 0 0,clip, width = .24\textwidth]
      {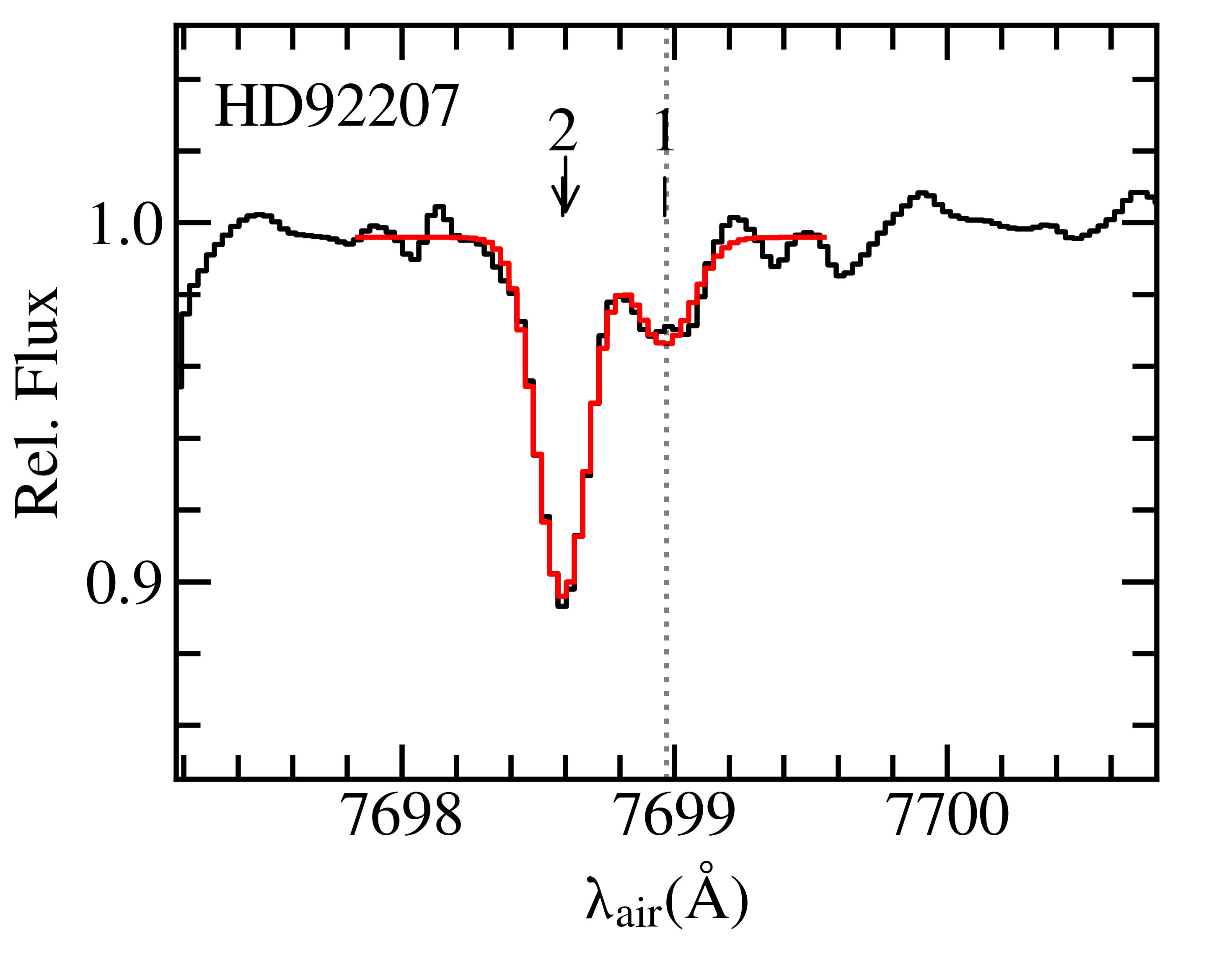}\label{fig:KI_HD92207}}
  \subfloat[]{
      \includegraphics[trim=25 0 0 0,clip, width = .24\textwidth]
      {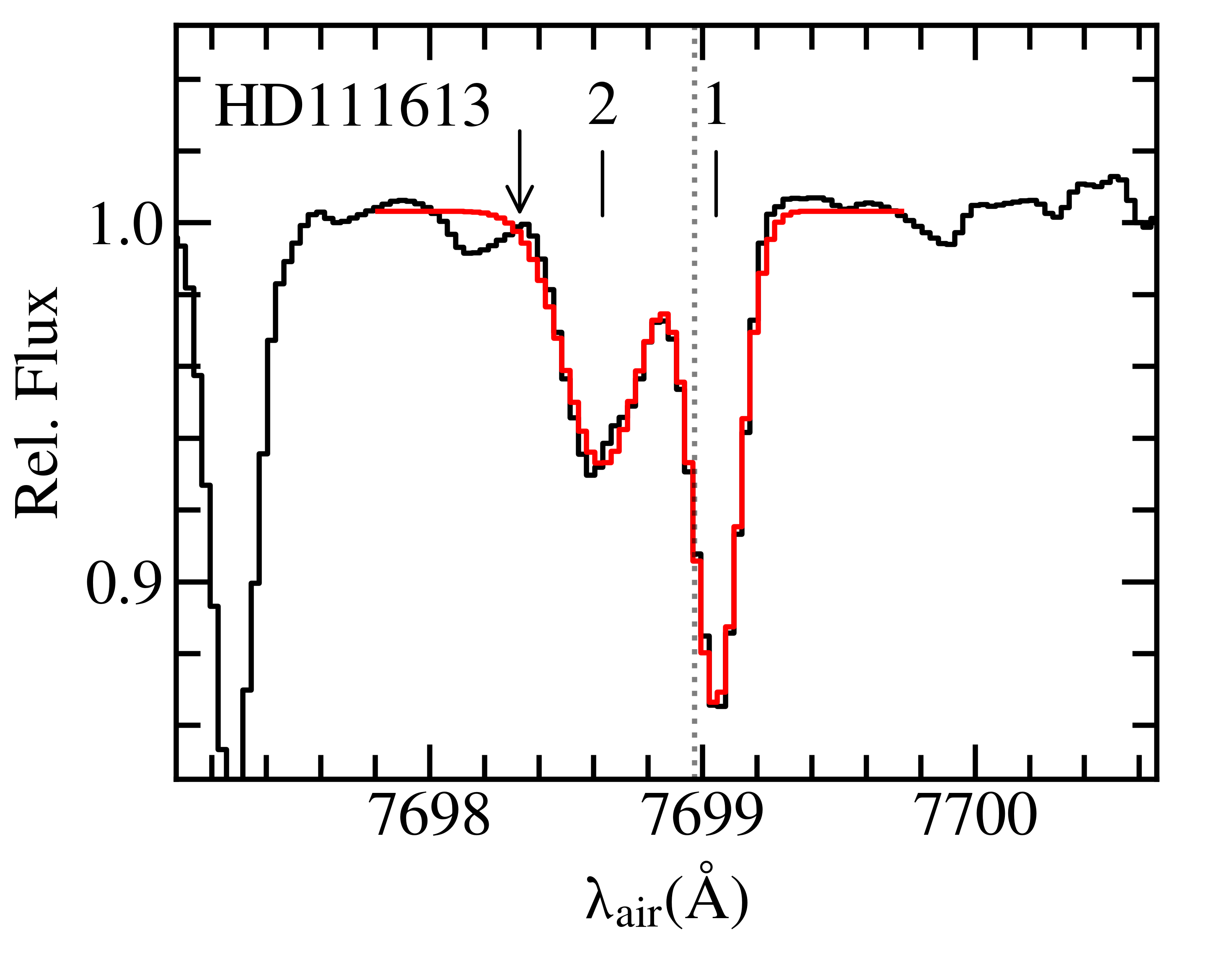}\label{fig:KI_HD111613}}\\[-3mm]
   \subfloat[]{
      \includegraphics[trim=25 0 0 0, width = .24\textwidth]
      {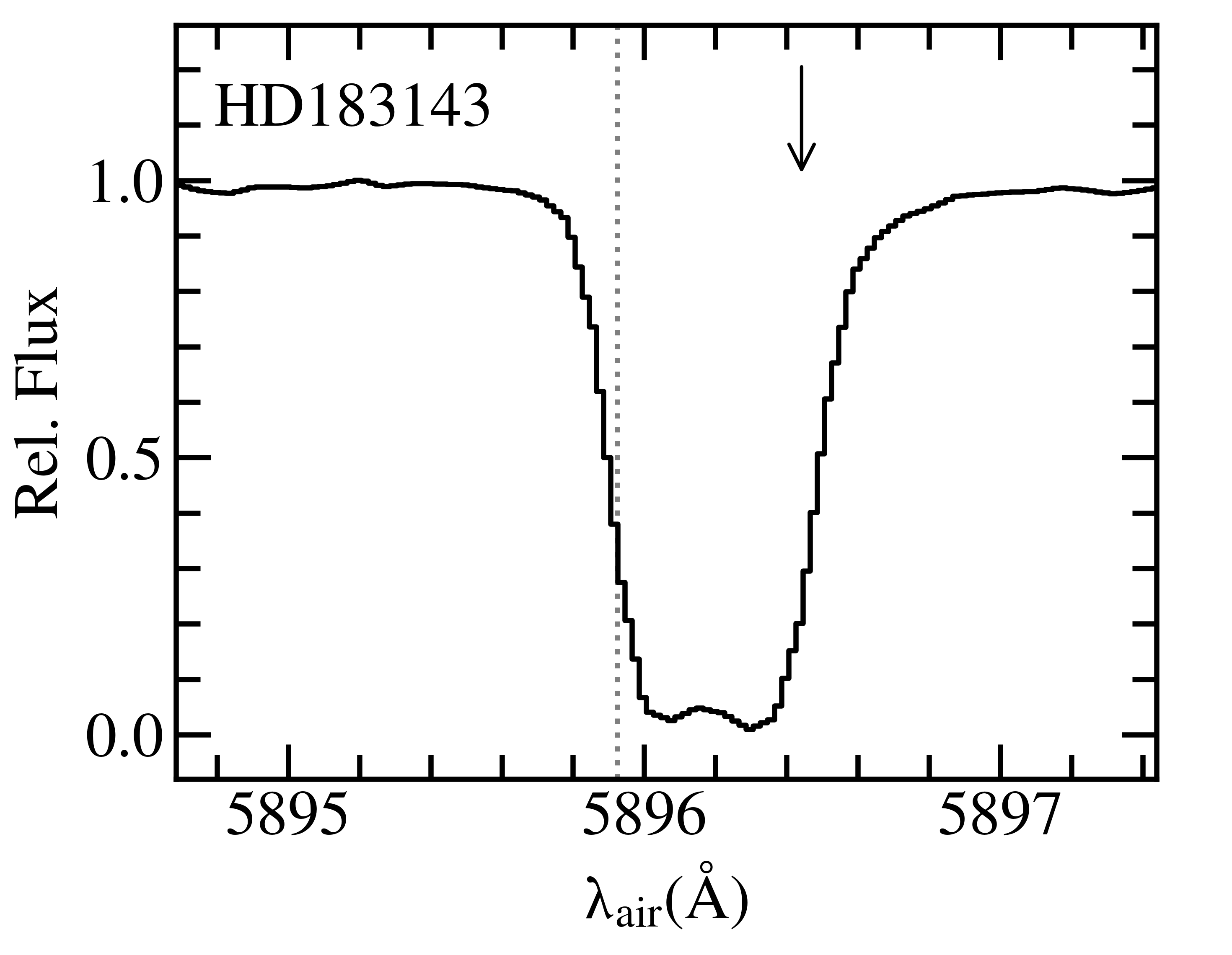}\label{fig:na_HD183143}}
  \subfloat[]{
      \includegraphics[trim=25 0 0 0,clip, width = .24\textwidth]
      {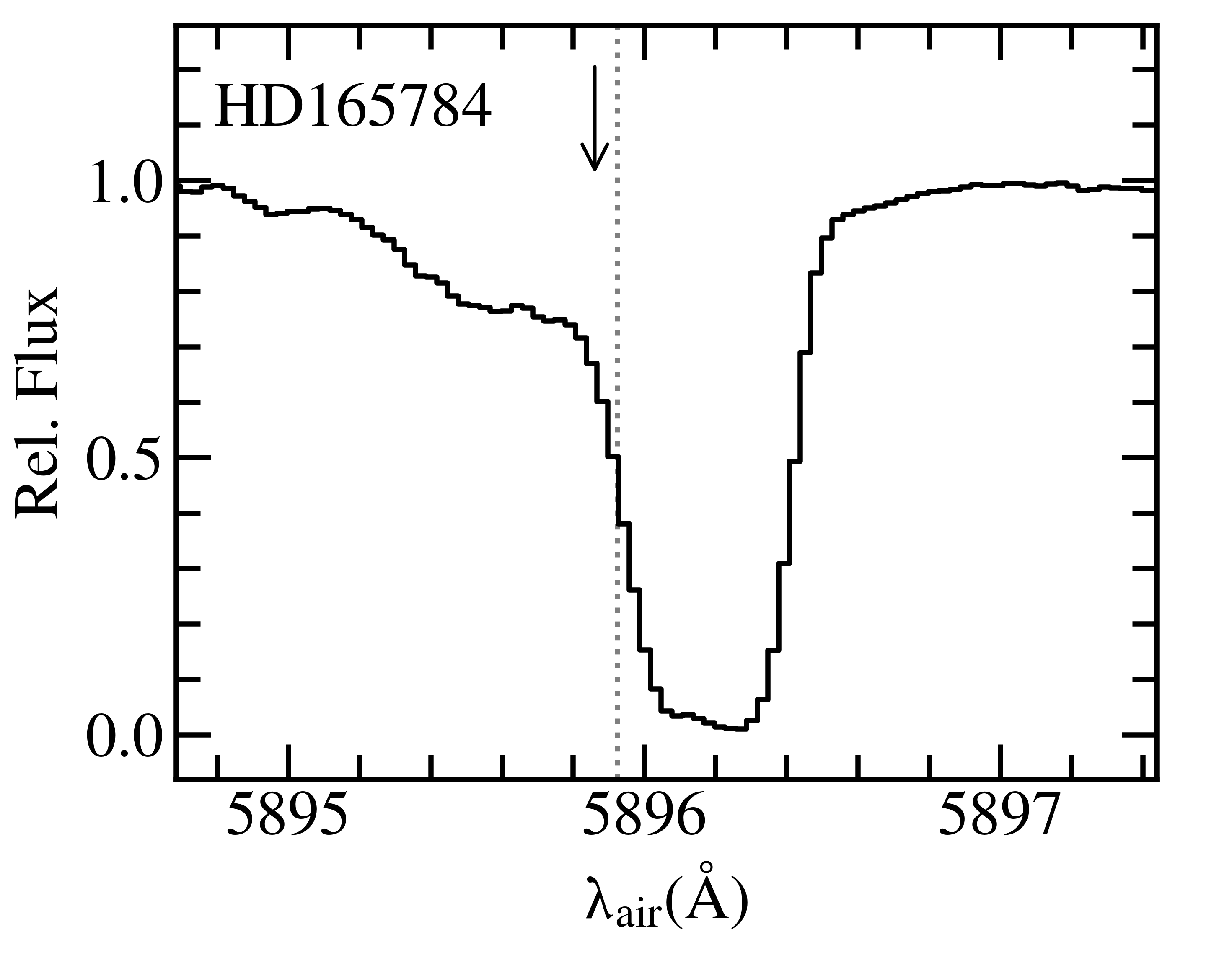}\label{fig:na_HD165784}}
  \subfloat[]{
      \includegraphics[trim=25 0 0 0,clip, width = .24\textwidth]
      {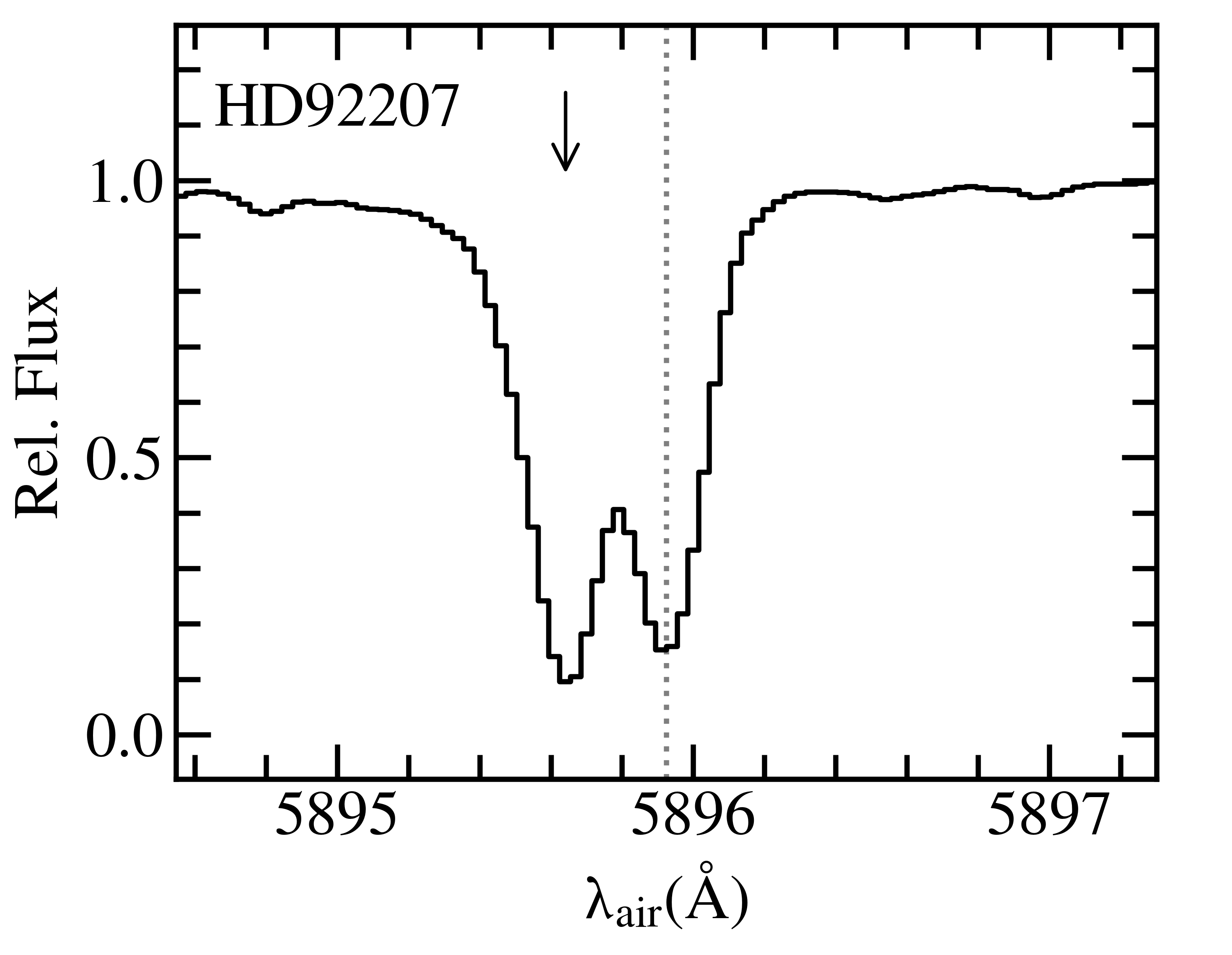}\label{fig:na_HD92207}}
  \subfloat[]{
      \includegraphics[trim=25 0 0 0,clip, width = .24\textwidth]
      {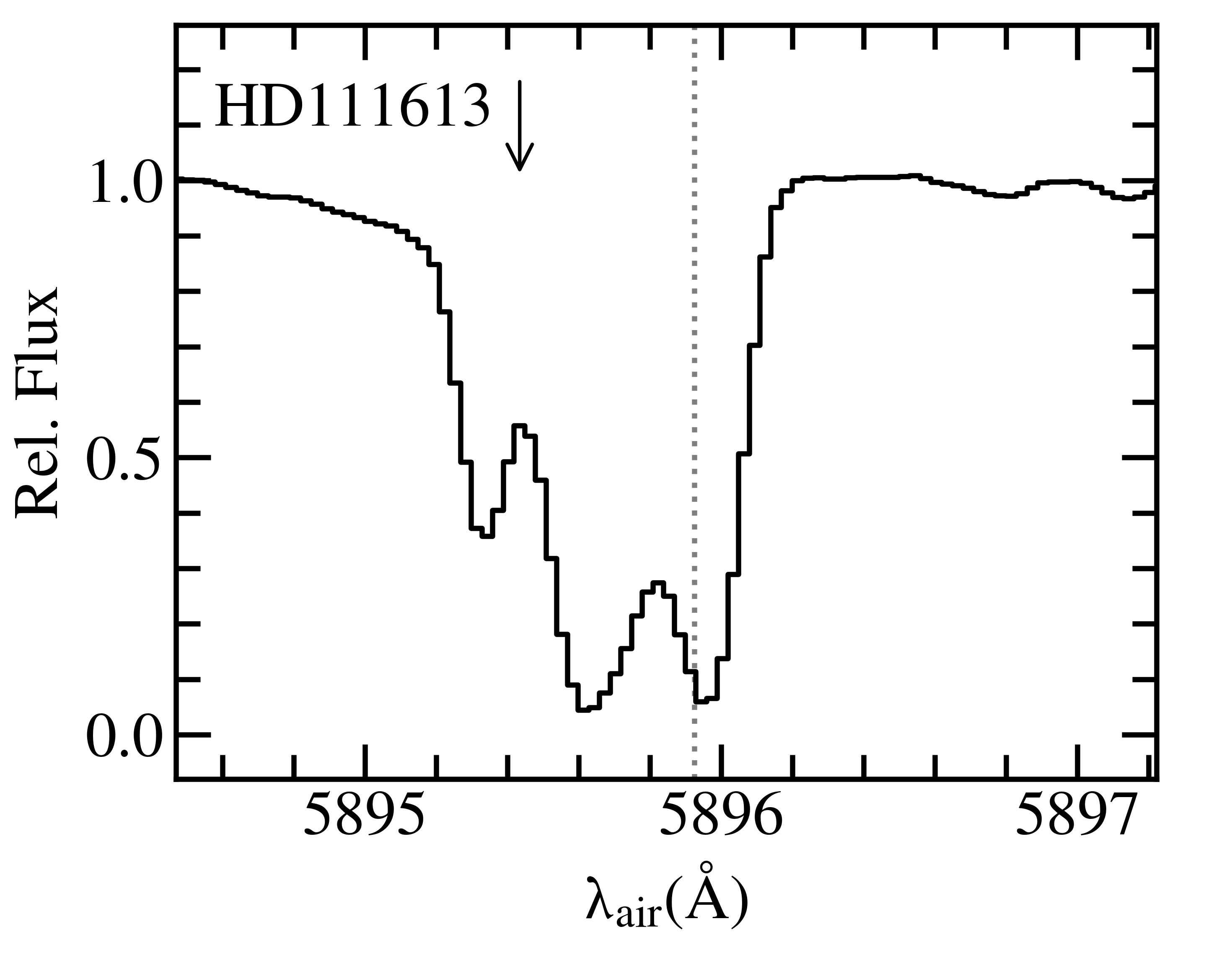}\label{fig:na_HD111613}}

  \caption{Spectra of atomic ISM lines.
 Upper panels: Observed spectra (black) of the interstellar 
  \ion{K}{i} line at
  $\lambda_\mathrm{air}$\,=\,7698.965\,{\AA} fitted with a two-Gaussian absorption 
  model (red). Component numbers are indicated. The vertical dotted lines represent the laboratory wavelengths 
  of the \ion{K}{i} lines. 
  The strong line at about $\lambda_\mathrm{air}$\,=\,9697.3\,{\AA} in panel \ref{fig:KI_HD111613} is a telluric O$_2$ line.
  Lower panels: Observed spectra (black) of the \ion{Na}{i} D$_1$ line at
  $\lambda_\mathrm{air}$\,=\,5895.924\,\AA\ (vertical dotted).
  For the cooler stars, HD~165784 and HD~111613, the interstellar \ion{Na}{i} D$_1$ lines are blended with a stellar component.
  The third \ion{Na}{i} component at $\sim$5895.3\,{\AA} in panel \ref{fig:na_HD111613} is very weak compared to the other two and barely present in \ion{K}{i}, such that it is neglected for the $RV$ analysis as little DIB absorption can be expected to originate from it.
  The position of the \ion{K}{i} and \ion{Na}{i} line in the stellar rest frame is indicated by a vertical arrow.
  Wavelengths are in the LSR.}
  \label{fig:KI_fits}
   
  \end{figure*}

    \begin{table*}
            \caption{Velocities and equivalent widths of the interstellar
            \ion{K}{i} $\lambda$7698.965\,\AA\ absorption components.}
        \centering
                {\small
\begin{tabular}{lrrrrrrrrr}
\hline\hline
Object & $RV_{1,\mathrm{bc}}$ & $RV_{2,\mathrm{bc}}$ & $RV_{1,\mathrm{LSR}}$ & $RV_{2,\mathrm{LSR}}$ & $|\Delta RV|$ & $RV_\mathrm{star,bc}$ & & $EW_1$ & $EW_2$\\
\cline{2-7} \cline{9-10}
 & \multicolumn{6}{c}{km\,s$^{-1}$} & & \multicolumn{2}{c}{m{\AA}}\\
\hline
HD 183143  &   $-10.6$  &     4.0  & 5.9 & 20.5 &   14.6  & 9.8 & &  119  &      100\\
HD 165784  &    $-3.1$  &    7.7   & 9.7 & 20.5 &    10.8  & $-16.0$ & & 105  &   63\\
HD 92207   &    8.5  &    $-6.1$   & $-0.2$ & $-14.8$ &  14.6  & $-5.7$ & & 8  &   23\\
HD 111613  &   7.0   &    $-$9.2  & 3.1 & $-13.1$ & 16.2   & $-21.0$ & & 28   &    24\\
HD 87737   &   ...   &     ...      &   ...  &   ...  &   ...    & 2.44 & & ... & ...\\
\hline
\end{tabular}
        \label{tab:rv_ism}
        }
    \end{table*}
    
    We used the non-LTE radiative transfer code {\sc Skirt} \citep{baes2011,camps2015} to model dust
    absorption in a 3D Monte Carlo simulation. The code allows different species, such as  graphitic and silicate dust, and 
    polycyclic aromatic hydrocarbons (PAHs) to be considered in detail. The HD~183143 extinction curve can be reproduced closely using a combination of
    the CCM extinction law and an additional PAH absorption component, for a combined $E(B-V)$\,=\,1.22 and $R_V$\,=\,3.3
    (see Fig.~\ref{fig:sed_pah}). 
    The peculiar ISM is represented by a slab with uniform PAH density extending over 2\,pc 
    along the sight line, approximated to the observed geometry (see below).
    The scattering process is modelled by the Henyey-Greenstein phase function 
    \citep{1941ApJ....93...70H}, which does not consider polarization.
    The optical properties are those of neutral PAH carbonaceous grains according to \cite{2001ApJ...554..778L}.
    Calorimetric properties are approximated by the analytical enthalpy prescription for 
    graphite \citep{2001ApJ...554..778L} with a bulk mass density set to the standard value 
    of 2240 kg m$^{-3}$ for graphite grains.
    The grain size distribution is adapted from model BARE\_GR\_S of 
    \cite{2004ApJS..152..211Z} with a minimum grain size of $a_\mathrm{min} = 3.5\,$\AA\ and a maximum grain size of 
    $a_\mathrm{max} = 50\,$\AA. We normalised the overall material density to an optical depth of
    $\tau(2175\,$\AA$) = 3.2$, which is equivalent to a PAH mass column density of $6.05 \cdot 10^{-5}$\,kg\,m$^{-2}$.

    The search for a source of PAH-rich material resulted in finding a bubble in the W3 and W4 bands with lower luminosity  
    than the ambient ISM in the general direction towards HD~183143, which is enclosed by a more luminous shell.
    Figure~\ref{fig:ir_bubble} shows images of this shell in the WISE 3 and 4 bands along with the extents of its major and minor axes.
    The sight line to HD~183143 intersects this shell tangentially, providing a potentially 
    significant column mass of absorbing material at optical--NIR wavelengths. The object at the centre of the bubble is the 
    $H$\,$\approx$19.8\,mag IR source \object{IRAS 19255+1816}, for which the only additional data available in the 
    literature are flux measurements with the Infrared Astronomical Satellite \citep[IRAS,][]{HeWa88}, which indicate a rise in flux towards 100\,$\mu$m.

    The shell is likely associated with the +4\,km\,s$^{-1}$ (barycentric) absorption component in the spectrum of HD~183143, which shows CN
    \citep{Hobbsetal09}. This hints at a carbon-rich circumstellar envelope (i.e. to a carbon-star nature for IRAS~19255+1816).
    The shell has a semi-major axis extension of $a$\,$\approx$\,6.7\,arcmin and a semi-minor axis 
    of $b$\,$\approx$\,5.2\,arcmin. At a tentative distance of $d_\mathrm{shell}$\,$\sim$\,2\,kpc this would translate to 
    $a$\,$\approx$\,3.9\,pc and $b$\,$\approx$\,3\,pc. The luminosities of the shell in the W3 and W4 bands also support a 
    localised PAH overabundance in this sight line because these band wavelengths match those of PAH emission bands
    \citep{2001ApJ...554..778L, 2014A&A...561A..82S}, as well as the improved fit of the extinction curve (Fig. \ref{fig:sed_pah}). So if some 
    DIBs show stronger normalised equivalent widths in this sight line compared to other DIBs, those DIBs should have a higher probability of having PAHs as carriers.
    
    It would, in general, be very interesting to observe more sight lines with PAH overabundances 
    to circumvent the degeneracy of those effects when only looking at a single example.
    For this particular shell it would be valuable to test other sight lines that pass the shell tangentially.
    Unfortunately, all candidate stars are much fainter, with little information available in the literature, and the 
    few brighter among them are low-luminosity foreground stars.
    Finally, it would be beneficial to investigate further the supposed central object of the 
    circumstellar shell, IRAS~19255+1816, to constrain its nature.

        \section{Atomic interstellar lines: Cloud velocities}\label{sec:atomic_lines}
    The sight lines were analysed for multiple cloud components and radial velocities $RV$ as 
    these are important details for the interpretation of the measured DIB profiles. All sight lines towards 
    reddened target stars show two
    interstellar cloud components, which were consequently fitted by a two-Gaussian model to the 
    interstellar \ion{K}{i} absorption line at $\lambda_\mathrm{air}$\,=\,7698.965\,{\AA} 
    \citep{NIST_ASD}\footnote{The component of the \ion{K}{i} resonance doublet at 
    $\lambda_\mathrm{air}$\,=\,7664.899\,{\AA} is often contaminated 
    by the telluric A band of O$_2$ \citep[e.g.][]{Kimeswengeretal21}.}. The fitting results are summarised in 
    Table~\ref{tab:rv_ism} and displayed in Fig.~\ref{fig:KI_fits}.
    We always label the local component with the lower absolute $RV$ as component 1 
    and the other component as component 2.
    For each component $RV$s are given in the barycentric rest frame $RV_\mathrm{bc}$ and in the local standard of rest 
    (LSR) $RV_\mathrm{LSR}$, with the transformation made using the solar peculiar velocity from \cite{2010schoenrich}.
    Furthermore, the velocity difference between the two components $|\Delta RV|$ and the $RV$ of the background stars
    from our own measurements is given in Table~\ref{tab:rv_ism}. The last two entries are the $EW$s of the \ion{K}{i} 
    absorption components. The $|\Delta RV|$
    is smaller than 17\,km\,s$^{-1}$ for all sight lines, so we consider all structures within DIB profiles 
    on a scale larger than 17\,km\,s$^{-1}$ as intrinsic band profile characteristics. 
    Figure~\ref{fig:KI_fits} (upper row) shows the Gaussian fits of the \ion{K}{i} lines, along with the respective \ion{Na}{i}
    D$_1$ lines (lower row), which were not fitted due to their strong saturation. The absorption components are labelled.
    We note that for HD~87737 there is no interstellar \ion{K}{i} component observable because the star is located 
    at a high Galactic latitude and it is relatively close to the Earth. 
  
    We attribute the \ion{K}{i} components to different spiral arms of the Milky Way.
    Component 1 is always the Local Arm component due to its lower absolute $RV$.
    We used the observed spiral structure of the Milky Way according to \cite{hou2014} to assign the second component; this is discussed in the following.
   
       The sight line to HD~183143 at Galactic longitude $\ell$\,$\approx$\,53{\degr} shows the second component 
    redshifted with respect to $RV_{1, \mathrm{LSR}}$. We attribute this component to a circumstellar shell 
    (Sect.~\ref{sec:shell_section}) that  is presumably close to the Sagittarius Arm. The assignment for HD~165784 at
    $\ell$\,$\approx$\,9{\degr} is similar, with a redshifted $RV_{2, \mathrm{LSR}}$ implying membership of 
    component 2 to the Sagittarius Arm.
\clearpage      

 \begin{figure*}[htb!]
  \centering
   \subfloat[DIB 9632]{
      \includegraphics[width = \textwidth/19*6, trim=0 20 0 20, clip]
      {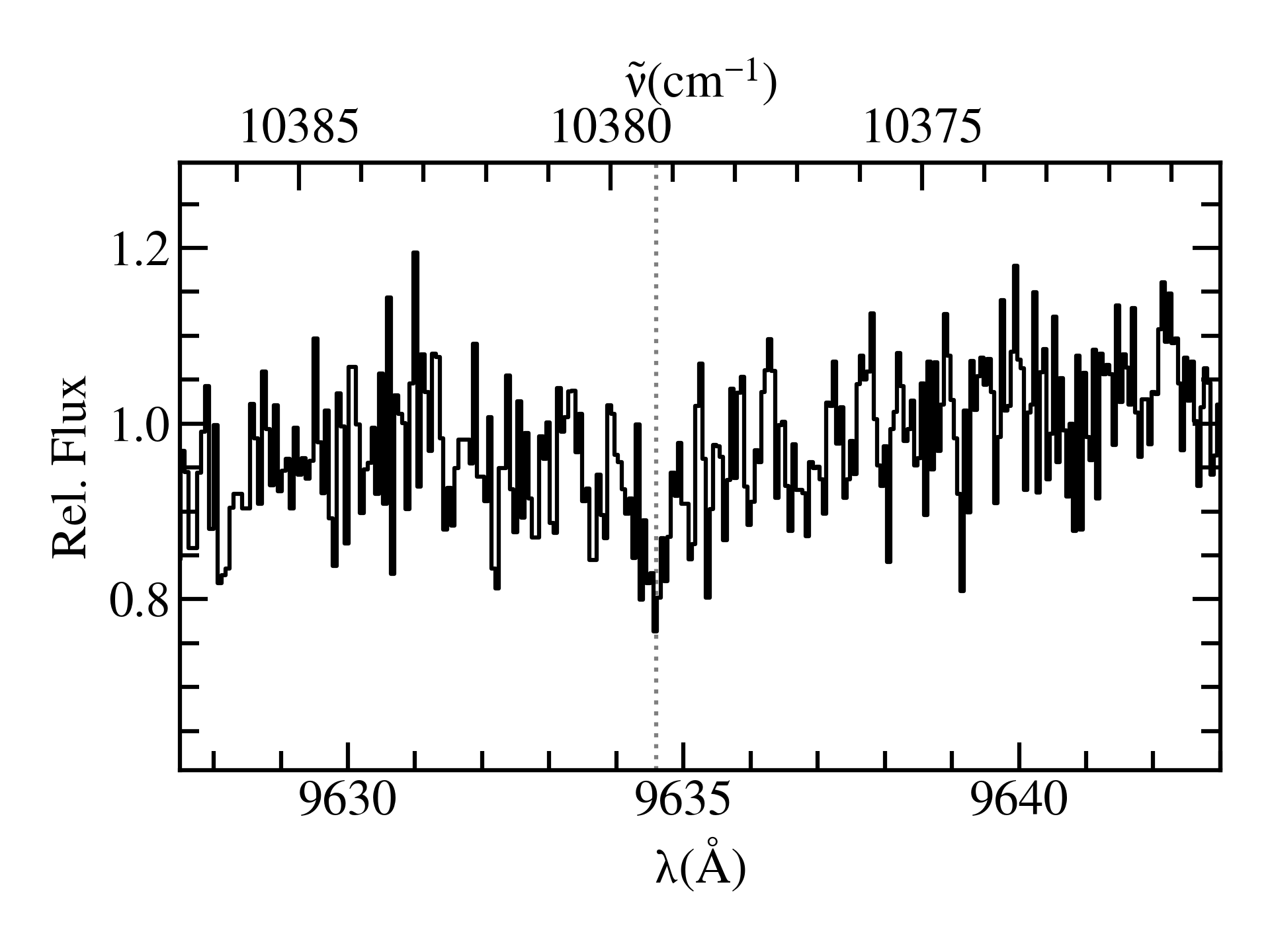}\label{fig:_hd183DIB9632}}
   \subfloat[DIB 9880]{
      \includegraphics[width = \textwidth/19*6, trim=0 20 0 20, clip]
      {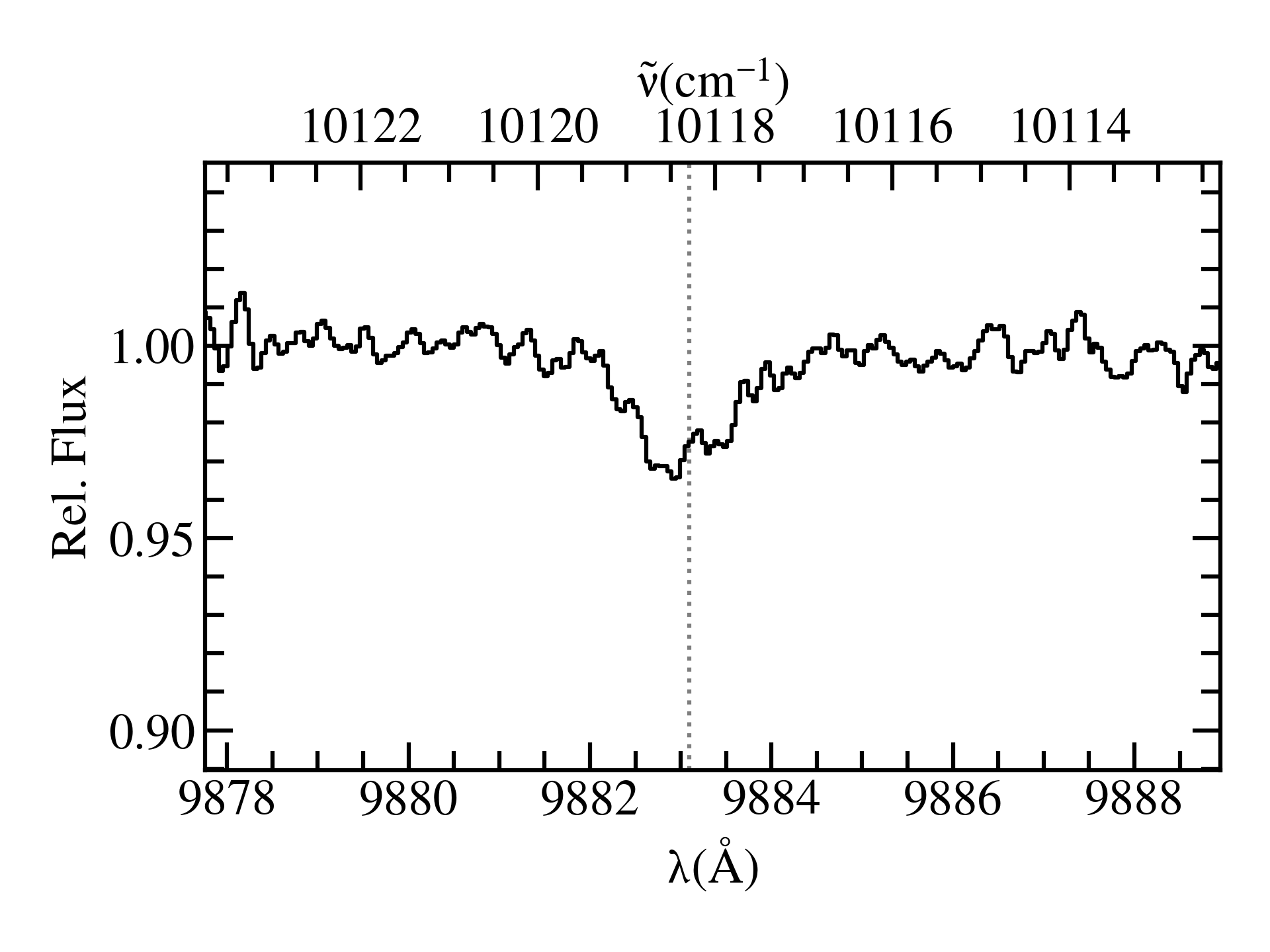}\label{fig:_hd183DIB9880}}
   \subfloat[DIB 10361]{
      \includegraphics[width = \textwidth/19*6, trim=0 20 0 20, clip]
      {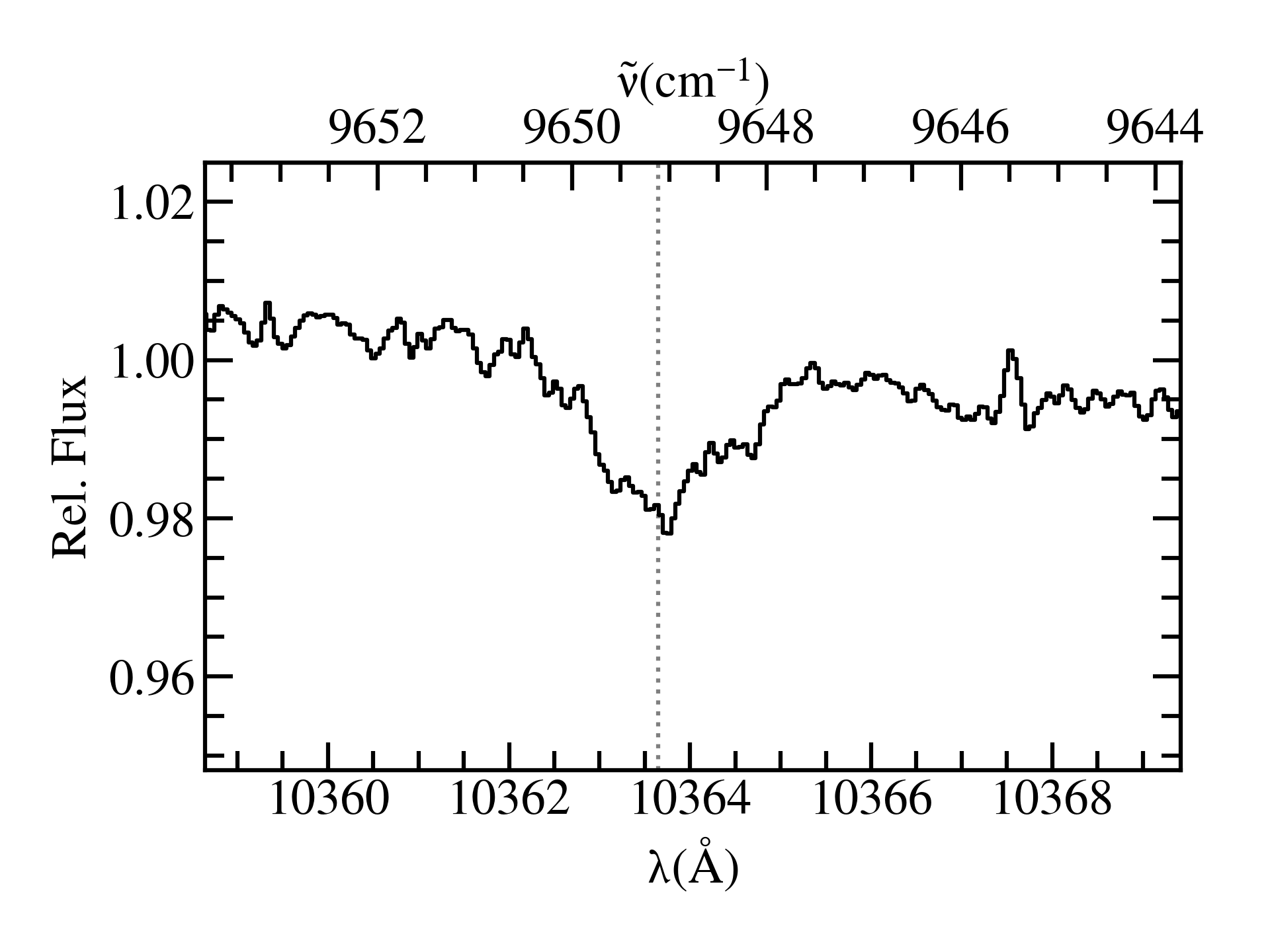}\label{fig:_hd183_10361}}\\[-4mm]
   \subfloat[DIB 10393]{
      \includegraphics[width = \textwidth/19*6, trim=0 20 0 20, clip]
      {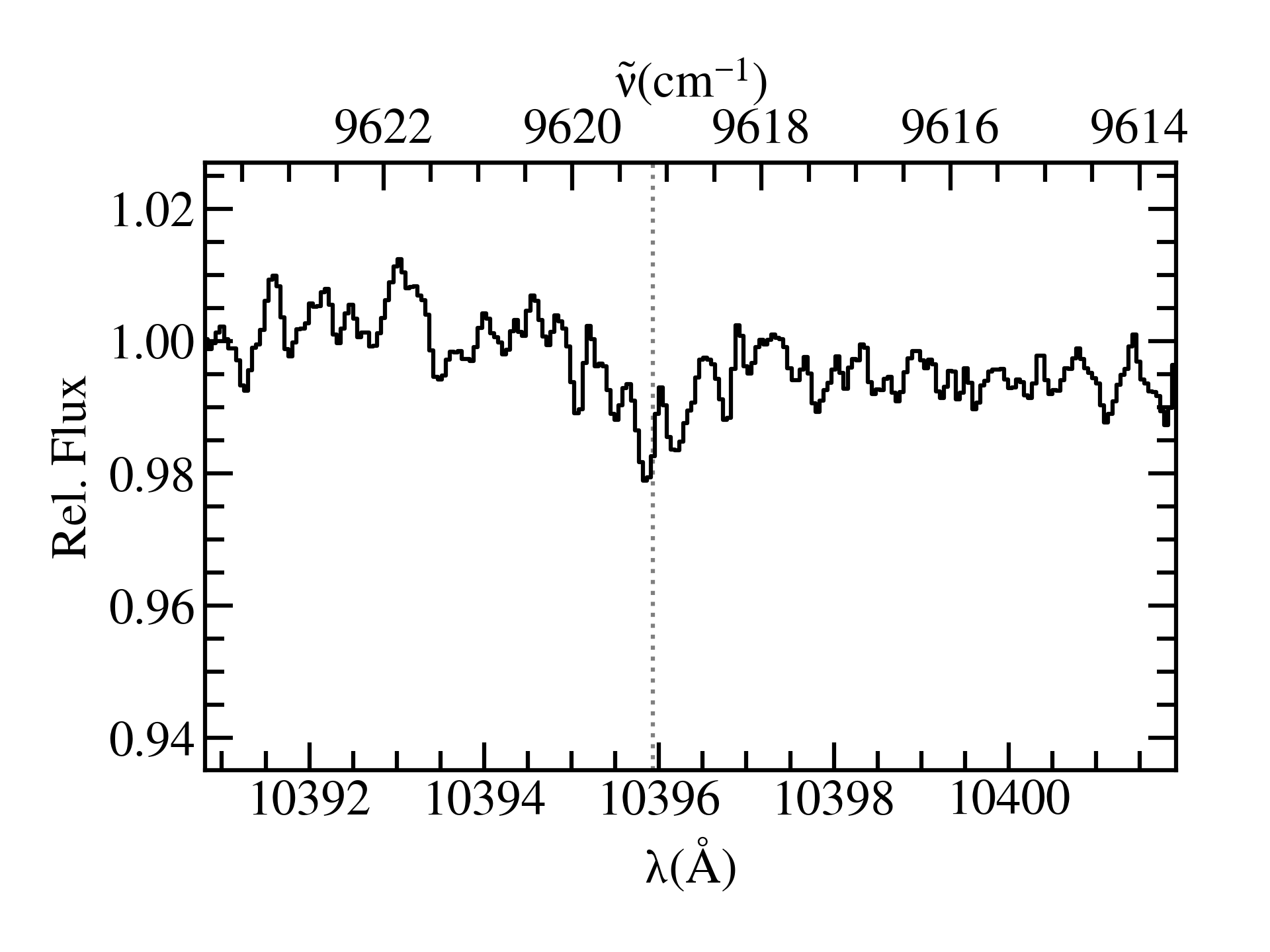}\label{fig:_hd183_10393}}
   \subfloat[DIB 10438]{
      \includegraphics[width = \textwidth/19*6, trim=0 20 0 10, clip]
      {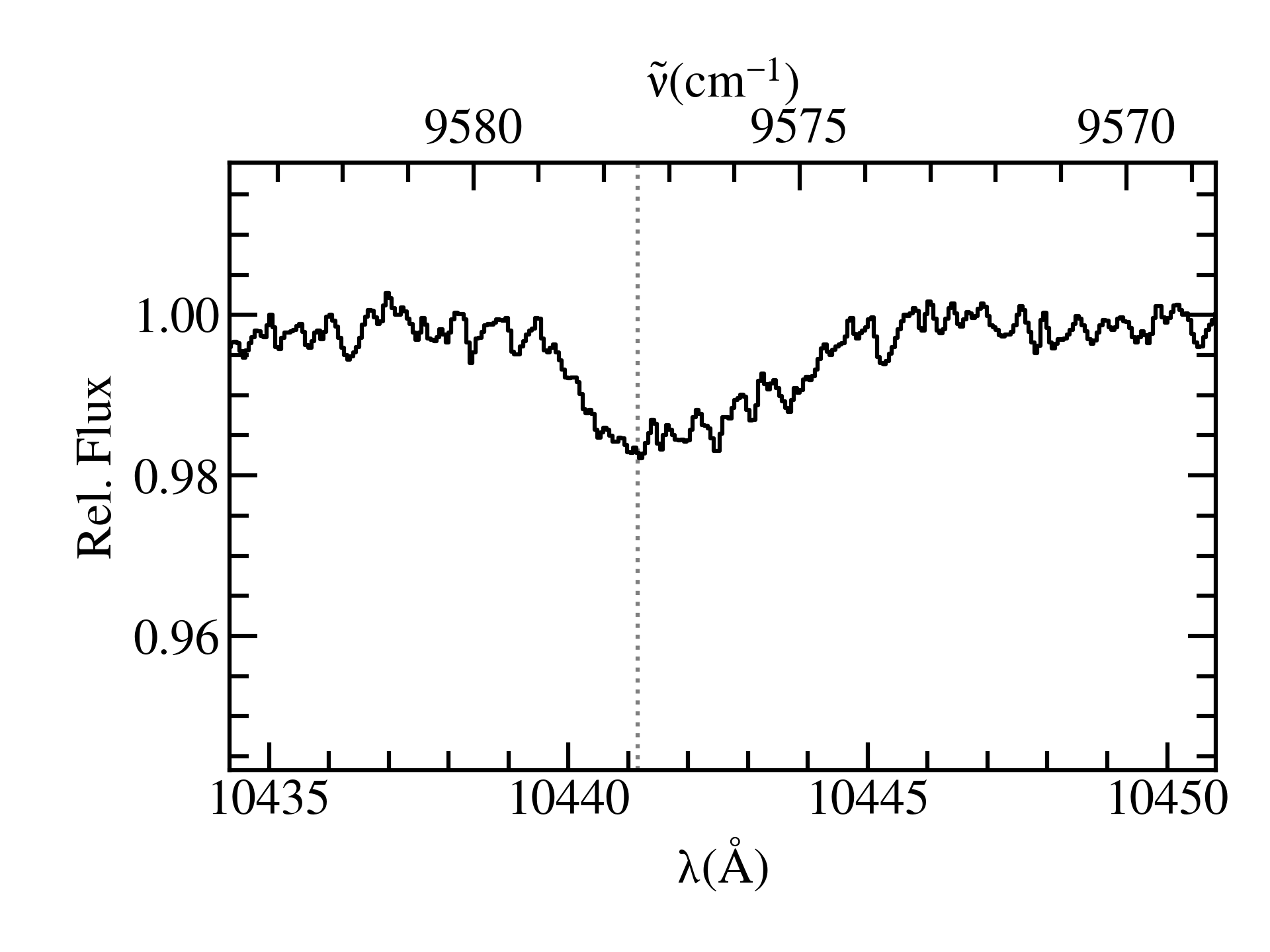}\label{fig:_hd183DIB10438}}
   \subfloat[DIB 10697]{
      \includegraphics[width = \textwidth/19*6, trim=0 20 0 10, clip]
      {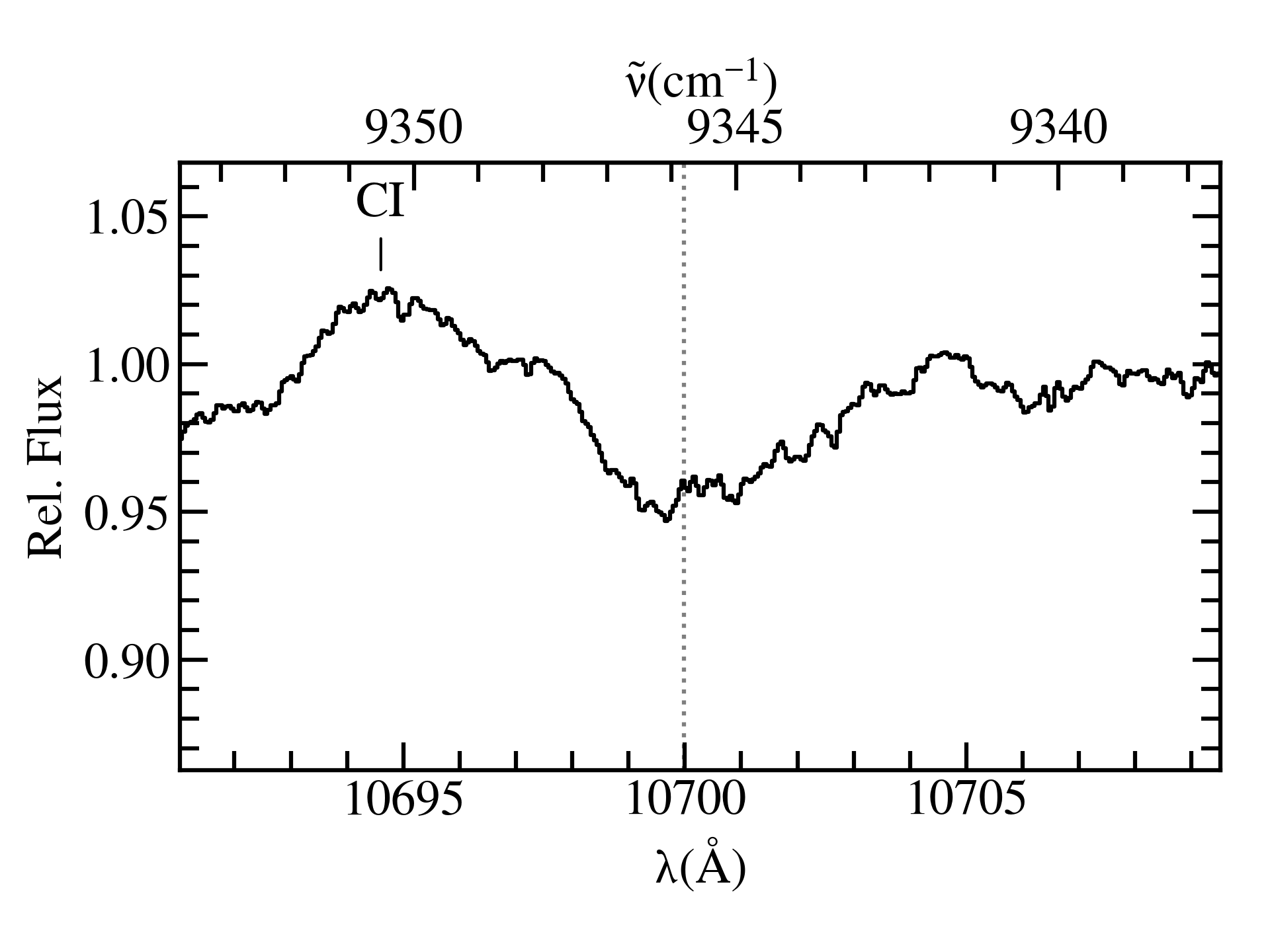}\label{fig:_hd183DIB10697}}\\[-4mm]
   \subfloat[DIB 10780]{
      \includegraphics[width = \textwidth/19*6, trim=0 20 0 10, clip]
      {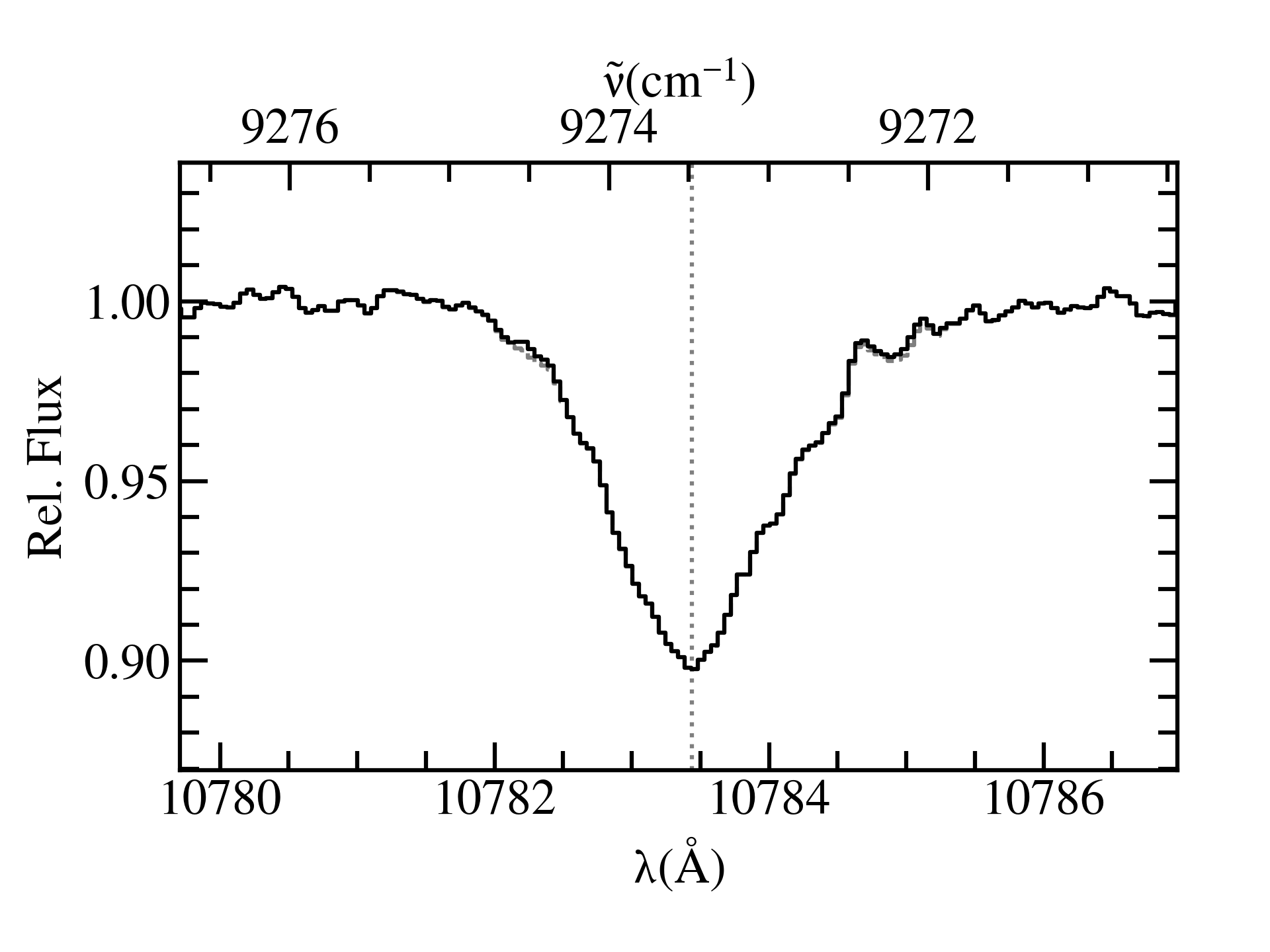}\label{fig:_hd183DIB10780}}
   \subfloat[DIB 10792]{
      \includegraphics[width = \textwidth/19*6, trim=0 20 0 10, clip]
      {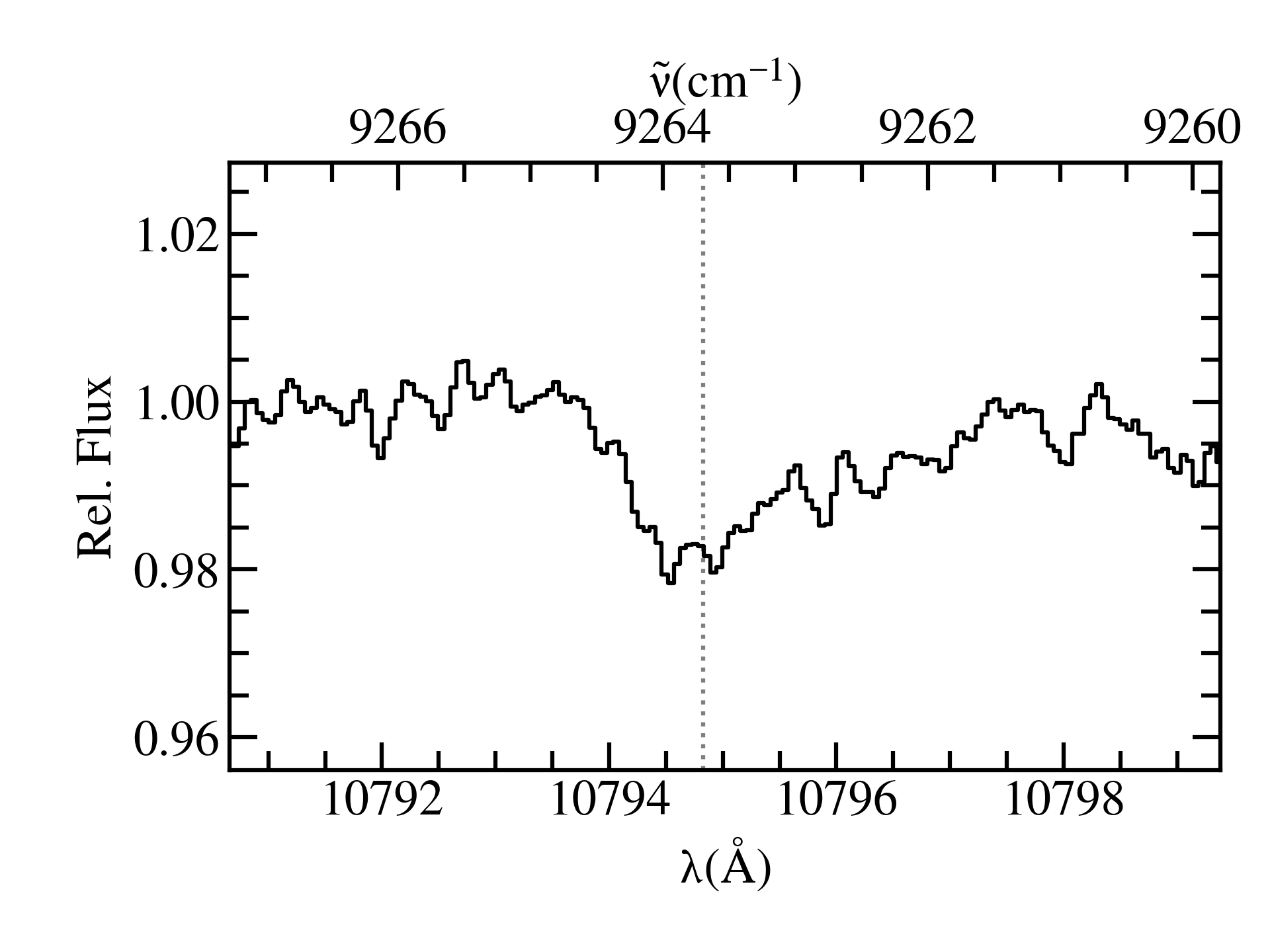}\label{fig:_hd183DIB10792}}
   \subfloat[DIB 11797]{
      \includegraphics[width = \textwidth/19*6, trim=0 20 0 10, clip]
      {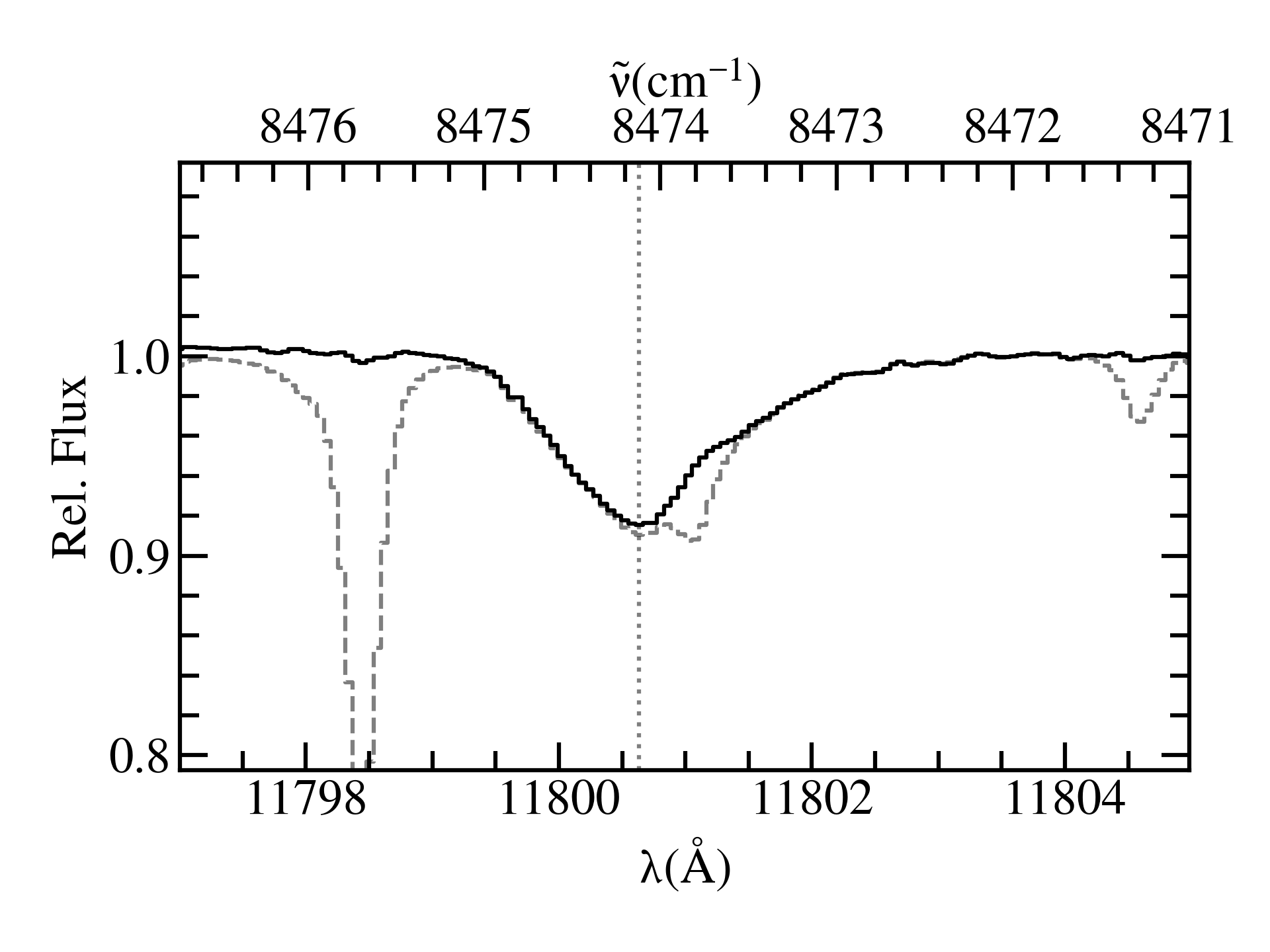}\label{fig:_hd183DIB11797}}\\[-4mm]
   \subfloat[DIB 12293]{
      \includegraphics[width = \textwidth/19*6, trim=0 20 0 10, clip]
      {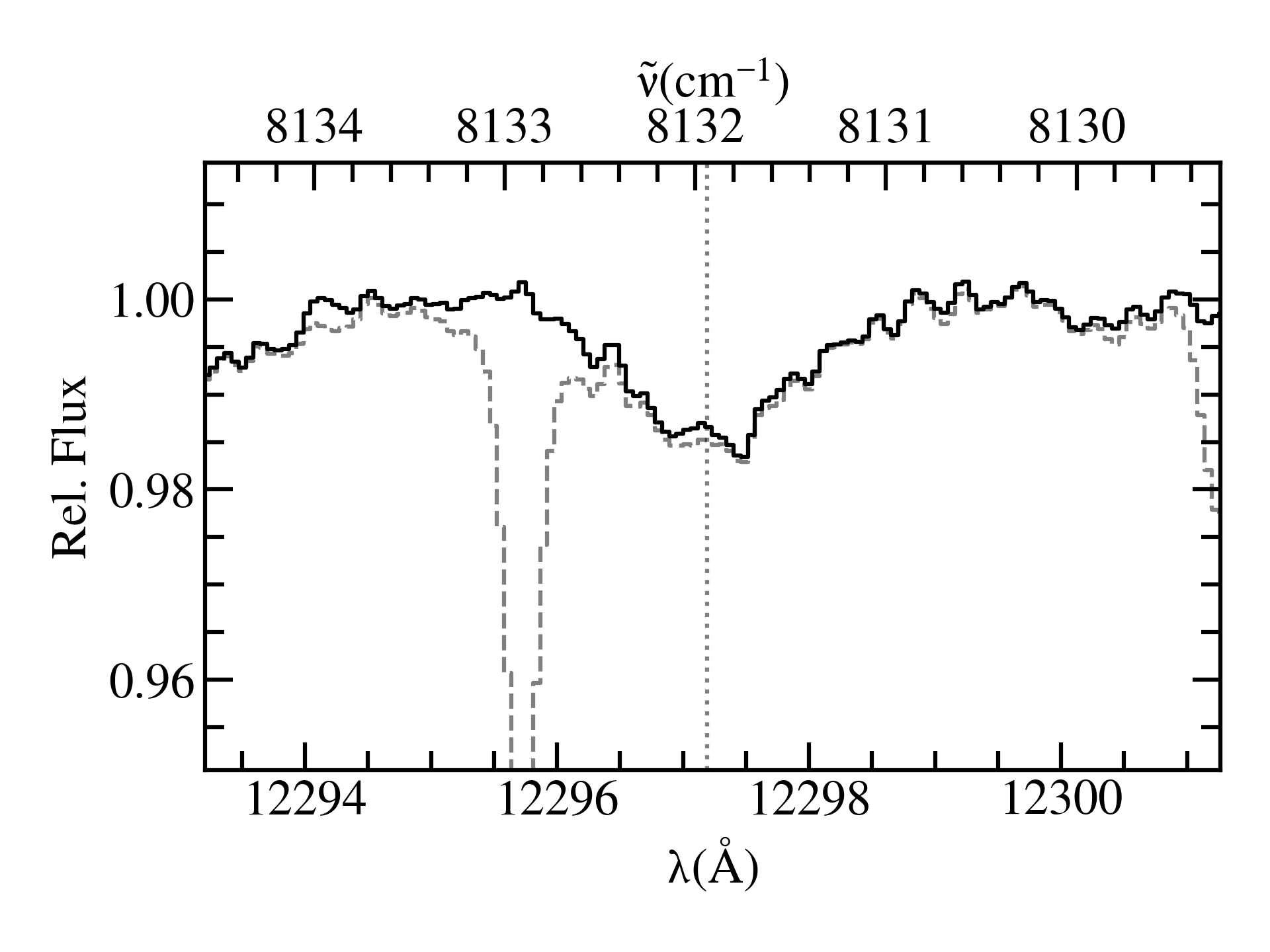}\label{fig:_hd183DIB12293}}
   \subfloat[DIB 12518]{
      \includegraphics[width = \textwidth/19*6, trim=0 20 0 10, clip]
      {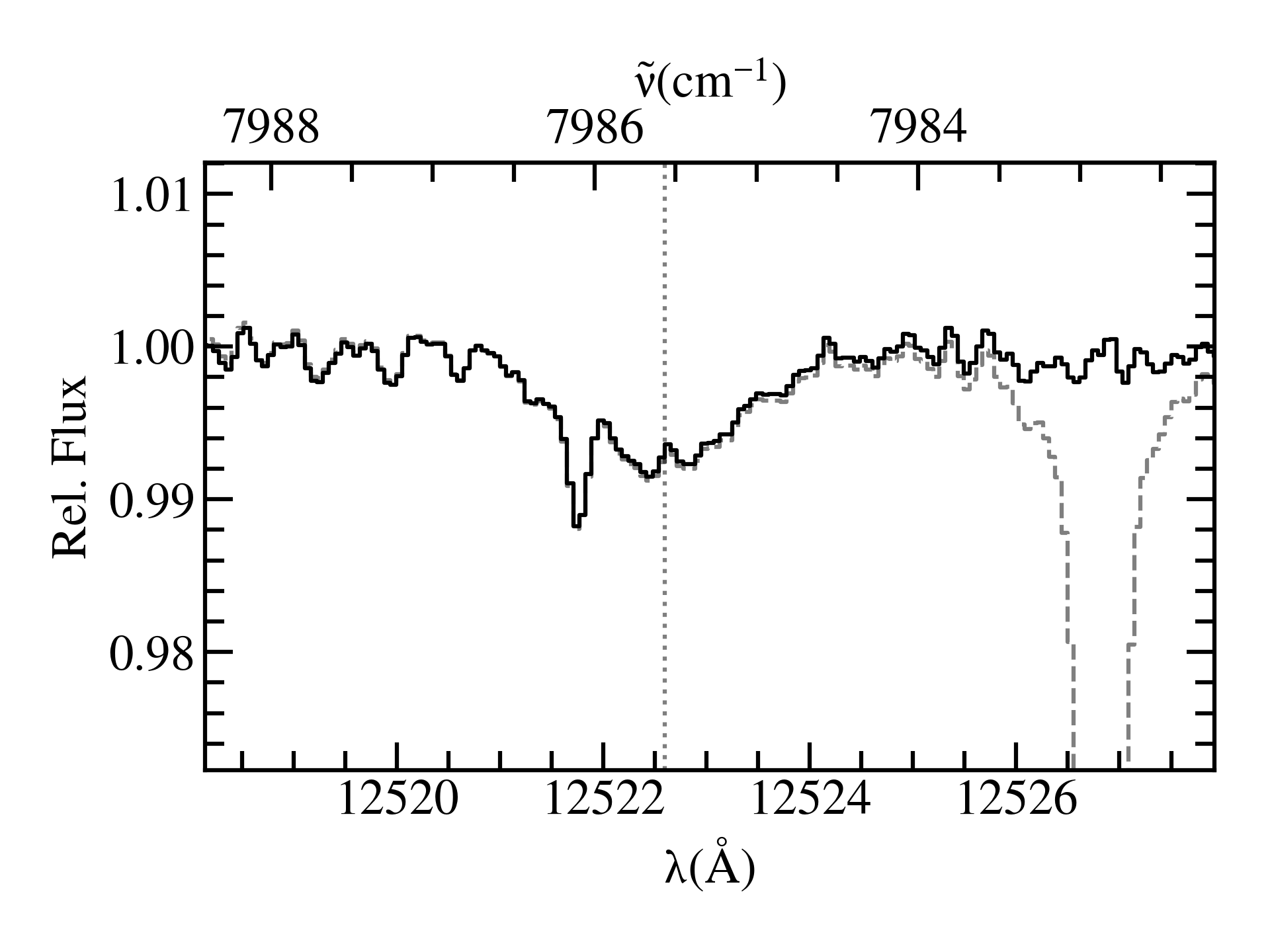}\label{fig:_hd183DIB12518}}
   \subfloat[DIB 12623]{
      \includegraphics[width = \textwidth/19*6, trim=0 20 0 10, clip]
      {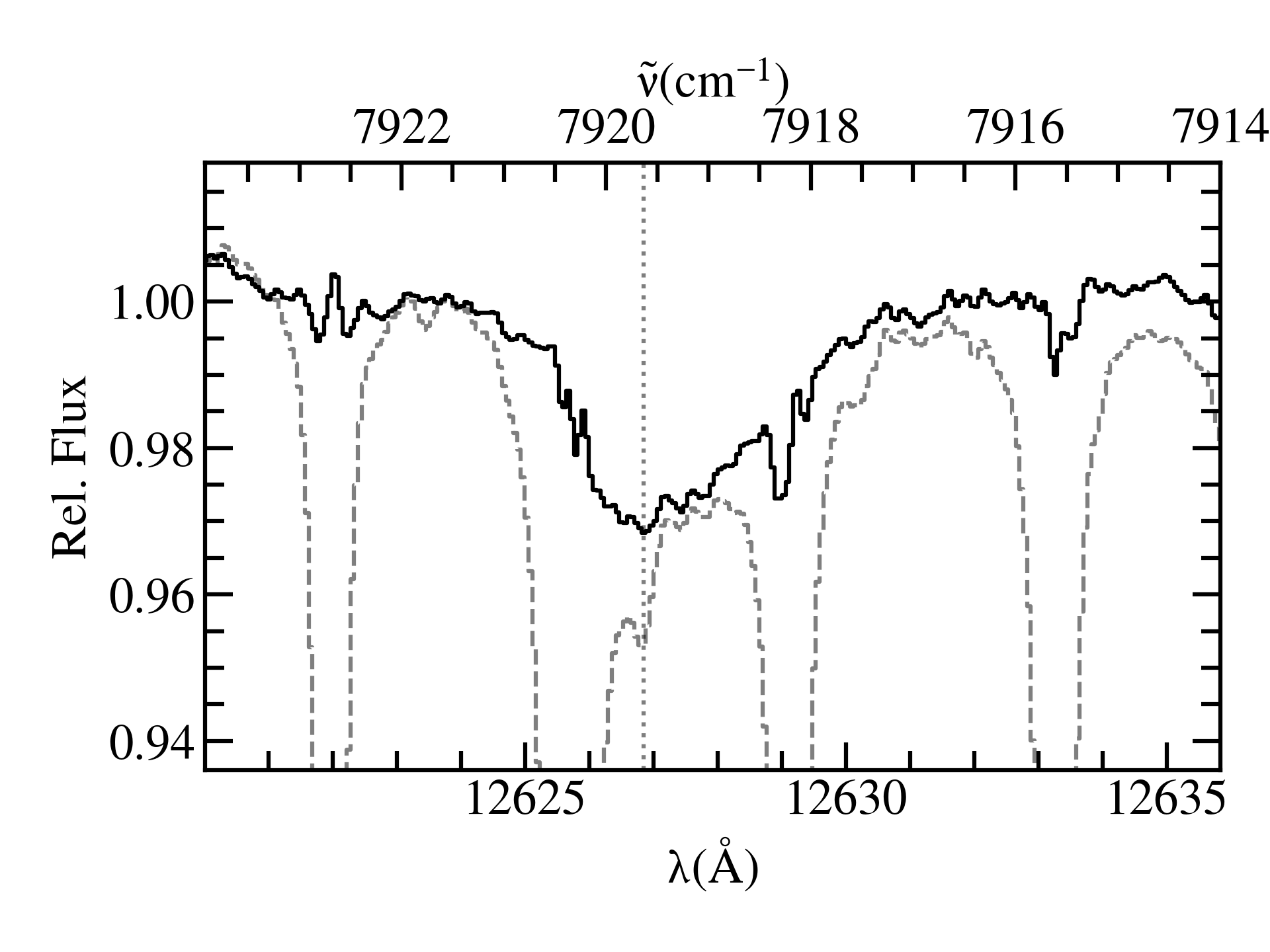}\label{fig:_hd183DIB12623}}\\[-4mm]
   \subfloat[DIB 12799]{
      \includegraphics[width = \textwidth/19*6, trim=0 20 0 10, clip]
      {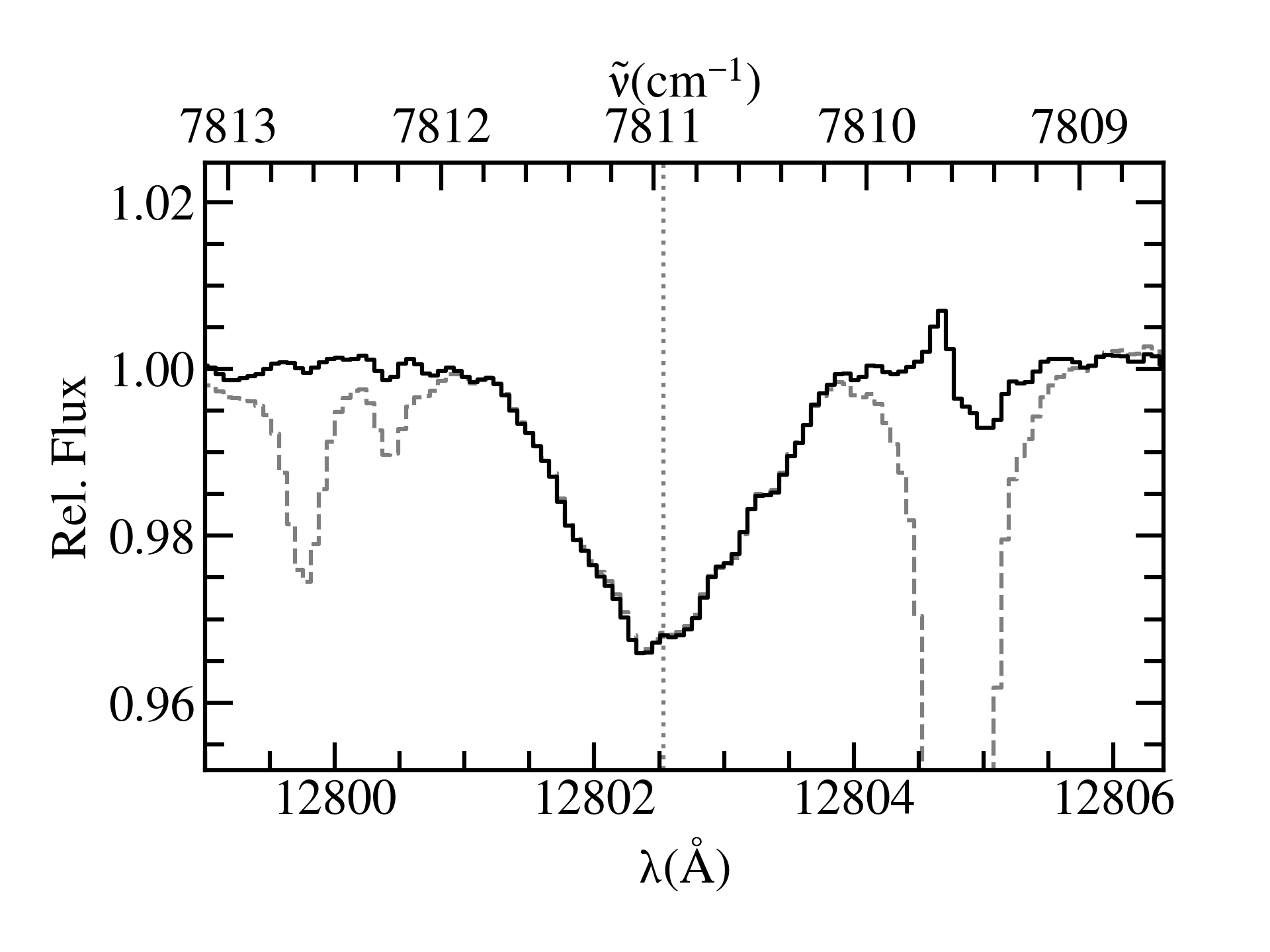}\label{fig:_hd183DIB12799}}
   \subfloat[DIB 13027]{
      \includegraphics[width = \textwidth/19*6, trim=0 20 0 10, clip]
      {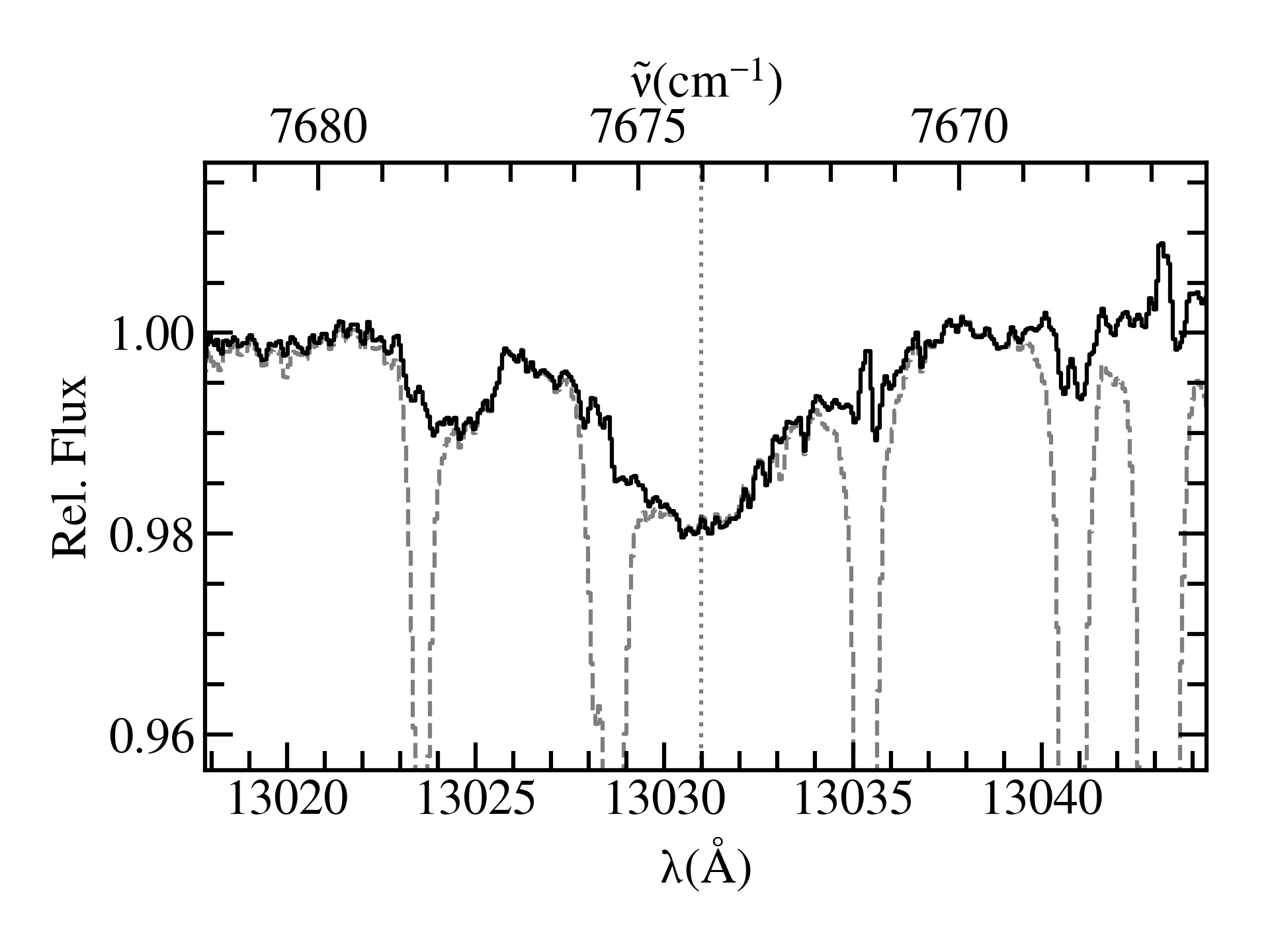}\label{fig:_hd183DIB13027}}
   \subfloat[DIB 13175]{
      \includegraphics[width = \textwidth/19*6, trim=0 20 0 10, clip]
      {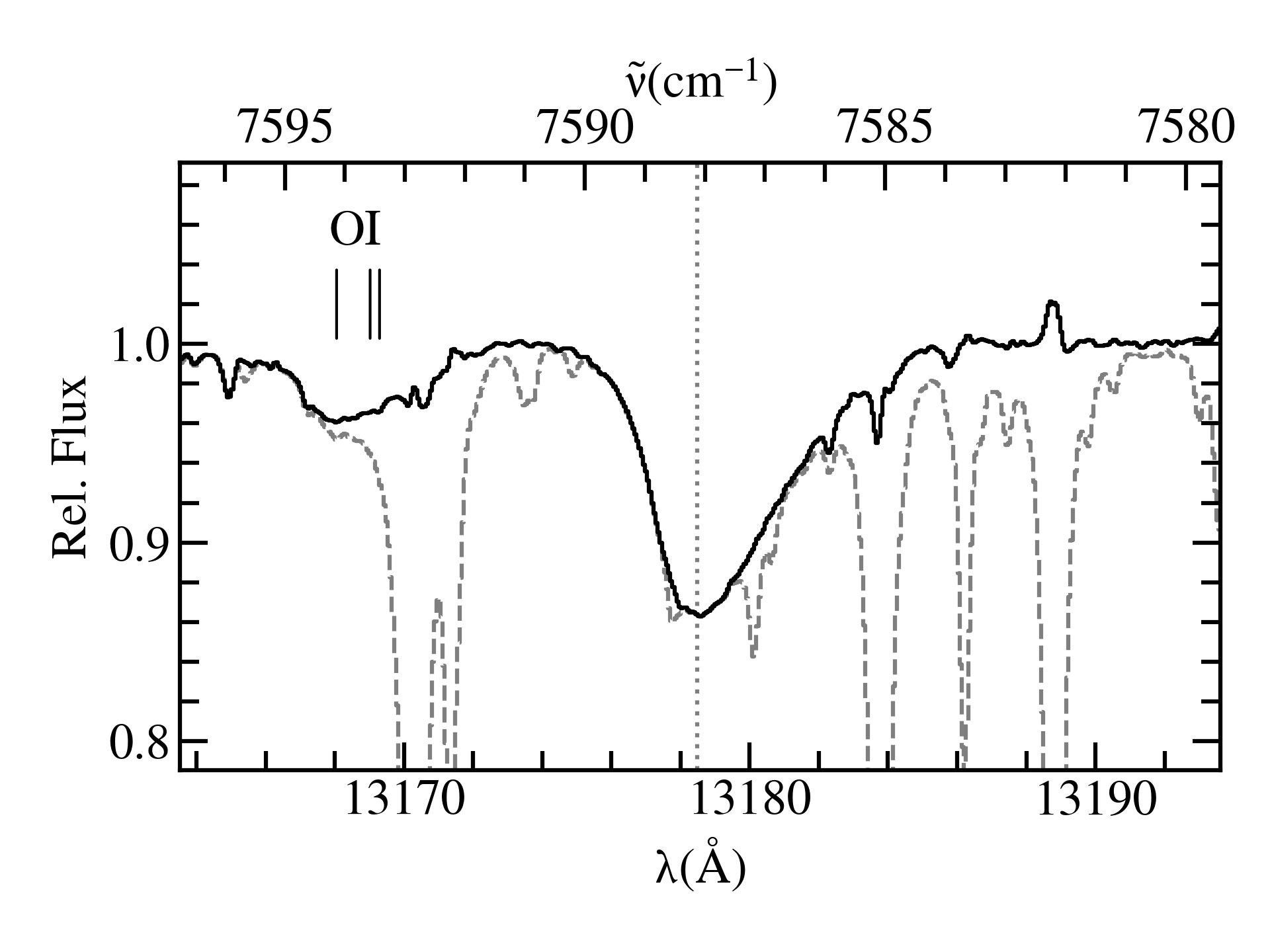}\label{fig:_hd183DIB13175}}
   \caption{Spectra of known DIBs in HD~183143 on the vacuum wavelength scale, in relative flux units. The central wavelength (see text) is indicated by a vertical dotted line. The spectrum before telluric correction is displayed as a grey dashed line in the cases where a correction was applied. Wavelengths are given in the barycentric rest frame.}
   \label{fig:known_dibs}
\end{figure*}     
\clearpage

    The sight line to HD~92207 penetrates the Carina Arm at a Galactic longitude of $\ell$\,$\approx$\,286{\degr}, so the Carina component $RV_{2, \mathrm{LSR}}$ is blueshifted with respect to the local component $RV_{1, \mathrm{LSR}}$, and is probably formed to a large degree in the IC~2599 \ion{H}{ii} region. 
    This hypothesis is supported by the good match of $RV_{2, \mathrm{bc}}$ with the radial velocity of the \ion{H}{ii} region $RV_{\ion{H}{ii}\mathrm{, bc}}$\,=\,$-$6.3$\pm$1\,km\,s$^{-1}$, obtained from hydrogen recombination lines \citep{1987caswell}, and by the high $R_V$ value found for this sight line.

    The sight line towards HD~111613 at $\ell$\,$\approx$303{\degr} shows a strong second \ion{K}{i} component, which is 
    attributed to the Carina Arm. There is a third weaker component blueward of the two strong ones, which can be seen 
    more clearly in the \ion{Na}{i} D$_1$ line (Fig. \ref{fig:na_HD111613}), but it is ignored for the purpose of this 
    study due to its weakness.

\section{DIB measurements \label{sec:dib_measurements}}
    Our characterisation of the DIBs concentrates on three aspects. The first two are the identification of 
    known and new DIBs,  and the measurement of standard DIB properties: wavelength $\lambda$, $EW$, full width at half maximum $FWHM$, and central 
    depth $A_\mathrm{c}$ (i.e. the classical data for DIB analyses). 
    The third aspect is the extraction of the DIB profiles to perform DIB profile analyses, which is rarely done in the
    literature, unless the DIBs show substructure (see above). 

    The first step in the analysis of the {\sc CRIRES} spectra is the identification of DIBs.
    We can build upon a collection of previously identified DIBs in the {\sc CRIRES} wavelength range from the work of 
    \cite{1990Natur.346..729J}, \cite{1994Natur.369..296F}, 
    \cite{2007A&A...465..993G}, \cite{geballe2011}, \cite{2014A&A...569A.117C}, and 
    \cite{2015ApJ...800..137H}. 
    Detailed line profiles for these known DIBs are displayed in Fig.~\ref{fig:known_dibs} for the
    standard star HD~183143. The telluric-corrected spectra and the 
    original spectra with telluric lines (where a correction was performed) are both shown.
    These are the highest resolution data that have been presented so far, and the cleanest in terms 
    of correction for telluric effects. However, the profiles still do not reflect the true DIB profiles as the sight line covers two absorption clouds (see the last section).
    The radial velocity difference of 14.6\,km\,s$^{-1}$ corresponds to a separation of the two absorption components by about 0.5 to 0.6\,{\AA} in the $YJ$ band, leading to some asymmetries (assuming an intrinsically symmetric DIB profile).
    However, several broad features like DIB $\lambda$10438, 10697, 12623, and 13175 (Figs.~\ref{fig:_hd183DIB10438}, \ref{fig:_hd183DIB10697}, \ref{fig:_hd183DIB12623}, and \ref{fig:_hd183DIB13175}) are clearly intrinsically asymmetric, with a steep blue drop and a slow redward rise.
    This band shape indicates an excitation of an electron, which changes the molecule's moment of inertia and results in the formation of a rotational branch head
    which leads to a short-wavelength spectral limit in the R branch \citep{2014IAUS..297...34S}.
    The detection of new DIBs is discussed in the next section.

    We measured the DIBs' central vacuum wavelength at their absorption minima. 
    If a DIB shows substructure with two minima, we chose the point of highest flux between the two 
    minima as the band centre, assuming a P and R branch to be present.
    An example can be seen  for HD~92207 in Fig.~\ref{fig:10780_11969_profile}, where the DIB centre 
    is located at $RV$\,=\,0\,km\,s$^{-1}$.
    In the case of newly detected DIBs, they are named after the approximate laboratory wavelength in air (see below).

    Equivalent widths of the DIBs were measured with a semi-automated approach.
    First, we selected two continuum regions with central wavelengths $\lambda_\mathrm{cont}$ around a DIB. 
    For each $\lambda_\mathrm{cont}$ a mean continuum flux was calculated from the 
    interval [$\lambda_\mathrm{cont} - d\lambda$, $\lambda_\mathrm{cont} + d\lambda$] 
    with an interval size of 8 or 20 bins. The continuum flux at the DIB position was then derived from a linear 
    interpolation between the two continuum regions in order to avoid tampering with the DIB profile.
    Those intervals were also used to determine the $S/N$ around the DIB.
    We did not fit predefined band profiles for $EW$ measurements. Commonly used profiles, for example Gaussians, do not 
    fit the band shapes well enough, which often show asymmetry or substructure. Instead, a direct integration 
    of the line between the two continuum regions was performed.
    Cosmics and residuals of telluric lines were excluded by eye for the $EW$ measurement.
    We calculated the errors $\sigma_{EW}$ of the equivalent width, using the method 
    of \cite{2011A&A...533A.129V}, 
    \begin{equation}
        \sigma_{EW} = \sqrt{2 \Delta\lambda\delta\lambda}/(S/N),
    \end{equation}
    where $\Delta\lambda$ is the wavelength range of the $EW$ measurement and $\delta\lambda$ 
    is the spectral dispersion. 
    If the telluric removal left noticeable residuals, we added an additional error on the order of the $EW$s of the telluric residuals.
    Finally, a third point was selected to mark the central band core with the maximum absorption $A_\mathrm{c}$, 
    which was selected by eye. Keeping both the continuum and the central depth fixed facilitated determining the $FWHM$  automatically.

    We used our $RV$ measurements of the interstellar \ion{K}{i} lines (Table~\ref{tab:rv_ism}) to estimate 
    the DIB rest-frame wavelengths in air ($\lambda_\mathrm{0, air}$). The DIB $EW$s and \ion{K}{i} $EW$s are 
    known to correlate \citep[][]{Galazutdinovetal04}, which leads to the conclusion that the DIB $RV$s should 
    relate closely to the $RV$s of the \ion{K}{i} components. Every sight line shows two resolved \ion{K}{i}
    components, but only one broadened DIB component, so that we chose the mean \ion{K}{i}-$RV$ as the estimated 
    DIB-$RV$. We used an unweighted mean because the correlation between the DIB $EW$s and \ion{K}{i} $EW$s is not very strong and should not be used to estimate DIB strengths. We calculated mean rest-frame wavelengths in air (Table \ref{tab:dib_list}), using the International 
    Astronomical Union (IAU) standard from \cite{1991ApJS...77..119M} for the vacuum-to-air wavelength conversion.
    The statistical error was calculated from the standard deviation of the observed wavelengths from the sample stars. We estimate the systematic error of the rest-frame wavelength to 0.3\,{\AA}, which corresponds 
    approximately to half of the velocity difference between the two \ion{K}{i} components. 
    
    All DIB identifications, both for known and for new DIBs, are summarised in Table~\ref{tab:dib_list}. The names of the
    DIBs are given as integers of approximate $\lambda_\mathrm{0,air}$, together with a 
    reference to the detection publication.
    For every sight line and DIB measurement, central wavelengths in vacuum
    $\lambda_\mathrm{vac}$ in the barycentric rest frame are tabulated, along with the $EW$,
    $A_\mathrm{c}$, and $FWHM$ values.   
    
    \begin{figure*}[ht!]
  \centering
 \subfloat[HD~183143]{
     \includegraphics[height=4.65cm]{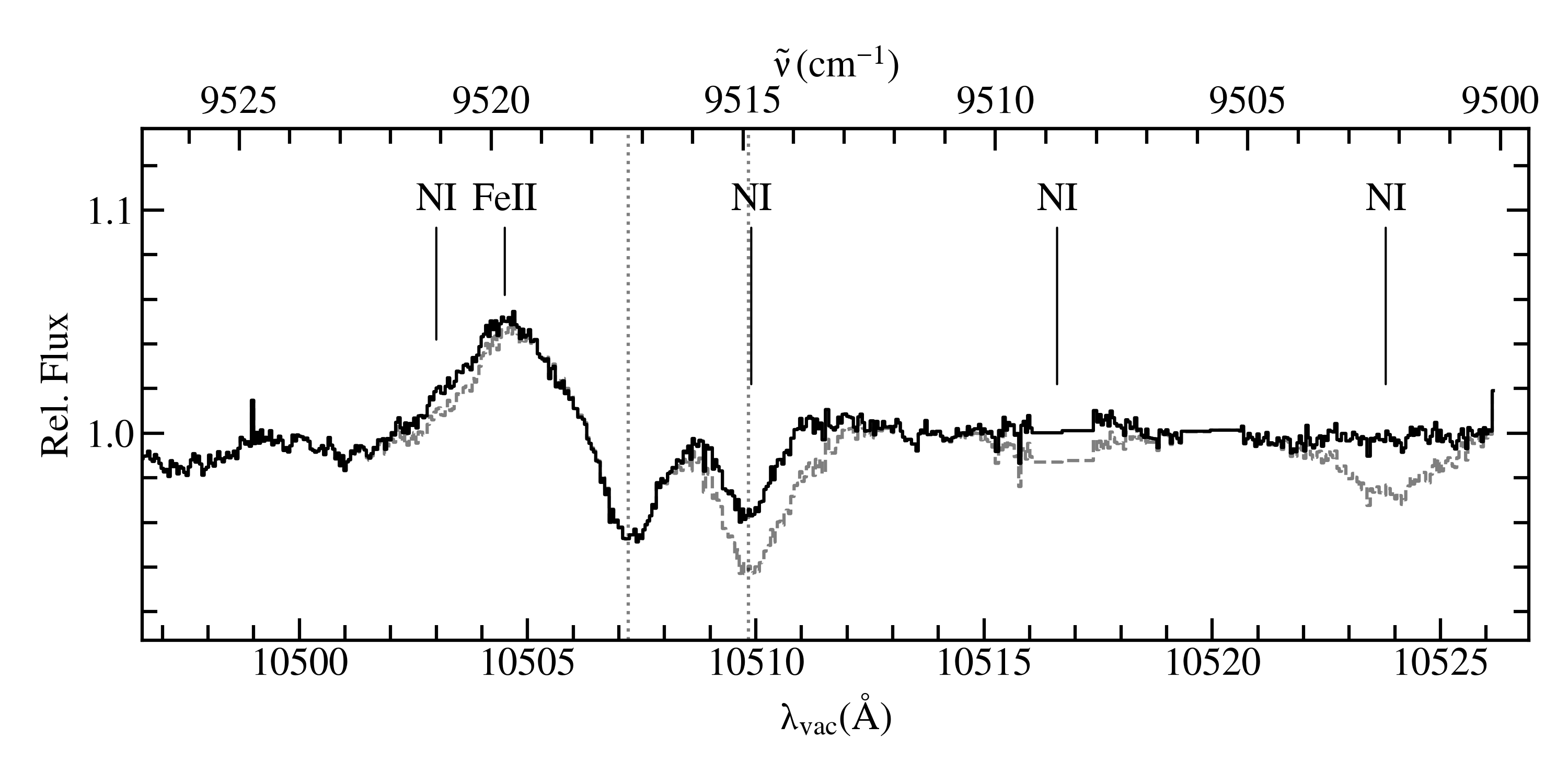}\label{fig:DIB10504_10507_HD183143}}
 \subfloat[HD~165784]{
     \includegraphics[height=4.65cm, trim=35 0 0 0, clip]{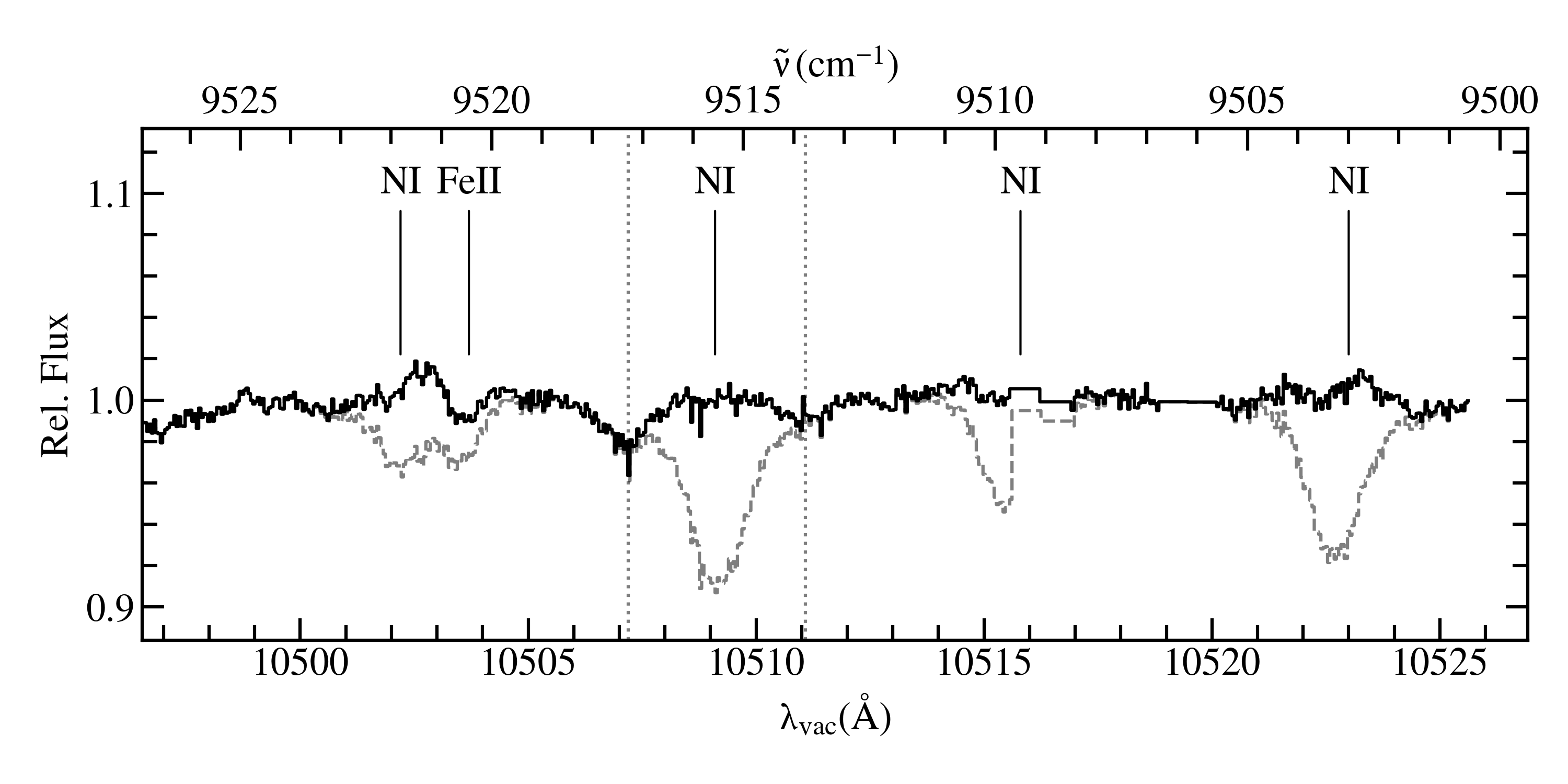}\label{fig:DIB10504_10507_HD165784}}\\[-1mm]
  \caption{Diffuse interstellar bands  $\lambda$10505 and 10507 in HD~183143 and HD~165784. The central DIB wavelengths are indicated by vertical dotted
   lines. The positions of stellar metal lines are also indicated. The spectrum corrected for \ion{N}{i} absorption lines
   (the photospheric \ion{Fe}{ii} emission line is not reproduced by the model) is shown in black, the spectrum 
   before the correction    in dashed grey.}
  \label{fig:comp_cox}
  \end{figure*}
  
    There are literature values for eight DIBs in HD~183143 
    by \cite{2014A&A...569A.117C}, based on X-Shooter observations on the ESO VLT, 
    at $R$\,$\approx$\,10\,000.
    Our $EW$ values agree with their measurements, with the exceptions of the DIBs $\lambda\lambda$10507, 10697,
    and 13175 shown in Fig.~\ref{fig:comp_cox} and Figs.~\ref{fig:_hd183DIB10697} and \ref{fig:_hd183DIB13175}.
    We measure an $EW$ for DIB 10507 of $EW(10507)$\,=\,41$\pm$10\,m{\AA}, which is much smaller than the 
    value of $EW_\mathrm{Cox}(10507)$\,=\,261\,m{\AA} (no errors given). 
    In Fig.~\ref{fig:comp_cox}
    we see that a stellar \ion{N}{i} line coincides with the DIB candidate 10507 from \cite{2014A&A...569A.117C}.
    A much weaker DIB remains in HD~183143 after the correction for the stellar component (the O-star background stars 
    of Cox et al., their Fig.~6, are unaffected by the \ion{N}{i} blend), the quality of which can be assessed by 
    comparison with the corrections of the other \ion{N}{i} lines of the multiplet in this wavelength range.
    In HD~183143, the DIB wavelength shifts by 1\,{\AA} relative to HD~165784, so it is likely that there is still a blend with an unknown stellar line, but our value is certainly a refined upper limit for $EW(10507)$.
    Our $EW$ for DIB 10697, $EW(10697)$\,=\,186$\pm$7\,m{\AA} is considerably smaller than 
    $EW_\mathrm{Cox}(10697)$\,=\,262$\pm$36\,m{\AA} because of a different continuum placement, compare our
    Fig.~\ref{fig:_hd183DIB10697} with their Fig.~6, where their continuum was located at about the half peak height of 
    the \ion{C}{i} emission line. Our $EW(13175)$\,=\,647$\pm$20\,m{\AA} is larger than 
    $EW_\mathrm{Cox}(13175)$\,=\,493$\pm$36\,m{\AA}.
    This difference is probably caused by a different continuum placement and the asymmetric DIB profile 
    which was fitted using a Gaussian by Cox et al., compare our Fig.~\ref{fig:_hd183DIB13175} with their Fig.~4.
    If Gaussians were used for measurements, a twin Gaussian would be much better suited for this DIB.
    Our $FWHM$ values of DIB $\lambda$13175 for HD~183143 and HD~165784 agree well with the literature values: \cite{2014A&A...569A.117C} measured $FWHM(13175)$\,=\,4.5$\pm$1.2\,{\AA}, \cite{1990Natur.346..729J} 4.0$\pm$0.5\,{\AA}, and Smoker et al. (2022, in prep.) a range from 3.6$\pm$0.1\,{\AA} to 5.3$\pm$0.1\,{\AA}.
    The $EW$ of the C$_{60}^+$ DIB $\lambda9632$, $EW(9632)$\,=\,316$\pm$99\,m{\AA} agrees well with $EW_\mathrm{Cox}(9632)$\,=\,263$\pm$3\,m{\AA}. The narrow absorption in the DIB centre (Fig. \ref{fig:_hd183DIB9632}) must not be confused with the broad DIB ($FWHM\,\approx\,2.5$\,{\AA}) and is probably caused by noise.

\begin{figure*}[t]
  \centering
   \subfloat[DIB 10125]{
      \includegraphics[width = \textwidth/19*6]
      {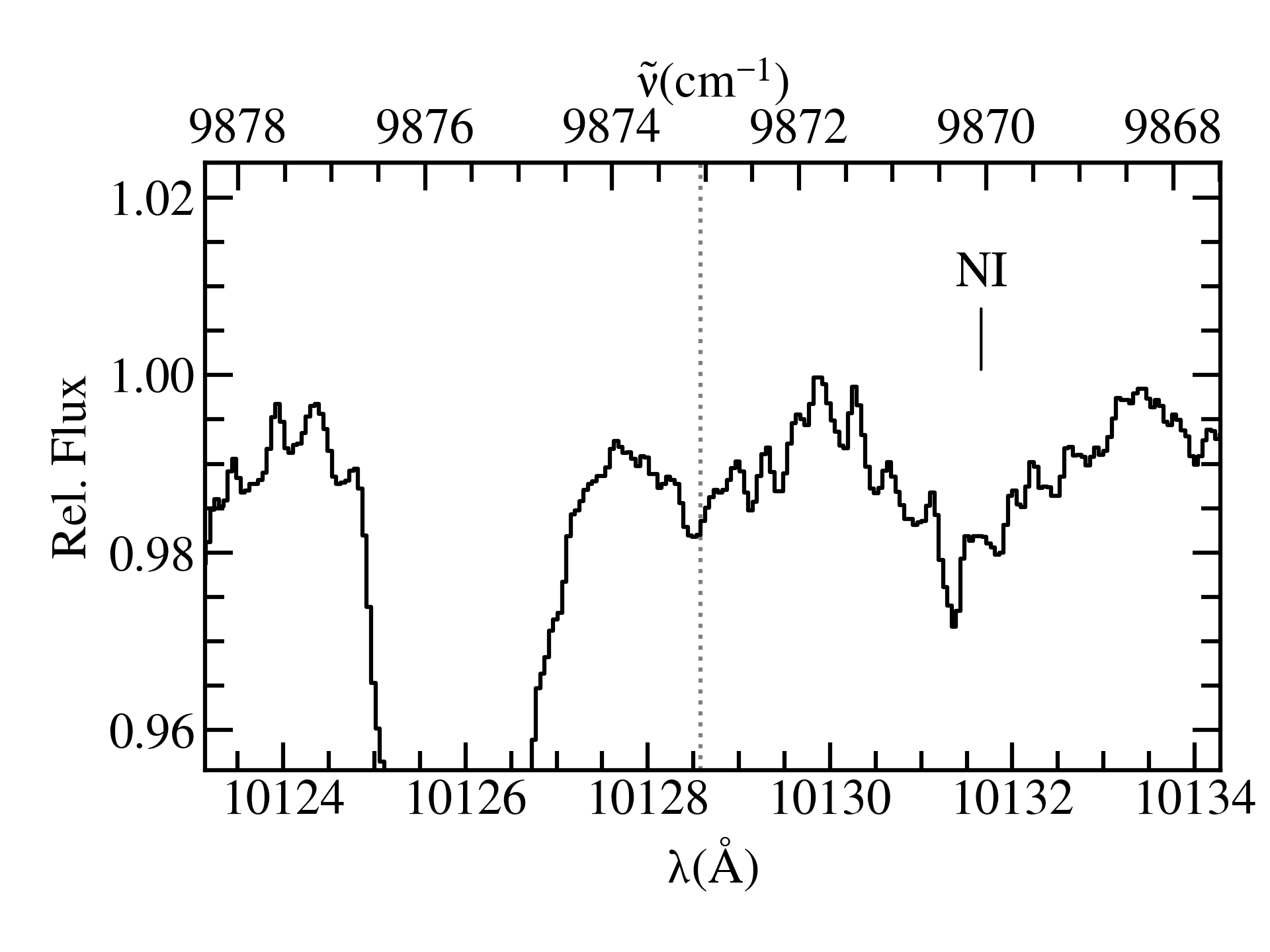}}
   \subfloat[DIB 10262]{
      \includegraphics[width = \textwidth/19*6]
      {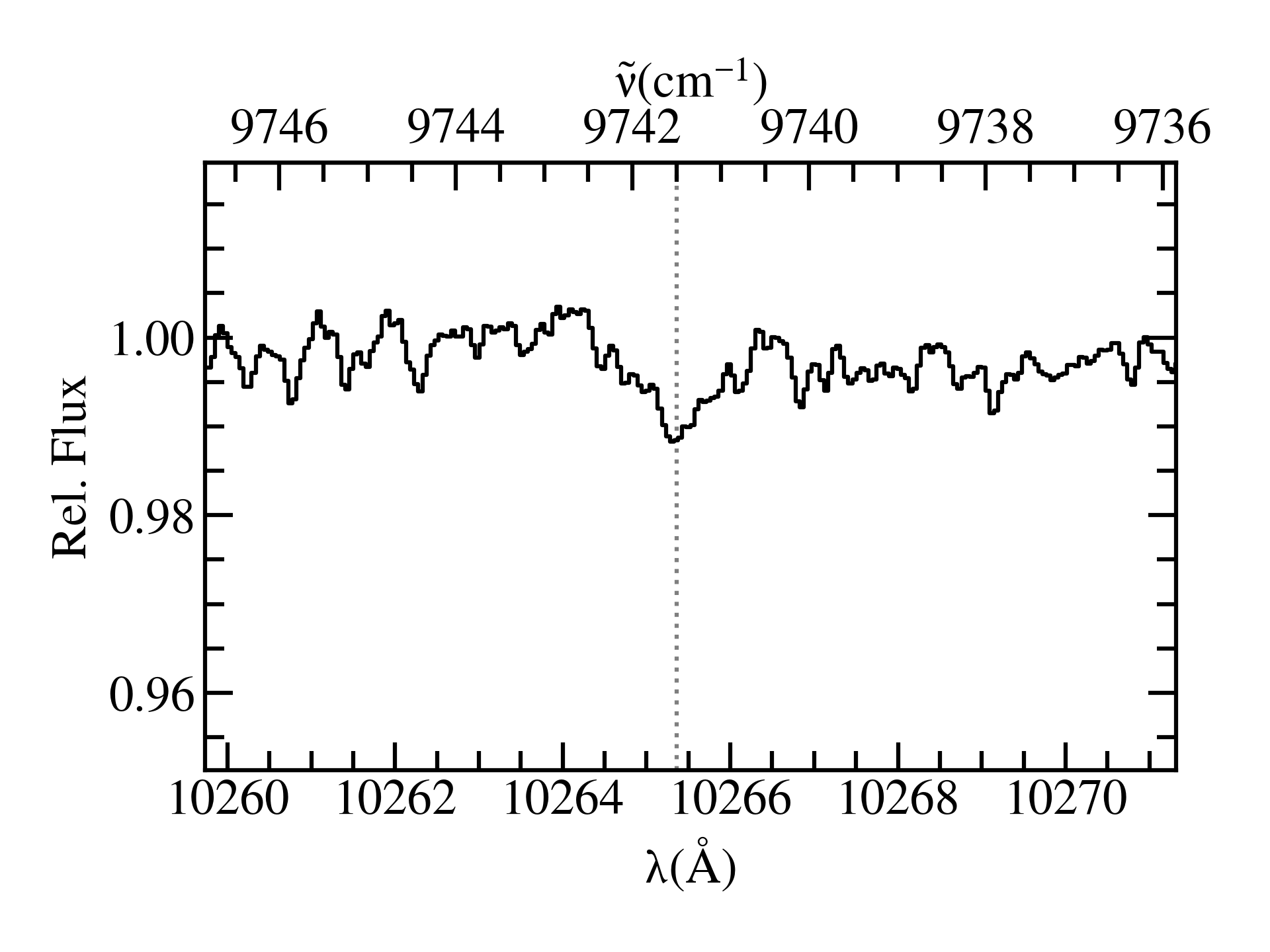}\label{fig:10263_hd183}}
   \subfloat[DIB 10735]{
      \includegraphics[width = \textwidth/19*6]
      {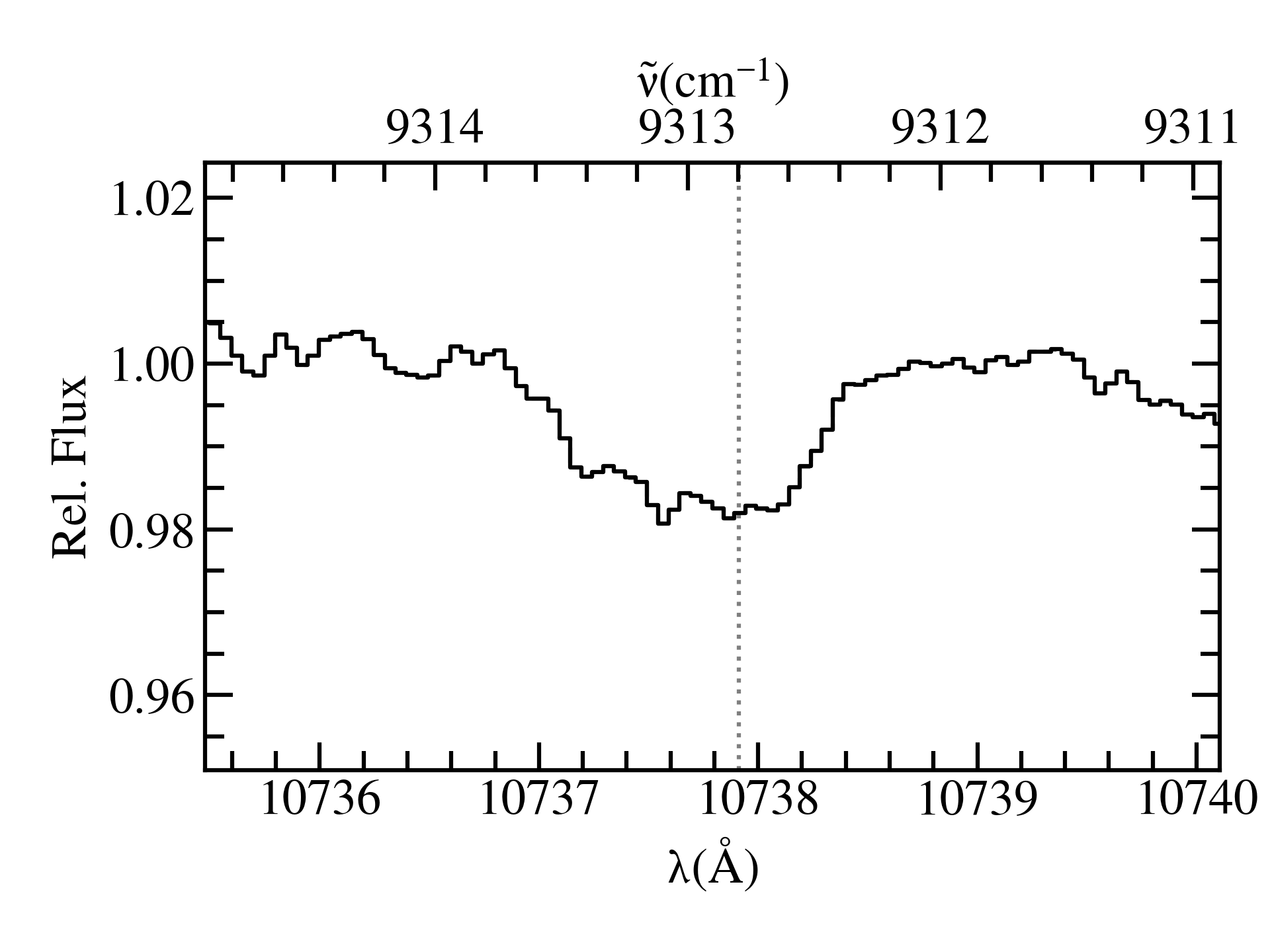}\label{fig:10734_hd183}}\\[-4mm]
   \subfloat[DIB 10884]{
      \includegraphics[width = \textwidth/19*6]
      {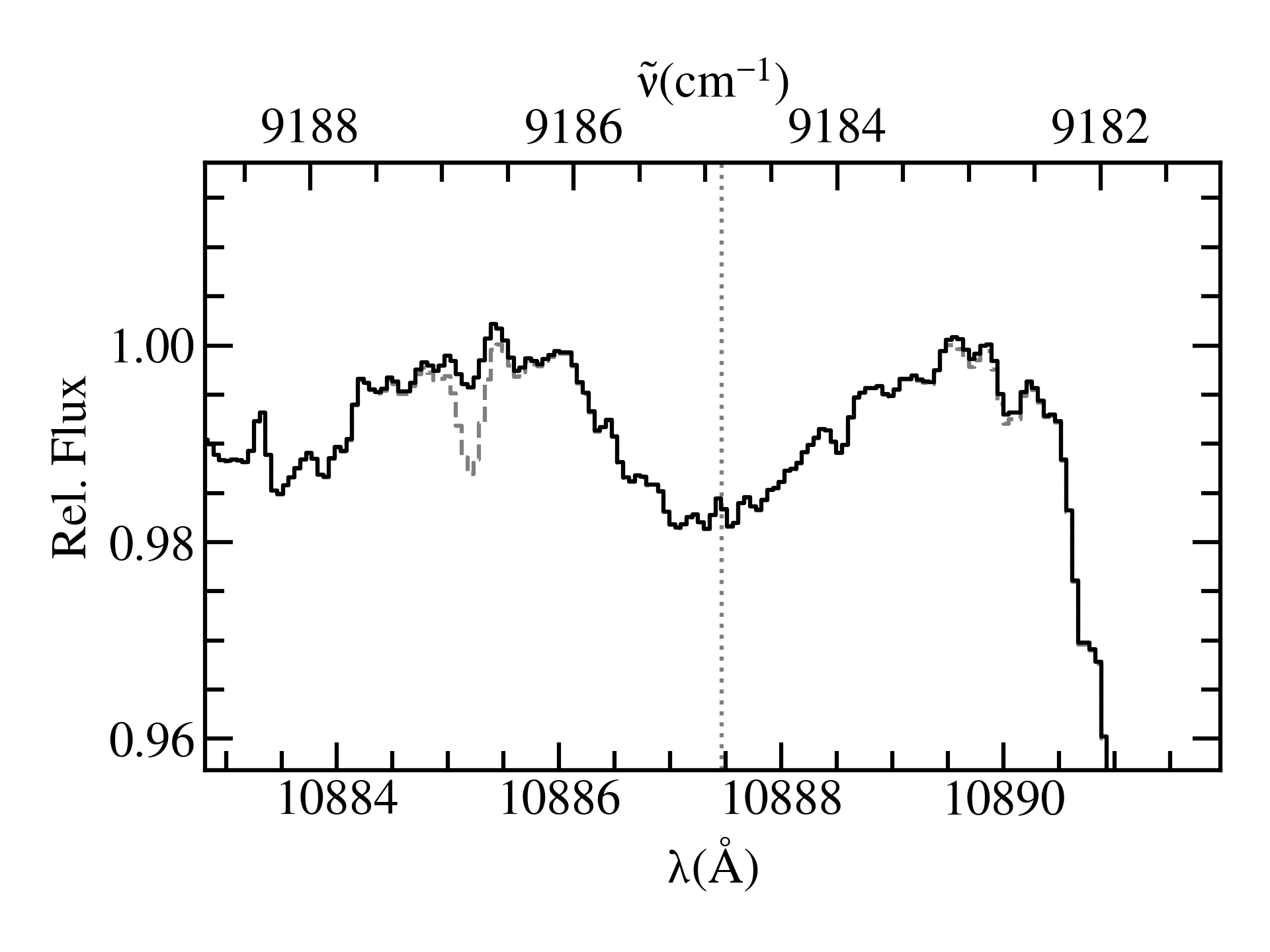}\label{fig:10884_hd183}}
   \subfloat[DIB 11048]{
      \includegraphics[width = \textwidth/19*6]
      {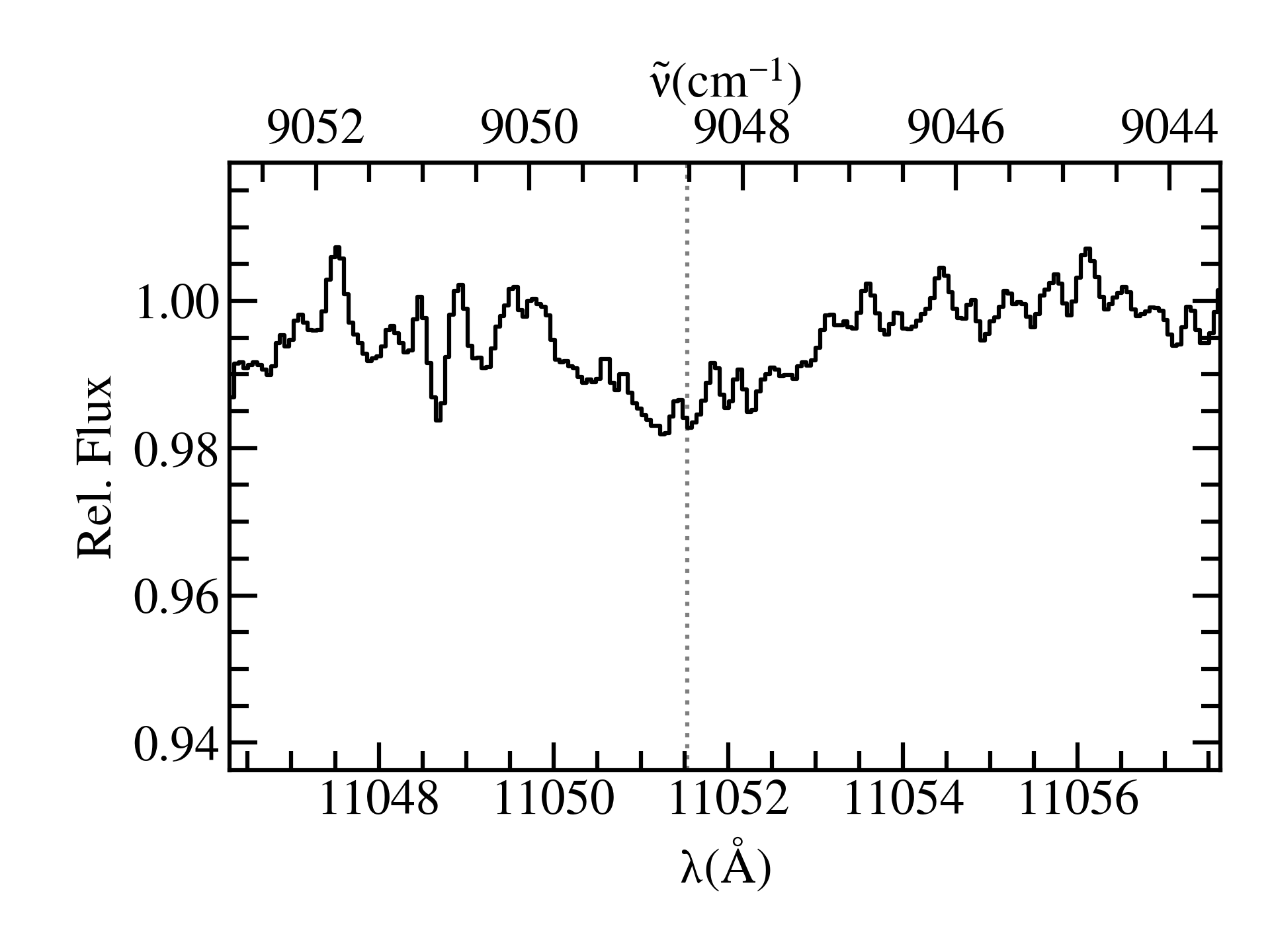}\label{fig:11048_hd183}}
   \subfloat[DIB 11695]{
      \includegraphics[width = \textwidth/19*6]
      {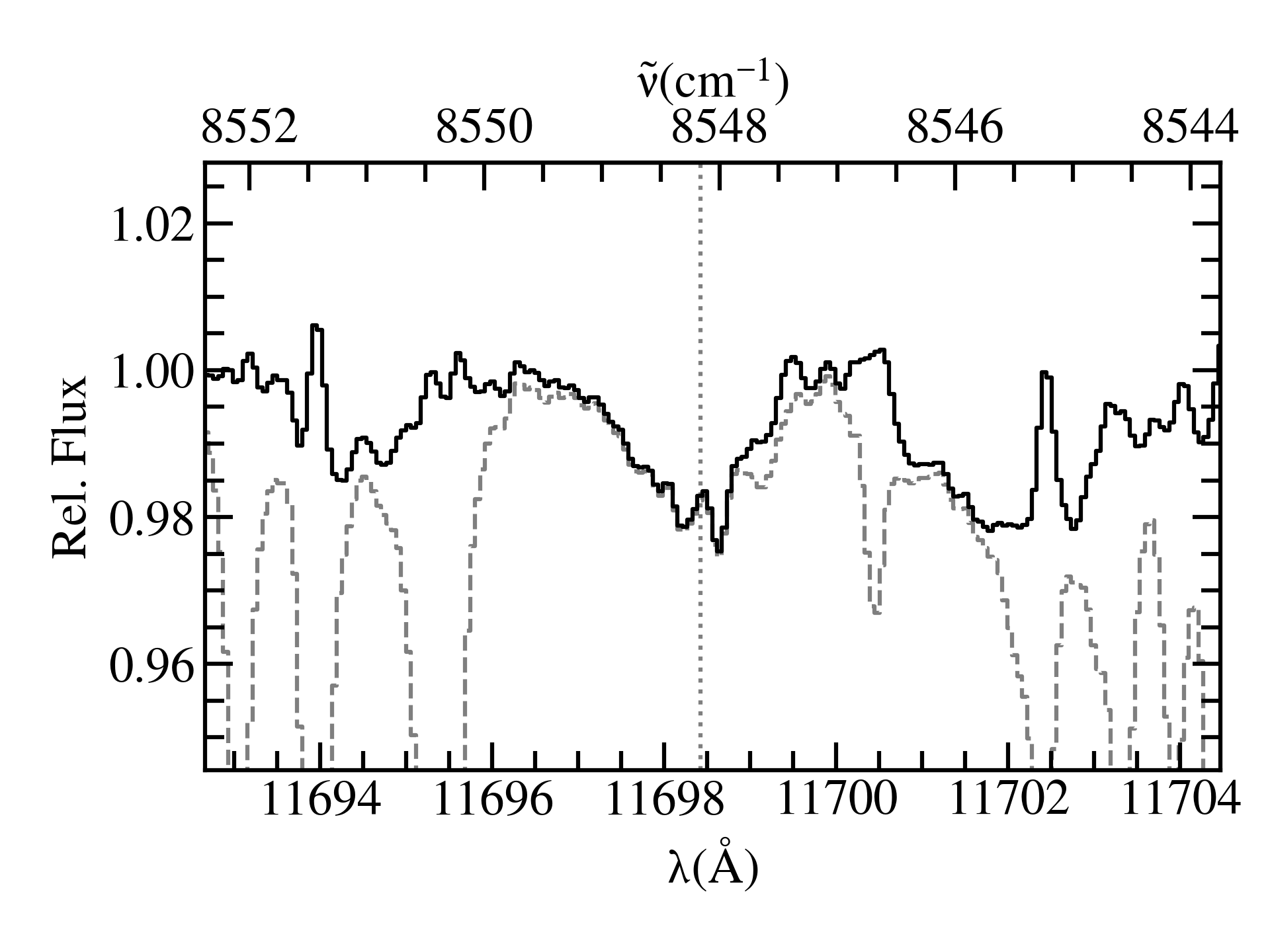}\label{fig:11695_hd183}}\\[-4mm]
   \subfloat[DIB 11699]{
      \includegraphics[width = \textwidth/19*6]
      {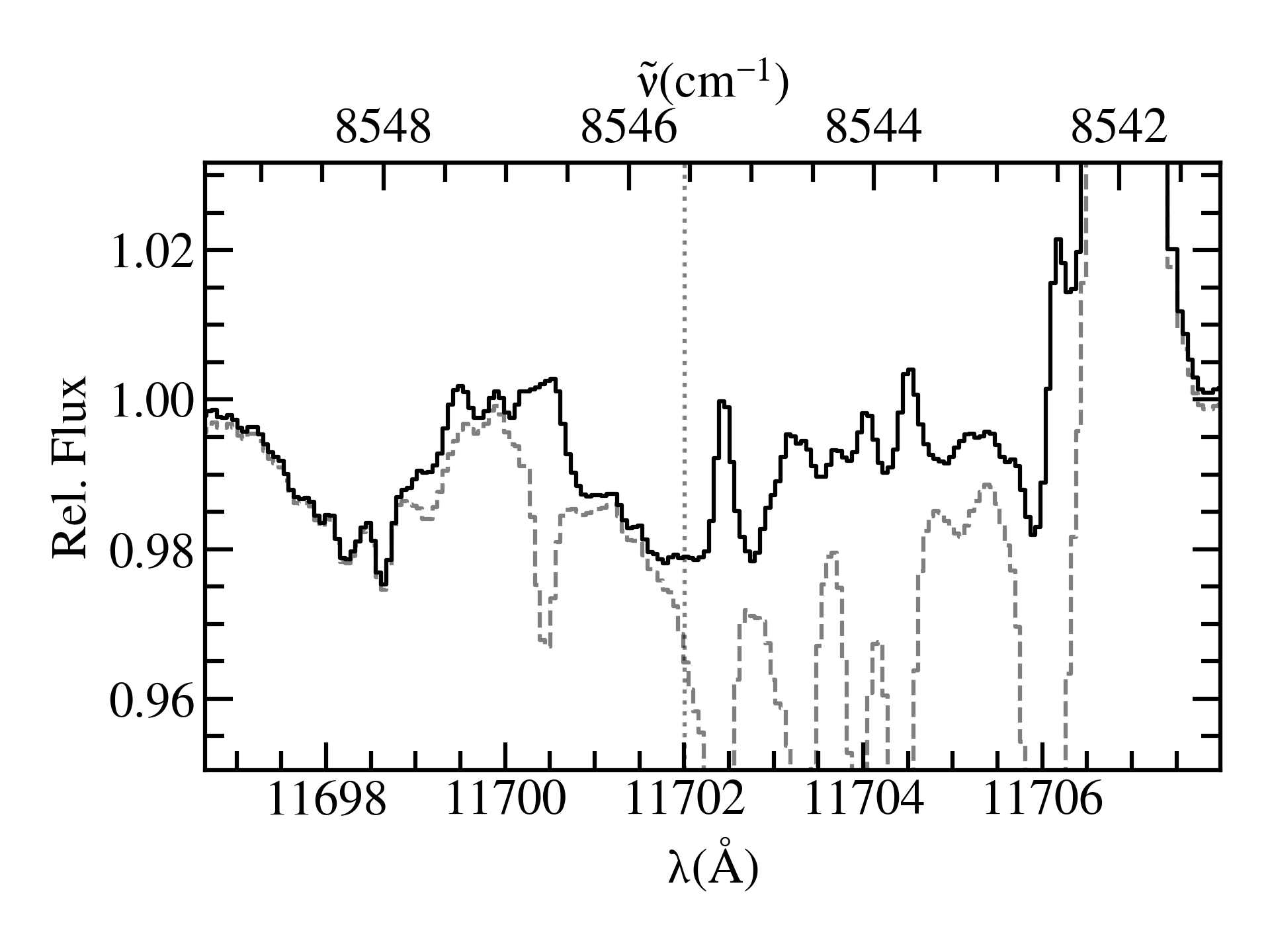}\label{fig:11699_hd183}}
   \subfloat[DIB 11721]{
      \includegraphics[width = \textwidth/19*6]
      {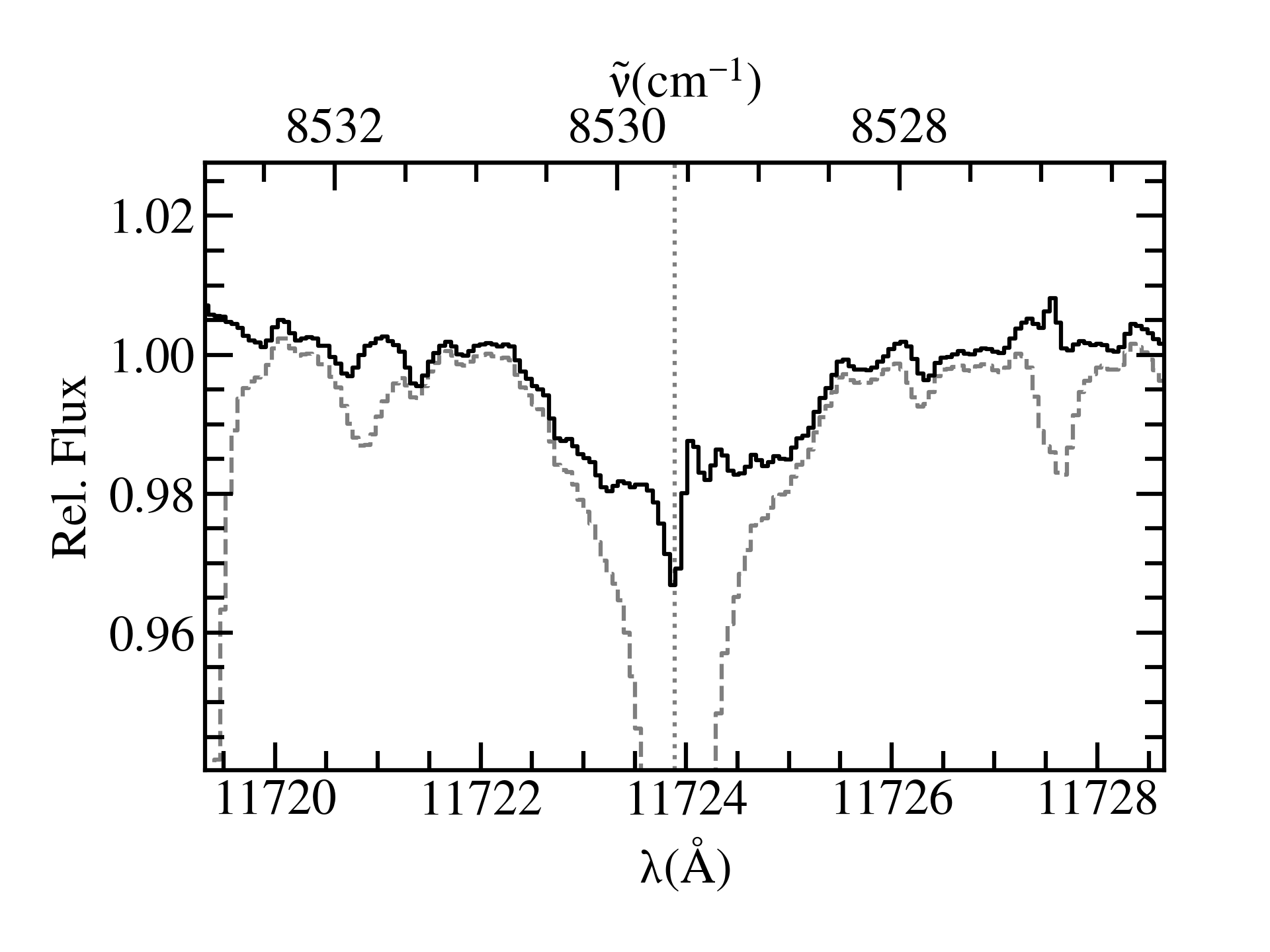}\label{fig:11721_hd183}}
   \subfloat[DIB 11792]{
      \includegraphics[width = \textwidth/19*6]
      {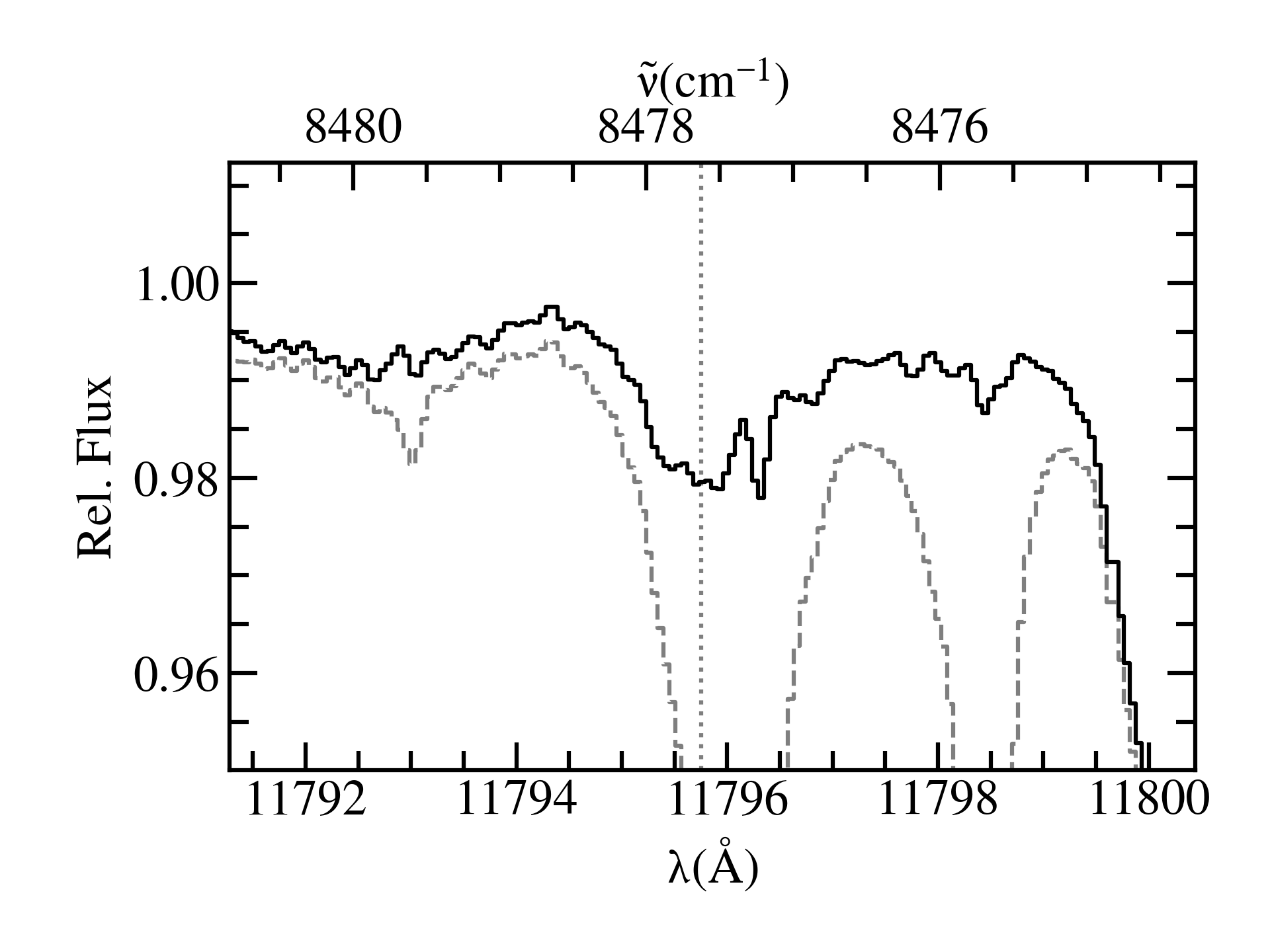}\label{fig:11792_hd183}}\\[-4mm]
   \subfloat[DIB 11970]{
      \includegraphics[width = \textwidth/19*6]
      {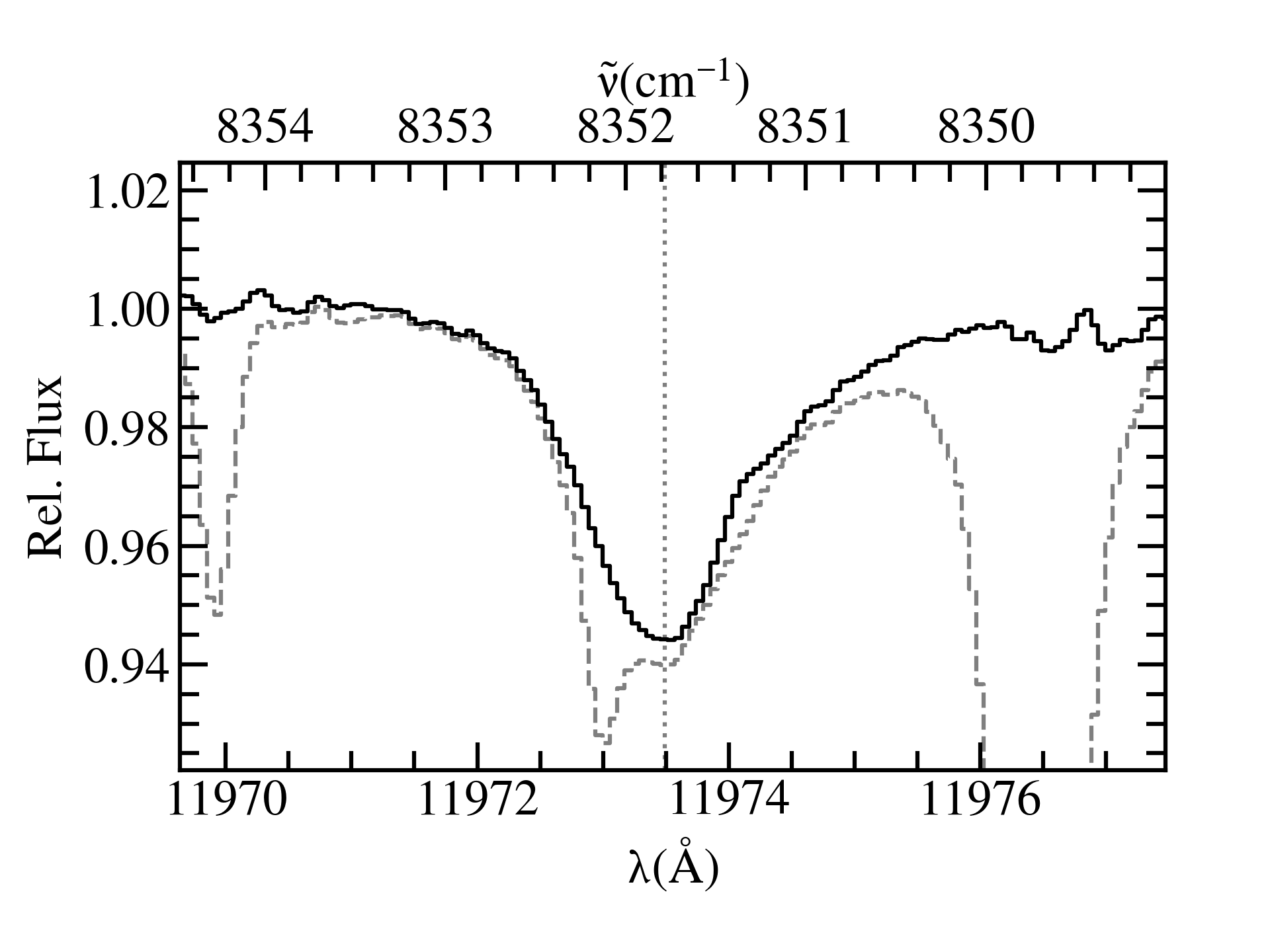}\label{fig:11969_hd183}}
   \subfloat[DIB 12222]{
      \includegraphics[width = \textwidth/19*6]
      {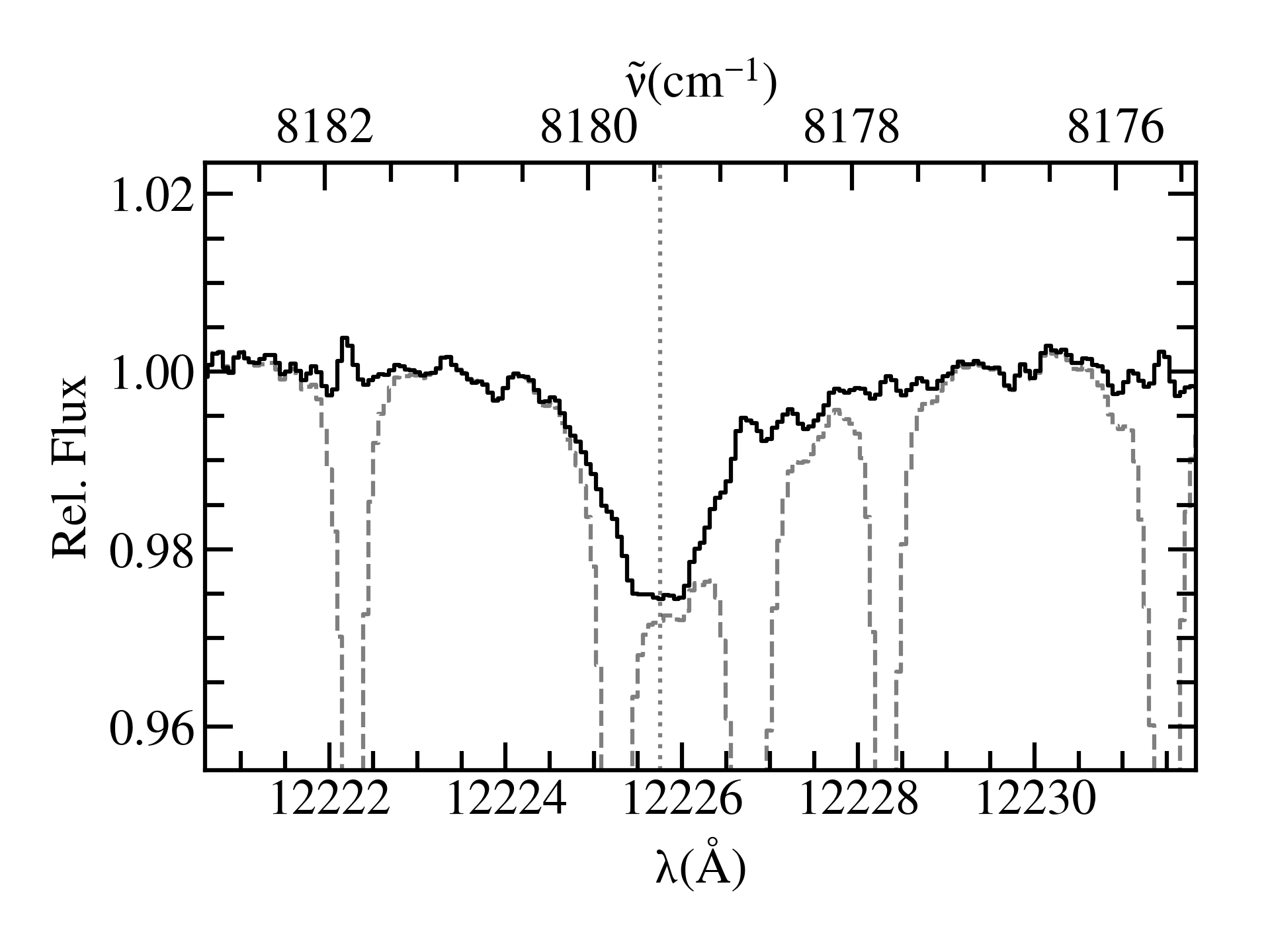}\label{fig:12222_hd183}}
   \subfloat[DIB 12838]{
      \includegraphics[width = \textwidth/19*6]
      {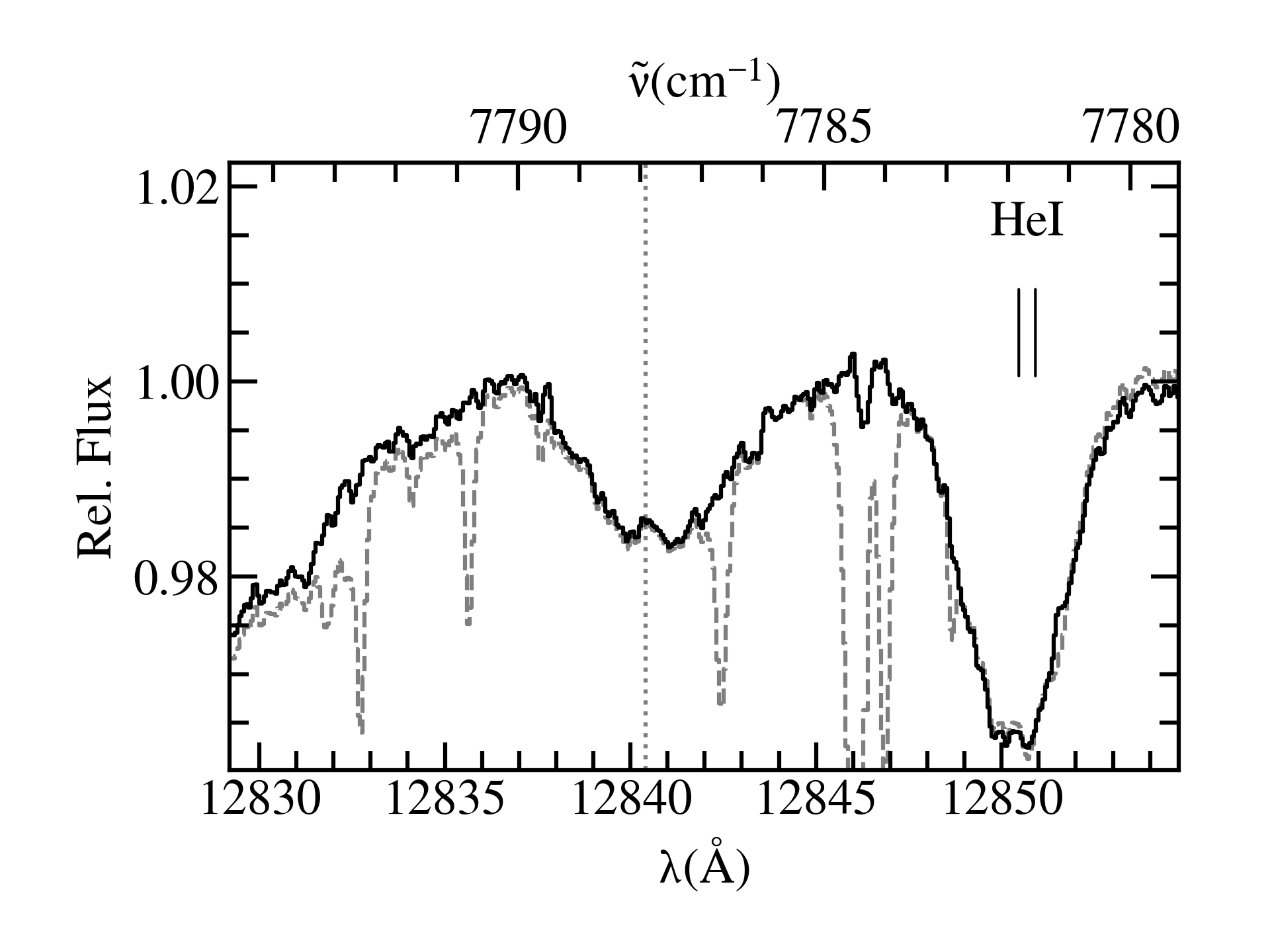}\label{fig:12837_hd183}}
   \caption{Normalised spectra of new DIB candidates in HD~183143 on the vacuum wavelength scale. The central wavelength is indicated by a dotted vertical line. The spectrum before telluric correction is displayed as a grey dashed line in the cases where a correction was applied. Wavelengths are given in the barycentric rest frame.} 
   \label{fig:dibs_hd183}
\end{figure*}

    \section{New NIR DIBs\label{sec:dib_candidates}}
    We identified a total of 12 new narrow DIBs in the NIR,  almost doubling the number of known DIBs in the 
    $YJ$ band;  about half of them were detected in  the two sight lines with highest reddening. 
    We compared their spectra to our unreddened 
    reference star HD~87737 and to the synthetic spectra of the background stars to exclude a stellar origin for the DIB candidate lines. 
    An overview of the new DIB line profiles in the DIB standard star HD~183143 is given in Fig.~\ref{fig:dibs_hd183} and
    correlations of $EW$ versus $E(B-V)$ for the target stars are shown in Fig.~\ref{fig:ebv_ew}, which clearly emphasise the need for additional data.
    Each new DIB candidate is briefly discussed in the following.\\[-8mm]
    \paragraph{$\lambda$10125.}
    This DIB is very weak 
    in HD~183143. It is located between a stellar \ion{C}{i} line at $\lambda_\mathrm{vac}$\,=\,10\,126.635\,\AA\ 
    and a stellar \ion{N}{i} line at $\lambda_\mathrm{vac}$\,=\,10\,131.055\,{\AA}. This DIB candidate 
    should be investigated further on sight lines with even higher reddening.\\[-8mm] 
    \paragraph{$\lambda$10262.}
    This is also a very weak DIB. 
    The $E(B-V) - EW$ correlation suggests that the normalised $EW$ of DIB $\lambda$10262 in
    HD~183143 is stronger than in HD~165784. This DIB is possibly more abundant per 
    unit reddening in the circumstellar shell in front of HD~183143 (see Sect.~\ref{sec:shell_section}).\\[-8mm]
    \paragraph{$\lambda$10735, $\lambda$10884, $\lambda$11048.}
    These DIBs are weak to moderately strong in HD~183143. All are unaffected by telluric absorption.
    \paragraph{$\lambda$11695.}
    This is a weak DIB, only detected in HD~165784 and HD~183143. The normalized $EW$ is 
    much stronger in HD~183143.
    Potentially, there are nearby \ion{Si}{i} lines; the strongest is located at $\lambda_\mathrm{vac}$\,=\,11\,703.442\,{\AA}. 
    There is a slightly stronger stellar \ion{Si}{i} line from the same multiplet at 
    $\lambda_\mathrm{vac}$\,=\,11\,644.128\,{\AA}, which may be used to cross-check the significance of a 
    possible blend with the DIB.
    However, in our background stars with $T_\mathrm{eff}$\,$>$\,9000\,K those lines are absent.\\[-8mm]
    \paragraph{$\lambda$11699.}
    This weak DIB is located in a region with weak telluric contamination. It is detected on three sight lines 
    and shows a tight $E(B-V) - EW$ relation.\\[-8mm]
    \paragraph{$\lambda$11721, $\lambda$11792.}
    Both DIB candidates are heavily blended with a telluric water vapour line. 
    However, the simultaneous detection of DIB $\lambda11721$ in four sight lines and of DIB $\lambda11792$
    in three sight lines, their good $E(B-V) - EW$-relations, and 
    their absence in the spectrum of HD~87737 all indicate that the feature is not an artefact 
    of the telluric correction.\\[-8mm]
    \paragraph{$\lambda$11970.}
    This DIB is very strong and rather broad, but its $EW$ can be substantially altered by a blend with a 
    stellar \ion{He}{i} line, increasingly for hotter stars (see Sect.~\ref{sec:stellar_blends}).
    It is also blended with telluric lines, in particular in the red wing.
    Its identification as a DIB is indisputable because of its strong profile and
    $A_\mathrm{c}$ correlation with the previously reported DIBs $\lambda\lambda$10780 and 11797.\\[-8mm]
    \paragraph{$\lambda$12222.}
    This weak DIB has several telluric lines in its vicinity and shows signs of a 
    PR branch in HD~165784, which has to be confirmed in other sight lines, preferably in a 
    hot ISM, such as \ion{H}{ii} regions.
    The available spectra of HD~92207 and HD~111613 do not cover this spectral region.\\[-8mm]
    \paragraph{$\lambda$12838.}
    This DIB is detected in four sightlines. The $E(B-V) - EW$ relation shows a linear
    trend. The red wing of the DIB is blended with a telluric line and has to be corrected carefully.
    We generally detect two single Gaussian band components for this DIB. This can be 
    interpreted either as two separate DIBs or one broader DIB with a pronounced P and R branch. We
    consider a PR branch hypothesis more likely because the separation is most pronounced in our 
    \ion{H}{ii} region sight line towards HD~92207. However, no definite conclusion can be drawn here
    because of the small sample size and  low $S/N$.

\begin{figure*}[t!]
    \subfloat[$\lambda10780$ (red), $\lambda11797$ (dashed black)]
      {\includegraphics[width=.295\textwidth]{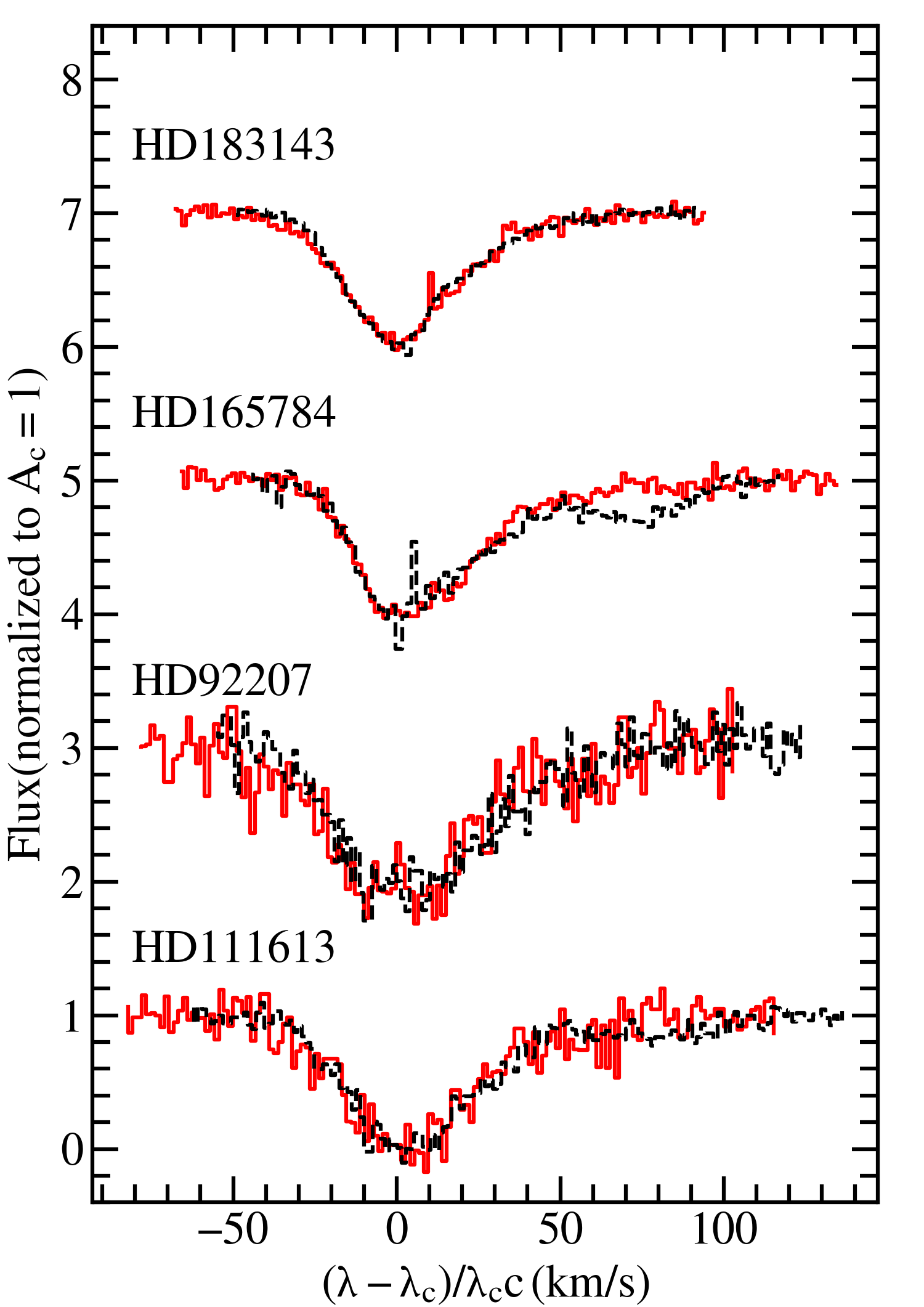}
      \label{fig:10780_11797_profile}}
    \subfloat[$\lambda10780$ (red), $\lambda11970$ (dashed black)]
       {\includegraphics[width=.295\textwidth]{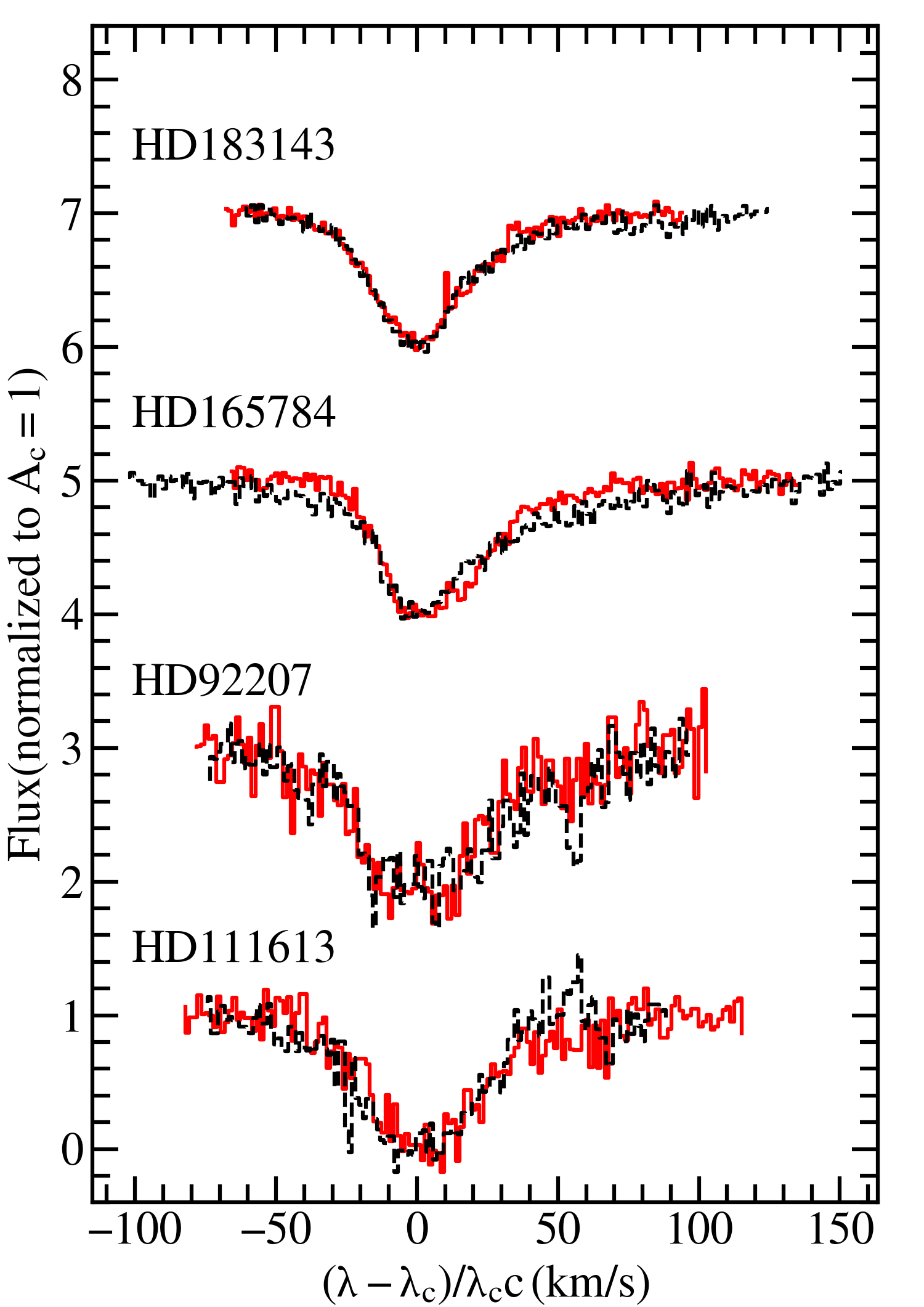}
       \label{fig:10780_11969_profile}}
    \subfloat[$\lambda11797$ (red), $\lambda11970$ (dashed black)]
      {\includegraphics[width=.295\textwidth]{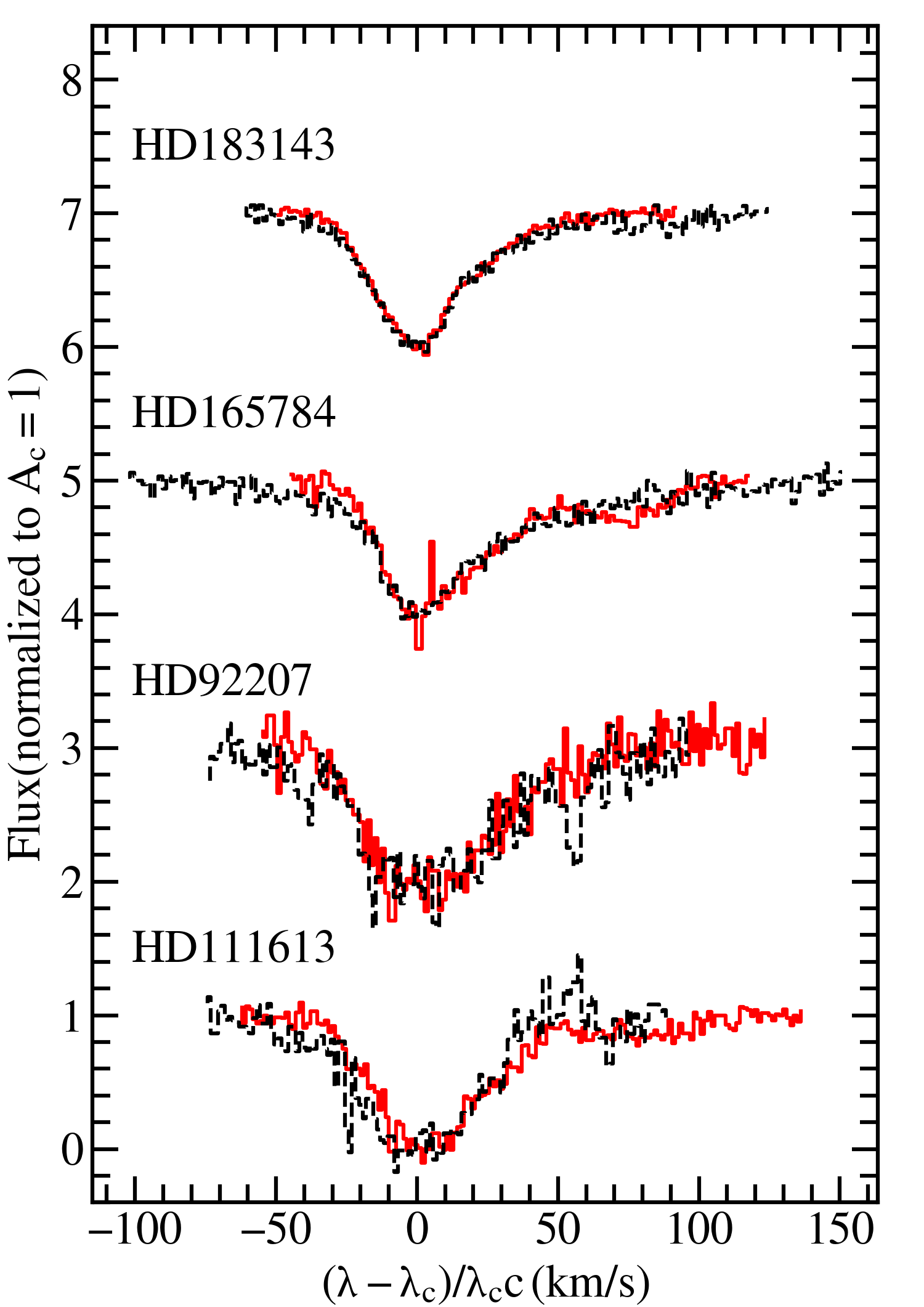}
      \label{fig:11797_11969_profile}}
   \caption{Profile comparisons for the members of the DIB profile family $\lambda10780$--11797--11970.} 
   \label{fig:profile_set_10780}
   \end{figure*}

    \section{DIB profile families\label{sec:profile_sets}}
\begin{figure*}[ht!]
   \centering
   \subfloat[]{\includegraphics[width = 0.26\textwidth]
      {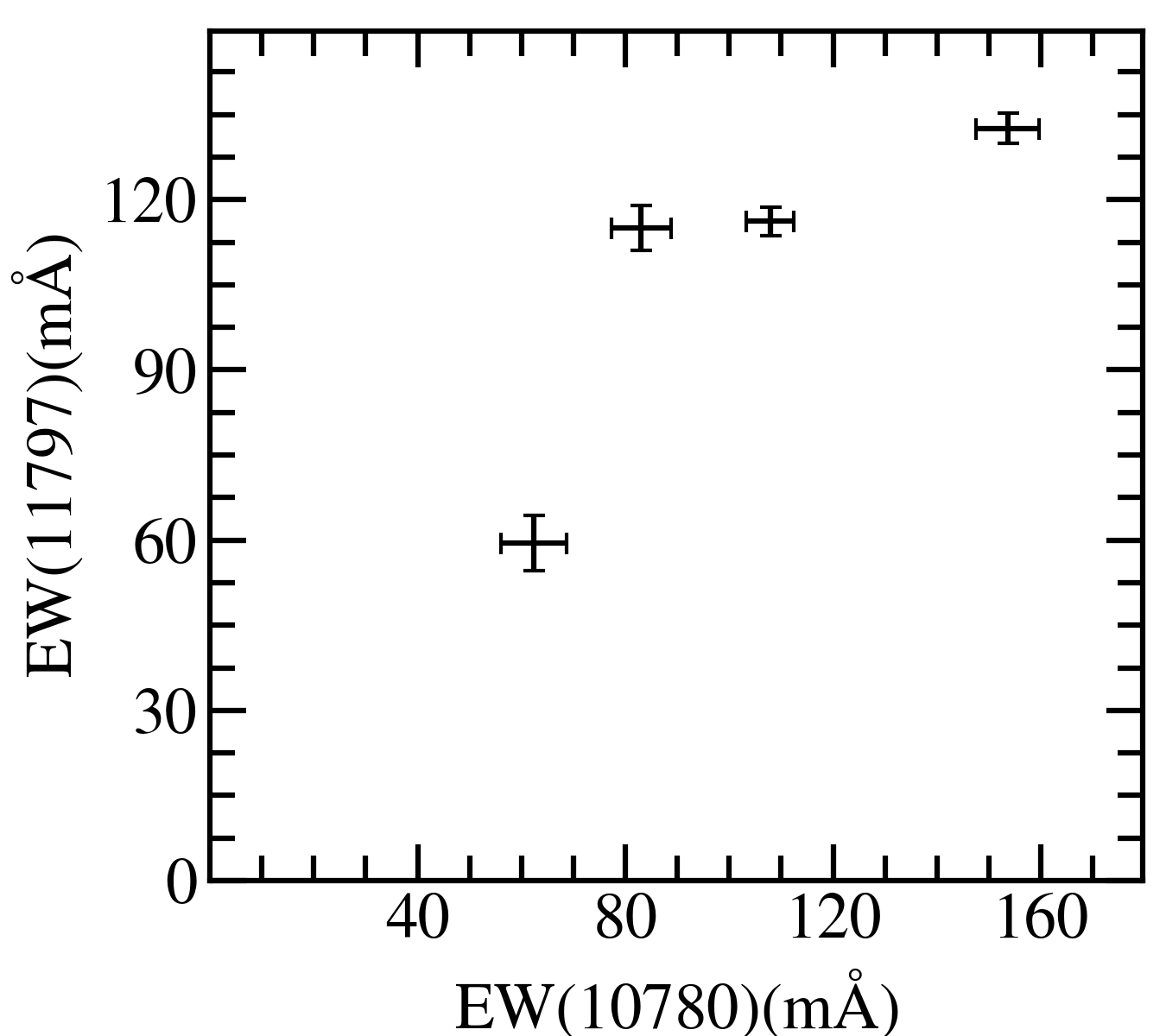}\label{fig:10780_11797_ew}}
   \subfloat[]{\includegraphics[width = 0.26\textwidth]
      {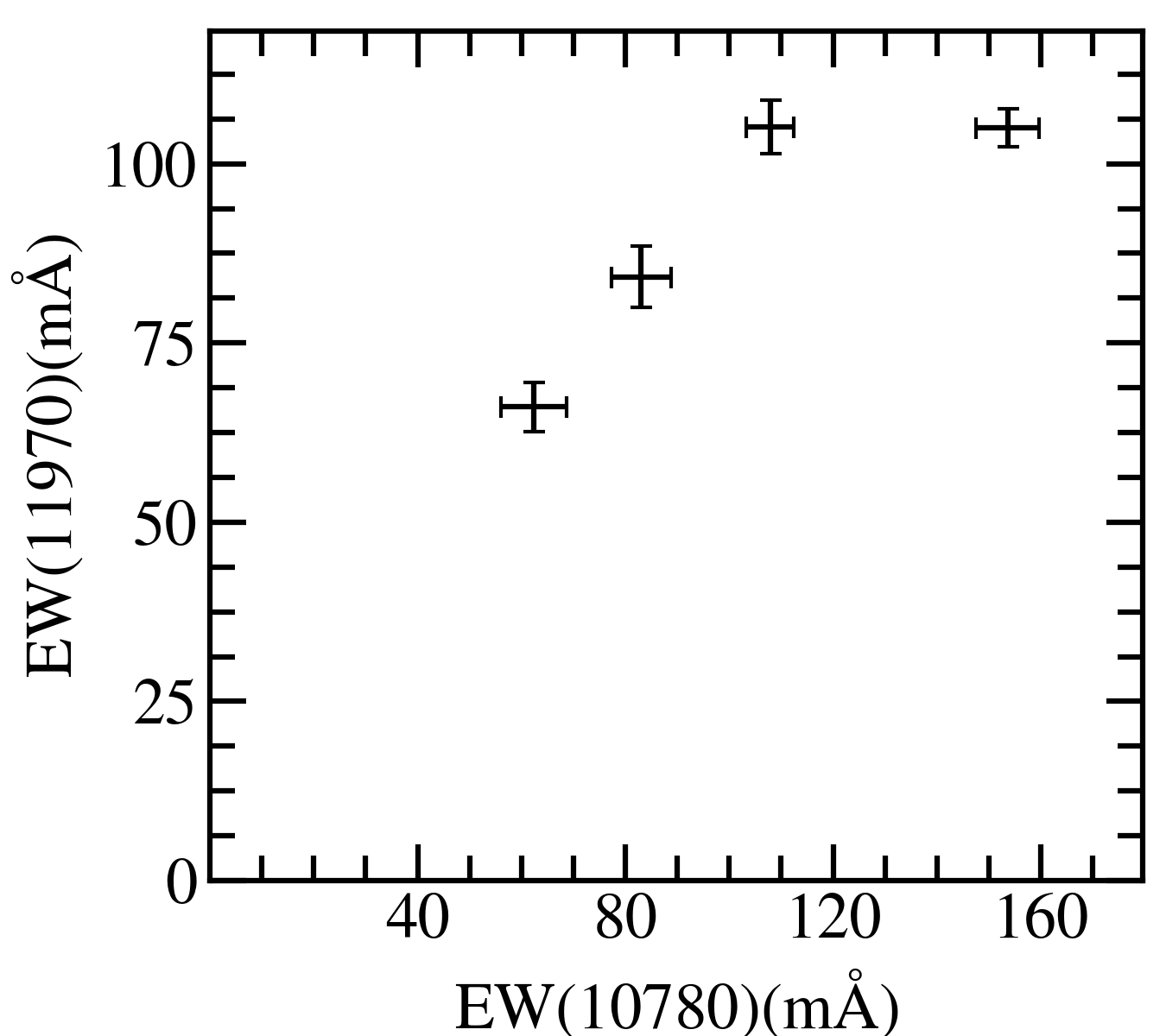}\label{fig:10780_11969_ew}}
   \subfloat[]{\includegraphics[width = 0.26\textwidth]
      {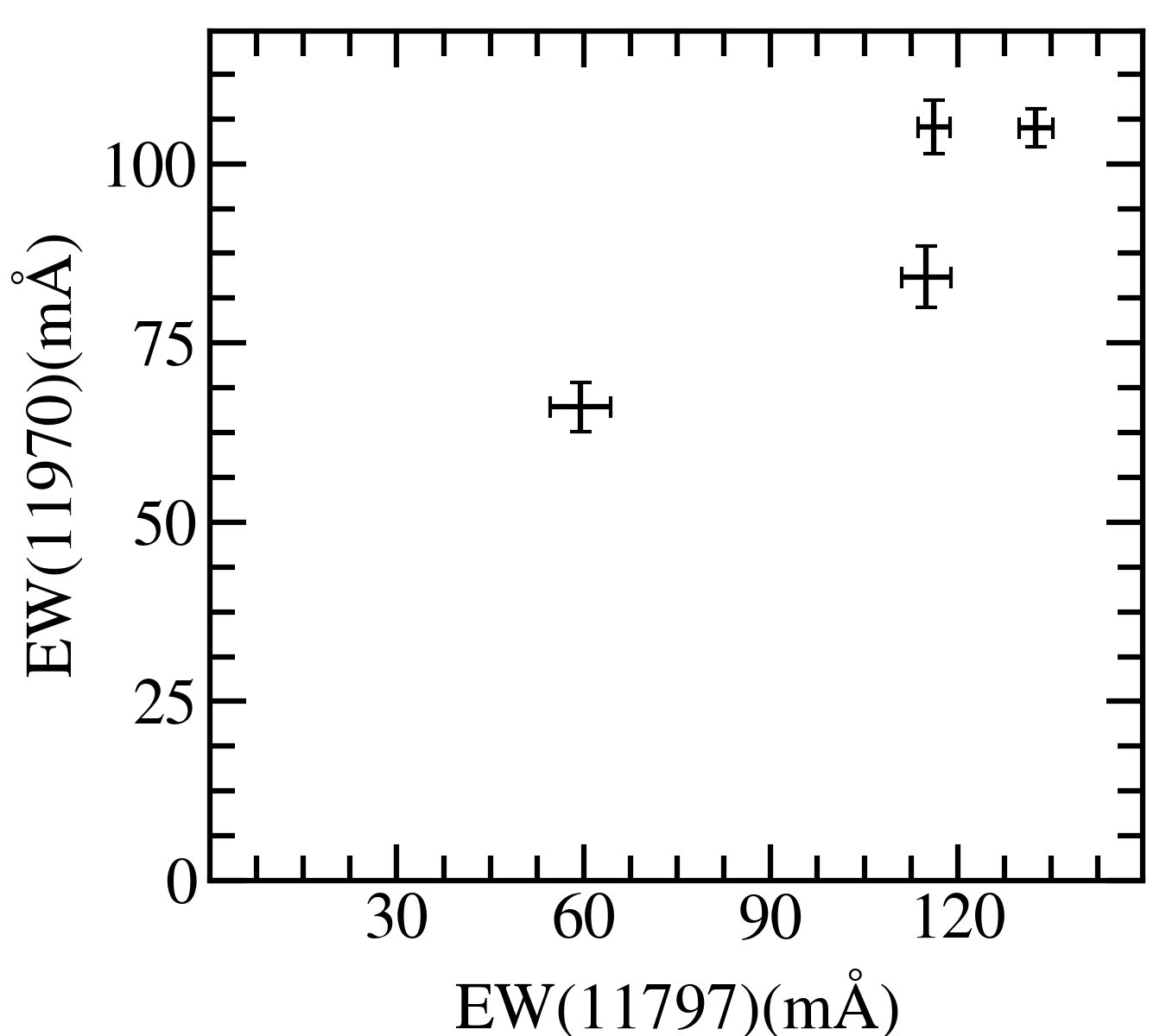}\label{fig:11797_11969_ew}}\\[-4mm]
   \subfloat[]{\includegraphics[width = 0.26\textwidth]
      {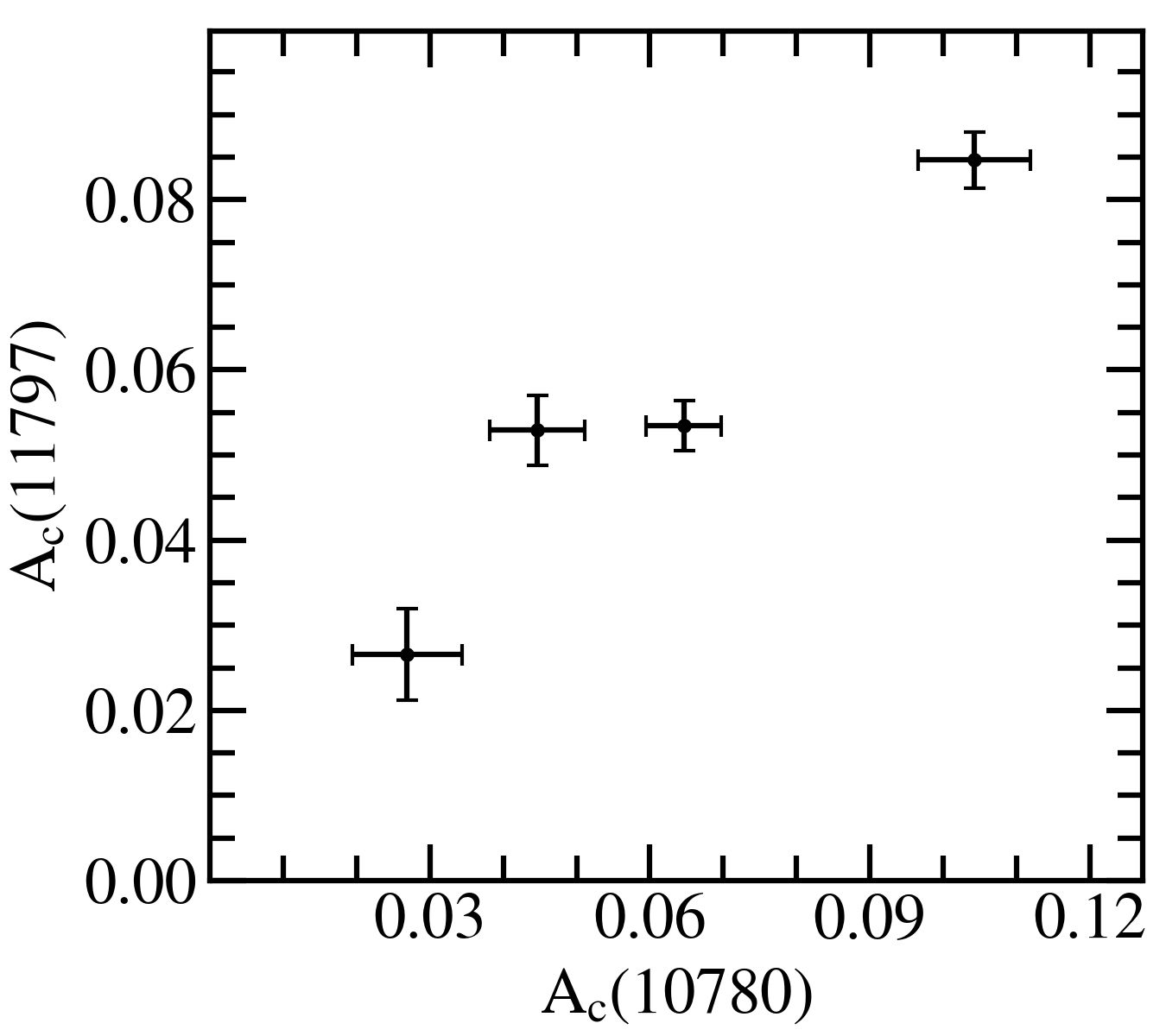}\label{fig:10780_11797_depth}}
   \subfloat[]{\includegraphics[width = 0.26\textwidth]
      {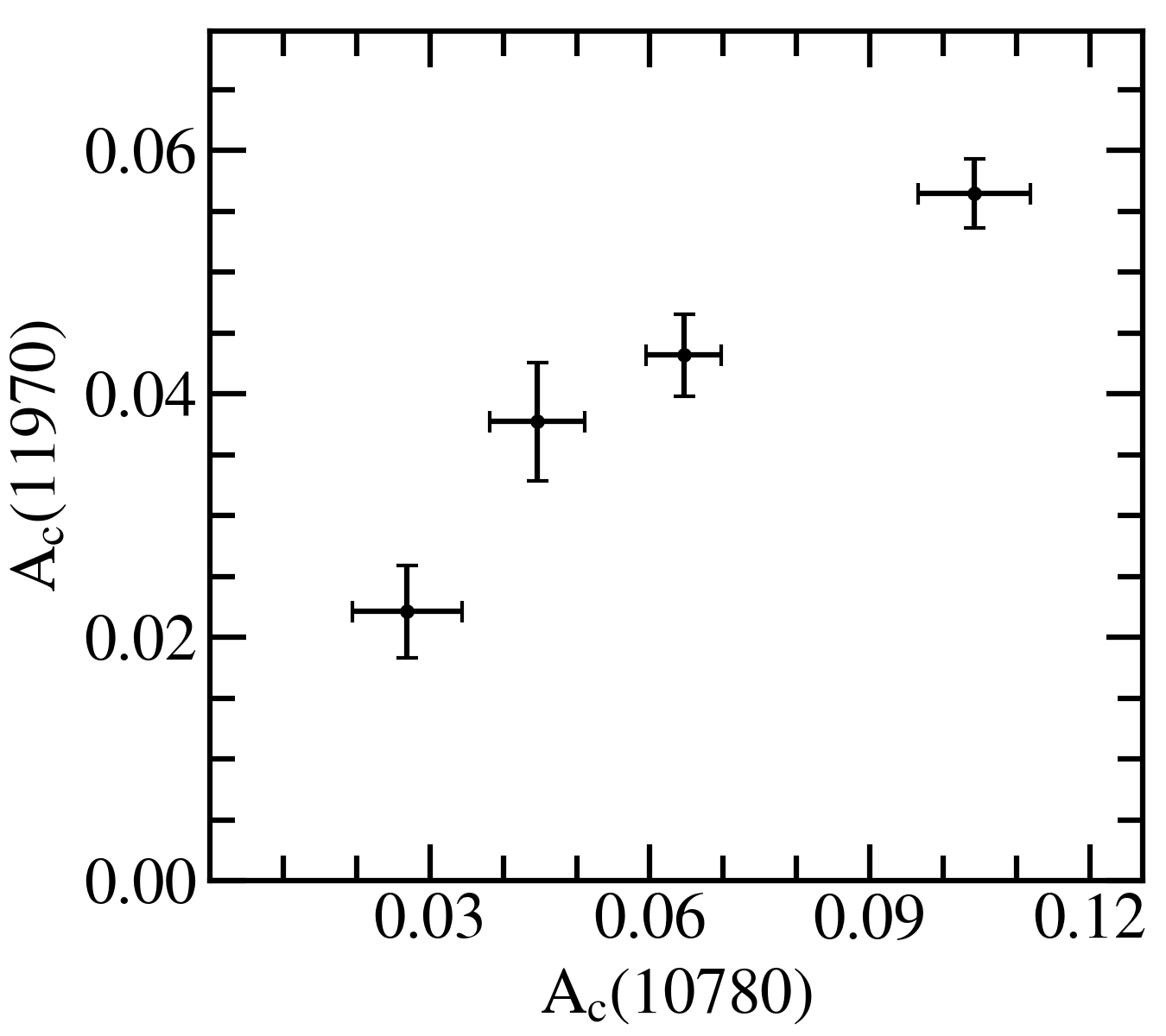}\label{fig:10780_11969_depth}}
   \subfloat[]{\includegraphics[width = 0.26\textwidth]
      {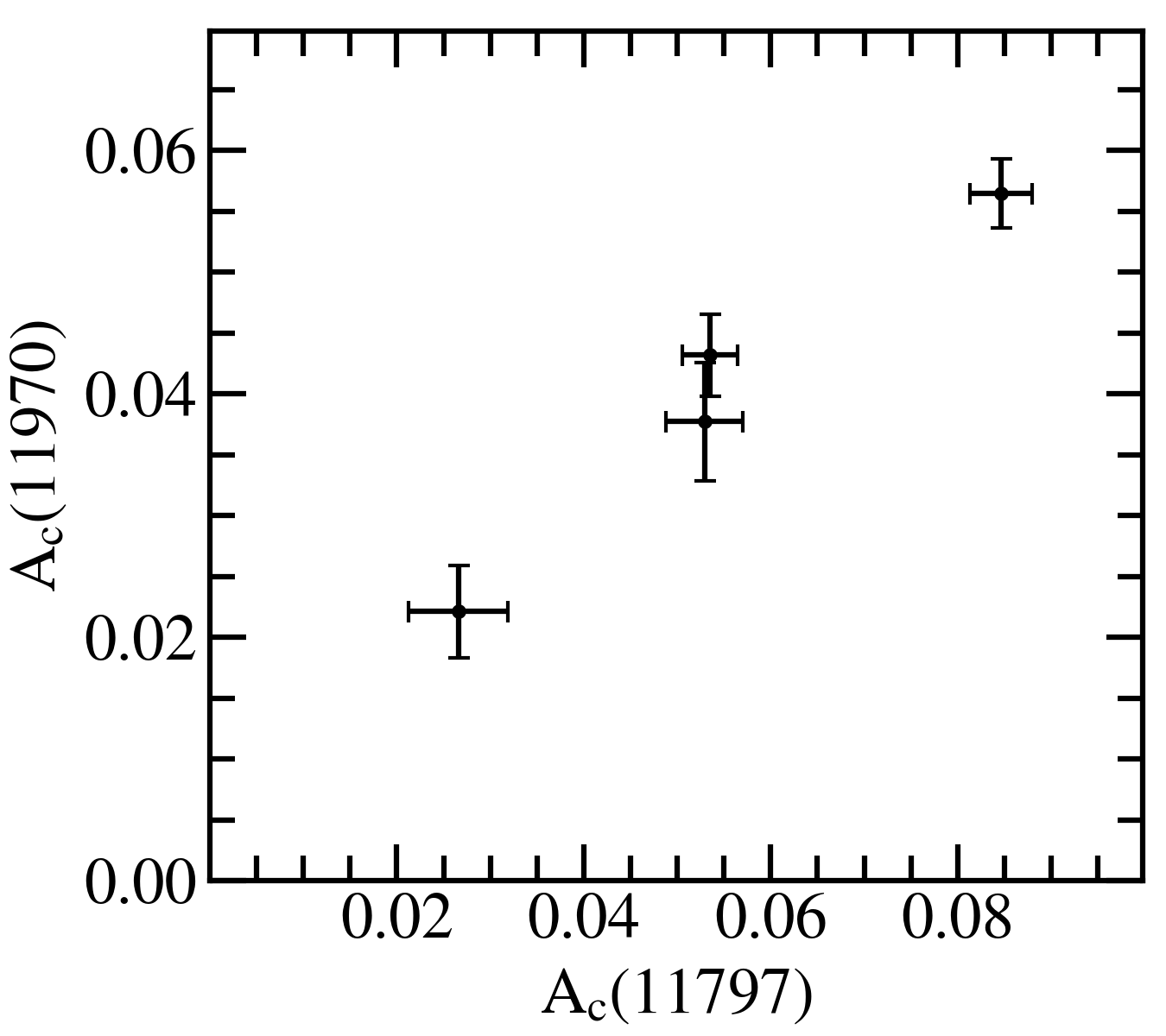}\label{fig:11797_11969_depth}}
   \caption{Plots of $EW$ correlation (top row) and $A_\mathrm{c}$ correlation (bottom row)  of the members of the DIB profile family $\lambda10780$--11797--11970. }
   \label{fig:profile_set_10780_ew_depth}
\end{figure*}
 
    We define a DIB profile family as a set of DIBs that show very similar profiles and 
    profile variations in all sight lines with detections of those DIBs. 
    Traditionally, DIBs are considered to have a common carrier if they show a strong
    correlation of their $EW$s \citep[e.g.][]{2022MNRAS.510.3546F}. But if those DIBs  are also part of the same DIB profile family, they 
    are more likely to have carriers with the same rotational constant and rotational 
    temperature \citep{2014IAUS..297...34S}. This implies that their carriers also occur in the same region along the 
    sight line, because they need the same environmental conditions and radial velocities 
    to exhibit the same profile variation.
    We therefore consider DIB profile families a good tool for establishing structural relations 
    between DIB carriers and to confirm $EW$-DIB families.

   \begin{figure*}
  \centering
  \subfloat[$\lambda10780$ -- HD~183143]{
      \includegraphics[trim=35 0 0 0, width = .2598\textwidth]
      {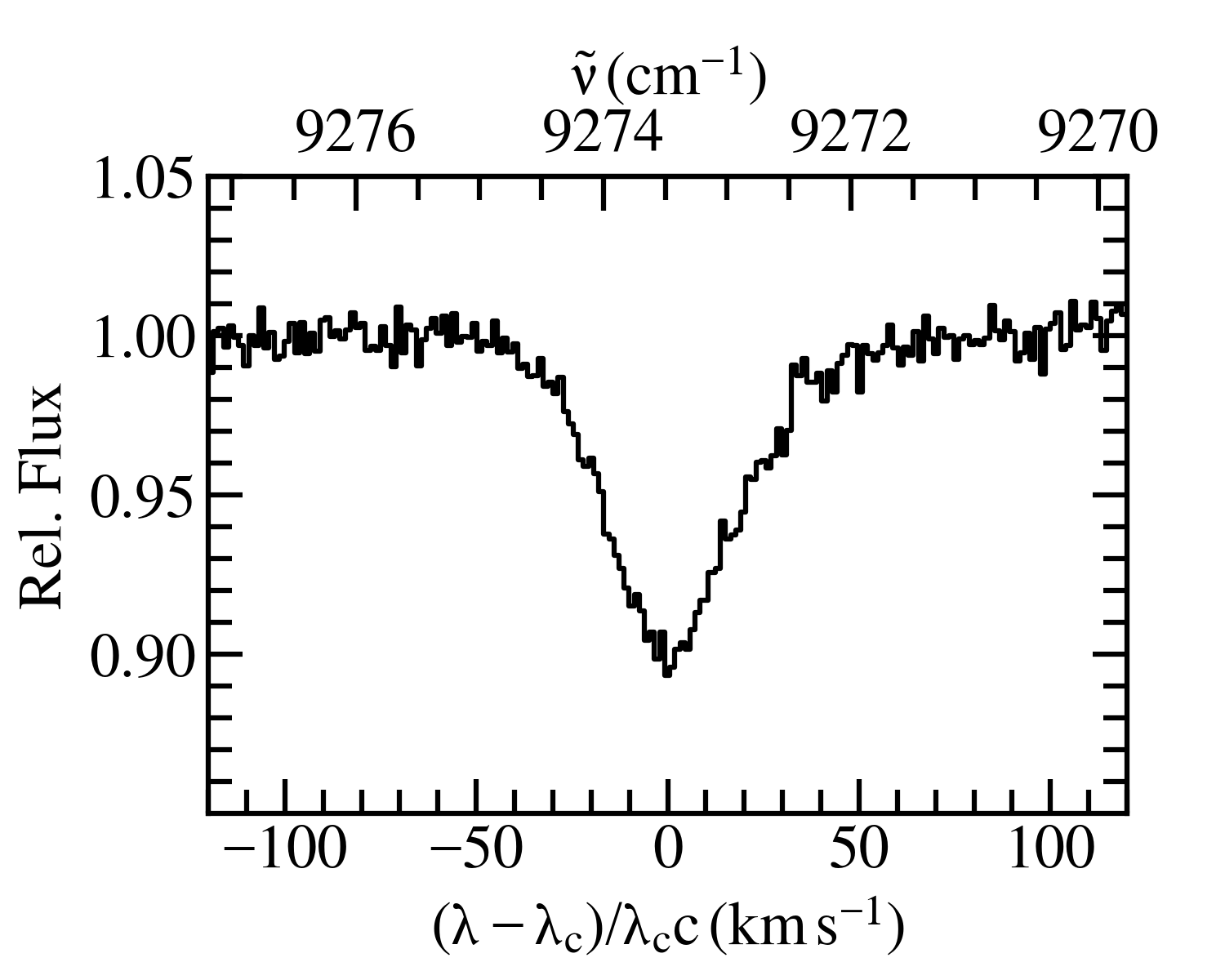}\label{fig:wavenumber_10780}}
  \subfloat[$\lambda11797$ -- HD~183143]{
      \includegraphics[trim=60 0 0 0,clip, width = .24\textwidth]
      {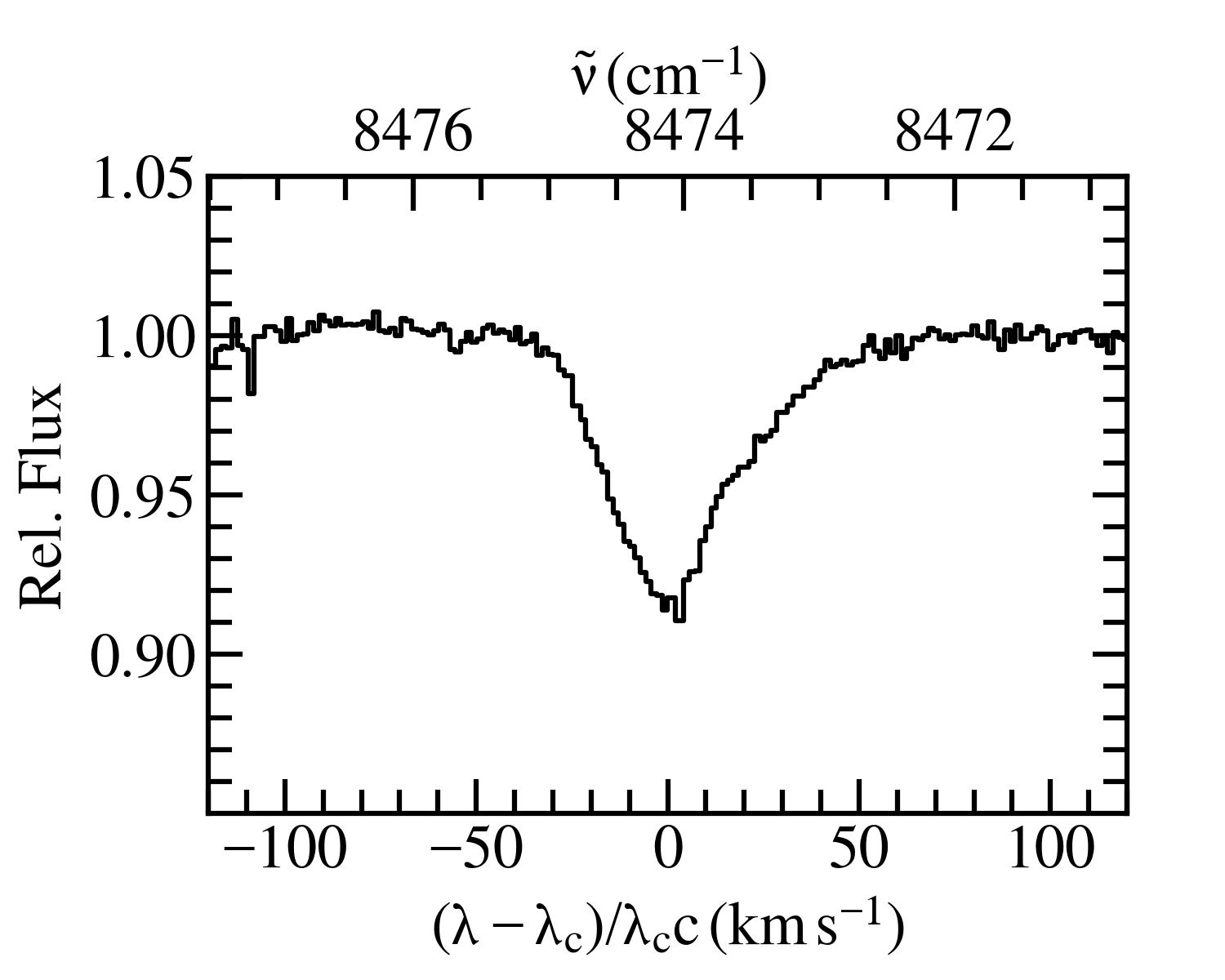}\label{fig:wavenumber_11797}}
  \subfloat[$\lambda11970$ -- HD~183143]{
      \includegraphics[trim=60 0 0 0,clip, width = .24\textwidth]
      {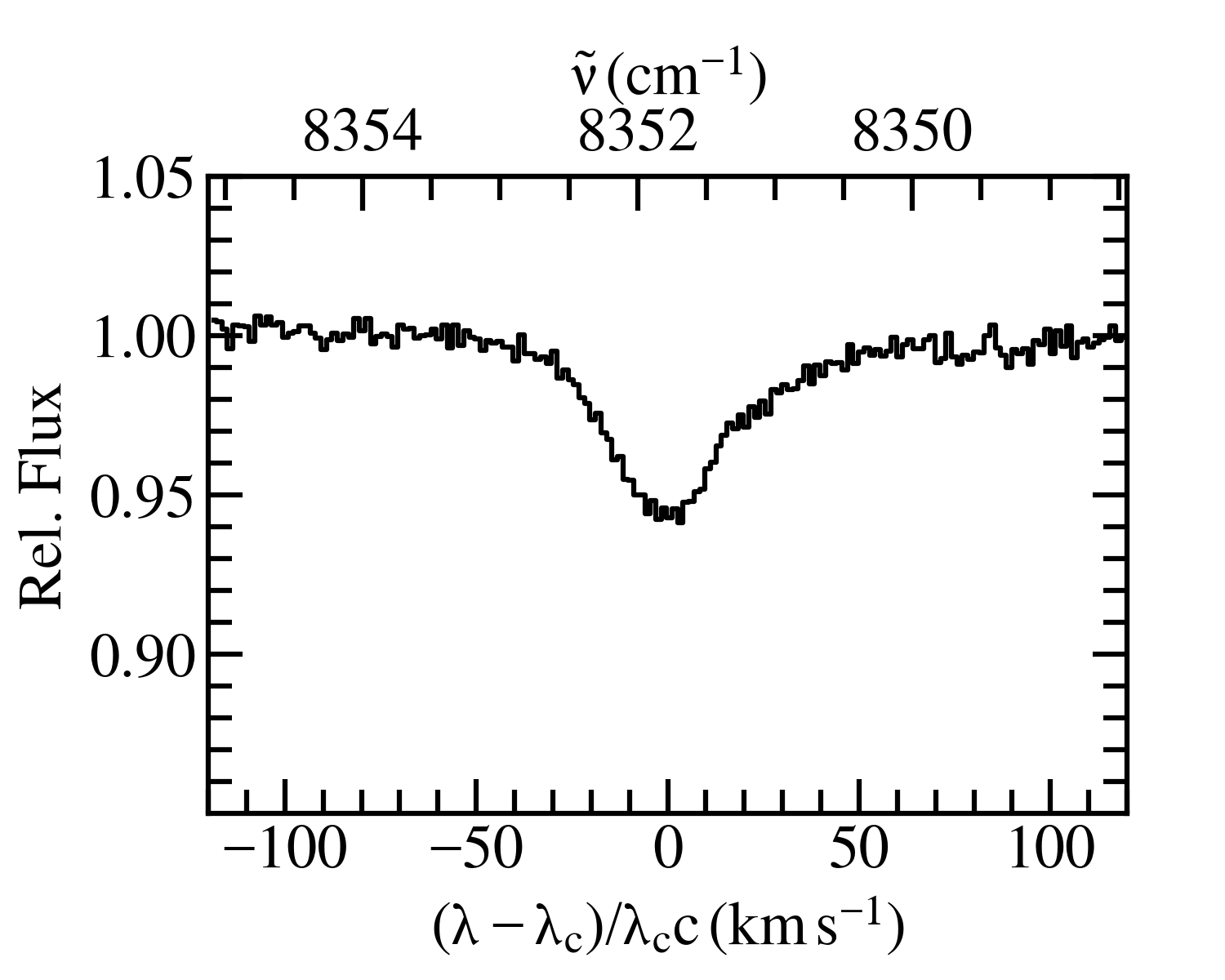}\label{fig:wavenumber_11969}}
  \subfloat[\ion{K}{i} -- HD~183143]{
      \includegraphics[trim=35 0 0 0,clip, width = .2598\textwidth]
      {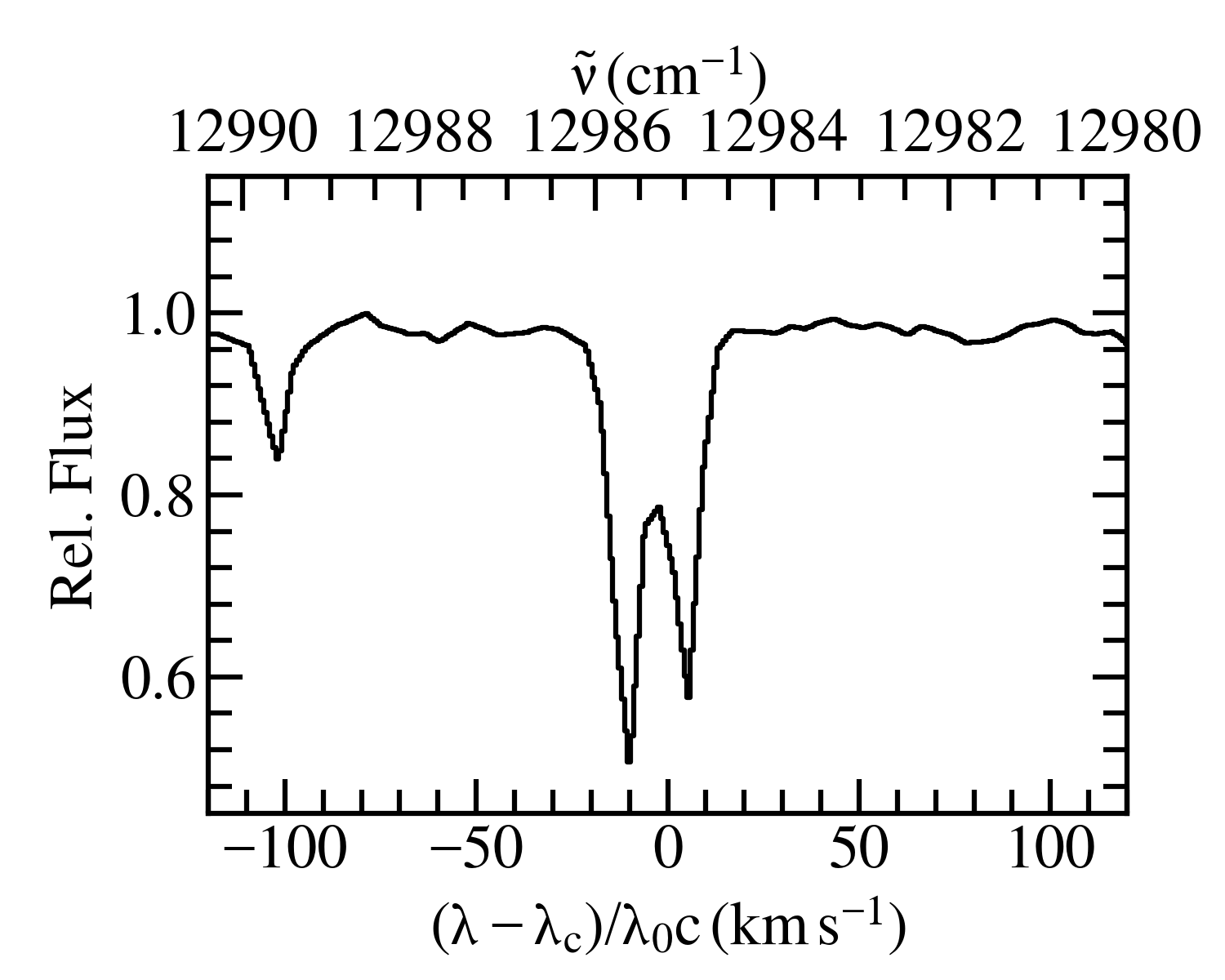}\label{fig:wavenumber_ki}}\\
  \subfloat[$\lambda10780$ -- HD~92207]{
      \includegraphics[trim=35 0 0 0, width = .2598\textwidth]
      {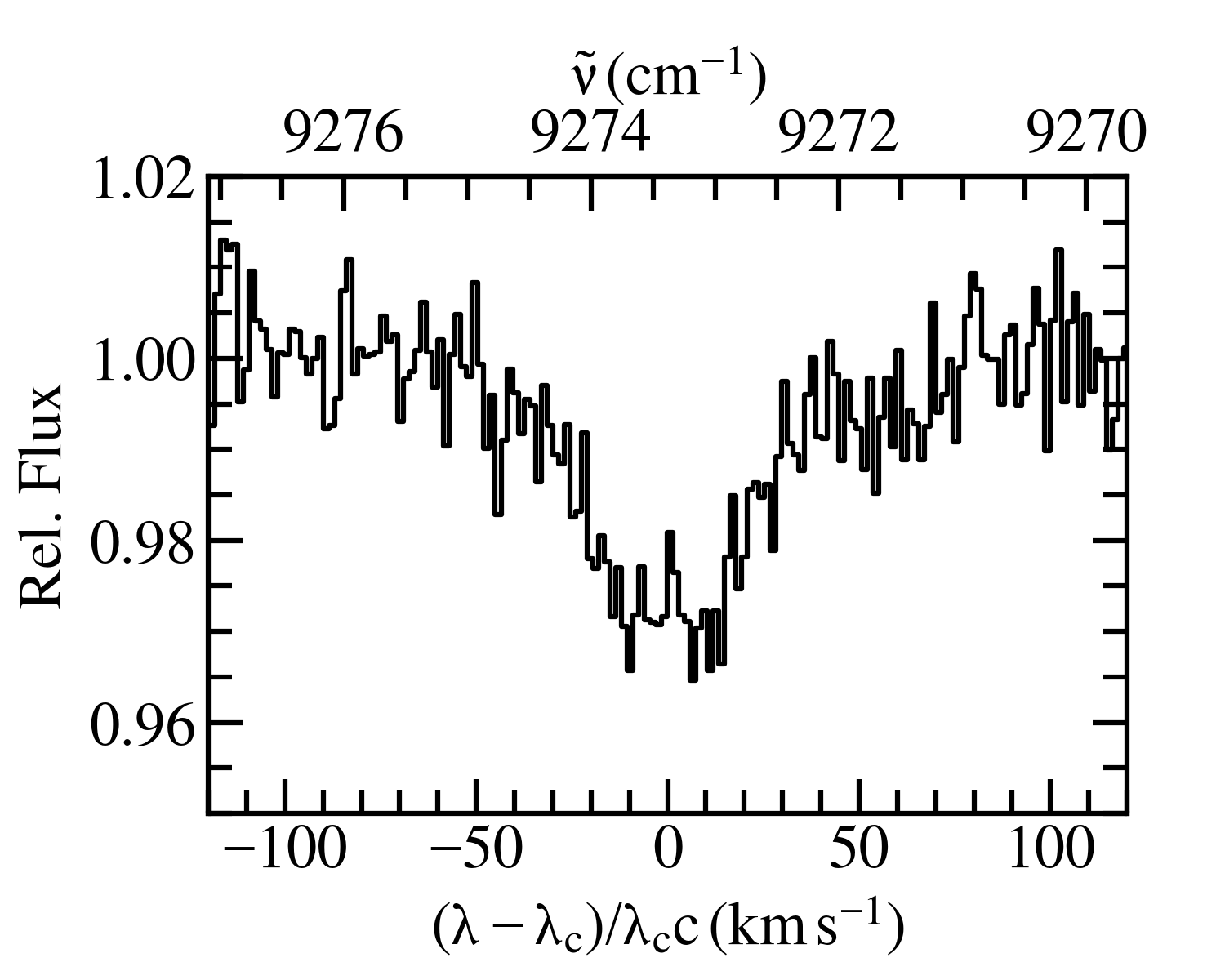}\label{fig:wavenumber_10780_hd922}}
  \subfloat[$\lambda11797$ -- HD~92207]{
      \includegraphics[trim=60 0 0 0,clip, width = .24\textwidth]
      {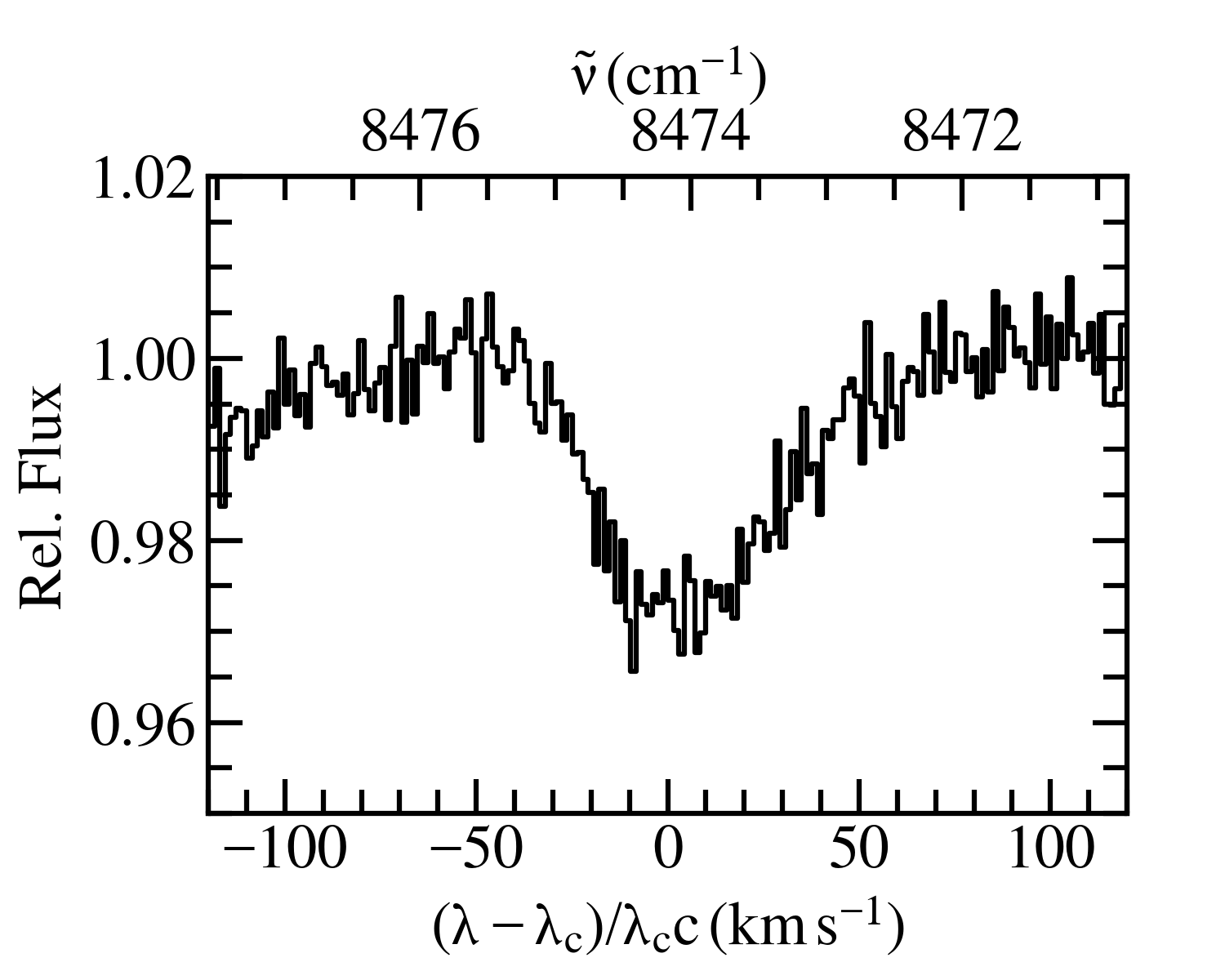}\label{fig:wavenumber_11797_hd922}}
  \subfloat[$\lambda11970$ -- HD~92207]{
      \includegraphics[trim=60 0 0 0,clip, width = .24\textwidth]
      {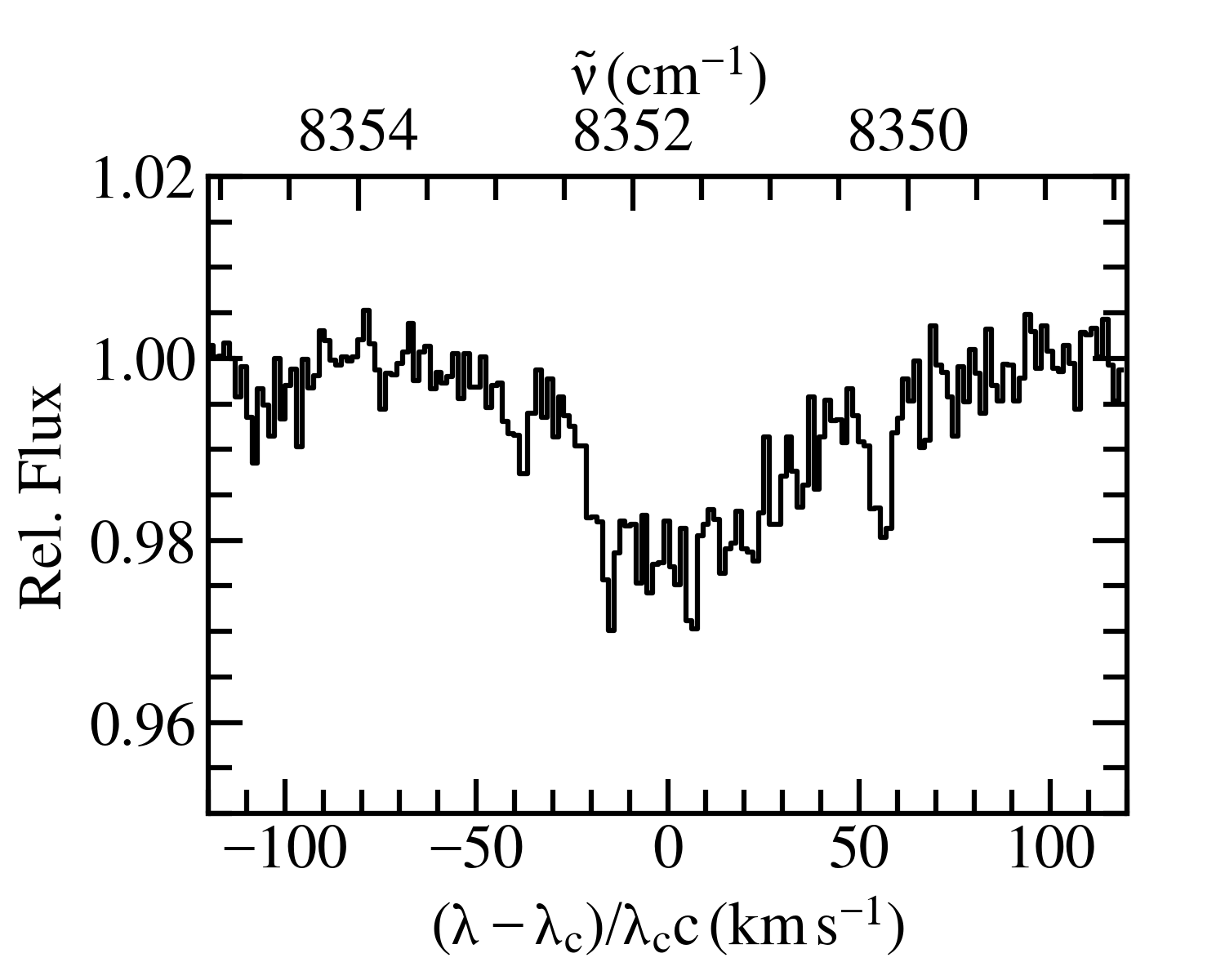}\label{fig:wavenumber_11969_hd922}}
  \subfloat[\ion{K}{i} -- HD~92207]{
      \includegraphics[trim=25 0 7 0,clip, width = .2598\textwidth]
      {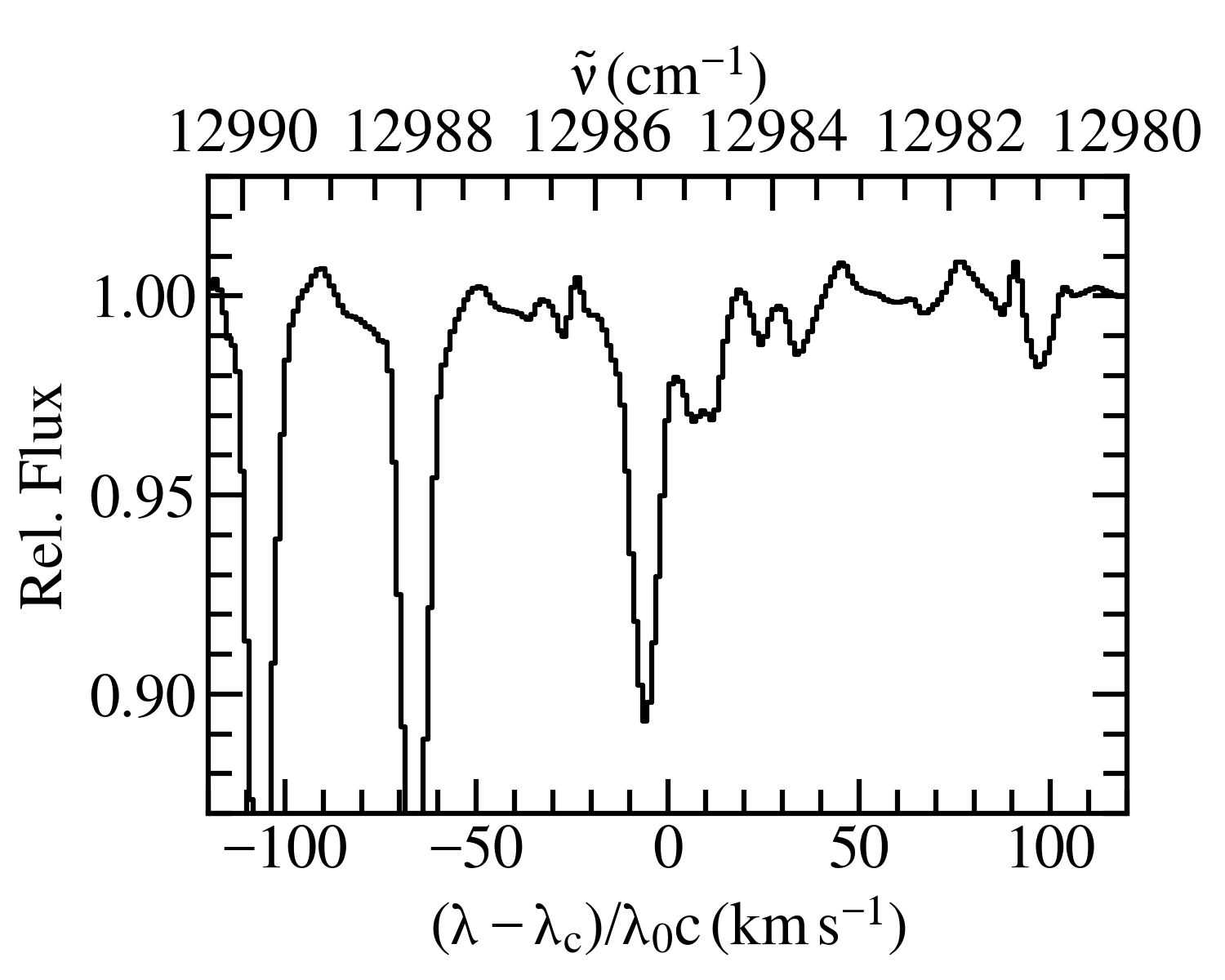}\label{fig:wavenumber_ki_hd922}}\\
  \caption{Profile plots of the DIB family $\lambda$10780--11797--11970 and interstellar \ion{K}{i} for the sight line of HD~183143 (top) and HD~92207 (bottom) in wavenumbers and radial velocity in the barycentric rest frame.} 
  \label{fig:profile_wavenumber}
  \end{figure*}    
  
    In this study the $EW$ analysis is secondary to the profile analysis because of our 
    small sample size. Consequently, if we want to verify whether two DIBs have the same or correlated 
    carriers, and we have measurements of both DIBs in only up to four sight lines, $EW$ 
    correlation gives us little information about family membership because we only have 
    up to four data points. Generally, the measurement of the equivalent width of a DIB can be subject to uncertainty
    because of a multitude of reasons: different (subjective) choice of the continuum, inaccurately treated stellar 
    or telluric blends, cosmics or deficiencies of the detector.
    Additional profile analysis can show us the exact nature of these deviations in the spectrum.
    Most of those deviations give rise to narrower features than the DIBs themselves and they are easily 
    distinguishable from DIBs (see e.g.~Fig.~\ref{fig:10780_11969_profile}). Thus, a profile analysis is 
    more robust against deviations than $EW$ correlations.

    Keeping this in mind, we compared all the DIB profiles from the present work, including already-known DIBs, in pairs  for a given sight line,
    after normalising each DIB to a central depth of $A_\mathrm{c}$\,=\,1. Good agreement was found for three DIBs.
    The starting point were the two strong DIBs $\lambda\lambda$10780 and 11797, for which  
    \cite{2015ApJ...800..137H} found a very strong $EW$ correlation with a correlation coefficient of 
    $r$\,=\,0.97(17) and an $EW$ ratio of nearly 1:1. This DIB pair is detected in four sight lines 
    (Fig. \ref{fig:10780_11797_profile}). 
    There is a clear uniform variation of the profiles: in HD~165784 the blue wing is 
    steeper and the core is broader than in HD~183143 and HD~111613. In HD~92207 the DIB core seems to be broader 
    than in all other sight lines. The matching profiles in all four sight 
    lines leads us to confirm the assumption that this pair is in the same DIB family.
    The DIB $\lambda$11970 was added as a third member, where a pairwise comparison with the other two normalised DIBs is shown in the other 
    panels of Fig.~\ref{fig:profile_set_10780}. Close agreement is also found in these cases, establishing the three
    DIBs as members of a profile family. We note that 
    DIB $\lambda$11970 had to be corrected for the stellar \ion{He}{i} component in some of the target stars (see 
    Sect.~\ref{sec:stellar_blends}) and for telluric absorption. For this DIB there are still some artefacts of a 
    corrected telluric line at $\sim$50\,km\,s$^{-1}$ visible in Fig.~\ref{fig:10780_11969_profile} and 
    Fig.~\ref{fig:11797_11969_profile}, in the form of weak emission in HD~111613 or as absorption in HD~92207.
    
    During an $EW$ correlation analysis the telluric artefacts cause a scatter in the otherwise linear 
    $EW$ correlation and weaken the relation between the DIBs $\lambda\lambda$10780, 11797, and 11970 (see 
    Fig.~\ref{fig:profile_set_10780_ew_depth}). Profile analysis circumvents this effect because the residuals due to 
    blending lines can clearly be identified during profile comparison. 
    For DIBs of the same profile family, $A_\mathrm{c}$ is a much better parameter to study the correlation of DIB strengths than $EW$s (Fig.~\ref{fig:profile_set_10780_ew_depth}).

    The continuum-normalized profiles of the profile family $\lambda$10780--11797--11970 are displayed in
    Fig.~\ref{fig:profile_wavenumber} along with the \ion{K}{i} line in velocity 
    (and wavenumber) 
    space for the 
    sight line towards HD~183143 and to HD~92207. We can see that 
    $A_\mathrm{c}$ decreases with increasing $\lambda$ for HD~183143, as do the $EW$s. A similar behaviour is found 
    for HD~165784 at 2/3 of the extinction of HD~183143.
    On the other hand, the ratios of $EW$s and $A_\mathrm{c}$ for the first two DIB profile family members
    are consistent with unity for the other two less reddened stars, while DIB $\lambda$11970 is a bit weaker.
    The comparison with the \ion{K}{i} line in Fig.~\ref{fig:profile_wavenumber} allows the impact of the ISM cloud structure on the DIB structure to be 
    estimated.
    
    To conclude, this DIB profile set is highly interesting because of the matching profiles. 
    If this can be confirmed with additional sight lines, it could be a 
    good starting point for laboratory studies to find a matching DIB carrier.

\section{Summary and conclusions}
   We have introduced a combined ISM sight line and stellar analysis approach that  involves a non-LTE analysis of 
   high-resolution optical and near-IR spectra of the background stars. A subsequent comparison of the model and
   observed SED facilitates a thorough characterisation of the sight line reddening and reddening law, which is more 
   robust than traditional approaches. Using telluric and stellar spectrum
   correction, 12 new DIB candidates were found in high spectral resolution observations of the $YJ$ band with 
   CRIRES, nearly doubling the known DIBs at these wavelengths. At the same time, this facilitated the characterisation of some known near-IR DIBs to be improved. Moreover, a first DIB profile family was established comprising
   the DIBs $\lambda$10780--11797--11970,
   which show remarkably similar profiles, suggesting that they emerge from the same carrier. Two of the DIBs had already been suggested to form a DIB family by
   \citet{2015ApJ...800..137H}. A match to the full SED
   of the DIB standard star HD~183143 including the unusual UV absorption bump was achieved for the first time, 
   requiring an extra PAH component that we attribute to a circumstellar shell on the sight line towards HD~183143.
   
   We have shown here the convincing potential of old CRIRES data for DIB studies in the near-IR. The extended 
   simultaneous wavelength coverage of the upgraded CRIRES \citep{Dornetal14} will allow systematic studies of the known
   and newly found near-IR DIBs to be performed in a much more time-efficient manner. New diagnostic links such as
   the DIB profile family established here plus recent improvements in laboratory techniques may steadily support the 
   way to further DIB carrier identifications.

\begin{acknowledgements}
   We would like to thank the referee for useful comments that helped to improve the paper.
   A.E., D.W. and N.P. gratefully acknowledge support by the Austrian Science Fund FWF
   within the DK-ALM (W1259-N27). M.F. and N.P. acknowledge financial support by the 
   Deutsche Forschungsgemeinschaft, DFG, project number PR 685/3-1.
   Based on observations collected at the European Southern Observatory under ESO programmes
   62.H-0176, 079.D-0810(A) and 093.C-0380(A) and on observations collected at the 
   Centro Astron\'omico Hispano Alem\'an at Calar Alto (CAHA), operated jointly by the 
   Max-Planck Institut f\"ur  Astronomie and the Instituto de Astrof\'isica de Andaluc\'ia 
   (CSIC), proposal H2001-2.2-011. Travel of N.P. to the Calar Alto Observatory was supported by the 
   DFG under grant PR 685/1-1.
   
   This research has made use of "Aladin sky atlas" developed at CDS, Strasbourg
   Observatory, France \citep{2000A&AS..143...33B}.
   
   This publication makes use of data products from the Wide-field Infrared Survey Explorer,
   which is a joint project of the University of California, Los Angeles, and the Jet 
   Propulsion Laboratory/California Institute of Technology, funded by the National 
   Aeronautics and Space Administration.
   
   This publication makes use
   of data products from the Two Micron All Sky Survey, which is a joint project
   of the University of Massachusetts and the Infrared Processing and Analysis
   Center/California Institute of Technology, funded by the National Aeronautics
   and Space Administration and the National Science Foundation.

   This work has made use of data from the European Space Agency (ESA) mission
   {\it Gaia} (\url{https://www.cosmos.esa.int/gaia}), processed by the {\it Gaia}
   Data Processing and Analysis Consortium (DPAC,
   \url{https://www.cosmos.esa.int/web/gaia/dpac/consortium}). Funding for the DPAC
   has been provided by national institutions, in particular the institutions
   participating in the {\it Gaia} Multilateral Agreement.
\end{acknowledgements}


%
   \bibliography{bibliography} 

\begin{thebibliography}{99}
\expandafter\ifx\csname natexlab\endcsname\relax\def\natexlab#1{#1}\fi

\bibitem[{{Baes} {et~al.}(2011){Baes}, {Verstappen}, {De Looze}, {Fritz},
  {Saftly}, {Vidal P{\'e}rez}, {Stalevski}, \& {Valcke}}]{baes2011}
{Baes}, M., {Verstappen}, J., {De Looze}, I., {et~al.} 2011, \apjs, 196, 22

\bibitem[{{Becker}(1998)}]{Becker98}
{Becker}, S.~R. 1998, ASP Conf.~Ser., 131, 137

\bibitem[{{Bisht} {et~al.}(2021){Bisht}, {Zhu}, {Yadav}, {Ganesh}, {Rangwal},
  {Durgapal}, {Sariya}, \& {Jiang}}]{Bishtetal21}
{Bisht}, D., {Zhu}, Q., {Yadav}, R.~K.~S., {et~al.} 2021, \mnras, 503, 5929

\bibitem[{{Bonnarel} {et~al.}(2000){Bonnarel}, {Fernique}, {Bienaym{\'e}},
  {Egret}, {Genova}, {Louys}, {Ochsenbein}, {Wenger}, \&
  {Bartlett}}]{2000A&AS..143...33B}
{Bonnarel}, F., {Fernique}, P., {Bienaym{\'e}}, O., {et~al.} 2000, \aaps, 143,
  33

\bibitem[{{Bresolin} {et~al.}(2001){Bresolin}, {Kudritzki}, {Mendez}, \&
  {Przybilla}}]{Bresolinetal01}
{Bresolin}, F., {Kudritzki}, R.-P., {Mendez}, R.~H., \& {Przybilla}, N. 2001,
  \apjl, 548, L159

\bibitem[{{Butler} \& {Giddings}(1985)}]{daresbury1985}
{Butler}, K. \& {Giddings}, J.~R. 1985, in Newsletter on Analysis of
  Astronomical Spectra, 9 (Univ. London)

\bibitem[{{Cami} {et~al.}(1997){Cami}, {Sonnentrucker}, {Ehrenfreund}, \&
  {Foing}}]{1997Cami}
{Cami}, J., {Sonnentrucker}, P., {Ehrenfreund}, P., \& {Foing}, B.~H. 1997,
  \aap, 326, 822

\bibitem[{{Campbell} {et~al.}(2015){Campbell}, {Holz}, {Gerlich}, \&
  {Maier}}]{2015Natur.523..322C}
{Campbell}, E.~K., {Holz}, M., {Gerlich}, D., \& {Maier}, J.~P. 2015, \nat,
  523, 322

\bibitem[{{Camps} \& {Baes}(2015)}]{camps2015}
{Camps}, P. \& {Baes}, M. 2015, Astronomy and Computing, 9, 20

\bibitem[{{Cardelli} {et~al.}(1989){Cardelli}, {Clayton}, \&
  {Mathis}}]{cardelli1989}
{Cardelli}, J.~A., {Clayton}, G.~C., \& {Mathis}, J.~S. 1989, \apj, 345, 245

\bibitem[{{Caswell} \& {Haynes}(1987)}]{1987caswell}
{Caswell}, J.~L. \& {Haynes}, R.~F. 1987, \aap, 171, 261

\bibitem[{{Clough} {et~al.}(1992){Clough}, {Iacono}, \&
  {Moncet}}]{1992JGR....9715761C}
{Clough}, S.~A., {Iacono}, M.~J., \& {Moncet}, J.-L. 1992, \jgr, 97, 15,761

\bibitem[{{Clough} {et~al.}(1981){Clough}, {Kneizys}, {Rothman}, \&
  {Gallery}}]{1981SPIE..277..152C}
{Clough}, S.~A., {Kneizys}, F.~X., {Rothman}, L.~S., \& {Gallery}, W.~O. 1981,
  Proc. SPIE, 277, 152

\bibitem[{{Cordiner} {et~al.}(2008{\natexlab{a}}){Cordiner}, {Cox}, {Trundle},
  {Evans}, {Hunter}, {Przybilla}, {Bresolin}, \& {Salama}}]{Cordineretal08a}
{Cordiner}, M.~A., {Cox}, N.~L.~J., {Trundle}, C., {et~al.} 2008{\natexlab{a}},
  \aap, 480, L13

\bibitem[{{Cordiner} {et~al.}(2019){Cordiner}, {Linnartz}, {Cox}, {Cami},
  {Najarro}, {Proffitt}, {Lallement}, {Ehrenfreund}, {Foing}, {Gull}, {Sarre},
  \& {Charnley}}]{2019ApJ...875L..28C}
{Cordiner}, M.~A., {Linnartz}, H., {Cox}, N.~L.~J., {et~al.} 2019, \apjl, 875,
  L28

\bibitem[{{Cordiner} {et~al.}(2008{\natexlab{b}}){Cordiner}, {Smith}, {Cox},
  {Evans}, {Hunter}, {Przybilla}, {Bresolin}, \& {Sarre}}]{Cordineretal08b}
{Cordiner}, M.~A., {Smith}, K.~T., {Cox}, N.~L.~J., {et~al.}
  2008{\natexlab{b}}, \aap, 492, L5

\bibitem[{{Cox} {et~al.}(2017){Cox}, {Cami}, {Farhang}, {Smoker},
  {Monreal-Ibero}, {Lallement}, {Sarre}, {Marshall}, {Smith}, {Evans}, {Royer},
  {Linnartz}, {Cordiner}, {Joblin}, {van Loon}, {Foing}, {Bhatt}, {Bron},
  {Elyajouri}, {de Koter}, {Ehrenfreund}, {Javadi}, {Kaper}, {Khosroshadi},
  {Laverick}, {Le Petit}, {Mulas}, {Roueff}, {Salama}, \& {Spaans}}]{Coxetal17}
{Cox}, N. L.~J., {Cami}, J., {Farhang}, A., {et~al.} 2017, \aap, 606, A76

\bibitem[{{Cox} {et~al.}(2014){Cox}, {Cami}, {Kaper}, {Ehrenfreund}, {Foing},
  {Ochsendorf}, {van Hooff}, \& {Salama}}]{2014A&A...569A.117C}
{Cox}, N.~L.~J., {Cami}, J., {Kaper}, L., {et~al.} 2014, \aap, 569, A117

\bibitem[{{Cutri} \& {et al.}(2012)}]{2012cutri}
{Cutri}, R.~M. \& {et al.} 2012, VizieR Online Data Catalog, II/311

\bibitem[{{Cutri} {et~al.}(2003){Cutri}, {Skrutskie}, {van Dyk}, {Beichman},
  {Carpenter}, {Chester}, {Cambresy}, {Evans}, {Fowler}, {Gizis}, {Howard},
  {Huchra}, {Jarrett}, {Kopan}, {Kirkpatrick}, {Light}, {Marsh}, {McCallon},
  {Schneider}, {Stiening}, {Sykes}, {Weinberg}, {Wheaton}, {Wheelock}, \&
  {Zacarias}}]{2003cutri}
{Cutri}, R.~M., {Skrutskie}, M.~F., {van Dyk}, S., {et~al.} 2003, VizieR Online
  Data Catalog, II/246

\bibitem[{{Dorn} {et~al.}(2014){Dorn}, {Anglada-Escude}, {Baade}, {Bristow},
  {Follert}, {Gojak}, {Grunhut}, {Hatzes}, {Heiter}, {Hilker}, {Ives}, {Jung},
  {K{\"a}ufl}, {Kerber}, {Klein}, {Lizon}, {Lockhart}, {L{\"o}winger},
  {Marquart}, {Oliva}, {Origlia}, {Pasquini}, {Paufique}, {Piskunov}, {Pozna},
  {Reiners}, {Smette}, {Smoker}, {Seemann}, {Stempels}, \&
  {Valenti}}]{Dornetal14}
{Dorn}, R.~J., {Anglada-Escude}, G., {Baade}, D., {et~al.} 2014, The Messenger,
  156, 7

\bibitem[{{Ehrenfreund} {et~al.}(2002){Ehrenfreund}, {Cami},
  {Jim{\'e}nez-Vicente}, {Foing}, {Kaper}, {van der Meer}, {Cox},
  {D'Hendecourt}, {Maier}, {Salama}, {Sarre}, {Snow}, \&
  {Sonnentrucker}}]{Ehrenfreundetal02}
{Ehrenfreund}, P., {Cami}, J., {Jim{\'e}nez-Vicente}, J., {et~al.} 2002, \apjl,
  576, L117

\bibitem[{{Elyajouri} \& {Lallement}(2019)}]{2019apogee}
{Elyajouri}, M. \& {Lallement}, R. 2019, \aap, 628, A67

\bibitem[{{Elyajouri} {et~al.}(2018){Elyajouri}, {Lallement}, {Cox}, {Cami},
  {Cordiner}, {Smoker}, {Farhang}, {Sarre}, \&
  {Linnartz}}]{2018A&A...616A.143E}
{Elyajouri}, M., {Lallement}, R., {Cox}, N.~L.~J., {et~al.} 2018, \aap, 616,
  A143

\bibitem[{{Fan} {et~al.}(2019){Fan}, {Hobbs}, {Dahlstrom}, {Welty}, {York},
  {Rachford}, {Snow}, {Sonnentrucker}, {Baskes}, \&
  {Zhao}}]{2019ApJ...878..151F}
{Fan}, H., {Hobbs}, L.~M., {Dahlstrom}, J.~A., {et~al.} 2019, \apj, 878, 151

\bibitem[{{Fan} {et~al.}(2022){Fan}, {Schwartz}, {Farhang}, {Cox},
  {Ehrenfreund}, {Monreal-Ibero}, {Foing}, {Salama}, {Kulik}, {MacIsaac}, {van
  Loon}, \& {Cami}}]{2022MNRAS.510.3546F}
{Fan}, H., {Schwartz}, M., {Farhang}, A., {et~al.} 2022, \mnras, 510, 3546

\bibitem[{{Firnstein} \& {Przybilla}(2012)}]{2012A&A...543A..80F}
{Firnstein}, M. \& {Przybilla}, N. 2012, \aap, 543, A80

\bibitem[{{Fitzpatrick} \& {Massa}(2005)}]{FeMa05}
{Fitzpatrick}, E.~L. \& {Massa}, D. 2005, \aj, 130, 1127

\bibitem[{{Foing} \& {Ehrenfreund}(1994)}]{1994Natur.369..296F}
{Foing}, B.~H. \& {Ehrenfreund}, P. 1994, \nat, 369, 296

\bibitem[{{Gaia Collaboration} {et~al.}(2021){Gaia Collaboration}, {Brown},
  {Vallenari}, {Prusti}, {de Bruijne}, {Babusiaux}, {Biermann}, {Creevey},
  {Evans}, {Eyer}, {Hutton}, {Jansen}, {Jordi}, {Klioner}, {Lammers},
  {Lindegren}, {Luri}, {Mignard}, {Panem}, {Pourbaix}, {Randich}, {Sartoretti},
  {Soubiran}, {Walton}, {Arenou}, {Bailer-Jones}, {Bastian}, {Cropper},
  {Drimmel}, {Katz}, {Lattanzi}, {van Leeuwen}, {Bakker}, {Cacciari},
  {Casta{\~n}eda}, {De Angeli}, {Ducourant}, {Fabricius}, {Fouesneau},
  {Fr{\'e}mat}, {Guerra}, {Guerrier}, {Guiraud}, {Jean-Antoine Piccolo},
  {Masana}, {Messineo}, {Mowlavi}, {Nicolas}, {Nienartowicz}, {Pailler},
  {Panuzzo}, {Riclet}, {Roux}, {Seabroke}, {Sordo}, {Tanga}, {Th{\'e}venin},
  {Gracia-Abril}, {Portell}, {Teyssier}, {Altmann}, {Andrae}, {Bellas-Velidis},
  {Benson}, {Berthier}, {Blomme}, {Brugaletta}, {Burgess}, {Busso}, {Carry},
  {Cellino}, {Cheek}, {Clementini}, {Damerdji}, {Davidson}, {Delchambre},
  {Dell'Oro}, {Fern{\'a}ndez-Hern{\'a}ndez}, {Galluccio}, {Garc{\'\i}a-Lario},
  {Garcia-Reinaldos}, {Gonz{\'a}lez-N{\'u}{\~n}ez}, {Gosset}, {Haigron},
  {Halbwachs}, {Hambly}, {Harrison}, {Hatzidimitriou}, {Heiter},
  {Hern{\'a}ndez}, {Hestroffer}, {Hodgkin}, {Holl}, {Jan{\ss}en}, {Jevardat de
  Fombelle}, {Jordan}, {Krone-Martins}, {Lanzafame}, {L{\"o}ffler}, {Lorca},
  {Manteiga}, {Marchal}, {Marrese}, {Moitinho}, {Mora}, {Muinonen}, {Osborne},
  {Pancino}, {Pauwels}, {Petit}, {Recio-Blanco}, {Richards}, {Riello},
  {Rimoldini}, {Robin}, {Roegiers}, {Rybizki}, {Sarro}, {Siopis}, {Smith},
  {Sozzetti}, {Ulla}, {Utrilla}, {van Leeuwen}, {van Reeven}, {Abbas}, {Abreu
  Aramburu}, {Accart}, {Aerts}, {Aguado}, {Ajaj}, {Altavilla}, {{\'A}lvarez},
  {{\'A}lvarez Cid-Fuentes}, {Alves}, {Anderson}, {Anglada Varela}, {Antoja},
  {Audard}, {Baines}, {Baker}, {Balaguer-N{\'u}{\~n}ez}, {Balbinot}, {Balog},
  {Barache}, {Barbato}, {Barros}, {Barstow}, {Bartolom{\'e}}, {Bassilana},
  {Bauchet}, {Baudesson-Stella}, {Becciani}, {Bellazzini}, {Bernet}, {Bertone},
  {Bianchi}, {Blanco-Cuaresma}, {Boch}, {Bombrun}, {Bossini}, {Bouquillon},
  {Bragaglia}, {Bramante}, {Breedt}, {Bressan}, {Brouillet}, {Bucciarelli},
  {Burlacu}, {Busonero}, {Butkevich}, {Buzzi}, {Caffau}, {Cancelliere},
  {C{\'a}novas}, {Cantat-Gaudin}, {Carballo}, {Carlucci}, {Carnerero},
  {Carrasco}, {Casamiquela}, {Castellani}, {Castro-Ginard}, {Castro Sampol},
  {Chaoul}, {Charlot}, {Chemin}, {Chiavassa}, {Cioni}, {Comoretto}, {Cooper},
  {Cornez}, {Cowell}, {Crifo}, {Crosta}, {Crowley}, {Dafonte}, {Dapergolas},
  {David}, {David}, {de Laverny}, {De Luise}, {De March}, {De Ridder}, {de
  Souza}, {de Teodoro}, {de Torres}, {del Peloso}, {del Pozo}, {Delbo},
  {Delgado}, {Delgado}, {Delisle}, {Di Matteo}, {Diakite}, {Diener},
  {Distefano}, {Dolding}, {Eappachen}, {Edvardsson}, {Enke}, {Esquej}, {Fabre},
  {Fabrizio}, {Faigler}, {Fedorets}, {Fernique}, {Fienga}, {Figueras},
  {Fouron}, {Fragkoudi}, {Fraile}, {Franke}, {Gai}, {Garabato},
  {Garcia-Gutierrez}, {Garc{\'\i}a-Torres}, {Garofalo}, {Gavras}, {Gerlach},
  {Geyer}, {Giacobbe}, {Gilmore}, {Girona}, {Giuffrida}, {Gomel}, {Gomez},
  {Gonzalez-Santamaria}, {Gonz{\'a}lez-Vidal}, {Granvik},
  {Guti{\'e}rrez-S{\'a}nchez}, {Guy}, {Hauser}, {Haywood}, {Helmi}, {Hidalgo},
  {Hilger}, {H{\l}adczuk}, {Hobbs}, {Holland}, {Huckle}, {Jasniewicz},
  {Jonker}, {Juaristi Campillo}, {Julbe}, {Karbevska}, {Kervella}, {Khanna},
  {Kochoska}, {Kontizas}, {Kordopatis}, {Korn}, {Kostrzewa-Rutkowska},
  {Kruszy{\'n}ska}, {Lambert}, {Lanza}, {Lasne}, {Le Campion}, {Le Fustec},
  {Lebreton}, {Lebzelter}, {Leccia}, {Leclerc}, {Lecoeur-Taibi}, {Liao},
  {Licata}, {Lindstr{\o}m}, {Lister}, {Livanou}, {Lobel}, {Madrero Pardo},
  {Managau}, {Mann}, {Marchant}, {Marconi}, {Marcos Santos}, {Marinoni},
  {Marocco}, {Marshall}, {Martin Polo}, {Mart{\'\i}n-Fleitas}, {Masip},
  {Massari}, {Mastrobuono-Battisti}, {Mazeh}, {McMillan}, {Messina},
  {Michalik}, {Millar}, {Mints}, {Molina}, {Molinaro}, {Moln{\'a}r},
  {Montegriffo}, {Mor}, {Morbidelli}, {Morel}, {Morris}, {Mulone}, {Munoz},
  {Muraveva}, {Murphy}, {Musella}, {Noval}, {Ord{\'e}novic}, {Orr{\`u}},
  {Osinde}, {Pagani}, {Pagano}, {Palaversa}, {Palicio}, {Panahi}, {Pawlak},
  {Pe{\~n}alosa Esteller}, {Penttil{\"a}}, {Piersimoni}, {Pineau}, {Plachy},
  {Plum}, {Poggio}, {Poretti}, {Poujoulet}, {Pr{\v{s}}a}, {Pulone}, {Racero},
  {Ragaini}, {Rainer}, {Raiteri}, {Rambaux}, {Ramos}, {Ramos-Lerate}, {Re
  Fiorentin}, {Regibo}, {Reyl{\'e}}, {Ripepi}, {Riva}, {Rixon}, {Robichon},
  {Robin}, {Roelens}, {Rohrbasser}, {Romero-G{\'o}mez}, {Rowell}, {Royer},
  {Rybicki}, {Sadowski}, {Sagrist{\`a} Sell{\'e}s}, {Sahlmann}, {Salgado},
  {Salguero}, {Samaras}, {Sanchez Gimenez}, {Sanna}, {Santove{\~n}a},
  {Sarasso}, {Schultheis}, {Sciacca}, {Segol}, {Segovia}, {S{\'e}gransan},
  {Semeux}, {Shahaf}, {Siddiqui}, {Siebert}, {Siltala}, {Slezak}, {Smart},
  {Solano}, {Solitro}, {Souami}, {Souchay}, {Spagna}, {Spoto}, {Steele},
  {Steidelm{\"u}ller}, {Stephenson}, {S{\"u}veges}, {Szabados}, {Szegedi-Elek},
  {Taris}, {Tauran}, {Taylor}, {Teixeira}, {Thuillot}, {Tonello}, {Torra},
  {Torra}, {Turon}, {Unger}, {Vaillant}, {van Dillen}, {Vanel}, {Vecchiato},
  {Viala}, {Vicente}, {Voutsinas}, {Weiler}, {Wevers}, {Wyrzykowski}, {Yoldas},
  {Yvard}, {Zhao}, {Zorec}, {Zucker}, {Zurbach}, \& {Zwitter}}]{Gaia21}
{Gaia Collaboration}, {Brown}, A.~G.~A., {Vallenari}, A., {et~al.} 2021, \aap,
  649, A1

\bibitem[{{Gaia Collaboration} {et~al.}(2016){Gaia Collaboration}, {Prusti},
  {de Bruijne}, {Brown}, {Vallenari}, {Babusiaux}, {Bailer-Jones}, {Bastian},
  {Biermann}, {Evans}, {Eyer}, {Jansen}, {Jordi}, {Klioner}, {Lammers},
  {Lindegren}, {Luri}, {Mignard}, {Milligan}, {Panem}, {Poinsignon},
  {Pourbaix}, {Randich}, {Sarri}, {Sartoretti}, {Siddiqui}, {Soubiran},
  {Valette}, {van Leeuwen}, {Walton}, {Aerts}, {Arenou}, {Cropper}, {Drimmel},
  {H{\o}g}, {Katz}, {Lattanzi}, {O'Mullane}, {Grebel}, {Holland}, {Huc},
  {Passot}, {Bramante}, {Cacciari}, {Casta{\~n}eda}, {Chaoul}, {Cheek}, {De
  Angeli}, {Fabricius}, {Guerra}, {Hern{\'a}ndez}, {Jean-Antoine-Piccolo},
  {Masana}, {Messineo}, {Mowlavi}, {Nienartowicz}, {Ord{\'o}{\~n}ez-Blanco},
  {Panuzzo}, {Portell}, {Richards}, {Riello}, {Seabroke}, {Tanga},
  {Th{\'e}venin}, {Torra}, {Els}, {Gracia-Abril}, {Comoretto},
  {Garcia-Reinaldos}, {Lock}, {Mercier}, {Altmann}, {Andrae}, {Astraatmadja},
  {Bellas-Velidis}, {Benson}, {Berthier}, {Blomme}, {Busso}, {Carry},
  {Cellino}, {Clementini}, {Cowell}, {Creevey}, {Cuypers}, {Davidson}, {De
  Ridder}, {de Torres}, {Delchambre}, {Dell'Oro}, {Ducourant}, {Fr{\'e}mat},
  {Garc{\'\i}a-Torres}, {Gosset}, {Halbwachs}, {Hambly}, {Harrison}, {Hauser},
  {Hestroffer}, {Hodgkin}, {Huckle}, {Hutton}, {Jasniewicz}, {Jordan},
  {Kontizas}, {Korn}, {Lanzafame}, {Manteiga}, {Moitinho}, {Muinonen},
  {Osinde}, {Pancino}, {Pauwels}, {Petit}, {Recio-Blanco}, {Robin}, {Sarro},
  {Siopis}, {Smith}, {Smith}, {Sozzetti}, {Thuillot}, {van Reeven}, {Viala},
  {Abbas}, {Abreu Aramburu}, {Accart}, {Aguado}, {Allan}, {Allasia},
  {Altavilla}, {{\'A}lvarez}, {Alves}, {Anderson}, {Andrei}, {Anglada Varela},
  {Antiche}, {Antoja}, {Ant{\'o}n}, {Arcay}, {Atzei}, {Ayache}, {Bach},
  {Baker}, {Balaguer-N{\'u}{\~n}ez}, {Barache}, {Barata}, {Barbier}, {Barblan},
  {Baroni}, {Barrado y Navascu{\'e}s}, {Barros}, {Barstow}, {Becciani},
  {Bellazzini}, {Bellei}, {Bello Garc{\'\i}a}, {Belokurov}, {Bendjoya},
  {Berihuete}, {Bianchi}, {Bienaym{\'e}}, {Billebaud}, {Blagorodnova},
  {Blanco-Cuaresma}, {Boch}, {Bombrun}, {Borrachero}, {Bouquillon}, {Bourda},
  {Bouy}, {Bragaglia}, {Breddels}, {Brouillet}, {Br{\"u}semeister},
  {Bucciarelli}, {Budnik}, {Burgess}, {Burgon}, {Burlacu}, {Busonero}, {Buzzi},
  {Caffau}, {Cambras}, {Campbell}, {Cancelliere}, {Cantat-Gaudin}, {Carlucci},
  {Carrasco}, {Castellani}, {Charlot}, {Charnas}, {Charvet}, {Chassat},
  {Chiavassa}, {Clotet}, {Cocozza}, {Collins}, {Collins}, {Costigan}, {Crifo},
  {Cross}, {Crosta}, {Crowley}, {Dafonte}, {Damerdji}, {Dapergolas}, {David},
  {David}, {De Cat}, {de Felice}, {de Laverny}, {De Luise}, {De March}, {de
  Martino}, {de Souza}, {Debosscher}, {del Pozo}, {Delbo}, {Delgado},
  {Delgado}, {di Marco}, {Di Matteo}, {Diakite}, {Distefano}, {Dolding}, {Dos
  Anjos}, {Drazinos}, {Dur{\'a}n}, {Dzigan}, {Ecale}, {Edvardsson}, {Enke},
  {Erdmann}, {Escolar}, {Espina}, {Evans}, {Eynard Bontemps}, {Fabre},
  {Fabrizio}, {Faigler}, {Falc{\~a}o}, {Farr{\`a}s Casas}, {Faye}, {Federici},
  {Fedorets}, {Fern{\'a}ndez-Hern{\'a}ndez}, {Fernique}, {Fienga}, {Figueras},
  {Filippi}, {Findeisen}, {Fonti}, {Fouesneau}, {Fraile}, {Fraser}, {Fuchs},
  {Furnell}, {Gai}, {Galleti}, {Galluccio}, {Garabato}, {Garc{\'\i}a-Sedano},
  {Gar{\'e}}, {Garofalo}, {Garralda}, {Gavras}, {Gerssen}, {Geyer}, {Gilmore},
  {Girona}, {Giuffrida}, {Gomes}, {Gonz{\'a}lez-Marcos},
  {Gonz{\'a}lez-N{\'u}{\~n}ez}, {Gonz{\'a}lez-Vidal}, {Granvik}, {Guerrier},
  {Guillout}, {Guiraud}, {G{\'u}rpide}, {Guti{\'e}rrez-S{\'a}nchez}, {Guy},
  {Haigron}, {Hatzidimitriou}, {Haywood}, {Heiter}, {Helmi}, {Hobbs},
  {Hofmann}, {Holl}, {Holland}, {Hunt}, {Hypki}, {Icardi}, {Irwin}, {Jevardat
  de Fombelle}, {Jofr{\'e}}, {Jonker}, {Jorissen}, {Julbe}, {Karampelas},
  {Kochoska}, {Kohley}, {Kolenberg}, {Kontizas}, {Koposov}, {Kordopatis},
  {Koubsky}, {Kowalczyk}, {Krone-Martins}, {Kudryashova}, {Kull}, {Bachchan},
  {Lacoste-Seris}, {Lanza}, {Lavigne}, {Le Poncin-Lafitte}, {Lebreton},
  {Lebzelter}, {Leccia}, {Leclerc}, {Lecoeur-Taibi}, {Lemaitre}, {Lenhardt},
  {Leroux}, {Liao}, {Licata}, {Lindstr{\o}m}, {Lister}, {Livanou}, {Lobel},
  {L{\"o}ffler}, {L{\'o}pez}, {Lopez-Lozano}, {Lorenz}, {Loureiro},
  {MacDonald}, {Magalh{\~a}es Fernandes}, {Managau}, {Mann}, {Mantelet},
  {Marchal}, {Marchant}, {Marconi}, {Marie}, {Marinoni}, {Marrese},
  {Marschalk{\'o}}, {Marshall}, {Mart{\'\i}n-Fleitas}, {Martino}, {Mary},
  {Matijevi{\v{c}}}, {Mazeh}, {McMillan}, {Messina}, {Mestre}, {Michalik},
  {Millar}, {Miranda}, {Molina}, {Molinaro}, {Molinaro}, {Moln{\'a}r},
  {Moniez}, {Montegriffo}, {Monteiro}, {Mor}, {Mora}, {Morbidelli}, {Morel},
  {Morgenthaler}, {Morley}, {Morris}, {Mulone}, {Muraveva}, {Musella},
  {Narbonne}, {Nelemans}, {Nicastro}, {Noval}, {Ord{\'e}novic},
  {Ordieres-Mer{\'e}}, {Osborne}, {Pagani}, {Pagano}, {Pailler}, {Palacin},
  {Palaversa}, {Parsons}, {Paulsen}, {Pecoraro}, {Pedrosa}, {Pentik{\"a}inen},
  {Pereira}, {Pichon}, {Piersimoni}, {Pineau}, {Plachy}, {Plum}, {Poujoulet},
  {Pr{\v{s}}a}, {Pulone}, {Ragaini}, {Rago}, {Rambaux}, {Ramos-Lerate},
  {Ranalli}, {Rauw}, {Read}, {Regibo}, {Renk}, {Reyl{\'e}}, {Ribeiro},
  {Rimoldini}, {Ripepi}, {Riva}, {Rixon}, {Roelens}, {Romero-G{\'o}mez},
  {Rowell}, {Royer}, {Rudolph}, {Ruiz-Dern}, {Sadowski}, {Sagrist{\`a}
  Sell{\'e}s}, {Sahlmann}, {Salgado}, {Salguero}, {Sarasso}, {Savietto},
  {Schnorhk}, {Schultheis}, {Sciacca}, {Segol}, {Segovia}, {Segransan},
  {Serpell}, {Shih}, {Smareglia}, {Smart}, {Smith}, {Solano}, {Solitro},
  {Sordo}, {Soria Nieto}, {Souchay}, {Spagna}, {Spoto}, {Stampa}, {Steele},
  {Steidelm{\"u}ller}, {Stephenson}, {Stoev}, {Suess}, {S{\"u}veges}, {Surdej},
  {Szabados}, {Szegedi-Elek}, {Tapiador}, {Taris}, {Tauran}, {Taylor},
  {Teixeira}, {Terrett}, {Tingley}, {Trager}, {Turon}, {Ulla}, {Utrilla},
  {Valentini}, {van Elteren}, {Van Hemelryck}, {van Leeuwen}, {Varadi},
  {Vecchiato}, {Veljanoski}, {Via}, {Vicente}, {Vogt}, {Voss}, {Votruba},
  {Voutsinas}, {Walmsley}, {Weiler}, {Weingrill}, {Werner}, {Wevers},
  {Whitehead}, {Wyrzykowski}, {Yoldas}, {{\v{Z}}erjal}, {Zucker}, {Zurbach},
  {Zwitter}, {Alecu}, {Allen}, {Allende Prieto}, {Amorim},
  {Anglada-Escud{\'e}}, {Arsenijevic}, {Azaz}, {Balm}, {Beck}, {Bernstein},
  {Bigot}, {Bijaoui}, {Blasco}, {Bonfigli}, {Bono}, {Boudreault}, {Bressan},
  {Brown}, {Brunet}, {Bunclark}, {Buonanno}, {Butkevich}, {Carret}, {Carrion},
  {Chemin}, {Ch{\'e}reau}, {Corcione}, {Darmigny}, {de Boer}, {de Teodoro}, {de
  Zeeuw}, {Delle Luche}, {Domingues}, {Dubath}, {Fodor}, {Fr{\'e}zouls},
  {Fries}, {Fustes}, {Fyfe}, {Gallardo}, {Gallegos}, {Gardiol}, {Gebran},
  {Gomboc}, {G{\'o}mez}, {Grux}, {Gueguen}, {Heyrovsky}, {Hoar}, {Iannicola},
  {Isasi Parache}, {Janotto}, {Joliet}, {Jonckheere}, {Keil}, {Kim},
  {Klagyivik}, {Klar}, {Knude}, {Kochukhov}, {Kolka}, {Kos}, {Kutka}, {Lainey},
  {LeBouquin}, {Liu}, {Loreggia}, {Makarov}, {Marseille}, {Martayan},
  {Martinez-Rubi}, {Massart}, {Meynadier}, {Mignot}, {Munari}, {Nguyen},
  {Nordlander}, {Ocvirk}, {O'Flaherty}, {Olias Sanz}, {Ortiz}, {Osorio},
  {Oszkiewicz}, {Ouzounis}, {Palmer}, {Park}, {Pasquato}, {Peltzer}, {Peralta},
  {P{\'e}turaud}, {Pieniluoma}, {Pigozzi}, {Poels}, {Prat}, {Prod'homme},
  {Raison}, {Rebordao}, {Risquez}, {Rocca-Volmerange}, {Rosen}, {Ruiz-Fuertes},
  {Russo}, {Sembay}, {Serraller Vizcaino}, {Short}, {Siebert}, {Silva},
  {Sinachopoulos}, {Slezak}, {Soffel}, {Sosnowska}, {Strai{\v{z}}ys}, {ter
  Linden}, {Terrell}, {Theil}, {Tiede}, {Troisi}, {Tsalmantza}, {Tur},
  {Vaccari}, {Vachier}, {Valles}, {Van Hamme}, {Veltz}, {Virtanen}, {Wallut},
  {Wichmann}, {Wilkinson}, {Ziaeepour}, \& {Zschocke}}]{Gaia16}
{Gaia Collaboration}, {Prusti}, T., {de Bruijne}, J.~H.~J., {et~al.} 2016,
  \aap, 595, A1

\bibitem[{{Galazutdinov} {et~al.}(2017){Galazutdinov}, {Lee}, {Han}, {Lee},
  {Valyavin}, \& {Kre{\l}owski}}]{2017MNRAS.467.3099G}
{Galazutdinov}, G.~A., {Lee}, J.-J., {Han}, I., {et~al.} 2017, \mnras, 467,
  3099

\bibitem[{{Galazutdinov} {et~al.}(2004){Galazutdinov}, {Manic{\`o}},
  {Pirronello}, \& {Kre{\l}owski}}]{Galazutdinovetal04}
{Galazutdinov}, G.~A., {Manic{\`o}}, G., {Pirronello}, V., \& {Kre{\l}owski},
  J. 2004, \mnras, 355, 169

\bibitem[{{Geballe} {et~al.}(2011){Geballe}, {Najarro}, {Figer},
  {Schlegelmilch}, \& {de La Fuente}}]{geballe2011}
{Geballe}, T.~R., {Najarro}, F., {Figer}, D.~F., {Schlegelmilch}, B.~W., \& {de
  La Fuente}, D. 2011, \nat, 479, 200

\bibitem[{{Giddings}(1981)}]{giddings1981}
{Giddings}, J.~R. 1981, PhD thesis, (Univ. London)

\bibitem[{{Gonz{\'a}lez} {et~al.}(2019){Gonz{\'a}lez}, {Briquet}, {Przybilla},
  {Nieva}, {De Cat}, {Saesen}, {Hubrig}, {Thoul}, {P{\'a}pics}, {Palaversa},
  {Naef}, {Neveu-Van Malle}, {J{\"a}rvinen}, {Pollard}, {Kilmartin}, {Mowlavi},
  \& {Butler}}]{Gonzalezetal19}
{Gonz{\'a}lez}, J.~F., {Briquet}, M., {Przybilla}, N., {et~al.} 2019, \aap,
  626, A94

\bibitem[{{Gonz{\'a}lez} {et~al.}(2017){Gonz{\'a}lez}, {Hubrig}, {Przybilla},
  {Carroll}, {Nieva}, {Ilyin}, {J{\"a}rvinen}, {Morel}, {Sch{\"o}ller},
  {Castro}, {Barb{\'a}}, {de Koter}, {Schneider}, {Kholtygin}, {Butler},
  {Veramendi}, {Langer}, \& {BOB Collaboration}}]{Gonzalezetal17}
{Gonz{\'a}lez}, J.~F., {Hubrig}, S., {Przybilla}, N., {et~al.} 2017, \mnras,
  467, 437

\bibitem[{{Groh} {et~al.}(2007){Groh}, {Damineli}, \&
  {Jablonski}}]{2007A&A...465..993G}
{Groh}, J.~H., {Damineli}, A., \& {Jablonski}, F. 2007, \aap, 465, 993

\bibitem[{{Hamano} {et~al.}(2015){Hamano}, {Kobayashi}, {Kondo}, {Ikeda},
  {Nakanishi}, {Yasui}, {Mizumoto}, {Matsunaga}, {Fukue}, {Mito}, {Yamamoto},
  {Izumi}, {Nakaoka}, {Kawanishi}, {Kitano}, {Otsubo}, {Kinoshita},
  {Kobayashi}, \& {Kawakita}}]{2015ApJ...800..137H}
{Hamano}, S., {Kobayashi}, N., {Kondo}, S., {et~al.} 2015, \apj, 800, 137

\bibitem[{{Heckman} \& {Lehnert}(2000)}]{HeLe00}
{Heckman}, T.~M. \& {Lehnert}, M.~D. 2000, \apj, 537, 690

\bibitem[{{Heger}(1922)}]{Heger22}
{Heger}, M.~L. 1922, Lick Observatory Bulletin, 10, 141

\bibitem[{{Helou} \& {Walker}(1988)}]{HeWa88}
{Helou}, G. \& {Walker}, D.~W., eds. 1988, {Infrared Astronomical Satellite
  (IRAS) Catalogs and Atlases.Volume 7: The Small Scale Structure Catalog.},
  Vol.~7

\bibitem[{{Henyey} \& {Greenstein}(1941)}]{1941ApJ....93...70H}
{Henyey}, L.~G. \& {Greenstein}, J.~L. 1941, \apj, 93, 70

\bibitem[{{Herbig}(1995)}]{1995ARA&A..33...19H}
{Herbig}, G.~H. 1995, \araa, 33, 19

\bibitem[{{Hobbs} {et~al.}(2009){Hobbs}, {York}, {Thorburn}, {Snow}, {Bishof},
  {Friedman}, {McCall}, {Oka}, {Rachford}, {Sonnentrucker}, \&
  {Welty}}]{Hobbsetal09}
{Hobbs}, L.~M., {York}, D.~G., {Thorburn}, J.~A., {et~al.} 2009, \apj, 705, 32

\bibitem[{{Hou} \& {Han}(2014)}]{hou2014}
{Hou}, L.~G. \& {Han}, J.~L. 2014, \aap, 569, A125

\bibitem[{{Joblin} {et~al.}(1990){Joblin}, {Maillard}, {D'Hendecourt}, \&
  {L{\'e}ger}}]{1990Natur.346..729J}
{Joblin}, C., {Maillard}, J.~P., {D'Hendecourt}, L., \& {L{\'e}ger}, A. 1990,
  \nat, 346, 729

\bibitem[{{Jung}(2014)}]{crires_pipeline_man}
{Jung}, Y. 2014, CRIRES Pipeline User Manual, VLT-MAN-ESO-019500-4406 (ESO)

\bibitem[{{Junkkarinen} {et~al.}(2004){Junkkarinen}, {Cohen}, {Beaver},
  {Burbidge}, {Lyons}, \& {Madejski}}]{Junkkarinenetal04}
{Junkkarinen}, V.~T., {Cohen}, R.~D., {Beaver}, E.~A., {et~al.} 2004, \apj,
  614, 658

\bibitem[{{Kaufer} {et~al.}(1999){Kaufer}, {Stahl}, {Tubbesing},
  {N{\o}rregaard}, {Avila}, {Francois}, {Pasquini}, \&
  {Pizzella}}]{Kauferetal99}
{Kaufer}, A., {Stahl}, O., {Tubbesing}, S., {et~al.} 1999, The Messenger, 95, 8

\bibitem[{{Kaufer} {et~al.}(1996){Kaufer}, {Stahl}, {Wolf}, {Gaeng},
  {Gummersbach}, {Kovacs}, {Mandel}, \& {Szeifert}}]{Kauferetal96}
{Kaufer}, A., {Stahl}, O., {Wolf}, B., {et~al.} 1996, \aap, 305, 887

\bibitem[{{K\"aufl} {et~al.}(2004){K\"aufl}, {Ballester}, {Biereichel},
  {Delabre}, {Donaldson}, {Dorn}, {Fedrigo}, {Finger}, {Fischer}, {Franza},
  {Gojak}, {Huster}, {Jung}, {Lizon}, {Mehrgan}, {Meyer}, {Moorwood}, {Pirard},
  {Paufique}, {Pozna}, {Siebenmorgen}, {Silber}, {Stegmeier}, \&
  {Wegerer}}]{Kaeufletal04}
{K\"aufl}, H.-U., {Ballester}, P., {Biereichel}, P., {et~al.} 2004, Proc. SPIE,
  5492, 1218

\bibitem[{{Kausch} {et~al.}(2015){Kausch}, {Noll}, {Smette}, {Kimeswenger},
  {Barden}, {Szyszka}, {Jones}, {Sana}, {Horst}, \& {Kerber}}]{Kauschetal15}
{Kausch}, W., {Noll}, S., {Smette}, A., {et~al.} 2015, \aap, 576, A78

\bibitem[{{Kimeswenger} {et~al.}(2021){Kimeswenger}, {Rainer}, {Przybilla}, \&
  {Kausch}}]{Kimeswengeretal21}
{Kimeswenger}, S., {Rainer}, M., {Przybilla}, N., \& {Kausch}, W. 2021, \aj,
  161, 66

\bibitem[{Kramida {et~al.}(2019)Kramida, {Yu.~Ralchenko}, Reader, \& {and NIST
  ASD Team}}]{NIST_ASD}
Kramida, A., {Yu.~Ralchenko}, Reader, J., \& {and NIST ASD Team}. 2019, {NIST
  Atomic Spectra Database (ver. 5.7.1), [Online]. Available:
  {\tt{https://physics.nist.gov/asd}} [2020, August 31]. National Institute of
  Standards and Technology, Gaithersburg, MD.}

\bibitem[{{Krelowski} \& {Walker}(1987)}]{1987Krelowski}
{Krelowski}, J. \& {Walker}, G.~A.~H. 1987, \apj, 312, 860

\bibitem[{{Kudritzki} {et~al.}(2014){Kudritzki}, {Urbaneja}, {Bresolin},
  {Hosek}, \& {Przybilla}}]{Kudritzkietal14}
{Kudritzki}, R.-P., {Urbaneja}, M.~A., {Bresolin}, F., {Hosek}, Matthew~W., J.,
  \& {Przybilla}, N. 2014, \apj, 788, 56

\bibitem[{{Kudritzki} {et~al.}(2008){Kudritzki}, {Urbaneja}, {Bresolin}, \&
  {Przybilla}}]{Kudritzkietal08}
{Kudritzki}, R.~P., {Urbaneja}, M.~A., {Bresolin}, F., \& {Przybilla}, N. 2008,
  Physica Scripta Volume T, 133, 014039

\bibitem[{{Kudritzki} {et~al.}(2013){Kudritzki}, {Urbaneja}, {Gazak}, {Macri},
  {Hosek}, {Bresolin}, \& {Przybilla}}]{Kudritzkietal13}
{Kudritzki}, R.-P., {Urbaneja}, M.~A., {Gazak}, Z., {et~al.} 2013, \apjl, 779,
  L20

\bibitem[{{Kurucz}(1993)}]{kur1993b}
{Kurucz}, R. 1993, CD-ROM No. 13 (Cambridge, Mass: SAO)

\bibitem[{{Lallement} {et~al.}(2018){Lallement}, {Cox}, {Cami}, {Smoker},
  {Farhang}, {Elyajouri}, {Cordiner}, {Linnartz}, {Smith}, {Ehrenfreund}, \&
  {Foing}}]{2018A&A...614A..28L}
{Lallement}, R., {Cox}, N.~L.~J., {Cami}, J., {et~al.} 2018, \aap, 614, A28

\bibitem[{{Li} \& {Draine}(2001)}]{2001ApJ...554..778L}
{Li}, A. \& {Draine}, B.~T. 2001, \apj, 554, 778

\bibitem[{{Massa} {et~al.}(1983){Massa}, {Savage}, \&
  {Fitzpatrick}}]{Massaetal83}
{Massa}, D., {Savage}, B.~D., \& {Fitzpatrick}, E.~L. 1983, \apj, 266, 662

\bibitem[{{Maza} {et~al.}(2014){Maza}, {Nieva}, \& {Przybilla}}]{Mazaetal14}
{Maza}, N.~L., {Nieva}, M.-F., \& {Przybilla}, N. 2014, \aap, 572, A112

\bibitem[{McCall {et~al.}(2010)McCall, Drosback, Thorburn, York, Friedman,
  Hobbs, Rachford, Snow, Sonnentrucker, \& Welty}]{McCall_2009}
McCall, B.~J., Drosback, M.~M., Thorburn, J.~A., {et~al.} 2010, \apj, 708, 1628

\bibitem[{{Mermilliod}(1997)}]{Mermilliod97}
{Mermilliod}, J.~C. 1997, VizieR Online Data Catalog, II/168

\bibitem[{{Morton}(1991)}]{1991ApJS...77..119M}
{Morton}, D.~C. 1991, \apjs, 77, 119

\bibitem[{{Nieva} \& {Przybilla}(2007)}]{NiPr07}
{Nieva}, M.~F. \& {Przybilla}, N. 2007, \aap, 467, 295

\bibitem[{{Nieva} \& {Przybilla}(2012)}]{NiPr12}
{Nieva}, M.~F. \& {Przybilla}, N. 2012, \aap, 539, A143

\bibitem[{{Nieva} \& {Przybilla}(2014)}]{NiPr14}
{Nieva}, M.-F. \& {Przybilla}, N. 2014, \aap, 566, A7

\bibitem[{{Omont}(2016)}]{Omont16}
{Omont}, A. 2016, \aap, 590, A52

\bibitem[{{Omont} {et~al.}(2019){Omont}, {Bettinger}, \&
  {T{\"o}nshoff}}]{Omontetal19}
{Omont}, A., {Bettinger}, H.~F., \& {T{\"o}nshoff}, C. 2019, \aap, 625, A41

\bibitem[{{Pfeiffer} {et~al.}(1998){Pfeiffer}, {Frank}, {Baumueller},
  {Fuhrmann}, \& {Gehren}}]{Pfeifferetal98}
{Pfeiffer}, M.~J., {Frank}, C., {Baumueller}, D., {Fuhrmann}, K., \& {Gehren},
  T. 1998, \aaps, 130, 381

\bibitem[{{Przybilla}(2005)}]{Przybilla05}
{Przybilla}, N. 2005, \aap, 443, 293

\bibitem[{{Przybilla} \& {Butler}(2001)}]{PrBu01}
{Przybilla}, N. \& {Butler}, K. 2001, \aap, 379, 955

\bibitem[{{Przybilla} \& {Butler}(2004)}]{PrBu04}
{Przybilla}, N. \& {Butler}, K. 2004, \apj, 609, 1181

\bibitem[{{Przybilla} {et~al.}(2001{\natexlab{a}}){Przybilla}, {Butler},
  {Becker}, \& {Kudritzki}}]{Przybillaetal01a}
{Przybilla}, N., {Butler}, K., {Becker}, S.~R., \& {Kudritzki}, R.~P.
  2001{\natexlab{a}}, \aap, 369, 1009

\bibitem[{{Przybilla} {et~al.}(2006){Przybilla}, {Butler}, {Becker}, \&
  {Kudritzki}}]{2006A&A...445.1099P}
{Przybilla}, N., {Butler}, K., {Becker}, S.~R., \& {Kudritzki}, R.~P. 2006,
  \aap, 445, 1099

\bibitem[{{Przybilla} {et~al.}(2000){Przybilla}, {Butler}, {Becker},
  {Kudritzki}, \& {Venn}}]{Przybillaetal00}
{Przybilla}, N., {Butler}, K., {Becker}, S.~R., {Kudritzki}, R.~P., \& {Venn},
  K.~A. 2000, \aap, 359, 1085

\bibitem[{{Przybilla} {et~al.}(2001{\natexlab{b}}){Przybilla}, {Butler}, \&
  {Kudritzki}}]{Przybillaetal01b}
{Przybilla}, N., {Butler}, K., \& {Kudritzki}, R.~P. 2001{\natexlab{b}}, \aap,
  379, 936

\bibitem[{{Przybilla} {et~al.}(2016){Przybilla}, {Fossati}, {Hubrig}, {Nieva},
  {J{\"a}rvinen}, {Castro}, {Sch{\"o}ller}, {Ilyin}, {Butler}, {Schneider},
  {Oskinova}, {Morel}, {Langer}, {de Koter}, \& {BOB
  Collaboration}}]{Przybillaetal16}
{Przybilla}, N., {Fossati}, L., {Hubrig}, S., {et~al.} 2016, \aap, 587, A7

\bibitem[{{Przybilla} {et~al.}(2021){Przybilla}, {Fossati}, \&
  {Jeffery}}]{Przybillaetal21}
{Przybilla}, N., {Fossati}, L., \& {Jeffery}, C.~S. 2021, \aap, 654, A119

\bibitem[{{Przybilla} {et~al.}(2008){Przybilla}, {Nieva}, {Tillich}, {Heber},
  {Butler}, \& {Brown}}]{Przybillaetal08}
{Przybilla}, N., {Nieva}, M.~F., {Tillich}, A., {et~al.} 2008, \aap, 488, L51

\bibitem[{{Ram{\'\i}rez-Tannus} {et~al.}(2018){Ram{\'\i}rez-Tannus}, {Cox},
  {Kaper}, \& {de Koter}}]{2018A&A...620A..52R}
{Ram{\'\i}rez-Tannus}, M.~C., {Cox}, N.~L.~J., {Kaper}, L., \& {de Koter}, A.
  2018, \aap, 620, A52

\bibitem[{{Rothman} {et~al.}(2005){Rothman}, {Jacquemart}, {Barbe}, {Chris
  Benner}, {Birk}, {Brown}, {Carleer}, {Chackerian}, {Chance}, {Coudert},
  {Dana}, {Devi}, {Flaud}, {Gamache}, {Goldman}, {Hartmann}, {Jucks}, {Maki},
  {Mandin}, {Massie}, {Orphal}, {Perrin}, {Rinsland}, {Smith}, {Tennyson},
  {Tolchenov}, {Toth}, {Vander Auwera}, {Varanasi}, \&
  {Wagner}}]{2005JQSRT..96..139R}
{Rothman}, L.~S., {Jacquemart}, D., {Barbe}, A., {et~al.} 2005, \jqsrt, 96, 139

\bibitem[{{Sarre}(2014)}]{2014IAUS..297...34S}
{Sarre}, P.~J. 2014, in The Diffuse Interstellar Bands, ed. J.~{Cami} \&
  N.~L.~J. {Cox}, Vol. 297, 34--40

\bibitem[{{Sarre} {et~al.}(1995){Sarre}, {Miles}, {Kerr}, {Hibbins}, {Fossey},
  \& {Somerville}}]{1995MNRAS.277L..41S}
{Sarre}, P.~J., {Miles}, J.~R., {Kerr}, T.~H., {et~al.} 1995, \mnras, 277, L41

\bibitem[{{Schiller} \& {Przybilla}(2008)}]{SchPr08}
{Schiller}, F. \& {Przybilla}, N. 2008, \aap, 479, 849

\bibitem[{{Sch{\"o}nrich} {et~al.}(2010){Sch{\"o}nrich}, {Binney}, \&
  {Dehnen}}]{2010schoenrich}
{Sch{\"o}nrich}, R., {Binney}, J., \& {Dehnen}, W. 2010, \mnras, 403, 1829

\bibitem[{{Seifahrt} {et~al.}(2010){Seifahrt}, {K{\"a}ufl}, {Z{\"a}ngl},
  {Bean}, {Richter}, \& {Siebenmorgen}}]{Seifahrtetal10}
{Seifahrt}, A., {K{\"a}ufl}, H.~U., {Z{\"a}ngl}, G., {et~al.} 2010, \aap, 524,
  A11

\bibitem[{{Siebenmorgen} {et~al.}(2014){Siebenmorgen}, {Voshchinnikov}, \&
  {Bagnulo}}]{2014A&A...561A..82S}
{Siebenmorgen}, R., {Voshchinnikov}, N.~V., \& {Bagnulo}, S. 2014, \aap, 561,
  A82

\bibitem[{{Skrutskie} {et~al.}(2006){Skrutskie}, {Cutri}, {Stiening},
  {Weinberg}, {Schneider}, {Carpenter}, {Beichman}, {Capps}, {Chester},
  {Elias}, {Huchra}, {Liebert}, {Lonsdale}, {Monet}, {Price}, {Seitzer},
  {Jarrett}, {Kirkpatrick}, {Gizis}, {Howard}, {Evans}, {Fowler}, {Fullmer},
  {Hurt}, {Light}, {Kopan}, {Marsh}, {McCallon}, {Tam}, {Van Dyk}, \&
  {Wheelock}}]{2006Skrutskie}
{Skrutskie}, M.~F., {Cutri}, R.~M., {Stiening}, R., {et~al.} 2006, \aj, 131,
  1163

\bibitem[{{Smette} {et~al.}(2015){Smette}, {Sana}, {Noll}, {Horst}, {Kausch},
  {Kimeswenger}, {Barden}, {Szyszka}, {Jones}, {Gallenne}, {Vinther},
  {Ballester}, \& {Taylor}}]{molecfit}
{Smette}, A., {Sana}, H., {Noll}, S., {et~al.} 2015, \aap, 576, A77

\bibitem[{{Snow}(2014)}]{Snow14}
{Snow}, T.~P. 2014, in The Diffuse Interstellar Bands, ed. J.~{Cami} \&
  N.~L.~J. {Cox}, Vol. 297, 3--12

\bibitem[{{Spieler} {et~al.}(2017){Spieler}, {Kuhn}, {Postler}, {Simpson},
  {Wester}, {Scheier}, {Ubachs}, {Bacalla}, {Bouwman}, \&
  {Linnartz}}]{2017ApJ...846..168S}
{Spieler}, S., {Kuhn}, M., {Postler}, J., {et~al.} 2017, \apj, 846, 168

\bibitem[{{Tielens}(2014)}]{Tielens14}
{Tielens}, A.~G.~G.~M. 2014, in The Diffuse Interstellar Bands, ed. J.~{Cami}
  \& N.~L.~J. {Cox}, Vol. 297, 399--411

\bibitem[{{Vos} {et~al.}(2011){Vos}, {Cox}, {Kaper}, {Spaans}, \&
  {Ehrenfreund}}]{2011A&A...533A.129V}
{Vos}, D.~A.~I., {Cox}, N.~L.~J., {Kaper}, L., {Spaans}, M., \& {Ehrenfreund},
  P. 2011, \aap, 533, A129

\bibitem[{{Wesselius} {et~al.}(1982){Wesselius}, {van Duinen}, {de Jonge},
  {Aalders}, {Luinge}, \& {Wildeman}}]{1982A&AS...49..427W}
{Wesselius}, P.~R., {van Duinen}, R.~J., {de Jonge}, A.~R.~W., {et~al.} 1982,
  \aaps, 49, 427

\bibitem[{{Zubko} {et~al.}(2004){Zubko}, {Dwek}, \&
  {Arendt}}]{2004ApJS..152..211Z}
{Zubko}, V., {Dwek}, E., \& {Arendt}, R.~G. 2004, \apjs, 152, 211

\end{thebibliography}
   \bibliographystyle{aa} 

%
\begin{appendix}
\setcounter{section}{1}
\setcounter{figure}{0}

   \begin{landscape}

    \begin{table}
    \begin{flushleft}{\large\textsf{\textbf{Appendix A: DIB measurements}}}\end{flushleft}
    \caption{Characterisation of the detected DIBs and new DIB candidates.}
    \setlength{\tabcolsep}{1.1mm}
    \resizebox{\columnwidth}{!}{
        {\small
    \begin{tabular}{lcc@{\extracolsep{4pt}}rrrcrrrcrrrcrrrc@{}}
    
\hline\hline
DIB & Ref. &                  Lab frame & \multicolumn{4}{c}{HD 183143} & \multicolumn{4}{c}{HD 165784} & \multicolumn{4}{c}{HD 92207} & \multicolumn{4}{c}{HD 111613} \\
\cline{4-7} \cline{8-11} \cline{12-15} \cline{16-19}
{} &   \  & $\lambda_\mathrm{0,air}$ & $\lambda_\mathrm{vac}$ &           $EW\ \ $ &    $A_\mathrm{c}\ \ $ & $FWHM$ & $\lambda_\mathrm{vac}$ &           $EW\ \ $ &   $A_\mathrm{c}\ \ $ & $FWHM$ & $\lambda_\mathrm{vac}$ &          $EW\ \ $ &   $A_\mathrm{c}\ \ $ & $FWHM$ & $\lambda_\mathrm{vac}$ &           $EW\ \ $ &   $A_\mathrm{c}\ \ $ & $FWHM$ \\
{} &   \  &                      (\AA) &                  (\AA) &       (m\AA)$\ \ $ &            $(\%)\ \ $ &  (\AA) &                  (\AA) &       (m\AA)$\ \ $ &           $(\%)\ \ $ &  (\AA) &                  (\AA) &      (m\AA)$\ \ $ &           $(\%)\ \ $ &  (\AA) &                  (\AA) &       (m\AA)$\ \ $ &           $(\%)\ \ $ &  (\AA) \\
\midrule
9632\tablefootmark{a}  &    2 &   $9632.6\,\pm\,0.8\ \ $ &                 9634.6 &    $316\,\pm\,99$ &     $12.0\,\pm\,11.6$ &    2.7 &                 9635.8 &      $77\,\pm\,14$ &  $3.9\,\pm\,2.1\ \ $ &    2.5 &                    ... &         ...$\ \ $ &            ...$\ \ $ &    ... &                    ... &          ...$\ \ $ &            ...$\ \ $ &    ... \\
9880  &    5 &   $~\,9880.4\,\pm\,0.1\ \ $ &                 9883.1 &   $44\,\pm\,6\ \ $ &   $2.9\,\pm\,0.9\ \ $ &    1.3 &                 9883.2 &   $21\,\pm\,3\ \ $ &  $1.2\,\pm\,0.3\ \ $ &    1.4 &                    ... &         ...$\ \ $ &            ...$\ \ $ &    ... &                    ... &          ...$\ \ $ &            ...$\ \ $ &    ... \\
10125 &    6 &  $10125.1\,\pm\,1.1\ \ $ &                10128.6 &    $9\,\pm\,3\ \ $ &   $1.0\,\pm\,0.7\ \ $ &    0.6 &                10127.2 &    $3\,\pm\,2\ \ $ &  $0.5\,\pm\,0.6\ \ $ &    0.6 &                    ... &         ...$\ \ $ &            ...$\ \ $ &    ... &                    ... &          ...$\ \ $ &            ...$\ \ $ &    ... \\
10262 &    6 &  $10262.5\,\pm\,0.2\ \ $ &                10265.4 &    $9\,\pm\,3\ \ $ &   $1.1\,\pm\,0.5\ \ $ &    0.8 &                10265.3 &    $5\,\pm\,3\ \ $ &  $0.6\,\pm\,0.5\ \ $ &    1.2 &                    ... &         ...$\ \ $ &            ...$\ \ $ &    ... &                    ... &          ...$\ \ $ &            ...$\ \ $ &    ... \\
10361 &    4 &  $10361.0\,\pm\,0.1\ \ $ &                10363.6 &   $30\,\pm\,3\ \ $ &   $1.9\,\pm\,0.4\ \ $ &    1.3 &                10363.9 &   $24\,\pm\,5\ \ $ &  $1.1\,\pm\,0.6\ \ $ &    1.9 &                    ... &         ...$\ \ $ &            ...$\ \ $ &    ... &                    ... &          ...$\ \ $ &            ...$\ \ $ &    ... \\
10393 &    4 &  $10392.5\,\pm\,1.5\ \ $ &                10395.9 &   $17\,\pm\,4\ \ $ &   $1.8\,\pm\,0.8\ \ $ &    1.0 &                10393.7 &   $18\,\pm\,7\ \ $ &  $1.3\,\pm\,1.2\ \ $ &    1.6 &                10396.3 &  $18\,\pm\,5\ \ $ &  $0.5\,\pm\,0.7\ \ $ &    1.5 &                    ... &          ...$\ \ $ &            ...$\ \ $ &    ... \\
10438 &    4 &  $10438.4\,\pm\,0.1\ \ $ &                10441.2 &   $64\,\pm\,4\ \ $ &   $1.7\,\pm\,0.4\ \ $ &    3.7 &                10441.3 &   $32\,\pm\,4\ \ $ &  $0.9\,\pm\,0.4\ \ $ &    3.3 &                    ... &         ...$\ \ $ &            ...$\ \ $ &    ... &                    ... &          ...$\ \ $ &            ...$\ \ $ &    ... \\
10504\tablefootmark{b,c} &    4 &  $10504.3\,\pm\,0.2\ \ $ &                10507.2 &   $50\,\pm\,4\ \ $ &   $4.1\,\pm\,0.8\ \ $ &    1.1 &                10507.2 &   $28\,\pm\,3\ \ $ &  $2.2\,\pm\,0.4\ \ $ &    1.1 &                    ... &         ...$\ \ $ &            ...$\ \ $ &    ... &                    ... &          ...$\ \ $ &            ...$\ \ $ &    ... \\
10507\tablefootmark{c} &    4 &  $10507.6\,\pm\,0.7\ \ $ &                10509.8 &   $41\,\pm\,10$ &   $3.4\,\pm\,0.4\ \ $ &    1.2 &                10511.0 &   $11\,\pm\,2\ \ $ &  $1.0\,\pm\,0.4\ \ $ &    1.1 &                    ... &         ...$\ \ $ &            ...$\ \ $ &    ... &                    ... &          ...$\ \ $ &            ...$\ \ $ &    ... \\
10697 &    4 &  $10696.8\,\pm\,0.4\ \ $ &                10700.0 &  $186\,\pm\,7\ \ $ &   $4.6\,\pm\,0.8\ \ $ &    4.0 &                10699.5 &  $133\,\pm\,4\ \ $ &  $3.7\,\pm\,0.5\ \ $ &    3.9 &                10699.4 &  $94\,\pm\,8\ \ $ &  $2.0\,\pm\,0.6\ \ $ &    4.0 &                10700.1 &   $91\,\pm\,6\ \ $ &  $2.1\,\pm\,0.7\ \ $ &    4.4 \\
10735 &    6 &  $10734.7\,\pm\,0.5\ \ $ &                10737.9 &   $20\,\pm\,2\ \ $ &   $1.9\,\pm\,0.4\ \ $ &    1.2 &                10737.4 &   $13\,\pm\,2\ \ $ &  $1.0\,\pm\,0.4\ \ $ &    1.2 &                    ... &         ...$\ \ $ &            ...$\ \ $ &    ... &                    ... &          ...$\ \ $ &            ...$\ \ $ &    ... \\
10780 &    3 &  $10780.4\,\pm\,0.2\ \ $ &                10783.4 &  $154\,\pm\,6\ \ $ &  $10.4\,\pm\,0.8\ \ $ &    1.3 &                10783.3 &  $108\,\pm\,5\ \ $ &  $6.5\,\pm\,0.5\ \ $ &    1.5 &                10783.5 &  $62\,\pm\,6\ \ $ &  $2.7\,\pm\,0.7\ \ $ &    1.8 &                10783.3 &   $83\,\pm\,6\ \ $ &  $4.5\,\pm\,0.6\ \ $ &    1.7 \\
10792 &    3 &  $10791.9\,\pm\,0.1\ \ $ &                10794.8 &   $39\,\pm\,4\ \ $ &   $1.7\,\pm\,0.5\ \ $ &    2.0 &                10795.0 &   $48\,\pm\,5\ \ $ &  $1.3\,\pm\,0.6\ \ $ &    3.3 &                    ... &         ...$\ \ $ &            ...$\ \ $ &    ... &                10794.7 &   $39\,\pm\,5\ \ $ &  $1.4\,\pm\,0.6\ \ $ &    2.2 \\
10884 &    6 &  $10884.2\,\pm\,0.3\ \ $ &                10887.5 &   $38\,\pm\,2\ \ $ &   $1.8\,\pm\,0.3\ \ $ &    2.0 &                10887.3 &   $28\,\pm\,2\ \ $ &  $2.0\,\pm\,0.4\ \ $ &    1.4 &                10886.9 &  $11\,\pm\,4\ \ $ &  $0.7\,\pm\,0.5\ \ $ &    2.0 &                10887.3 &   $22\,\pm\,7\ \ $ &  $1.1\,\pm\,0.7\ \ $ &    2.1 \\
11048\tablefootmark{d} &    6 &  $11048.3\,\pm\,0.4\ \ $ &                11051.5 &   $39\,\pm\,4\ \ $ &   $1.7\,\pm\,0.6\ \ $ &    2.6 &                11051.7 &   $13\,\pm\,4\ \ $ &  $0.6\,\pm\,0.5\ \ $ &    2.2 &                11051.1 &  $19\,\pm\,4\ \ $ &  $0.9\,\pm\,0.5\ \ $ &    1.5 &                11050.9 &   $16\,\pm\,3\ \ $ &  $0.9\,\pm\,0.4\ \ $ &    1.5 \\
11695 &    6 &  $11695.2\,\pm\,0.2\ \ $ &                11698.4 &      $32\,\pm\,11$ &   $2.1\,\pm\,1.8\ \ $ &    1.7 &                11698.4 &    $9\,\pm\,2\ \ $ &  $1.3\,\pm\,0.3\ \ $ &    0.8 &                    ... &         ...$\ \ $ &            ...$\ \ $ &    ... &                    ... &          ...$\ \ $ &            ...$\ \ $ &    ... \\
11699 &    6 &  $11698.7\,\pm\,0.3\ \ $ &                11702.0 &      $35\,\pm\,12$ &   $1.8\,\pm\,1.8\ \ $ &    2.1 &                11702.0 &   $28\,\pm\,5\ \ $ &  $1.2\,\pm\,0.7\ \ $ &    2.4 &                    ... &         ...$\ \ $ &            ...$\ \ $ &    ... &                11701.5 &   $19\,\pm\,4\ \ $ &  $1.1\,\pm\,0.5\ \ $ &    1.6 \\
11721 &    6 &  $11720.5\,\pm\,0.3\ \ $ &                11723.9 &   $46\,\pm\,3\ \ $ &   $2.2\,\pm\,0.4\ \ $ &    2.5 &                11723.5 &   $39\,\pm\,3\ \ $ &  $1.7\,\pm\,0.4\ \ $ &    2.3 &                11724.0 &  $28\,\pm\,4\ \ $ &  $1.0\,\pm\,0.5\ \ $ &    3.3 &                11723.6 &   $22\,\pm\,4\ \ $ &  $0.9\,\pm\,0.5\ \ $ &    2.5 \\
11792 &    6 &  $11792.7\,\pm\,0.1\ \ $ &                11795.8 &   $17\,\pm\,2\ \ $ &   $1.5\,\pm\,0.3\ \ $ &    1.1 &                11795.9 &   $10\,\pm\,2\ \ $ &  $0.9\,\pm\,0.3\ \ $ &    1.4 &                11796.1 &   $9\,\pm\,4\ \ $ &  $0.6\,\pm\,0.6\ \ $ &    1.4 &                    ... &          ...$\ \ $ &            ...$\ \ $ &    ... \\
11797\tablefootmark{e} &    1 &  $11797.3\,\pm\,0.2\ \ $ &                11800.6 &  $133\,\pm\,3\ \ $ &   $8.5\,\pm\,0.3\ \ $ &    1.5 &                11800.5 &  $116\,\pm\,3\ \ $ &  $5.3\,\pm\,0.3\ \ $ &    1.7 &                11800.6 &  $59\,\pm\,5\ \ $ &  $2.7\,\pm\,0.5\ \ $ &    2.2 &                11800.4 &  $115\,\pm\,4\ \ $ &  $5.3\,\pm\,0.4\ \ $ &    1.8 \\
11970\tablefootmark{f} &    6 &  $11970.1\,\pm\,0.2\ \ $ &                11973.5 &  $105\,\pm\,3\ \ $ &   $5.6\,\pm\,0.3\ \ $ &    1.4 &                11973.3 &  $105\,\pm\,4\ \ $ &  $4.3\,\pm\,0.3\ \ $ &    1.6 &                11973.5 &  $66\,\pm\,3\ \ $ &  $2.2\,\pm\,0.4\ \ $ &    2.2 &                11973.3 &   $84\,\pm\,4\ \ $ &  $3.8\,\pm\,0.5\ \ $ &    2.0 \\
12222 &    6 &  $12222.4\,\pm\,0.2\ \ $ &                12225.8 &   $45\,\pm\,2\ \ $ &   $2.6\,\pm\,0.3\ \ $ &    1.5 &                12225.7 &   $28\,\pm\,3\ \ $ &  $1.7\,\pm\,0.4\ \ $ &    1.6 &                    ... &         ...$\ \ $ &            ...$\ \ $ &    ... &                    ... &          ...$\ \ $ &            ...$\ \ $ &    ... \\
12293 &    5 &  $12293.9\,\pm\,0.1\ \ $ &                12297.2 &   $24\,\pm\,2\ \ $ &   $1.4\,\pm\,0.3\ \ $ &    1.7 &                12297.3 &   $12\,\pm\,3\ \ $ &  $0.9\,\pm\,0.4\ \ $ &    1.5 &                    ... &         ...$\ \ $ &            ...$\ \ $ &    ... &                    ... &          ...$\ \ $ &            ...$\ \ $ &    ... \\
12518 &    5 &  $12519.4\,\pm\,0.1\ \ $ &                12522.6 &   $16\,\pm\,2\ \ $ &   $0.8\,\pm\,0.3\ \ $ &    2.0 &                12523.0 &   $30\,\pm\,4\ \ $ &  $0.5\,\pm\,0.3\ \ $ &    3.8 &                    ... &         ...$\ \ $ &            ...$\ \ $ &    ... &                    ... &          ...$\ \ $ &            ...$\ \ $ &    ... \\
12623 &    5 &  $12623.6\,\pm\,0.2\ \ $ &                12626.8 &  $105\,\pm\,3\ \ $ &   $3.1\,\pm\,0.3\ \ $ &    3.4 &                12627.3 &   $56\,\pm\,2\ \ $ &  $1.3\,\pm\,0.2\ \ $ &    3.7 &                    ... &         ...$\ \ $ &            ...$\ \ $ &    ... &                    ... &          ...$\ \ $ &            ...$\ \ $ &    ... \\
12799 &    5 &  $12799.3\,\pm\,0.4\ \ $ &                12802.5 &   $50\,\pm\,4\ \ $ &   $3.4\,\pm\,0.5\ \ $ &    1.5 &                12802.5 &   $18\,\pm\,3\ \ $ &  $1.5\,\pm\,0.5\ \ $ &    1.3 &                12803.4 &  $21\,\pm\,5\ \ $ &  $1.2\,\pm\,0.6\ \ $ &    2.1 &                12802.6 &   $33\,\pm\,3\ \ $ &  $1.5\,\pm\,0.3\ \ $ &    2.2 \\
12838 &    6 &  $12837.6\,\pm\,0.9\ \ $ &                12840.4 &   $72\,\pm\,3\ \ $ &   $1.4\,\pm\,0.3\ \ $ &    4.6 &                12840.7 &   $35\,\pm\,6\ \ $ &  $0.9\,\pm\,0.6\ \ $ &    3.7 &                12842.5 &  $28\,\pm\,3\ \ $ &  $0.6\,\pm\,0.3\ \ $ &    5.8 &                12840.6 &   $23\,\pm\,4\ \ $ &  $0.8\,\pm\,0.5\ \ $ &    2.5 \\
13027 &    4 &  $13027.0\,\pm\,0.7\ \ $ &                13031.0 &  $134\,\pm\,3\ \ $ &   $1.9\,\pm\,0.2\ \ $ &    4.7 &                13030.2 &   $48\,\pm\,6\ \ $ &  $1.2\,\pm\,0.4\ \ $ &    5.6 &                    ... &         ...$\ \ $ &            ...$\ \ $ &    ... &                    ... &          ...$\ \ $ &            ...$\ \ $ &    ... \\
13175 &    1 &  $13175.1\,\pm\,0.1\ \ $ &                13178.5 &  $647\,\pm\,20$ &  $13.7\,\pm\,0.2\ \ $ &    4.1 &                13178.9 &  $466\,\pm\,21$ &  $9.0\,\pm\,0.5\ \ $ &    4.2 &                    ... &         ...$\ \ $ &            ...$\ \ $ &    ... &                    ... &          ...$\ \ $ &            ...$\ \ $ &    ... \\
\bottomrule
\end{tabular}
}
    \label{tab:dib_list}
    }
    \tablebib{
    (1) \citet{1990Natur.346..729J}; 
    (2) \citet{1994Natur.369..296F}; 
    (3) \citet{2007A&A...465..993G}; 
    (4) \citet{2014A&A...569A.117C}; 
    (5) \citet{2015ApJ...800..137H}; 
    (6) This work
    }
    \tablefoot{All wavelengths are measured in the barycentric rest frame. The rest wavelengths are given in Col. 3 with a statistical error. Rest wavelengths have a systematic error of $\pm\ 0.3\,\AA$ due to the double \ion{K}{i} components (see Section \ref{sec:dib_measurements}). Non-detection of DIBs are a result either of
    a strength too low for detection or of missing spectral coverage.
    Stellar blends:
    \tablefoottext{a}{\ion{Mg}{ii}, $\lambda_\mathrm{air}$\,=\,9631.891, 9632.430\,{\AA}} 
    \tablefoottext{b}{\ion{Fe}{ii}, $\lambda_\mathrm{air}$\,=\,10501.504\,{\AA}}
    \tablefoottext{c}{\ion{N}{i}, $\lambda_\mathrm{air}$=10506.997\,{\AA}}
    \tablefoottext{d}{\ion{He}{i}, $\lambda_\mathrm{air}$\,=\,11044.983\,{\AA}}
    \tablefoottext{e}{\ion{C}{i}, $\lambda_\mathrm{air}$\,=\,11801.098\,{\AA}}
    \tablefoottext{f}{\ion{He}{i}, $\lambda_\mathrm{air}$\,=\,11969.060, 11969.464\,{\AA}.}
    }
    \end{table}
    \end{landscape}

\setcounter{section}{2}


\setcounter{figure}{0}
\begin{figure*}[b]
\begin{flushleft}{\large\textsf{\textbf{Appendix B: E(B-V)--EW relations}}}\end{flushleft}
  \centering
   \subfloat{
      \includegraphics[width = \textwidth/19*6]
      {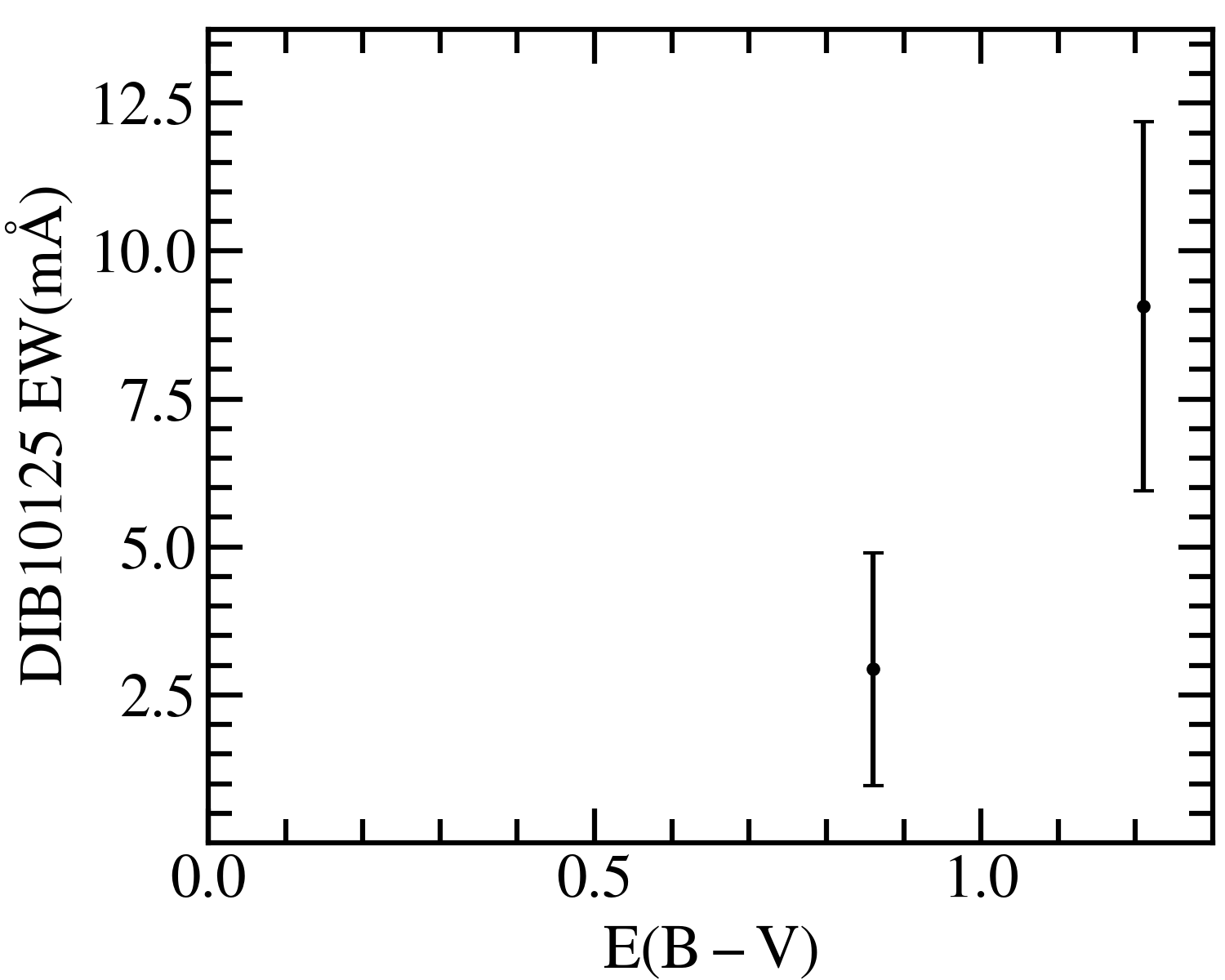}}
   \subfloat{
      \includegraphics[width = \textwidth/19*6]
      {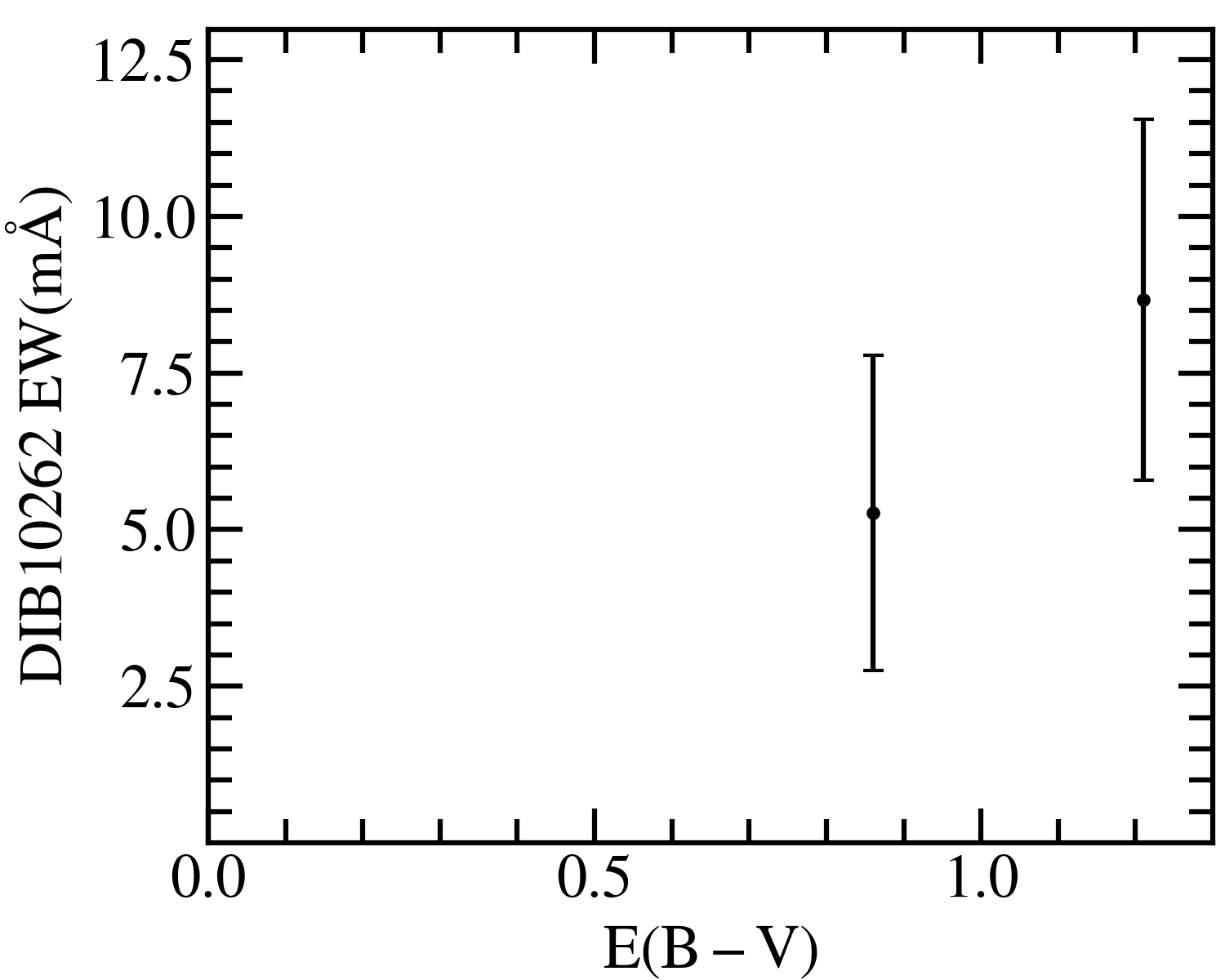}\label{fig:10263_ebv}}
   \subfloat{
      \includegraphics[width = \textwidth/19*6]
      {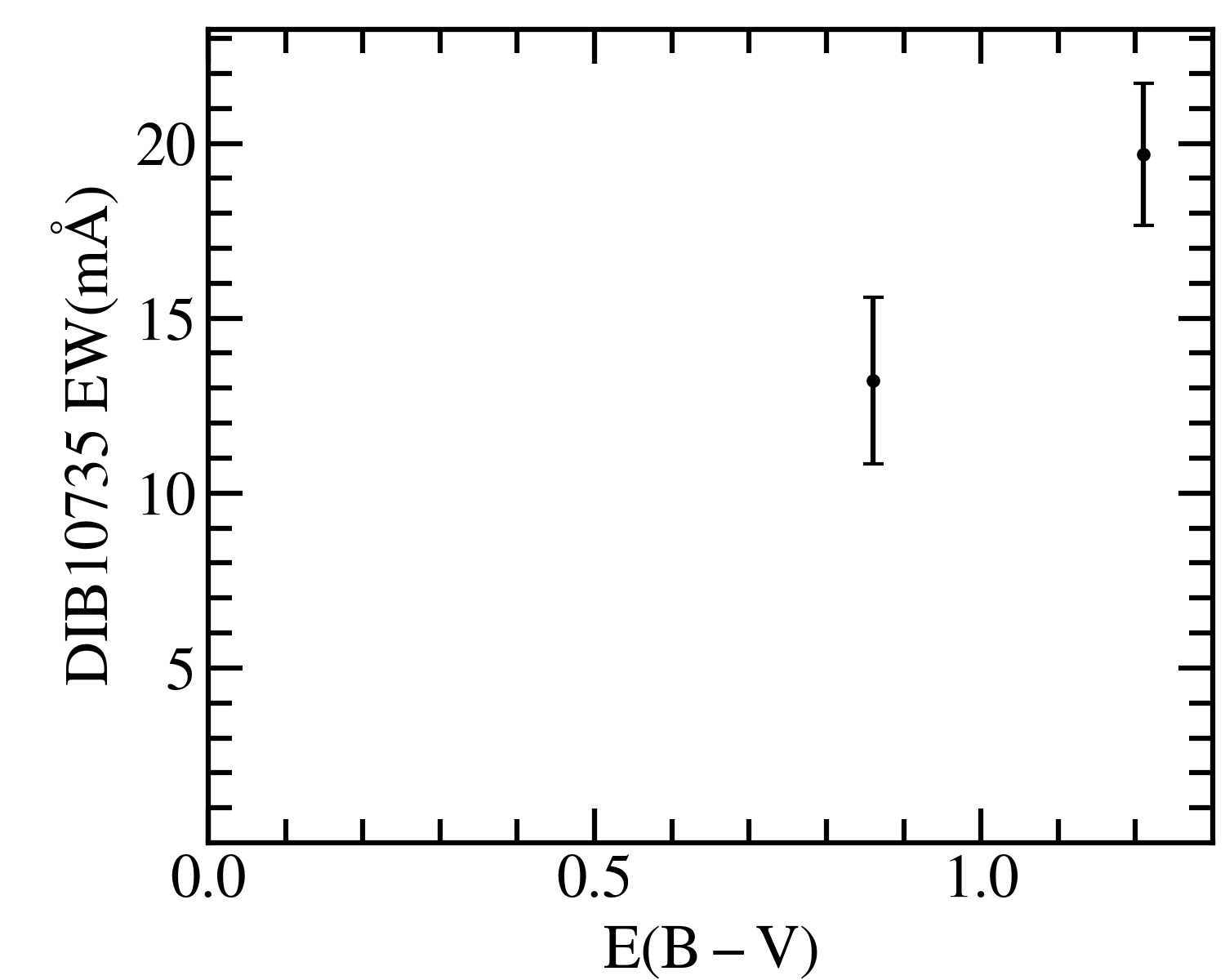}\label{fig:10734_ebv}}\\
   \subfloat{
      \includegraphics[width = \textwidth/19*6]
      {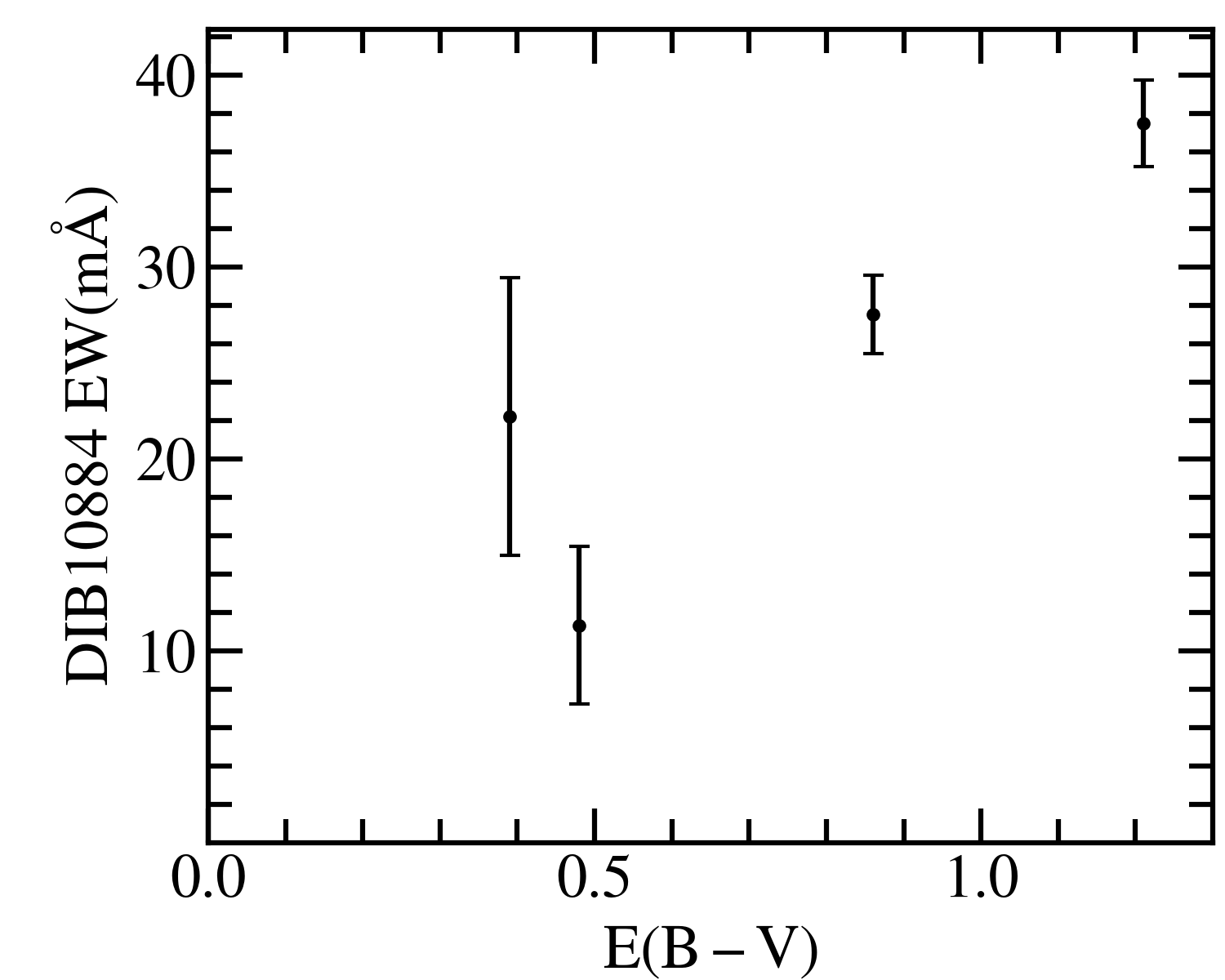}\label{fig:10884_ebv}}
   \subfloat{
      \includegraphics[width = \textwidth/19*6]
      {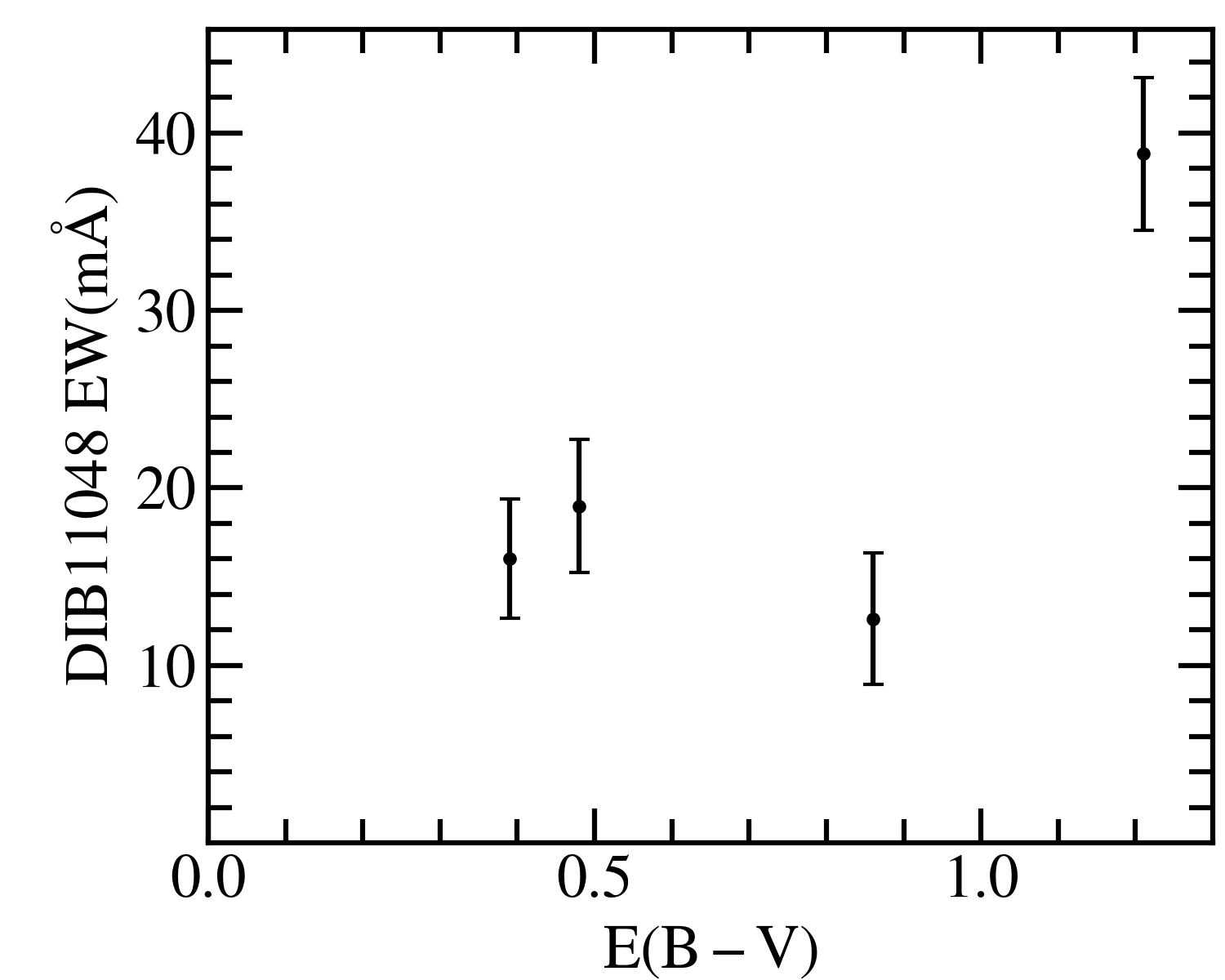}\label{fig:11048_ebv}}
   \subfloat{
      \includegraphics[width = \textwidth/19*6]
      {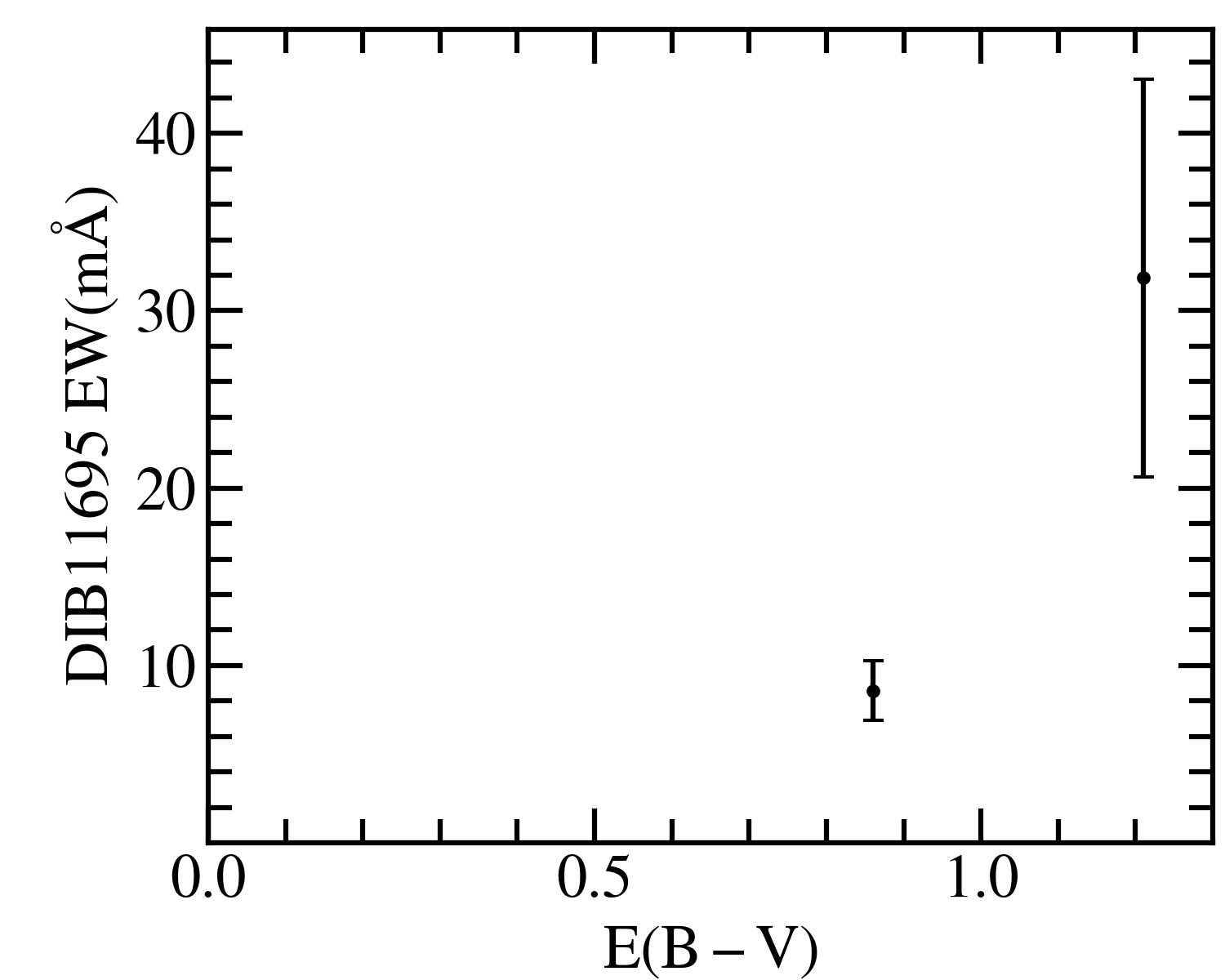}\label{fig:11695_ebv}}\\
   \subfloat{
      \includegraphics[width = \textwidth/19*6]
      {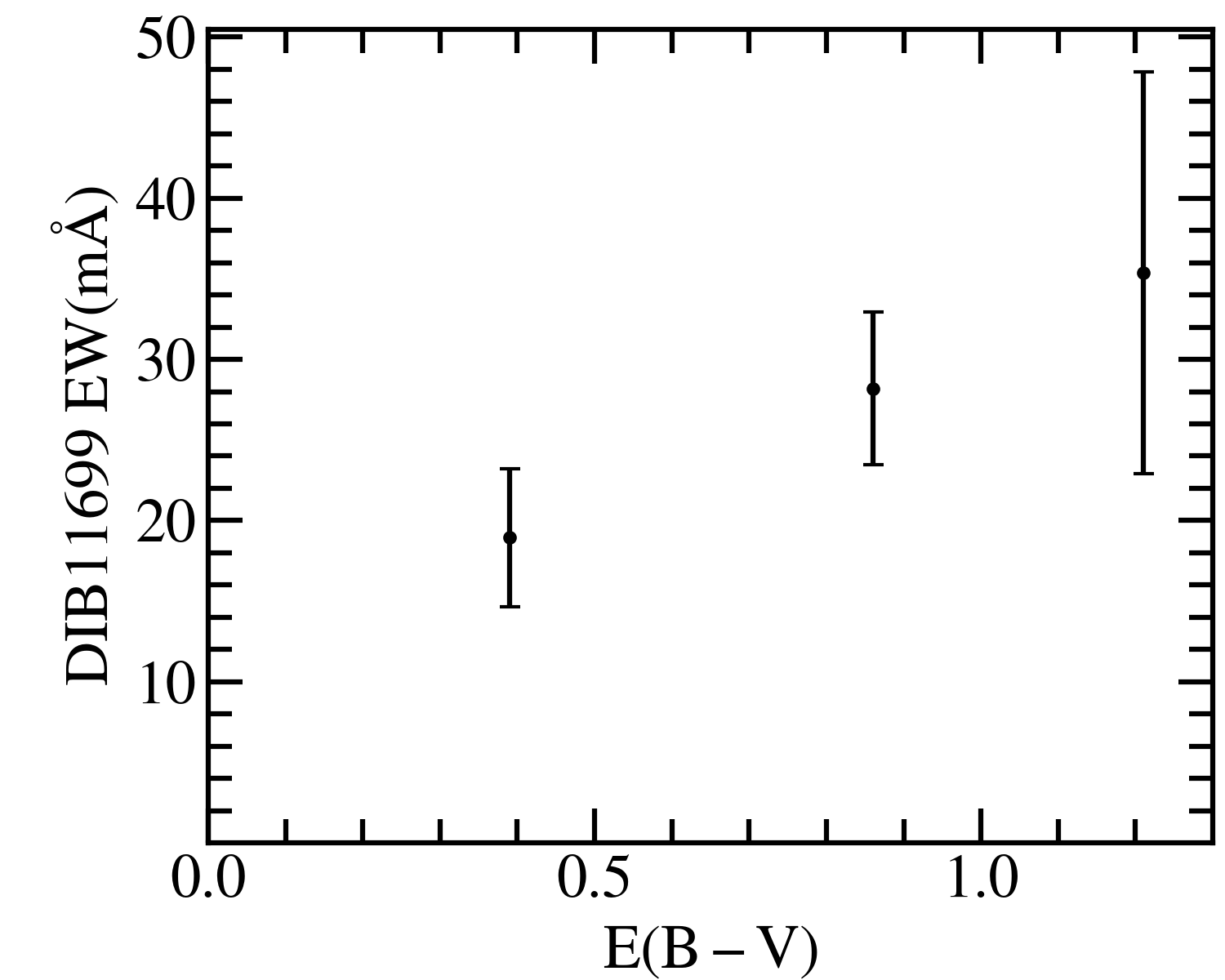}\label{fig:11699_ebv}}
   \subfloat{
      \includegraphics[width = \textwidth/19*6]
      {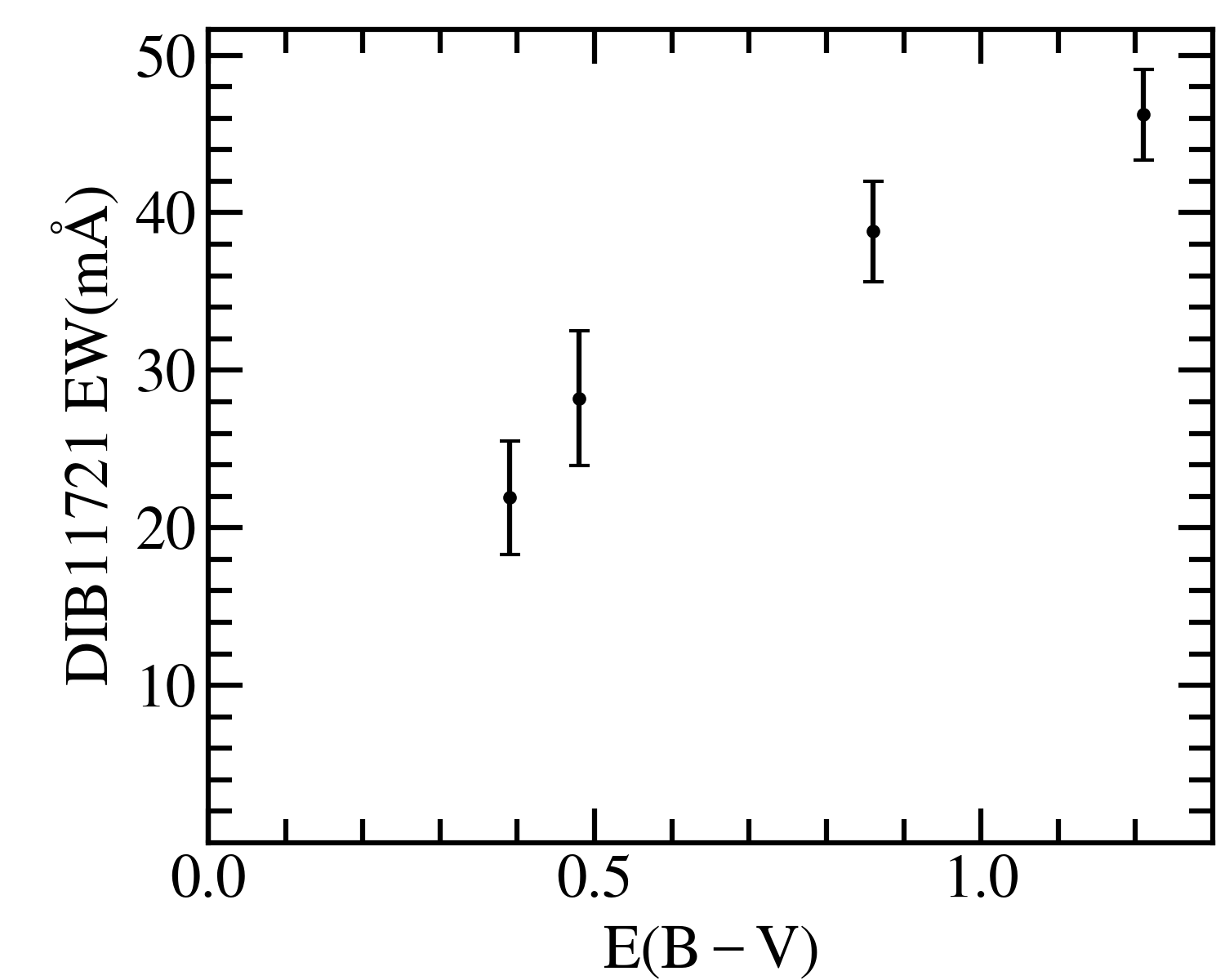}\label{fig:11721_ebv}}
   \subfloat{
      \includegraphics[width = \textwidth/19*6]
      {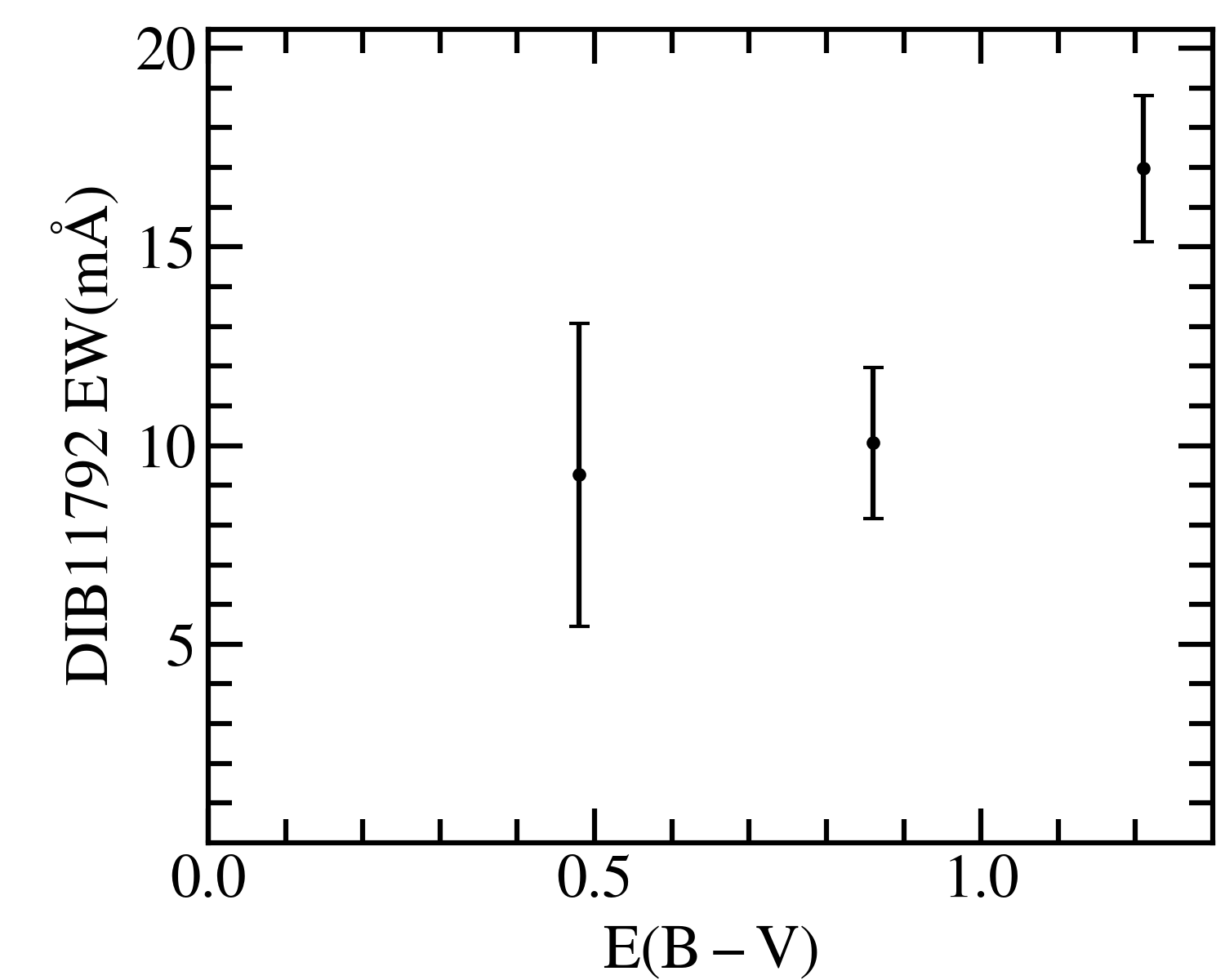}\label{fig:11792_ebv}}\\
   \subfloat{
      \includegraphics[width = \textwidth/19*6]
      {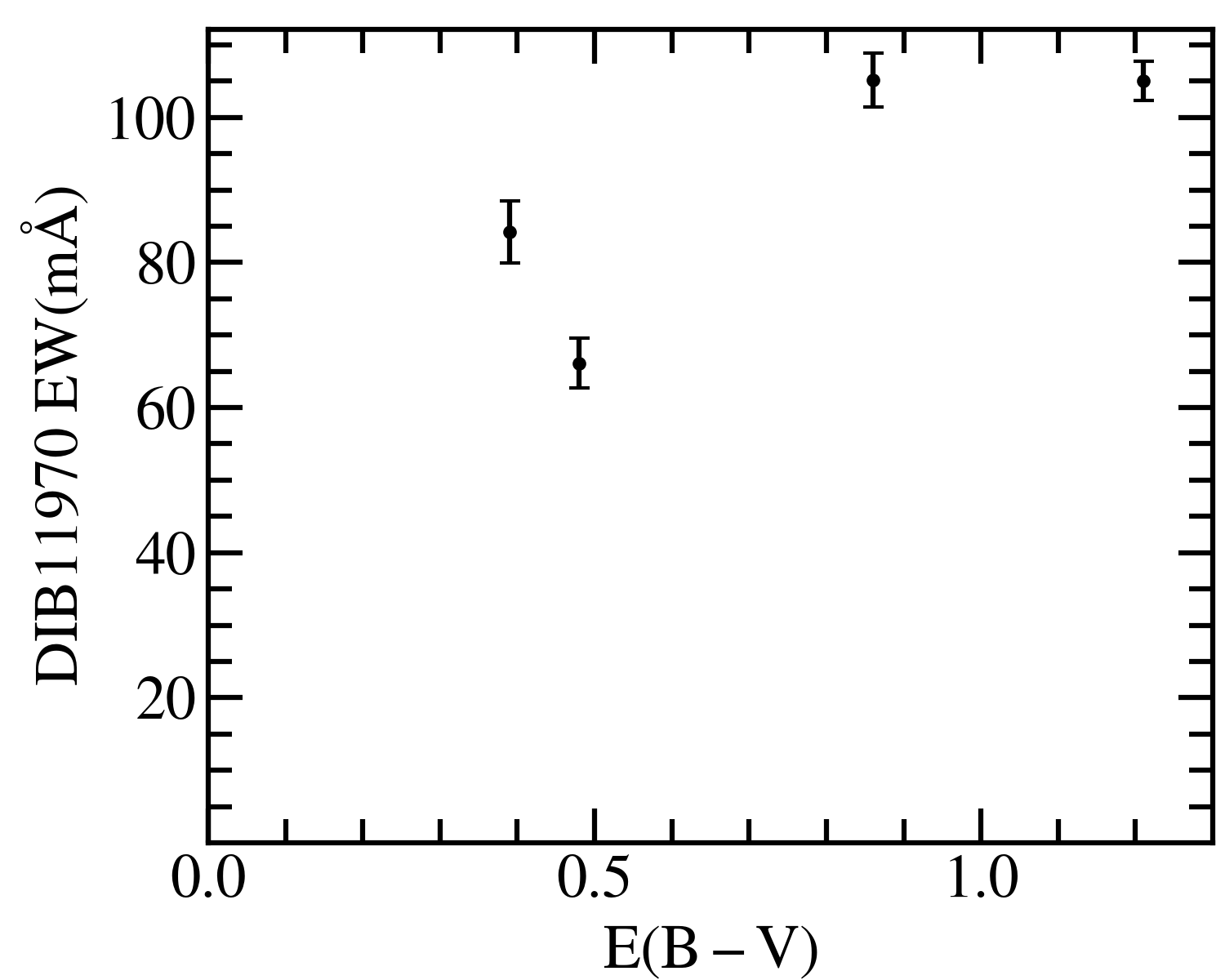}\label{fig:11969_ebv}}
   \subfloat{
      \includegraphics[width = \textwidth/19*6]
      {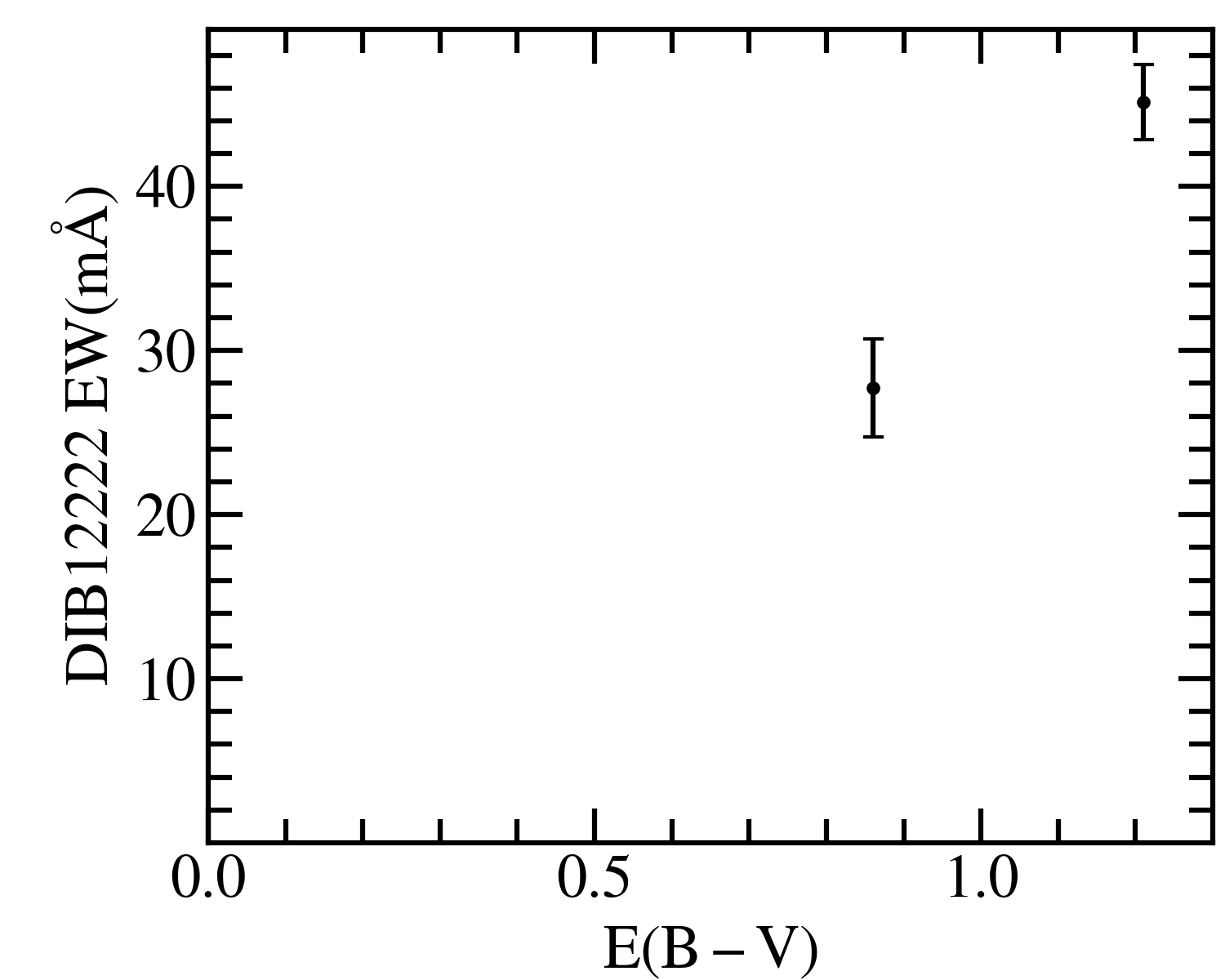}\label{fig:12222_ebv}}
   \subfloat{
      \includegraphics[width = \textwidth/19*6]
      {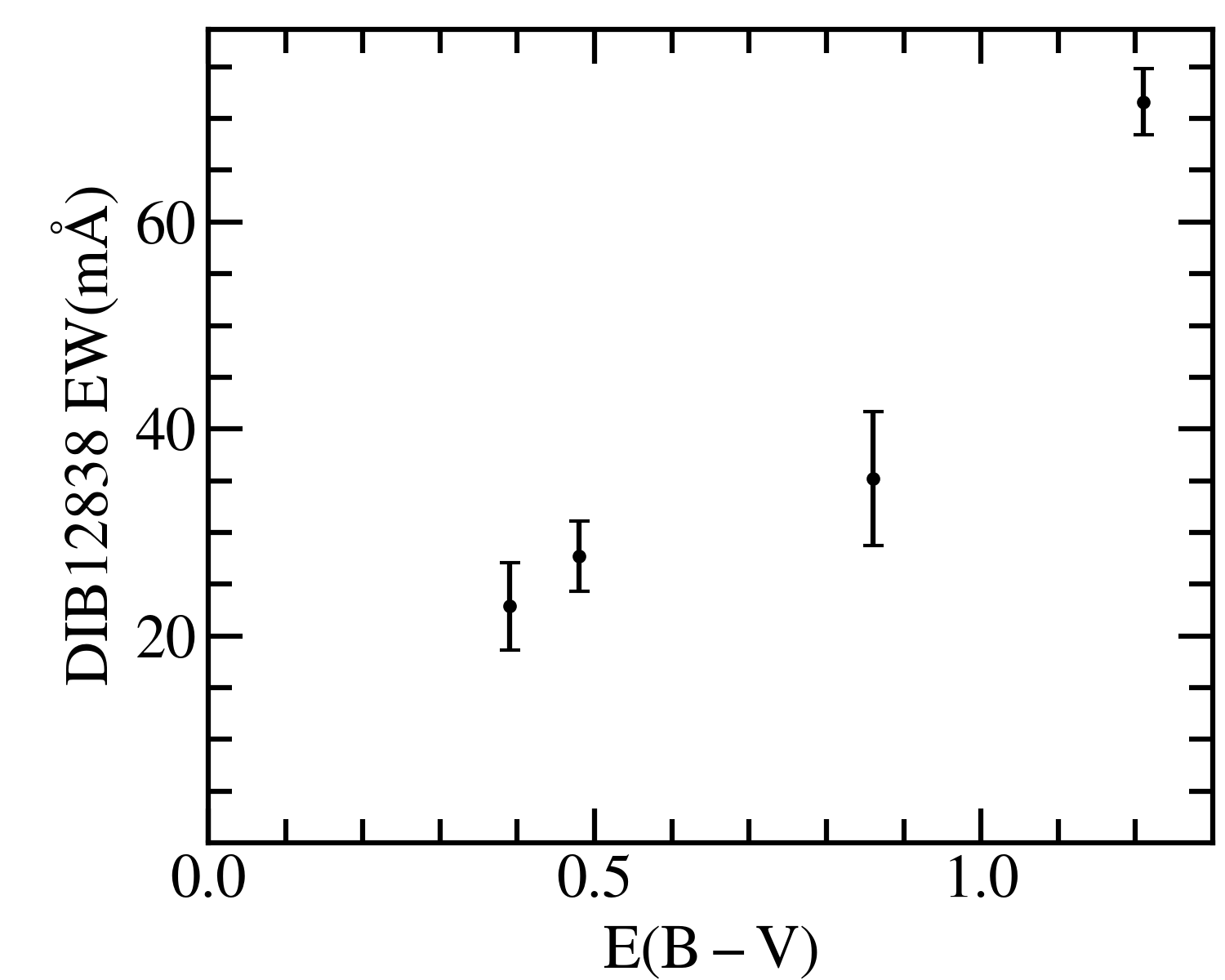}\label{fig:12837_ebv}}
   \caption{Correlations of the equivalent widths of the newly found NIR DIB candidates with $E(B-V)$. Reddening values of the four DIB target stars are the following: $E(B-V)$\,=\,0.39 (HD~111613), 0.48 (HD~92207), 0.86 (HD~165784), and 1.22 (HD~183143).}
   \label{fig:ebv_ew}
\end{figure*}

\end{appendix}
\end{document}